\documentclass[12pt,oneside,english,french]{book}
\usepackage[T1]{fontenc}
\usepackage[utf8]{inputenc}
\usepackage{float}
\usepackage{textcomp}
\usepackage{enumitem}
\usepackage{amsmath}
\usepackage{amssymb}
\usepackage{cancel}
\usepackage{graphicx}
\usepackage{wasysym}

\makeatletter

\newcommand{\noun}[1]{\textsc{#1}}
\providecommand{\tabularnewline}{\\}

\usepackage{fixltx2e}
\usepackage[french]{babel}
\usepackage{graphicx,amsmath}
\usepackage{epstopdf}
\usepackage{epsf,color}
\usepackage[mathscr]{eucal}
\usepackage{amsmath}
\usepackage{slashed}
\usepackage{placeins}
\usepackage{microtype}
\usepackage{amssymb,amsfonts}
\usepackage[nohints]{minitoc}
\usepackage{graphicx}
\usepackage{tikz}
\usepackage{dsfont}
\usepackage{txfonts}
\usepackage{emptypage}
\usepackage{dsfont}
\definecolor{Mygreen}{rgb}{0.00, 0.41, 0.27}
\definecolor{Mypink}{rgb}{1.0, 0.0, 0.5}
\definecolor{Myblue}{rgb}{0.00, 0.227, 0.51}
\definecolor{Myred}{rgb}{0.80, 0.2, 0.0}
\usepackage[colorlinks,citecolor=blue,linkcolor=Myblue,urlcolor=blue]{hyperref}
\usepackage{float}
\usepackage{bm}
\usepackage{color}
\usepackage{indentfirst}
\usepackage{import}
\usepackage{multicol}
\usepackage[T1]{fontenc}
\usepackage[utf8]{inputenc}
\usepackage[utf8]{luainputenc}
\usepackage[pages=some]{background}
\usepackage{lipsum}
\usepackage{enumitem}
\usepackage[normalem]{ulem}
\usepackage{fancyhdr}
\usepackage{array}
\usepackage{pdfpages}
\usepackage{titlesec}
\fancyheadoffset[RE,LO]{0\textwidth}
\usepackage{fix-cm} 
\makeatletter
\newcommand\MHUGE{\@setfontsize\Huge{32}{41}} 
\newcommand\HUGE{\@setfontsize\Huge{38}{47}} 
\newcommand\GHUGE{\@setfontsize\Huge{60}{70}} 
\makeatother
\AtBeginDocument{\mathcode`v=\varv}
\newcolumntype{K}[1]{>{\centering\arraybackslash}p{#1}}

\titleformat{\chapter}[display]
  {\bfseries\Huge}
  {\filleft{\color[rgb]{0.03,0.21,0.27}\MHUGE\MakeUppercase{\chaptertitlename}\vspace{1cm}~~\GHUGE\thechapter}}
  {1ex}
  {\filleft}
  [\vspace{2ex}%
\titlerule]

\usepackage{geometry}
\addto{\captionsfrench}{
  
}

\usepackage{listings,lstautogobble}
\def\simlt{\lower.5ex\hbox{$\; \buildrel < \over \sim \;$}}
\def\simgt{\lower.5ex\hbox{$\; \buildrel > \over \sim \;$}}

\makeatletter
\AtBeginDocument{%
  \expandafter\renewcommand\expandafter\subsection\expandafter
    {\expandafter\@fb@secFB\subsection}%
  \newcommand\@fb@secFB{\FloatBarrier
    \gdef\@fb@afterHHook{\@fb@topbarrier \gdef\@fb@afterHHook{}}}%
  \g@addto@macro\@afterheading{\@fb@afterHHook}%
  \gdef\@fb@afterHHook{}%
}
\makeatother

\makeatother

\usepackage{babel}
\makeatletter
\addto\extrasfrench{%
   \providecommand{\og}{\leavevmode\flqq~}%
   \providecommand{\fg}{\ifdim\lastskip>\z@\unskip\fi~\frqq}%
}

\makeatother
\begin{document}

\includepdf{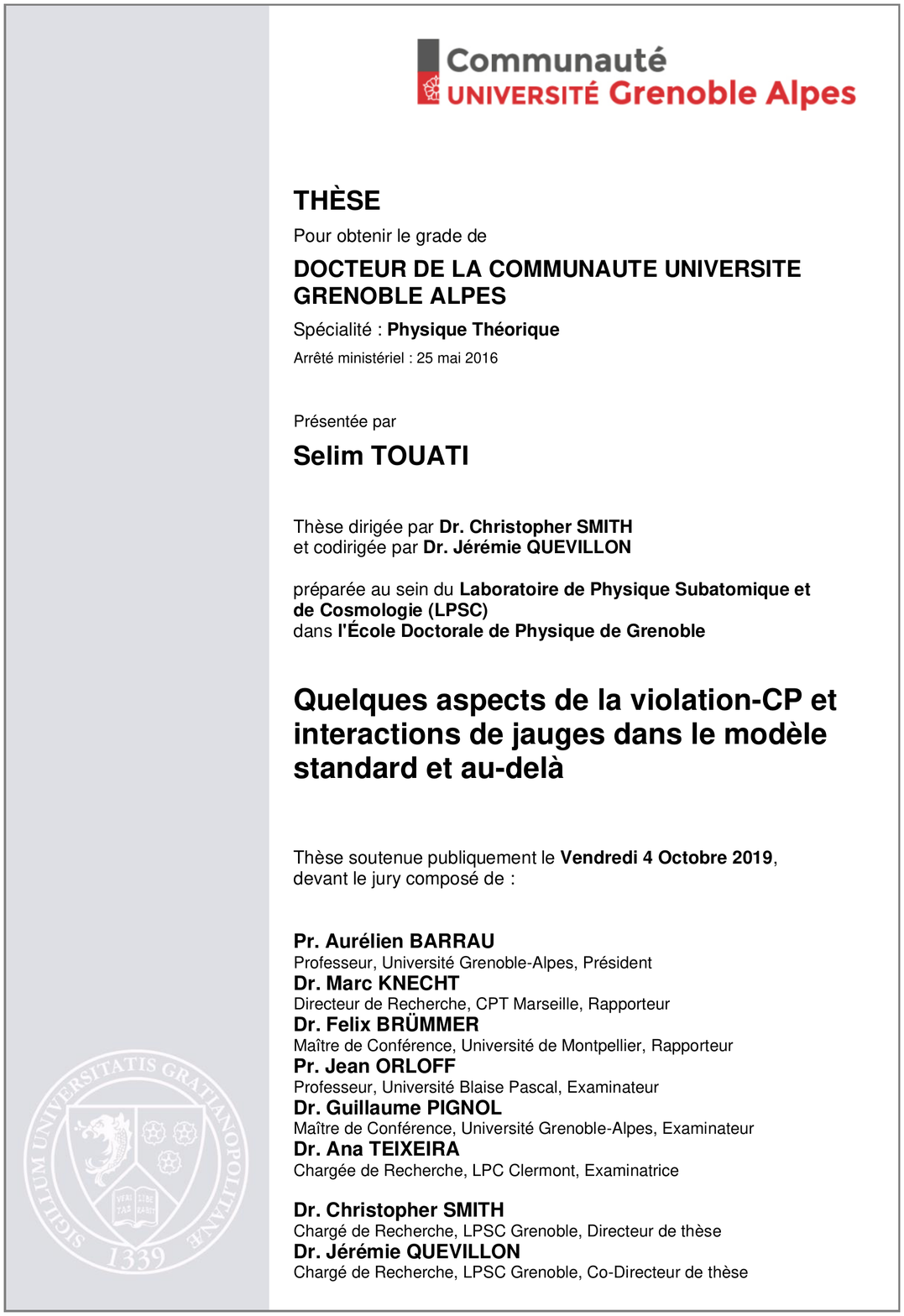}

\newpage\null\thispagestyle{empty}\newpage

\frontmatter

\chapter*{Remerciements}

\markboth{{Remerciements}}{Remerciements}

Je voudrais commencer par remercier Aurélien Barrau pour m’avoir écouté
et fait confiance il y a 4 ans en me donnant l’opportunité de poursuivre
mes études de physique fondamentale à l’Université Joseph Fourier
à Grenoble. J’ai beaucoup apprécié ton geste, merci pour ton humanité.\newline

Je tiens à remercier Arnaud Lucotte, directeur du Laboratoire de Physique
Subatomique et de Cosmologie (LPSC), pour m’avoir accueilli dans cet
établissement où j’ai pu réaliser ma thèse dans d’excellentes conditions.
Je remercie aussi l’ensemble des membres du LPSC. Merci à David Maurin
pour le travail qu’il fait pour les doctorants au Comité de Suivi
de Thèse. Merci à Marie-Hélène Genest pour avoir été en charge de
mon suivi individuel.\newline

Je remercie également l’ensemble des membres de mon jury de thèse.
Merci à Felix Brümmer et à Marc Knecht pour avoir accepté d’être les
rapporteurs de mon manuscrit ainsi que pour avoir contribué à améliorer
sa qualité. Merci à Aurélien Barrau, Jean Orloff, Guillaume Pignol
et Ana Teixeira pour leur disponibilité et pour l’intérêt qu’ils ont
porté à ce travail.\newline

Je tiens à remercier vivement mon directeur de thèse Christopher Smith
pour m'avoir accordé sa confiance, supervisé ma thèse ainsi que pour
la grande liberté qu’il m’a donnée. Merci pour ta disponibilité, ta
patience et pour tous tes précieux conseils. Ta rigueur, ton esprit
critique et ton honnêteté scientifique m’ont marqué. Merci de m’avoir
montré ce qu’est le métier de théoricien et que l’on ne peut vraiment
comprendre quelque chose qu’en la faisant réellement. 

Un énorme merci à Jérémie Quevillon, codirecteur de cette thèse depuis
son arrivée au labo. Outre le lien qui nous unit en tant que Mineurs,
travailler avec toi a été un plaisir ! Merci de m’avoir initié au
monde des théories effectives, merci pour ton dynamisme, ta disponibilité
et pour ta bienveillance. J’espère avoir la chance de travailler de
nouveau avec toi par la suite.\newline

Je souhaite remercier chaleureusement l’ensemble des membres du groupe
Théorie du LPSC pour leur accueil. Faire ma thèse avec vous a vraiment
été super ! J'ai pu apprendre beaucoup aussi bien sur le plan scientifique
que sur le fonctionnement d’une équipe de recherche. Je remercie l’ensemble
des chercheurs permanents. Merci à Sabine Kraml pour faire vivre le
groupe en dehors du labo en organisant les différents diners avec
les visiteurs ou encore les sorties en montagne. Merci à Ingo Schienbein
pour les différentes notions de physique des particules qu’il m’a
enseigné en tant que professeur ainsi que pour son ouverture d’esprit
en tant que personne, c'est toujours un plaisir de parler avec toi.
J’ai une pensée également pour Marianne Mangin-Brinet qui j’espère,
reviendra dans le groupe au plus vite. Je remercie également l’ensemble
des postdoctorants et doctorants du groupe Théorie que j’ai pu croiser
durant ma thèse. Merci à Kentarou, Olek, Kimiko, Uschi, Kseniia, Humberto,
Pierre-Henri et Rola. Vous êtes tous des personnes incroyables par
lesquelles je me suis énormément enrichit humainement. Merci à mon
ami Mohammad pour sa grande gentillesse ainsi que pour tous les moments
que nous avions partagés, que ce soit au labo ou pendant les multiples
ballades le long de la rivière. I would like to thank Hoa for all
the exchanges we had and for having accompanied me through the numerous
weekend sessions in the lab during the summer when I was writing the
manuscript. Thankfully, there was the legendary “cool room”. Good
luck for your Ph.D ! Enfin, je souhaite bonne chance à tous les nouveaux
arrivants dans le groupe.\newline

Je souhaite remercier aussi tous les collègues du premier étage, en
particulier le groupe UCN pour avoir partagé de nombreuses pauses
café ensemble. Merci en particulier à Benoît Clément et à Guillaume
Pignol pour leur enthousiasme et leur pédagogie quand il s’agit d’expliquer
une notion, toujours avec humour et dans une ambiance conviviale.
J’apprécie beaucoup les échanges avec vous. Après vos trolls sur les
théoriciens, j’ai finalement appris à utiliser la machine à café.
Merci à Sébastien Curtoni pour sa gentillesse et pour être toujours
disposé à parler avec joie de son travail en physique médicale. Merci
à Mahfoud pour sa positivité en toutes circonstances et de toujours
donner le sourire aux personnes qu’il croise dans les couloirs avec
ses nombreuses blagues.\newline

Mes années passées à Grenoble n’auraient pas été ce qu’elles sont
sans la présence de Thomas Meideck et de Léonard Aubry. Merci à vous
pour toutes les discussions intéressantes qu’on a eues sur des sujets
très divers mais toujours passionnants, les parties de Go jouées au
club ou ailleurs, les débats politiques animés ainsi que pour les
nombreuses réunions dans notre QG préféré, la Nat’, où l’on y faisait
un peu tout cela à la fois. C’est toujours un plaisir de vous voir
!\newline

Je remercie vivement mon partenaire de la première heure, Killian
Martineau, qui a toujours été là depuis mes tout débuts à Grenoble.
Depuis le binôme choc en PSA, on a continué à évoluer ensemble et
il ne se passe pas un jour sans qu’on ait une interaction. Merci pour
ton estime à mon égard et pour avoir toujours cru en moi, je ne sais
pas si tu es naïf ou fou (en fait si je sais) mais une chose est certaine,
tu es un ami formidable ! Nos petites chamailleries et nos éternels
débats sur le statut des mathématiques en physique, sur nos visions
différentes de la recherche ou encore sur la manière de procéder dans
un système conformiste témoignent de nos styles a priori très différents
mais comme le yin et le yang, complémentaires. A chaque fois que l’on
s’est associé et que l’on a usé de cette complémentarité, on a tout
déchiré ! Merci pour tous les échanges qu’on a eus et je suis sûr
qu’il y en aura d’autres dans le futur. Je te souhaite beaucoup de
succès dans ta carrière de physicien qui démarre, tu le mérites !
Peut-être que le binôme choc se reformera un jour, qui sait ;)\newline

Je remercie également tous les doctorants que j’ai pu croiser durant
mon passage au LPSC et avec qui j’ai pu interagir ou partager des
moments. Merci à tous les membres du BIDUL pour les évènements qu’ils
organisent et le travail qu’ils font pour les doctorants.\newline

Merci à tous les enseignants-chercheurs qui m’ont fait confiance et
donné l’opportunité d’enseigner dans leurs Unités d’Enseignements
respectives. Merci à Olivier Jacquin, Thierry Dombre et Arnaud Blaise.
Je tiens à remercier tout particulièrement Laurent Derome qui m’a
donné la chance d’effectuer des travaux dirigés de relativité restreinte.
C’était stimulant et pédagogiquement très formateur. J’ai beaucoup
apprécié de travailler avec toi. Merci aux étudiants pour leur curiosité,
leur intérêt pour la physique fondamentale et pour m’avoir poussé
dans mes retranchements avec les nombreux paradoxes à résoudre qu’ils
m’ont offert. Grâce à eux, j’ai aussi pu progresser.\newline

Je souhaite remercier également tous les professeurs que j’ai eu durant
mes études, surtout ceux dont la passion transparaît et qui respectent
sincèrement leurs étudiants, en passant par les classes prépas, l’École
des Mines, l’École Normale Supérieure et le M2 PSA à l’Université
Joseph Fourier. Celui qui m’a sans doute le plus marqué est Pierre
Salati, qui m’a enseigné la relativité générale à l’ENS et la théorie
quantique des champs à l’UJF, tant il honore ce magnifique métier.
Je le remercie pour son dévouement envers l’enseignement et ses étudiants
ainsi que pour son génie pédagogique. Je n’ai jamais vu un professeur
y mettre autant du sien en 12 ans d’études ! Tous les étudiants qui
ont eu la chance de l’avoir voient de quoi je parle.\newline

Je remercie aussi tous les tuteurs que j’ai eu durant les précédents
stages de recherche que j’ai pu effectuer au cours de mes études,
dans des thématiques très variées de physique théorique. Merci à András
Borbély et Frédéric Gruy à l’Ecole des Mines de Saint-Etienne, merci
à Terai-sensei à Osaka City University et merci à Thomas Buchert au
CRAL à Lyon. J’ai beaucoup appris avec chacun de vous ! \newline

De manière plus personnelle, je souhaite remercier deux personnes
qui ont joué un rôle clef dans mon parcours. Merci à Sadok Kebaier,
mon professeur particulier au lycée, pour avoir participé à déclencher
ma passion pour les sciences dures. Merci à mon ami Zak et à « son
» fameux théorème pour m’avoir donné le goût de la théorie. Cette
rencontre a été vraiment déterminante pour moi et dans la direction
que j’ai prise par la suite.\newline

Enfin, c’est tout naturellement que j’exprime mes plus profonds remerciements
à ma famille, pour leur soutien inconditionnel malgré le chemin tortueux
que j’ai emprunté et qui m’a mené à cette thèse. Merci à ma grand-mère
sans qui rien de tout cela n’aurait été possible, merci à ma mère
pour l’éducation qu’elle m’a donnée et pour m’avoir toujours soutenu
quelles que soient les directions prises, merci à mes petites sœurs.
J’ai une pensée pour mon grand-père, dont le souhait était de me voir
réussir au baccalauréat, j’espère lui faire honneur. Je remercie également
tous mes amis proches, ou plutôt ma deuxième famille, qui ont toujours
été là pour moi. Ils se reconnaitront. Merci en particulier à ceux
qui sont venus de loin (certains d’un autre continent) pour assister
à ma soutenance. Je dédie ce manuscrit à Omar, qui nous a quittés
en cours de route.

\newpage\null\thispagestyle{empty}\newpage

\dominitoc
\tableofcontents 
\newpage

\setcounter{page}{1}

\setlength{\parindent}{16pt}

\mainmatter

\chapter*{Introduction
\addstarredchapter{Introduction}
\markboth{{Introduction}}{Introduction}}

Il y a 2600 ans, avec la naissance de la \emph{philosophia naturalis}\footnote{Expression en latin signifiant \og philosophie naturelle \fg{},
la branche de la philosophie dédiée à l'étude objective de la nature
et de l'univers physique.}, les penseurs grecs présocratiques se sont mis à chercher une essence
première en toute chose. Autrement dit, ils ont commencé à s'interroger
sur la constitution du monde dans lequel ils vivaient. Quoi de plus
naturel. Cette démarche marque le début d'une grande aventure pour
l'humanité, peut être même la plus grande que l'Homme n'ait jamais
entrepris. Elle a commencé avec \noun{Thalès de Milet} il y a 27 siècles
et n'a jamais cessé depuis. Au cours des siècles, cette quête a connu
de nombreux rebondissements.

La question fondamentale qui guide cette aventure est simple: de quoi
le monde est-il fait ? 

L'une des premières tentative de réponse a été apportée par \noun{Empèdocle}.
Pour lui, la Nature était composée de quatre éléments: la Terre, l'Air,
le Feu et l'Eau et toutes choses étaient faites de différentes combinaisons
de ces quatre éléments. Cette théorie a été complétée par \noun{Aristote}
qui y rajouta deux forces, la gravité qui faisait tomber les éléments
Terre et Eau et la légèreté qui faisait monter l'Air et le Feu. Il
y avait déjà là un besoin de théoriser seulement par la pensée et
le modèle imaginé par ces deux penseurs aboutit à une séparation entre
le contenu en matière d'une part et les forces qui s'y appliquent
d'autre part qui est toujours d'usage aujourd'hui. 

Par ailleurs, une autre interrogation occupait les esprits curieux
à cette époque. Il s'agissait de savoir si la matière était indéfiniment
sécable ou pas. Pour \noun{Aristote} et sa théorie des quatre éléments,
la matière était continue, c'est à dire que l'on pouvait diviser un
bout de matière en bouts plus petits, et ce indéfiniment. Une autre
école de pensée en Grèce antique dont le représentant le plus célèbre
est \noun{Démocrite} formulait l'hypothèse atomiste. Pour lui, la
matière était composée de grains élémentaires appelés \emph{atomes}\footnote{Du grec signifiant insécable.},
il y en avait plusieurs types et les différentes manières de les assembler
expliquaient la diversité des choses matérielles. De plus, \noun{Démocrite}
postulait également l'existence du vide, car selon lui, sans vide
aucun mouvement de matière n'est possible. Le débat entre les deux
écoles de pensée, continue et atomiste, s'est poursuivi sans jamais
qu'un des deux camps ne puissent vraiment l'emporter par des preuves
solides. 

Il faudra attendre 23 siècles pour que le jeune \noun{Einstein} dans
le premier de ses trois articles de 1905 (quelques semaines avant
celui sur la relativité restreinte) vienne trancher. Il s'appuie sur
l'observation faite par \noun{Brown} quelques années auparavant de
la dérive d'un grain de pollen dans un fluide, ce dernier était animé
d'un mouvement permanent chaotique que \noun{Brown} ne parvient pas
à expliquer. \noun{Einstein} comprend alors que si la matière était
continue, c'est à dire formée par des éléments infiniment petits et
infiniment nombreux, alors en moyenne les collisions provenant de
toutes les directions s'équilibreraient à chaque instant et le grain
observé ne bougerait pas. Par conséquent, \noun{Einstein} déduit que
le fluide n'est pas continu mais composé de grains de dimensions finies,
les molécules. Ainsi, il explique le mouvement brownien attribué alors
au hasard comme résultant de collisions entre particules. Il calcule
même leurs tailles à partir du mouvement qui les animait et fournit
la preuve définitive de l'intuition de \noun{Démocrite}. Les atomistes
l'emportèrent. 

A ce stade, il était donc établit que la matière était composée d'atomes
et non continue, mais est-ce la fin de l'aventure initiale lancée
par \noun{Thalès }? Est-ce que ces atomes ne seraient-ils pas sécables
à leur tour ? A l'époque, certains physiciens soupçonnaient déjà que
ce soit le cas suite aux travaux d'un certain \noun{Thomson} à Cambridge
qui avait observé à l'aide d'un système similaire aux tubes cathodiques,
une particule de matière chargée négativement et de très faible masse
par rapport aux atomes, appelée \emph{électron}. Il a été réalisé
rapidement que ces électrons devaient provenir des atomes eux-mêmes.
Quelques années plus tard, en 1911, c'est un physicien britannique
du nom de \noun{Rutherford} qui est venu compléter la réponse en étudiant
les collisions de particules alpha (émises par des atomes radioactifs)
avec des atomes d'or. Il établit que les atomes étaient composés de
noyaux chargés positivement, autour desquels il y avait un certain
nombre d'électrons, les particules découvertes par \noun{Thomson}.
Les atomes possédaient donc une structure interne. Comme les noyaux
n'avaient pas tous la même charge, il a été alors considéré que ces
derniers étaient composés de particules portant une charge élémentaire
positive appelées \emph{protons}\footnote{Du grec signifiant \og premier \fg{}, car considéré comme fondamental.}
et leur nombre expliquait la différence entre les noyaux. Les constituants
fondamentaux de la matière étaient donc les électrons et les protons.

L'aventure principale aurait pu s'arrêter là, mais en 1932, \noun{Chadwick}
découvre que le noyau atomique contient une autre particule, de masse
à peu près égale à celle du proton mais de charge électrique nulle,
\emph{le neutron}. Il fallait donc l'inclure à la liste des briques
fondamentales de la matière contenant désormais: le proton, le neutron
et l'électron. Cet état de connaissance sur la composition de la matière
est resté stable pendant une trentaine d'années.

Cependant, même après toutes ces découvertes, l'esprit humain ne peut
pas s'empêcher de relancer l'interrogation initiale des philosophes
présocratiques. Est-ce que les protons, les neutrons et les électrons
sont insécables ou bien possèdent-ils eux aussi une sous-structure
interne ? C'est le jeu des poupées russes. A partir des années 1960s,
avec la construction des premiers grands collisionneurs de particules,
l'évolution des connaissances sur la matière s'est considérablement
accélérée et la liste des particules élémentaires s'est vue étoffée.

En effet, ce type d'expériences consiste à faire entrer en collision
à très grande vitesse, des protons entre eux ou avec des électrons.
Plus la vitesse des particules est élevée, ou plus généralement l'énergie
de la collision, plus petite sera la taille de la structure sondée
par l'expérience. Les collisions de protons à haute énergie ont révélé
que ces derniers avaient une sous-structure, ils étaient eux-mêmes
composés de \emph{quarks}. Au fur et à mesure que ces expériences
de collisions se poursuivaient, de nouvelles particules élémentaires
étaient découvertes. Par exemple, les quarks se déclinent en six types
différents, appelés \emph{saveurs}. Pour l'anecdote, ce terme a été
lancé par \noun{Gell-Mann} en 1971 à son étudiant de l'époque dans
un magasin de glace en Californie: \og Tout comme les glaces, les
quarks possèdent à la fois une couleur et une saveur \fg{}.

Aujourd'hui, quelle réponse pourrions-nous apporter à la question
posée par les penseurs présocratiques il y a 27 siècles sur la constitution
des choses matérielles ?

\begin{figure}[H]
\centering
\includegraphics[scale=0.6]{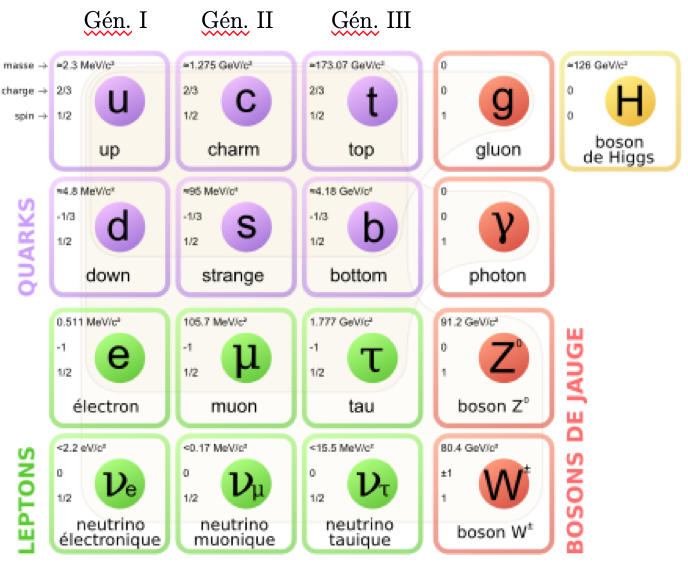}
\caption{Contenu en particules du Modèle Standard.}
\label{SMPcContent}
\end{figure}

Le modèle standard de la physique des particules, développé dans le
début des années 1970s, résume l'état des connaissances actuelles
sur les briques élémentaires qui constituent la matière (recensées
dans la figure \ref{SMPcContent}) ainsi que leurs interactions (hormis
la gravité). Il est le résultat d'un échange permanent entre théories
et expériences s'étalant sur une cinquantaine d'années et impliquant
un grand nombre de chercheurs. Il permet d'expliquer la diversité
des processus observés et est en accord avec toutes les données expérimentales
avec un niveau de précision inégalé dans l'histoire des sciences.
Le dernier épisode en date dans l'aventure a eu lieu le 4 Juillet
2012, avec la découverte du boson de Higgs par les collaborations
ATLAS et CMS, dernière pièce manquante du modèle standard qui vient
couronner le tout.

Cette thèse s'inscrit dans le cadre de la phénoménologie des particules
élémentaires.

Le \textbf{Chapitre \ref{Ch1}} est consacré au modèle standard de
la physique des particules. Il s'agit ici de poser le cadre théorique
nécessaire à l'étude des particules élémentaires et de leurs interactions.
Après avoir introduit le concept de théorie de jauge à travers l'exemple
de l'électrodynamique quantique pour le cas des théories abéliennes
puis des théories de Yang-Mills pour le cas non-abélien, nous établirons
le lagrangien du MS, en traitant les différentes pièces qui le composent
séparément. Cela consistera à traiter les différentes interactions
fondamentales, en passant par la chromodynamique quantique pour le
cas de l'interaction forte puis du modèle de Glashow-Salam-Weinberg
pour la théorie unifiée électrofaible. Ensuite, nous étudierons la
notion de brisure spontanée de symétrie électrofaible avec le mécanisme
de Higgs et les interactions de Yukawa qui permettent d'introduire
les masses dans le MS. Puis, nous nous pencherons sur le secteur de
la saveur du MS qui présente un interêt particulier dans cette thèse.
Enfin, nous passerons en revue les succès et les problèmes du MS.

Le \textbf{Chapitre \ref{Ch2}} traitera de quelques aspects de la
violation-$\mathcal{CP}$ dans le cadre du MS et au-delà. Tout d'abord,
nous étudierons les sources de violation-$\mathcal{CP}$ présentes
dans le MS, à la fois dans le secteur de l'interaction faible et celui
de l'interaction forte. Puis, nous étudierons la physique des neutrinos
et particulièrement le phénomène d'oscillation, d'abord d'un point
de vue théorique puis des faits expérimentaux qui l'ont confirmés.
Ce phénomène d'oscillation a permis d'établir au cours des 20 dernières
années que les neutrinos avaient une masse. Nous nous intéresserons
ensuite à quelques mécanismes de génération de masses des neutrinos,
qui seront utiles par la suite (chapitre 3). Pour clore ce chapitre,
nous considérerons un phénomène particulier violant la symétrie $\mathcal{CP}$,
le moment dipolaire électrique (EDM) des particules élémentaires,
nous décrirons certaines de ses propriétés ainsi que quelques aspects
expérimentaux concernant sa mesure pour enfin lister les différentes
contraintes établies et leurs impacts sur la recherche de nouvelle
physique.

Le \textbf{Chapitre \ref{Ch3}} est entièrement basé sur l'article
\cite{ST-EDM}. Dans cette étude, les structures de saveurs des EDMs
de quarks et de leptons dans le MS et au-delà seront scrutées en utilisant
des outils inspirés de \og Minimal Flavor Violation \fg{} (MFV).
Alors que les invariants de saveurs de type Jarlskog sont adéquats
pour estimer la violation-$\mathcal{CP}$ en présence de boucles de
fermions, d'autres structures non-invariantes émergent à partir de
processus de type rainbow. Le but est de construire systématiquement
ces dernières structures de saveurs dans le secteur des quarks et
celui des leptons, en considérant différents mécanismes de générations
de masses des neutrinos. Numériquement, nous montrerons que ces structures
sont généralement beaucoup plus grandes et pas nécessairement corrélées
avec les invariants de type Jarlskog. Finalement, le formalisme est
adapté pour traiter avec une troisième classes de structures de saveurs,
sensibles aux phases savoureuses $U(1)$. Nous utiliserons les méthodes
développées pour étudier l'impact de la violation-$\mathcal{CP}$
forte et la relation entre les phases de Majorana de neutrinos et
des interactions éventuelles violant les nombres baryoniques et/ou
leptoniques.

Le \textbf{Chapitre \ref{Ch4}} est dédié au concept de théorie effective
des champs (EFT). Nous commencerons dans un premier temps par introduire
la notion d'EFT et de poser le cadre théorique qui permet son expression.
Puis, dans un second temps, nous nous destinerons à la construction
d'une théorie effective pour bosons de jauges. Cette étude sera basée
sur l'article \cite{QST-GBEFT}. Après avoir introduit l'exemple pédagogique
que constitue le lagrangien d'Euler-Heisenberg pour l'étude des interactions
de photons, considéré comme l'archétype d'une EFT, nous généraliserons
leur résultat pour des bosons de jauge d'une symétrie générique, comme
par exemple $SU(2)$, $SU(3)$, plusieurs groupes de grandes unification
ou encore des symétries mixtes comme $U(1)\otimes SU(N)$ et $SU(M)\otimes SU(N)$.
En utilisant l'approche diagrammatique, nous effectuerons une procédure
de correspondance (matching) détaillée qui restera manifestement invariante
de jauge à toutes les étapes, mais ne reposant pas sur les équations
du mouvement et sera donc valide off-shell. Nous fournirons les expression
analytiques des coefficients de Wilson des opérateurs de dimensions
quatre, six et huit induits par des scalaires, des fermions et des
vecteurs siégeants dans des représentations génériques du groupe de
jauge. Ces expressions reposent sur une analyse des invariants quartiques
de Casimir, pour lesquels nous fournissons une revue avec les conventions
adoptées pour le calcul des diagrammes de Feynman. Finalement, nous
montrerons qu'à une boucle, certains opérateurs sont redondants quelque
soit la représentation ou le spin de la particule intégrée, réduisant
l'apparente complexité de la base d'opérateurs qui pourrait être construite
uniquement sur des arguments de symétrie.

\chapter{Le modèle standard de la physique des particules\label{Ch1}}

\minitoc

\section{Théories de jauges}

Une théorie de jauge est une théorie construite pour être invariante
sous certaines transformations internes qui n'agissent pas sur l'espace-temps,
appelée transformations de jauges. Ce principe d'invariance de jauge
est extrêmement puissant et permet de décrire les interactions entre
particules élémentaires. Une interaction donnée possède généralement
des symétries\footnote{Par symétrie, nous entendons un ensemble de transformations qui laissent
invariante l'interaction.} et la seule connaissance de ces dernières permet de décrire l'interaction
en question en contraignant sa structure grâce au formalisme des théories
de jauges. 

\subsection{Théories de jauges abéliennes: cas de l'électrodynamique quantique}

L'électrodynamique quantique (QED) est la version quantique de la
théorie de l'électromagnétisme. En théorie quantique des champs (TQC),
le champ électromagnétique est quantifiée et son quanta de champ est
le photon. Lorsque deux particules chargées électriquement interagissent,
elles le font par l'intermédiaire du champ quantique électromagnétique
qui contrairement à sa version classique, présente une structure granulaire.
Cette granularité du champ électromagnétique permet de décrire une
interaction élémentaire par un échange de grain qui compose le champ
véhiculant l'interaction. Ces quantas de champs sont des particules
appelées \emph{particules médiatrices} ou encore \emph{bosons de jauges}.
Dans le cas de le QED, la seule particule médiatrice est le photon.
Ainsi, la QED est donc la théorie des interactions entre fermions
et photons. Dans cette partie, nous allons exprimer cette théorie
dans un formalisme de théorie de jauge.

D'abord, le lagrangien\footnote{En fait, c'est la densité lagrangienne que nous appellerons abusivement
lagrangien.} de Dirac décrivant des fermions libres 
\begin{equation}
\mathcal{L}_{0}=\overline{\psi}(\gamma^{\mu}\partial_{\mu}-m)\psi
\end{equation}
possède une symétrie globale $U(1)$\footnote{Le groupe $U(1)$ est l'ensemble des nombres complexes de module unité.
Mathématiquement, $U(1)=\{e^{i\alpha}|\alpha\in\mathbb{R}\}$.}, c'est à dire qu'il est invariant sous l'action des éléments (transformations)
de ce groupe. Les transformations de $U(1)$ agissent uniquement sur
le champ fermionique de la manière suivante: 
\begin{equation}
\psi(x)\rightarrow\psi^{\prime}(x)=e^{i\alpha}\psi(x),
\end{equation}
avec $\alpha$ un paramètre arbitraire réel. La transformation est
dite globale car $\alpha$ ne dépend pas du point de l'espace-temps
$x$. 

Pour construire la QED, c'est à dire ajouter les interactions (médiées
par les photons) à la théorie des fermions libres précédente, il suffit
d'étendre la symétrie globale à une symétrie locale, c'est à dire
avec un paramètre qui dépend du point de l'espace-temps $\alpha\rightarrow\alpha(x)$,
autrement dit une fonction. Afin d'imposer l'invariance de $\mathcal{L}_{0}$
sous les transformations $U(1)$ locales $\psi(x)\rightarrow\psi^{\prime}(x)=e^{i\alpha(x)}\psi(x)$
dites transformations de jauges, il est nécessaire de remplacer la
dérivée partielle par une \emph{dérivée covariante} contenant un champ
vectoriel $A_{\mu}$
\begin{equation}
\partial_{\mu}\rightarrow D_{\mu}=\partial_{\mu}-ieA_{\mu},
\end{equation}
où $e$ représente la constante de couplage de l'interaction électromagnétique.
Dans la construction des théories de jauges, ce remplacement de dérivées
est un principe général, appelé \emph{principe de substitution minimale,}
qui permet d'introduire les interactions à partir de la théorie libre.
Ainsi, le lagrangien invariant de jauge locale sous $U(1)$ est donné
par
\begin{equation}
\mathcal{L}=\overline{\psi}(\gamma^{\mu}D_{\mu}-m)\psi.
\end{equation}
En développant la dérivée covariante, on obtient le terme d'interaction
$\mathcal{L}_{int}$ entre les fermions et le champ vectoriel médiateur
$A_{\mu}$ 
\begin{equation}
\mathcal{L}=\mathcal{L}_{0}+\underbrace{e\overline{\psi}\gamma^{\mu}\psi A_{\mu}}_{\mathcal{L}_{int}}.
\end{equation}

Pour résumer, afin d'introduire les interactions, nous sommes naturellement
amenés à introduire un champ vectoriel médiateur de l'interaction
électromagnétique pour appliquer le principe de substitution minimale.
Plus généralement, dans une théorie de jauge, l'application du principe
de substitution minimale permet d'introduire naturellement les bosons
de jauges nécessaires, véhiculant l'interaction. La nature et le nombre
de particules médiatrices sont dictés par la structure du groupe de
jauge sous lequel la théorie est invariante locale.

Notons que le champ vectoriel médiateur $A_{\mu}$ n'est pas dynamique,
car il lui manque un terme cinétique dans cette théorie. On ajoute
alors un terme cinétique à $A_{\mu}$ invariant de jauge locale sous
$U(1)$, déjà présent dans l'électrodynamique classique, à savoir
\begin{equation}
\mathcal{L}_{A}=-\frac{1}{4}F_{\mu\nu}F^{\mu\nu},
\end{equation}
où $F_{\mu\nu}=\partial_{\mu}A_{\nu}-\partial_{\nu}A_{\mu}$ est le
tenseur électromagnétique. L'équation du mouvement de $A_{\mu}$,
obtenue à partir de l'équation d'Euler-Lagrange appliquée au terme
cinétique $\mathcal{L}_{A}$, donne alors les équations de Maxwell
dans le vide. On peut alors complètement identifier $A_{\mu}$ au
photon et le lagrangien de QED est donc
\begin{equation}
\begin{aligned}\mathcal{L}_{QED} & =\overline{\psi}(\gamma^{\mu}D_{\mu}-m)\psi-\frac{1}{4}F_{\mu\nu}F^{\mu\nu}\\
 & =\mathcal{L}_{0}+\mathcal{L}_{int}+\mathcal{L}_{A}.
\end{aligned}
\end{equation}

\subsection{Théories de jauges non-abéliennes}

Les étapes de construction de la QED avec le groupe de jauge $U(1)$
peuvent être étendues à la construction de théories basées sur des
groupes de jauges non-abéliens, c'est à dire à des transformations
non-commutatives. Dans un premier temps, il s'agit d'identifier les
symétries globales du lagrangien libre. Puis, les interactions entrent
en action à l'aide du principe de substitution minimal, moyennant
l'introduction de nouveaux champs vectoriels de jauges médiateurs
de l'interaction. Enfin, les champs vectoriels sont rendus dynamiques
en leur ajoutant des termes cinétiques.

Considérons une théorie libre décrite par un multiplet de champs fermioniques
de masse $m$, $\Psi=(\psi_{1},\ldots,\psi_{n})^{T}$ ainsi que le
lagrangien suivant
\begin{equation}
\mathcal{L}_{0}=\overline{\Psi}(\gamma^{\mu}\partial_{\mu}-m)\Psi.\label{eq:L0nonAbel}
\end{equation}
Cette théorie possède une symétrie globale sous les transformations
\begin{equation}
\Psi(x)\rightarrow U(\alpha^{1},\ldots,\alpha^{N})\Psi(x),
\end{equation}
où $U$ sont des matrices unitaires $n\times n$ représentant un groupe
de Lie non-abélien de rang $N$, dépendant de $N$ paramètres réels
$\alpha^{1},\ldots,\alpha^{N}$. Notons que dans le contexte de ce
chapitre et plus généralement celui du modèle standard, les groupes
de Lie non-abéliens d'interêts seront $SU(2)$ et $SU(3)$, avec les
champs fermioniques siégeant dans les représentations fondamentales
respectives ($n=2$) et ($n=3$).

Une propriété cruciale des groupes de Lie simplement connexes est
que les éléments du groupe peuvent être exprimés comme une exponentielle
des éléments de l'algèbre associée. Une base de l'algèbre de Lie est
donnée par les générateurs. Par conséquent, tous les éléments de l'algèbre
peuvent être exprimés comme combinaison linéaire des générateurs.
Ainsi, les matrices $U$ peuvent être exprimées de la sorte
\begin{equation}
U(\alpha^{1},\ldots,\alpha^{N})=e^{i(\alpha^{1}T_{1}+\ldots+\alpha^{N}T_{N})},
\end{equation}
avec $T_{1},\ldots,T_{N}$ les générateurs du groupe de Lie, matrices
hermitiennes formant l'algèbre de Lie
\begin{equation}
\left[T_{a},T_{b}\right]=if_{abc}T_{c},
\end{equation}
où $f_{abc}$ sont les constantes de structures réelles caractéristiques
du groupe. On adopte la convention de normalisation des générateurs
\begin{equation}
Tr\left(T_{a}T_{b}\right)=\frac{1}{2}\delta_{ab}.
\end{equation}

Ensuite, la deuxième étape dans la construction de la théorie de jauge
consiste à étendre la symétrie globale précédente à une symétrie locale
en faisant dépendre du point d'espace-temps $x$ les paramètres $\{\alpha^{a},a=1\ldots N\}$,
c'est à dire en les élevant au statut de fonctions $\{\alpha^{a}(x),a=1\ldots N\}$.
Afin que le lagrangien (\ref{eq:L0nonAbel}) soit invariant sous les
transformations de jauges locales, on applique le principe de substitution
minimal en introduisant une dérivée covariante
\begin{equation}
\partial_{\mu}\rightarrow D_{\mu}=\partial_{\mu}\mathbb{I}-ig\mathbb{W}_{\mu},
\end{equation}
impliquant un champ de vecteur $\mathbb{W}_{\mu}$ ainsi qu'une constante
de couplage $g$. Comme la dérivée covariante agit sur les $n$-uplets
$\Psi=\left(\begin{array}{c}
\psi_{1}\\
\vdots\\
\psi_{n}
\end{array}\right)$, le champ de vecteur $\mathbb{W}_{\mu}$ est nécessairement une matrice
carrée de dimension $n$ et peut être développé avec les générateurs
comme suit
\begin{equation}
\mathbb{W}_{\mu}(x)=T_{a}W_{\mu}^{a}(x),
\end{equation}
où les $\{W_{\mu}^{a}(x),a=1\ldots N\}$ sont les $N$ champs de jauges
médiateurs de cette interaction. Par suite, les termes d'interactions
sont obtenus de la même manière que précédemment, c'est à dire en
développant la dérivée covariante dans le lagrangien (\ref{eq:L0nonAbel}).
En effet, 
\begin{equation}
\mathcal{L}_{0}\rightarrow\mathcal{L}=\mathcal{L}_{0}+\mathcal{L}_{int},
\end{equation}
avec le terme d'interaction
\begin{equation}
\mathcal{L}_{int}=g\overline{\Psi}\gamma^{\mu}\mathbb{W}_{\mu}\Psi=g\overline{\Psi}\gamma^{\mu}T_{a}\Psi W_{\mu}^{a},
\end{equation}
contenant les interactions de $N$ courants $j_{a}^{\mu}=g\overline{\Psi}\gamma^{\mu}T_{a}\Psi$
avec les champs de jauges $W_{\mu}^{a}$.

Les transformations de jauges locales qui laissent invariante la théorie
impliquent la matrice $U\equiv U(\alpha^{1}(x),\ldots,\alpha^{N}(x))$
et sont données par
\begin{equation}
\begin{aligned}\Psi & \rightarrow\Psi^{\prime}=U\Psi,\\
\mathbb{W}_{\mu} & \rightarrow\mathbb{W}_{\mu}^{\prime}=U\mathbb{W}_{\mu}U^{-1}-\frac{i}{g}(\partial_{\mu}U)U^{-1}.
\end{aligned}
\label{eq:NonAbelGaugeTransfo}
\end{equation}
Pour une transformation infinitésimale $\alpha^{a}(x)\ll1$, la transformation
de jauge pour le champ de vecteur $\mathbb{W}_{\mu}$ peut être écrite
en termes des champs de jauges $W_{\mu}^{a}$ comme ceci
\begin{equation}
W_{\mu}^{a}\rightarrow W_{\mu}^{\prime a}=W_{\mu}^{a}+\frac{1}{g}\partial_{\mu}\alpha^{a}+f_{abc}W_{\mu}^{b}\alpha^{c}.
\end{equation}
Remarquons que cette transformation diffère du cas abélien par son
dernier terme, nouveau, car d'origine purement non-abélienne.

Finalement, la dernière étape consiste à rendre dynamiques les champs
de jauges $W_{\mu}^{a}$ en leurs donnant un terme cinétique, obtenu
en généralisant le cas de l'électromagnétisme. En effet, le tenseur
électromagnétique $F_{\mu\nu}$ est généralisé comme suit
\begin{equation}
\mathbb{F}_{\mu\nu}=T_{a}F_{\mu\nu}^{a}=\partial_{\mu}\mathbb{W}_{\nu}-\partial_{\nu}\mathbb{W}_{\mu}-ig\left[\mathbb{W}_{\mu},\mathbb{W}_{\nu}\right],
\end{equation}
avec $N$ tenseurs de champs de jauges (équivalents au tenseur électromagnétique
pour le photon)
\begin{equation}
F_{\mu\nu}^{a}=\partial_{\mu}W_{\nu}^{a}-\partial_{\nu}W_{\mu}^{a}+gf_{abc}W_{\mu}^{b}W_{\nu}^{c}.
\end{equation}
Sous la transformation de jauge (\ref{eq:NonAbelGaugeTransfo}), le
tenseur $\mathbb{F}_{\mu\nu}$ se transforme selon
\begin{equation}
\mathbb{F}_{\mu\nu}\rightarrow\mathbb{F}_{\mu\nu}^{\prime}=U\mathbb{F}_{\mu\nu}U^{-1}.
\end{equation}
Afin de construire un invariant de jauge, on peut prendre la trace
du tenseur $\mathbb{F}_{\mu\nu}$ qui grâce à sa propriété de cyclicité
élimine les matrices $U$ induites par la transformation de jauge.
De façon explicite,
\begin{equation}
Tr\left(\mathbb{F}_{\mu\nu}^{\prime}\mathbb{F}^{\prime\mu\nu}\right)=Tr\left(U\mathbb{F}_{\mu\nu}U^{-1}U\mathbb{F}^{\mu\nu}U^{-1}\right)=Tr\left(U^{-1}U\mathbb{F}_{\mu\nu}U^{-1}U\mathbb{F}^{\mu\nu}\right)=Tr\left(\mathbb{F}_{\mu\nu}\mathbb{F}^{\mu\nu}\right),
\end{equation}
où l'on a utilisé l'unitarité de $U$,$U^{\dagger}U=\mathbb{I}$.
Comme le terme cinétique doit obligatoirement être invariant de jauge,
il s'ensuit que la version non-abélienne du terme cinétique pour les
champs de jauges $W_{\mu}^{a}$ est donc
\begin{equation}
\mathcal{L}_{W}=-\frac{1}{2}Tr\left(\mathbb{F}_{\mu\nu}\mathbb{F}^{\mu\nu}\right)=-\frac{1}{4}F_{\mu\nu}^{a}F^{a,\mu\nu}.
\end{equation}
La partie quadratique de ce terme cinétique décrit la propagation
libre des champs de jauges alors que les parties cubiques et quartiques
décrivent l'auto-interaction des champs de jauges, phénomène propre
aux théories de jauges non-abéliennes. Ces interactions des champs
de jauges avec eux mêmes sont complètement déterminées par la symétrie
de jauge. On a
\begin{equation}
\begin{aligned}\mathcal{L}_{W} & =-\frac{1}{4}(\partial_{\mu}W_{\nu}^{a}-\partial_{\nu}W_{\mu}^{a})(\partial^{\mu}W^{a,\nu}-\partial^{\nu}W^{a,\mu}) & \text{propagation des champs libres}\\
 & -\frac{g}{2}f_{abc}(\partial_{\mu}W_{\nu}^{a}-\partial_{\nu}W_{\mu}^{a})W^{b,\mu}W^{c,\nu} & \text{auto-interaction à 3 champs}\\
 & -\frac{g^{2}}{4}f_{abc}f_{ade}W_{\mu}^{b}W_{\nu}^{c}W^{d,\mu}W^{e,\nu}. & \text{auto-interaction à 4 champs}
\end{aligned}
\end{equation}
Pour finir, notons que les champs de jauges n'ont pas de masses. En
effet, des termes de masses du type $\frac{M^{2}}{2}W_{\mu}^{a}W^{a,\mu}$
ne sont pas invariants sous les transformations locales (\ref{eq:NonAbelGaugeTransfo})
et brise donc la symétrie de jauge.

\section{La chromodynamique quantique: théorie de l'interaction forte}

La chromodynamique quantique (QCD) est la théorie décrivant l'interaction
forte, cette \og force \fg{} qui permet la cohésion des noyaux atomiques
censés éclater sous l'effet des forces répulsives électromagnétiques
s'exerçant entre les protons. La QCD est une théorie de jauge non-abélienne
basée sur le groupe de jauge $SU(3)$, groupe de Lie spécial unitaire
de dimension 3. Les fermions assujettis à l'interaction forte sont
les quarks, qui composent les hadrons. La QCD est donc la théorie
des quarks en interactions fortes. Les quarks existent sous trois
états de \emph{couleurs} différents. La couleur est la \og charge
forte \fg{}, elle est à l'interaction forte ce que la charge électrique
est à l'interaction électromagnétique.

Les trois états de couleurs des quarks forment la représentation fondamentale
du groupe de jauge $SU(3)$. La représentation fondamentale étant
de dimension 3, les quarks sont représentés par des triplets de champs
fermioniques $\Psi=(q_{1},q_{2},q_{3})^{T}$ pour chaque type (saveur)
de quarks. Le groupe de jauge est alors noté $SU(3)_{c}$ avec l'indice
$c$ référant à \og couleur \fg{}. Le groupe de couleur $SU(3)_{c}$
a huit générateurs $\{T_{a}|a=1\ldots8\}$, qui s'expriment dans la
représentation fondamentale en termes des huit matrices de Gell-Mann
$3\times3$, $\{\lambda_{a}|a=1\ldots8\}$
\begin{equation}
T_{a}=\frac{1}{2}\lambda_{a},\hspace{1em}a=1\ldots8,
\end{equation}
Dans le formalisme des théories de jauges vu précédemment, les interactions
sont contenues dans la dérivée covariante. Pour la QCD, dans l'étape
de construction de la théorie de jauge, afin d'appliquer le principe
de substitution minimale sur la théorie libre, on est amené à introduire
huit champs vectoriels de jauges $\{G_{\mu}^{a}|a=1\ldots8\}$, médiateurs
de l'interaction forte, appelés \emph{les gluons}. Ainsi, la dérivée
covariante agissant sur le triplets de couleurs de quarks $\Psi$
et est donnée par
\begin{equation}
D_{\mu}=\partial_{\mu}-ig_{s}\frac{\lambda}{2}G_{\mu}^{a},
\end{equation}
De plus, les tenseurs de champs de jauges sont définis comme suit
\begin{equation}
G_{\mu\nu}^{a}=\partial_{\mu}G_{\nu}^{a}-\partial_{\nu}G_{\mu}^{a}+g_{s}f_{abc}G_{\mu}^{b}G_{\nu}^{c},
\end{equation}
où $g_{s}$ est la constante de couplage forte et est communément
exprimée en fonction de la constante de structure fine de l'interaction
forte $\alpha_{s}$ de la manière suivante $\alpha_{s}=\tfrac{g_{s}^{2}}{4\pi}$.

\begin{figure}
\centering
\includegraphics[width=0.6\textwidth]{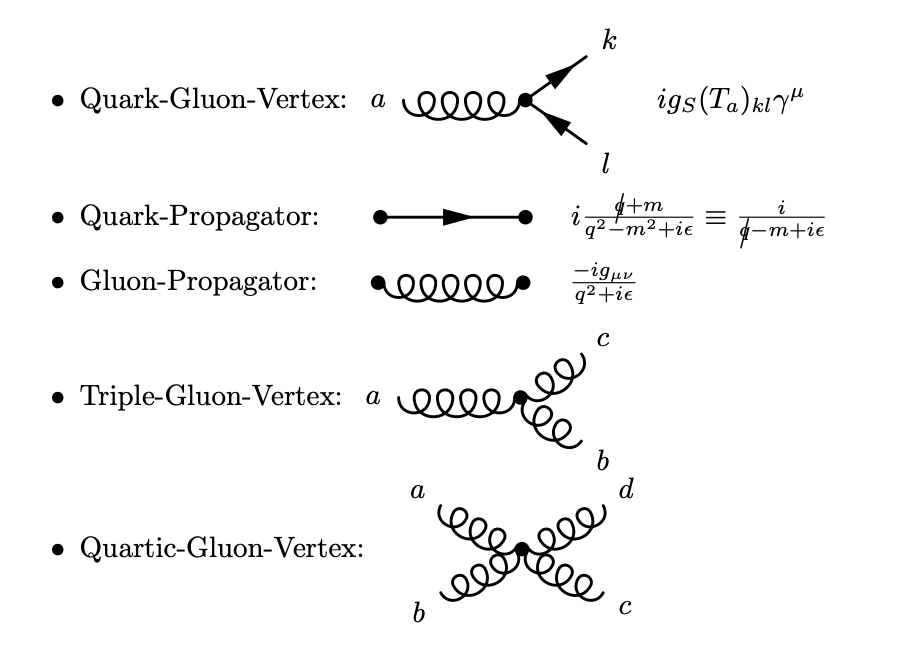}
\caption{Propagateurs et interactions en QCD}
\label{QCDFeynmanRules}
\end{figure}

Par suite, en utilisant les outils développés dans la partie précédente,
on écrit le lagrangien de QCD (pour une saveur donnée) de la manière
suivante
\begin{equation}
\begin{aligned}\mathcal{L}_{QCD} & =\overline{\Psi}(i\gamma^{\mu}D_{\mu}-m)\Psi+\mathcal{L}_{G}\\
 & =\overline{\Psi}(i\gamma^{\mu}\partial_{\mu}-m)\Psi+g_{s}\overline{\Psi}\gamma^{\mu}\frac{\lambda_{a}}{2}\Psi G_{\mu}^{a}-\frac{1}{4}G_{\mu\nu}^{a}G^{a,\mu\nu}.
\end{aligned}
\end{equation}
Ce lagrangien contient les interactions des courants de quarks et
des gluons ainsi que les auto-interactions à 3-gluons et 4-gluons.
Ces interactions élémentaires sont représentées en figure \ref{QCDFeynmanRules}
et leurs règles de Feynman sont données. Notons que pour obtenir le
lagrangien complet de QCD, on doit sommer le lagrangien précédent
(décrivant uniquement un type de quark) sur les 6 saveurs de quarks
du MS $q=u,d,c,s,b,t$, chacune représentées par un triplet de couleur
$\Psi_{q}$ avec leurs masses respectives $m_{q}$. Enfin, nous avons
volontairement omis le lagrangien de fixation de jauge sur lequel
on reviendra par la suite de façon plus générale.

\section{La théorie électrofaible de Glashow-Salam-Weinberg: vers le modèle
standard}

La théorie électrofaible \cite{GlashowSM,WeinbergSM}, proposée par
Glashow, Weinberg et Salam est la théorie unifiée de l'interaction
électromagnétique et de l'interaction faible\footnote{L'interaction faible est l'interaction responsable des désintégrations
nucléaires. Elle concerne tous les fermions.}. La théorie de Glashow-Salam-Weinberg (GSW) est la formulation la
plus complète de la théorie des interactions unifiées électrofaibles
et est en accord jusqu'à ce jour avec toutes les données expérimentales
d'origine électrofaible. A basse énergie, cette théorie reproduit
à la fois la QED et la théorie de Fermi des interactions faibles,
qui constituait déjà une très bonne description des interactions nucléaires
faibles à basses énergies. De plus, ce modèle est minimal dans le
sens où il contient le minimum de degrés de libertés nécessaires permettant
de décrire tous les résultats expérimentaux connus.

La théorie GSW est une théorie de jauge non-abélienne basée sur le
groupe de jauge $SU(2)_{L}\times U(1)_{Y}$. Au moment de l'élaboration
de la théorie GSW, les résultats de la désormais célèbre expérience
de Wu en 1956 \cite{WuEXP} qui a établi que l'interaction faible
violait la parité de façon maximale, était bien connus. Etant donné
que l'interaction faible ne couple que des particules de chiralités
gauches, la théorie GSW est chirale dans la mesure où elle traite
les composantes de chiralités gauches différemment de celles de chiralités
droites. L'indice $L$ dans le groupe de jauge de l'interaction faible
$SU(2)_{L}$ fait référence à la chiralité gauche (Left). Mathématiquement,
cela signifie que sous l'action du groupe de jauge $SU(2)_{L}$, les
fermions gauches et droits se transforment selon différentes représentations
du groupe.

\begin{figure}[h]
\centering
\includegraphics[width=0.6\textwidth]{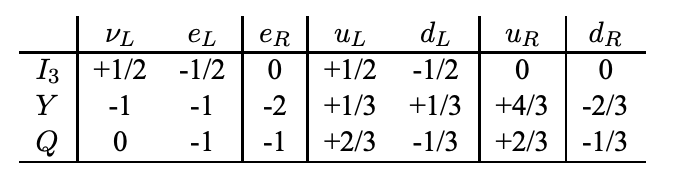}
\caption{Nombres quantiques d'isospin $I_{3}$ et d'hypercharge $Y$ pour les leptons et les quarks de chiralités gauche et droite ainsi que leurs charges électriques}
\label{QuantumNumbers}
\end{figure}

En effet, au sein de chaque génération, les fermions gauches sont
placés dans des doublets se transformant comme la représentation fondamentale
de $SU(2)_{L}$ et les fermions droits sont des singlets siégeant
dans la représentation triviale de $SU(2)_{L}$ car ils n'interagissent
pas. Les objets élémentaires de la théorie sont

\begin{eqnarray*}
\text{Leptons}: &  & \left(\begin{array}{c}
\nu_{e}\\
e
\end{array}\right)_{L},\left(\begin{array}{c}
\nu_{\mu}\\
\mu
\end{array}\right)_{L},\left(\begin{array}{c}
\nu_{\tau}\\
\tau
\end{array}\right)_{L},e_{R},\mu_{R},\tau_{R}\\
\text{Quarks}: &  & \left(\begin{array}{c}
u\\
d
\end{array}\right)_{L},\left(\begin{array}{c}
c\\
s
\end{array}\right)_{L},\left(\begin{array}{c}
t\\
b
\end{array}\right)_{L},u_{R},d_{R},c_{R},s_{R},t_{R},b_{R}.
\end{eqnarray*}
Ces derniers peuvent être classifiés par la donnée de trois nombres
quantiques, à savoir: l'isospin faible $I$, sa troisième composante
$I_{3}$ et l'hypercharge faible $Y$. Les champs de chiralité gauche
ont un isospin $I=\tfrac{1}{2}$ et forment donc des doublets alors
que les champs de chiralité droite possède un isospin nul $I=0$ et
sont alors des singlets. La relation de Gell-Mann-Nishijima lie ces
nombres quantiques à la charge électrique $Q$ de la manière suivante
\begin{equation}
Q=I_{3}+\frac{Y}{2}.\label{eq:GellMannNishijima}
\end{equation}
Les différents nombres quantiques pour les fermions élémentaires (quarks
et leptons) sont donnés pour la première génération dans le tableau
représenté en figure \ref{QuantumNumbers}. Ceux-ci sont identiques
pour les autres générations.

Ainsi, cette structure est incorporée dans un formalisme de théorie
de jauge avec des transformations locales du groupe $SU(2)_{L}\times U(1)_{Y}$.
Ce groupe de Lie non-simple possède quatre générateurs, trois pour
$SU(2)_{L}$, $\{T_{a}=I_{a}|a=1,2,3\}$ et un pour $U(1)_{Y},$ $T_{4}=Y$,
où les $I_{a}$ sont les opérateurs d'isospins et $Y$ l'hypercharge,
formant l'algèbre de Lie
\begin{equation}
\left[I_{a},I_{b}\right]=i\epsilon_{abc}I_{c},\hspace{1em}\left[I_{a},Y\right]=0.
\end{equation}

Le lagrangien complet de la théorie électrofaible est formé de quatre
composantes chacune traitant d'un secteur: jauge, Higgs, fermion et
Yukawa
\begin{equation}
\mathcal{L}_{EW}=\mathcal{L}_{G}+\mathcal{L}_{H}+\mathcal{L}_{F}+\mathcal{L}_{Y}.
\end{equation}
Dans la suite, nous allons nous pencher sur chacune de ces composantes.
Les lagrangiens de Higgs et de Yukawa seront traités dans la section
suivante consacrée à la brisure spontanée de symétrie.

\paragraph{Les champs de jauges $\mathcal{L}_{G}$\newline}

Chaque générateurs du groupe de jauge $SU(2)_{L}\times U(1)_{Y}$
est associé à un boson de jauge. Nous avons alors un triplet de champ
vectoriels $W_{\mu}^{1,2,3}$ associé aux $I_{1,2,3}$ ainsi qu'un
singlet vectoriel $B_{\mu}$ associé à $Y$. Ces derniers permettent
de définir les tenseurs de champs de jauges suivants
\begin{equation}
\begin{aligned}W_{\mu\nu}^{a} & =\partial_{\mu}W_{\nu}^{a}-\partial_{\nu}W_{\mu}^{a}+g_{2}\epsilon_{abc}W_{\mu}^{b}W_{\nu}^{c},\\
B_{\mu\nu} & =\partial_{\mu}B_{\nu}-\partial_{\nu}B_{\mu}.
\end{aligned}
\end{equation}
En raison du fait que le groupe de jauge soit semi-simple et qu'il
contienne deux facteurs, nous avons deux constantes de couplages indépendantes:
$g_{2}$ pour $SU(2)_{L}$ et $g_{1}$ pour $U(1)_{Y}$. A partir
des tenseurs de champs de jauges définis plus haut, on construit le
lagrangien de jauge suivant
\begin{equation}
\mathcal{L}_{G}=-\frac{1}{4}W_{\mu\nu}^{a}W^{\mu\nu,a}-\frac{1}{4}B_{\mu\nu}B^{\mu\nu},
\end{equation}
invariant de jauge local sous $SU(2)_{L}\times U(1)_{Y}$. Notons
que des termes de masses explicites pour les champs de jauges sont
interdits dans la mesure où ils brisent l'invariance de jauge. Les
masses sont données dans un second temps à l'aide du mécanisme de
Higgs que nous verrons plus loin.

\paragraph{Le lagrangien des fermions $\mathcal{L}_{F}$\newline}

Comme les fermions de chiralités différentes vivent dans des représentations
différentes, on se doit de distinguer les champs gauches et droits.
Pour cela, on utilise les projecteurs chiraux
\begin{equation}
\psi_{L}=\frac{1-\gamma^{5}}{2}\psi,\hspace{1em}\psi_{R}=\frac{1+\gamma^{5}}{2}\psi.
\end{equation}
Comme nous l'avons vu précédemment, les champs fermioniques gauches
de chaque génération $I\in\{1,2,3\}$ de quarks et de leptons sont
organisés dans des doublets de $SU(2)$, tandis que les champs droits
sont représentés par des singlets
\begin{equation}
\psi_{L}^{I}=\left(\begin{array}{c}
\psi_{L+}^{I}\\
\psi_{L-}^{I}
\end{array}\right),\hspace{1em}\psi_{R,\sigma}^{I},
\end{equation}
où l'indice $\sigma=\pm$ dénote les fermions de type up (+) et de
type down (-). Tous les multiplets sont des états propres de l'hypercharge
faible $Y$ en sorte que la relation de Gell-Mann-Nishijima (\ref{eq:GellMannNishijima})
s'applique. La dérivée covariante est différente pour les champs gauches
et droits car ils appartiennent à des représentations différentes
et ont donc des générateurs différents. Elle est donnée par
\begin{equation}
D_{\mu}^{L,R}=\partial_{\mu}-ig_{2}I_{a}^{L,R}W_{\mu}^{a}+ig_{1}\frac{Y}{2}B_{\mu},\hspace{1em}\text{avec}\hspace{1em}I_{a}^{L}=\frac{1}{2}\sigma_{a},\hspace{1em}I_{a}^{R}=0
\end{equation}
et induit les interactions entre les fermions et les champs de jauges
\begin{equation}
\mathcal{L}_{F}=\sum_{I}\overline{\psi}_{L}^{I}i\gamma^{\mu}D_{\mu}^{L}\psi_{L}^{I}+\sum_{I,\sigma}\overline{\psi}_{R,\sigma}^{I}i\gamma^{\mu}D_{\mu}^{R}\psi_{R,\sigma}^{I},
\end{equation}
où l'indice $I$ indexe les générations de particules.

\section{Brisure spontanée de la symétrie électrofaible}

Jusque là, nous avons un modèle qui décrit les interactions électrofaibles
entre particules élémentaires, le modèle GSW. Cependant, en raison
du fait que les termes de masses brisent l'invariance de jauge, nous
n'étions pas autorisés à en écrire et de ce fait aucune particule
n'est dotée de masse dans ce modèle. Or, nous savons que quasiment
toutes les particules ont une masse et il faut donc trouver un moyen
de les introduire dans le modèle GSW. La solution à ce problème réside
dans le concept de \emph{brisure spontanée de symétrie}, déjà bien
connu dans d'autres domaines de la physique comme en matière condensée
par exemple. Le mécanisme de brisure spontanée de la symétrie électrofaible
a été développé en 1964 indépendamment par Englert et Brout, par Higgs
et par Guralnik, Hagen et Kibble. Il est communément appelé simplement
\og mécanisme de Higgs \fg{} au détriment des autres physiciens.
Suite à la découverte en 2012 par l'expérience ATLAS au CERN \cite{HiggsAtlas}
du boson de Higgs prédit par ce mécanisme, le prix Nobel de physique
l'année suivante a été décerné à Brout, Englert et Higgs pour leur
prédiction alors confirmée.

\subsection{Le mécanisme de Higgs}

Le mécanisme de Higgs permet de donner une masse aux particules à
partir de la brisure spontanée de la symétrie électrofaible. Nous
allons décrire ici comment briser spontanément la symétrie $SU(2)_{L}\times U(1)_{Y}$
de sorte à conserver le sous-groupe de jauge électromagnétique $U(1)_{em}$
non brisé. Dans ce but, on introduit un doublet d'isospin faible de
champs scalaires complexes d'hypercharge $Y=1$
\begin{equation}
H(x)=\left(\begin{array}{c}
\phi^{+}(x)\\
\phi^{0}(x)
\end{array}\right).
\end{equation}
Ce champ est appelé \emph{champ de Higgs}. Le champ de Higgs est couplé
aux bosons de jauges par l'application du principe de substitution
minimal,
\begin{equation}
\mathcal{L}_{H}=(D_{\mu}H)^{\dagger}(D^{\mu}H)-V(H),\label{eq:LHiggs}
\end{equation}
avec une dérivée covariante
\begin{equation}
D_{\mu}=\partial_{\mu}-ig_{2}\frac{\sigma_{a}}{2}W_{\mu}^{a}+i\frac{g_{1}}{2}B_{\mu},
\end{equation}
où les $\{\sigma_{a}|a=1,2,3\}$ sont les matrices de Pauli. Le potentiel
$V(H)$, qui contient les auto-interactions du champ de Higgs, est
donné par
\begin{equation}
V(H)=-\mu^{2}H^{\dagger}H+\frac{\lambda}{4}(H^{\dagger}H)^{2},
\end{equation}
avec $\mu$ et $\lambda$ des constantes réelles.

Dans le vide, c'est à dire l'état de plus basse énergie, le potentiel
est minoré. Pour $\mu^{2}$ et $\lambda$ tout deux positifs, le minimum
n'est pas en $H=0$ mais plutôt réalisé par toutes les configurations
de champs non-nulles vérifiants $H^{\dagger}H=\tfrac{2\mu^{2}}{\lambda}$.
En sélectionnant la configuration réelle et neutre électriquement
($QH=0$)
\begin{equation}
Q=I_{3}+\frac{Y}{2}=\left(\begin{array}{cc}
1 & 0\\
0 & 0
\end{array}\right),
\end{equation}
on obtient alors la \emph{valeur moyenne dans le vide (VEV)
\begin{equation}
\langle0|H|0\rangle=\frac{1}{\sqrt{2}}\left(\begin{array}{c}
0\\
v
\end{array}\right)\hspace{1em}\text{avec}\hspace{1em}v=\frac{2\mu}{\sqrt{\lambda}}.
\end{equation}
}Bien que le lagrangien soit invariant de jauge sous le groupe $SU(2)_{L}\times U(1)_{Y}$,
le vide $\langle0|H|0\rangle$ quant à lui ne présente pas cette symétrie.
Dans ces conditions, on dit que la symétrie a été spontanément brisée.
Néanmoins, la configuration vide $\langle0|H|0\rangle$ est toujours
symétrique sous l'action du sous-groupe électromagnétique $U(1)_{em}$
qui est généré par l'opérateur charge électrique $Q$. Il s'ensuit
que la symétrie de jauge électromagnétique reste donc préservée.

Le champ de Higgs peut être réécrit de la manière suivante
\begin{equation}
H(x)=\left(\begin{array}{c}
\phi^{+}(x)\\
\frac{v+h(x)+i\chi(x)}{\sqrt{2}}
\end{array}\right),
\end{equation}
avec les composantes $\phi^{+}$, $h$ et$\chi$ ayant des valeurs
moyennes dans le vide nulles. En développant le potentiel $V(H)$
autour de la VEV en supposant que les autres composantes apportent
des petites perturbations autour de cette configuration du vide, on
obtient alors un terme de masse pour $h$ alors que $\phi^{+}$ et
$\chi$ restent sans masses. En exploitant la symétrie de jauge du
lagrangien, les composantes $\phi^{+}$ et $\chi$ peuvent être éliminées
par une transformation de jauge judicieuse, ce qui signifie que ce
sont des degrés de libertés non physiques (would-be Goldstone bosons).
En choisissant cette jauge particulière où $\phi^{+}=\chi=0$, appelée
\emph{jauge unitaire}, le champ de Higgs s'écrit simplement sous la
forme
\begin{equation}
H(x)=\frac{1}{\sqrt{2}}\left(\begin{array}{c}
0\\
v+h(x)
\end{array}\right).\label{eq:HiggsUnitaryGauge}
\end{equation}
En ce qui concerne le potentiel $V(H)$, il s'exprime dans cette jauge
unitaire en fonction de la VEV et du seul degré de liberté physique
$h(x)$ de la manière suivante
\begin{equation}
V(v,h)=\mu^{2}h^{2}+\frac{\mu^{2}}{v}h^{3}+\frac{\mu^{2}}{4v^{2}}h^{4}=\frac{m_{h}^{2}}{2}h^{2}+\frac{m_{h}^{2}}{2v}h^{3}+\frac{m_{h}^{2}}{8v^{2}}h^{4}.
\end{equation}
Le degré de liberté physique $h(x)$ est un champ réel décrivant une
particule scalaire neutre et massive, le célèbre \emph{boson de Higgs},
de masse
\begin{equation}
m_{h}=\sqrt{2}\mu,
\end{equation}
et d'auto-couplages cubiques et quartiques proportionnels à $m_{h}^{2}$.
De plus, les couplages aux champs de jauges sont obtenus à partir
du terme cinétique dans le lagrangien de Higgs (\ref{eq:LHiggs})
et contiennent les vertex trilinéaires $hWW$, $hZZ$ et quadrilinéaires
$hhWW$, $hhZZ$. 

Pour clore cette partie sur le mécanisme de Higgs, notons que ce dernier
permet aux bosons de jauges d'acquérir une masse à travers leurs couplages
au champ de Higgs contenus dans le terme cinétique. Cependant, les
fermions sont toujours de masses nulles. C'est le rôle de la dernière
composante non traitée du lagrangien $\mathcal{L}_{EW}$, formée des
interactions de Yukawa $\mathcal{L}_{Y}$, qui va permettre aux fermions
d'acquérir leurs masses.

\subsection{Masses des fermions : le secteur de Yukawa}

Afin de doter les fermions d'une masse, on introduit les interactions
de Yukawa couplant le champ de Higgs aux champs fermioniques. Pour
une génération de quarks et de leptons donnée, le lagrangien de Yukawa
invariant de jauge encapsulant ces interactions peut s'exprimer en
termes des doublets $L=(\nu_{L},e_{L})^{T}$, $Q=(u_{L},d_{L})^{T}$
et du champ de Higgs $H$ ainsi que son conjugué de charge $H^{c}=i\sigma_{2}H=(\phi^{0\ast},-\phi^{-})^{T}$,
où $\phi^{-}=(\phi^{+})^{\dagger}$,
\begin{equation}
\mathcal{L}_{Y}=-y_{e}\bar{L}He_{R}-y_{d}\bar{Q}Hd_{R}-y_{u}\bar{Q}H^{c}u_{R}+h.c.
\end{equation}
En explicitant ce lagrangien en fonction des différentes composantes
du champ de Higgs, on obtient
\begin{equation}
\begin{aligned}\mathcal{L}_{Y} & =-y_{e}(\bar{\nu}_{L}\phi^{+}e_{R}+\bar{e}_{R}\phi^{-}\nu_{L}+\bar{e}_{L}\phi^{0}e_{R}+\bar{e}_{R}\phi^{0\ast}e_{L})\\
 & =-y_{d}(\bar{u}_{L}\phi^{+}d_{R}+\bar{d}_{R}\phi^{-}u_{L}+\bar{d}_{L}\phi^{0}d_{R}+\bar{d}_{R}\phi^{0\ast}d_{L})\\
 & =-y_{u}(-\bar{u}_{R}\phi^{+}d_{L}+\bar{d}_{L}\phi^{-}u_{R}+\bar{u}_{R}\phi^{0}u_{L}+\bar{u}_{L}\phi^{0\ast}u_{R}).
\end{aligned}
\end{equation}
Par suite, en se plaçant dans la jauge unitaire (\ref{eq:HiggsUnitaryGauge}),
le lagrangien de Yukawa prend la forme plus simple
\begin{equation}
\mathcal{L}_{Y}=-\sum_{f}m_{f}\bar{\psi}_{f}\psi_{f}-\sum_{f}\frac{m_{f}}{v}\bar{\psi}_{f}\psi_{f}h,
\end{equation}
qui contient alors les termes de masses des fermions, avec des masses
$m_{f}$ dépendants de la VEV et de la constante de Yukawa associée
au fermion en question
\begin{equation}
m_{f}=y_{f}\frac{v}{\sqrt{2}}.
\end{equation}
Hormis les termes de masses, on trouve également dans $\mathcal{L}_{Y}$
les interactions entre les fermions massifs et le boson de Higgs dont
le couplage est proportionnel à la masse du fermion impliqué $m_{f}$.

Pour conclure, notons que le cas réaliste en présence de trois générations
permet à un nouveau phénomène d'émerger, le mélange des quarks. Ce
phénomène aura une répercussion sur les courants chargés de quarks
en interaction faible et permettra d'avoir de la violation-$\mathcal{CP}$
dans le MS. Nous traiterons ces aspects en détails dans la Section
\ref{Ch1Sec5} ainsi que dans le Chapitre \ref{Ch2} consacré à la
violation-$\mathcal{CP}$.

\subsection{Champs et paramètres physiques}

Le terme cinétique du lagrangien de Higgs (\ref{eq:LHiggs}) contenant
les interactions entre les champs de Higgs et de jauges permet d'avoir
des termes de masses pour les bosons vecteurs sous la forme non-diagonale
suivante
\begin{equation}
\frac{1}{2}\left(\frac{g_{2}}{2}v\right)^{2}(W_{1}^{2}+W_{2}^{2})+\frac{1}{2}\left(\frac{v}{2}\right)^{2}\left(W_{\mu}^{3},B_{\mu}\right)\left(\begin{array}{cc}
g_{2}^{2} & g_{1}g_{2}\\
g_{1}g_{2} & g_{1}^{2}
\end{array}\right)\left(\begin{array}{c}
W^{3,\mu}\\
B^{\mu}
\end{array}\right).\label{eq:BosonsMassTerms}
\end{equation}
Les champs physiques deviennent apparents en appliquant une rotation
aux champs $W_{\mu}^{a}$ et $B_{\mu}$ qui sont liés aux générateurs
de $SU(2)_{L}\times U(1)_{Y}$. On définit alors
\begin{equation}
W_{\mu}^{\pm}=\frac{1}{\sqrt{2}}(W_{\mu}^{1}\mp iW_{\mu}^{2})
\end{equation}
ainsi que
\begin{equation}
\left(\begin{array}{c}
Z_{\mu}\\
A_{\mu}
\end{array}\right)=\left(\begin{array}{cc}
\cos\theta_{W} & \sin\theta_{W}\\
-\sin\theta_{W} & \cos\theta_{W}
\end{array}\right)\left(\begin{array}{c}
W^{3,\mu}\\
B^{\mu}
\end{array}\right).
\end{equation}
Exprimé dans cette base, c'est à dire en termes des champs physiques,
le terme de masse précédent (\ref{eq:BosonsMassTerms}) prend une
forme diagonale
\begin{equation}
M_{W}^{2}W_{\mu}^{+}W^{-\mu}+\frac{1}{2}(A_{\mu},Z_{\mu})\left(\begin{array}{cc}
0 & 0\\
0 & M_{Z}^{2}
\end{array}\right)\left(\begin{array}{c}
A_{\mu}\\
Z_{\mu}
\end{array}\right),
\end{equation}
avec
\begin{equation}
M_{W}=\frac{1}{2}g_{2}v,\hspace{1em}M_{Z}=\frac{v}{2}\sqrt{g_{1}^{2}+g_{2}^{2}}.
\end{equation}
L'angle de la rotation effectuée $\theta_{W}$ est appelé \emph{angle
de Weinberg} et il est donné par
\begin{equation}
\cos\theta_{W}=\frac{g_{2}}{\sqrt{g_{1}^{2}+g_{2}^{2}}}=\frac{M_{W}}{M_{Z}}.
\end{equation}
En effectuant cette rotation sur les interactions du lagrangien des
fermions $\mathcal{L}_{F}$ et en identifiant $A_{\mu}$ au photon
se couplant à l'électron avec une constante de couplage $e$, on peut
relier la charge électrique et les constantes de couplages de jauges
de la façon suivante
\begin{equation}
e=\frac{g_{1}g_{2}}{\sqrt{g_{1}^{2}+g_{2}^{2}}},\hspace{1em}g_{2}=\frac{e}{\sin\theta_{W}},\hspace{1em}g_{1}=\frac{e}{\cos\theta_{W}}.
\end{equation}
Ainsi, ces relations nous permettent de remplacer les paramètres originaux
de la théorie $g_{1},g_{2},\lambda,\mu^{2},y_{f}$ par un jeu de paramètres
physiques équivalents $e,M_{W},M_{Z},m_{h},m_{f},V_{CKM}$\footnote{La matrice CKM traduit le mélange des quarks et sera définie plus
tard.}. L'interêt de ces nouveaux paramètres est qu'ils peuvent être mesurés
directement dans les expériences. A ce jour, tous ces paramètres ont
été mesurés expérimentalement.

Concernant les interactions des fermions avec les champs de jauges,
contenues dans le lagrangien $\mathcal{L}_{F}$, on peut également
les exprimer en fonction des champs et des paramètres physiques. Ils
se manifestent alors comme des couplages entre le courant électromagnétique
$J_{em}^{\mu}$, le courant faible neutre $J_{NC}^{\mu}$ et le courant
faible chargé $J_{CC}^{\mu}$ avec les champs vectoriels de jauges
correspondants
\begin{equation}
\mathcal{L}_{FG}=J_{em}^{\mu}A_{\mu}+J_{NC}^{\mu}Z^{\mu}+J_{CC}^{\mu}W_{\mu}^{+}+J_{CC}^{\mu\dagger}W_{\mu}^{-},
\end{equation}
où les différents courants sont donnés par
\begin{equation}
\begin{aligned}J_{em}^{\mu} & =-e\sum_{f}Q_{f}\bar{\psi}_{f}\gamma^{\mu}\psi_{f},\\
J_{NC}^{\mu} & =\frac{g_{2}}{2\cos\theta_{W}}\sum_{f}\bar{\psi}_{f}(C_{V}^{f}\gamma^{\mu}-C_{A}^{f}\gamma^{\mu}\gamma^{5})\psi_{f},\\
J_{CC}^{\mu} & =\frac{g_{2}}{\sqrt{2}}\left(\sum_{I=1,2,3}\bar{\nu}^{I}\gamma^{\mu}\frac{1-\gamma^{5}}{2}e^{I}+\sum_{I,J=1,2,3}\bar{u}^{I}\gamma^{\mu}\frac{1-\gamma^{5}}{2}V_{IJ}d^{J}\right),
\end{aligned}
\end{equation}
où $I,J$ sont des indices de saveurs, c'est à dire $(e^{1},e^{2},e^{3})=(e,\mu,\tau)$
et $(\nu^{1},\nu^{2},\nu^{3})=(\nu_{e},\nu_{\mu},\nu_{\tau})$. De
plus, les constantes de couplages intervenant dans les courants neutres
dépendent de la charge électrique du fermion $Q_{f}$ ainsi que de
son isospin $I_{3}^{f}$ comme suit
\begin{equation}
\begin{aligned}C_{v}^{f} & =I_{3}^{f}-2Q_{f}\sin^{2}\theta_{W},\\
C_{A}^{f} & =I_{3}^{f}.
\end{aligned}
\end{equation}

Pour conclure cette partie, nous allons réécrire le lagrangien de
jauge $\mathcal{L}_{G}$ en termes des champs physiques afin de révéler
la forme des auto-interactions de bosons de jauges. En effet, comme
nous l'avons vu auparavant, le caractère non-abélien du groupe de
jauge induit des auto-interactions des champs de jauges complètement
déterminées par la structure du groupe. En procédant, on obtient
\begin{equation}
\begin{aligned}\mathcal{L}_{G,self} & =e\left[W_{\mu\nu}^{+}W^{-\mu}A^{\nu}+W_{\mu}^{+}W_{\nu}^{-}F^{\mu\nu}+h.c.\right]\\
 & +e\cot\theta_{W}\left[W_{\mu\nu}^{+}W^{-\mu}Z^{\nu}+W_{\mu}^{+}W_{\nu}^{-}Z^{\mu\nu}+h.c.\right]\\
 & -\frac{e^{2}}{4\sin^{2}\theta_{W}}\left[(W_{\mu}^{-}W_{\nu}^{+}-W_{\nu}^{-}W_{\mu}^{+})W_{\mu}^{+}W_{\nu}^{-}+h.c.\right]\\
 & -\frac{e^{2}}{4}(W_{\mu}^{+}A_{\nu}-W_{\nu}^{+}A_{\mu})(W^{-\mu}A^{\nu}-W^{-\nu}A^{\mu})\\
 & -\frac{e^{2}}{4}\cot^{2}\theta_{W}(W_{\mu}^{+}Z_{\nu}-W_{\nu}^{+}Z_{\mu})(W^{-\mu}Z^{\nu}-W^{-\nu}Z^{\mu})\\
 & +\frac{e^{2}}{2}\cot\theta_{W}(W_{\mu}^{+}A_{\nu}-W_{\nu}^{+}A_{\mu})(W^{-\mu}Z^{\nu}-W^{-\nu}Z^{\mu})+h.c.
\end{aligned}
\end{equation}
Insistons sur le fait que dans le MS, ces constantes d'auto-couplages
sont uniquement déterminées par la symétrie de jauge, c'est à dire
par la structure du groupe de jauge. Il s'ensuit alors que toutes
déviations mesurées à ces constantes ne peut s'expliquer qu'en dehors
du MS, autrement dit par des processus de nouvelle physique ayant
lieu à des échelles d'énergies supérieures.

\section{Le secteur de la saveur du modèle standard \label{Ch1Sec5}}

Cette partie est largement inspirée de \cite{HDRChristopher}.

\subsection{Symétrie de saveur et paramètres libres dans le modèle standard}

Le modèle standard est une théorie de jauge basée sur le groupe $G_{SM}=SU(3)_{c}\times SU(2)_{L}\times U(1)_{Y}$,
où les groupes correspondent respectivement aux interactions forte
et électrofaible. Le lagrangien des fermions en interactions est le
suivant:
\begin{equation}
\mathcal{L}_{fermions}=\sum_{\psi,I}\bar{\psi}^{I}(i\cancel{D}_{\psi})\psi^{I},
\end{equation}
où la somme est effectuée sur tous les fermions et toutes les saveurs,
$\psi=Q,U,D,L,E$ (voir tableau \ref{TableauMS}) et $I=1,2,3$ est
l'indice de saveur (pour la I-ème génération). Les bosons de jauges
qui médient les différentes interactions sont contenus dans la dérivée
covariante:

\begin{equation}
D_{\psi}^{\mu}=\partial^{\mu}-ig_{s}T_{\psi}^{a}G_{a}^{\mu}-ig\overrightarrow{T_{\psi}}\centerdot\overrightarrow{W^{\mu}}-ig'\frac{Y_{\psi}}{2}B^{\mu},
\end{equation}
où $g_{s}$, $g$ et $g'$ désignent respectivement les constantes
de couplages des interactions forte, faible et d'hypercharge et où
$(T_{\psi}^{a},\overrightarrow{T_{\psi}},\frac{Y_{\psi}}{2})$ sont
les générateurs du groupe dans la représentation de $G_{SM}$ portée
par $\psi$. Les différents générateurs pour chaque particules et
chaque interactions sont donnés dans le tableau \ref{TableauMS}.

\begin{table}
\centering

\begin{tabular}{c|c|c|c|cccc}
\hline 
Matière $\psi$ & $\mathcal{B}$ & $\mathcal{L}$ & $G_{F}$ & $G_{SM}$: & $T_{\psi}^{a}$ & $\overrightarrow{T_{\psi}}$ & $Y_{\psi}/2$\tabularnewline
\hline 
\hline 
$Q=\left(\begin{array}{c}
u_{L}\\
d_{L}
\end{array}\right)$ & 1/3 & 0 & $(3,1,1,1,1)$ & $(3,2)_{+1/3}$ & $+\lambda^{a}/2$ & $\overrightarrow{\sigma}/2$ & +1/6\tabularnewline
$U=u_{R}^{\text{\ensuremath{\dagger}}}$ & -1/3 & 0 & $(1,3,1,1,1)$ & $(3,1)_{-4/3}$ & $-\lambda^{a}/2$ & 0 & -2/3\tabularnewline
$D=d_{R}^{\dagger}$ & -1/3 & 0 & $(1,1,3,1,1)$ & $(3,1)_{+2/3}$ & $-\lambda^{a}/2$ & 0 & +1/3\tabularnewline
\hline 
$L=\left(\begin{array}{c}
\nu_{L}\\
e_{L}
\end{array}\right)$ & 0 & 1 & $(1,1,1,3,1)$ & $(1,2)_{-1}$ & 0 & $\overrightarrow{\sigma}/2$ & -1/2\tabularnewline
$E=e_{R}^{\dagger}$ & 0 & -1 & $(1,1,1,1,3)$ & $(1,1)_{+2}$ & 0 & 0 & +1\tabularnewline
\hline 
\end{tabular}

\caption{La première colonne représente le contenu en matière du MS. $\mathcal{B}$
et $\mathcal{L}$ sont respectivement les nombres baryonique et leptonique.
On trouve les représentations des groupes de saveur $G_{F}$ et du
MS $G_{SM}$ sous lesquelles se transforment chaque $\psi$ ainsi
que les générateurs associés à $SU(3)_{C}$, $SU(2)_{L}$ et $U(1)_{Y}$.
Les $\lambda^{a}$ dénotent les matrices de Gell-Mann et $\protect\overrightarrow{\sigma}=(\sigma^{1},\sigma^{2},\sigma^{3})$
les matrices de Pauli. Notons que les fermions droits sont définis
à partir du conjugué de leurs spineurs de Weyl gauches. Par $U$,
on peut comprendre $u_{R}^{\dagger},\bar{u}_{R}$ ou $u_{R}^{C}$
qui se transforment tous sous la même représentation $G_{SM}\times G_{F}$.}

\label{TableauMS}
\end{table}

On remarque que la dérivée covariante ne dépend pas de l'indice de
saveur. Autrement dit, les interactions de jauges ne dépendent pas
de la saveur des fermions. Le lagrangien est invariant sous une permutation
des trois indices de saveur et plus généralement sous l'action d'une
transformation unitaire qui mélange les saveurs des différents champs
fermioniques. Pour chacun des cinq champs de matière $\psi=Q,U,D,L,E$
, il y a invariance (dans l'espace des saveurs) sous le groupe des
matrices unitaires $3\times3\::\:U(3)_{\psi}$:

$\forall\psi\in\{Q,U,D,L,E\},\forall g_{\psi}\in U(3),$

\begin{equation}
\psi^{I}\rightarrow\sum_{J}(g_{\psi})^{IJ}\psi^{J}:\mathcal{L}_{fermions}\rightarrow\sum_{\psi,I}\sum_{J,K}\bar{\psi}^{K}(g_{\psi}^{\dagger})^{KI}(i\cancel{D}_{\psi})(g_{\psi})^{IJ}\psi^{J}=\mathcal{L}_{fermions}.
\end{equation}
$g_{\psi}$ étant unitaire, on a bien $(g_{\psi}^{\dagger})^{IK}(g_{\psi})^{KJ}=\delta^{IJ}$.

Le secteur de jauge du MS est donc invariant sous l'action du groupe
de symétrie globale suivant, appelé \emph{groupe de saveur}:
\begin{equation}
G_{F}=U(3)^{5}=U(3)_{Q}\times U(3)_{U}\times U(3)_{D}\times U(3)_{L}\times U(3)_{E}.
\end{equation}

L'action de ce groupe sur les champs fermioniques de matière est définie
de la manière suivante : 
\begin{equation}
\begin{cases}
\psi\overset{G_{F}}{\rightarrow} & g_{\psi}\psi\hspace{1em}\text{pour les champs de chiralité gauche}\hspace{1em}\psi=Q,L\\
\psi\overset{G_{F}}{\rightarrow} & \psi g_{\psi}\hspace{1em}\text{pour les champs de chiralité droite}\hspace{1em}\psi=U,D,E
\end{cases},\text{où }g_{\psi}\in U(3)_{\psi}.
\end{equation}

Cependant, cette symétrie n'est pas exacte dans le MS. Elle est explicitement
brisée par le secteur de Yukawa (couplage fermion-fermion-Higgs).
\begin{equation}
\mathcal{L}_{Yukawa}=-U^{I}Y_{u}^{IJ}Q^{J}H^{\dagger C}-D^{I}Y_{d}^{IJ}Q^{J}H^{\dagger}-E^{I}Y_{e}^{IJ}L^{J}H^{\dagger}+h.c,\label{eq:6}
\end{equation}
où h.c dénote le conjugué hermitique de l'expression auquel il est
ajouté.

En effet, ces couplages de Yukawa brisent la symétrie de saveur $(G_{F})$
étant donné qu'ils couplent deux types de fermions différents. Sous
$G_{F}$, chaque fermion se transforme différemment (avec une matrice
unitaire différente). Notons que ce secteur est la source de la majorité
des paramètres libres du MS. Comme il y a trois générations de particules,
dans l'espace des saveurs $Y_{u,d,e}\in M_{3}(\mathbb{C})$ (matrices
$3\times3$ à coefficients complexes). Par conséquent, chacune des
matrices contient 9 paramètres complexes, soit $9\times2=18$ paramètres
réels. Au total, le secteur de Yukawa introduit $18\times3=54$ paramètres
réels. Heureusement, ils ne sont pas tous physiques et en redéfinissant
les champs de sorte à ce qu'ils absorbent certains paramètres, nous
nous retrouvons avec 13 paramètres libres pour le secteur de Yukawa.
Sachant qu'il y en a 19 dans le MS, cela représente encore environ
deux tiers des paramètres libres. 

Détaillons ceci du point de vue de la symétrie de saveur.

Tout d'abord, remarquons que $G_{F}$ n'est pas entièrement brisé
par les couplages de Yukawa. En effet, il reste des symétries résiduelles
qualifiées d'accidentelles dans le MS:
\begin{equation}
U(3)^{5}\rightarrow U(1)_{\mathcal{B}}\times U(1)_{\mathcal{L}_{e}}\times U(1)_{\mathcal{L}_{\mu}}\times U(1)_{\mathcal{L}_{\tau}},
\end{equation}
où $\mathcal{B}$ désigne le nombre baryonique et $\mathcal{L}_{e,\mu,\tau}$
les nombres leptoniques de saveurs. 

Comptons le nombre de paramètres physiques. Pour cela, il faut déterminer
le nombre de générateurs brisés et ensuite on peut absorber autant
de paramètres que de générateurs brisés en redéfinissant les champs.
Ainsi, nous avons donc : 

\begin{equation}
\text{\#paramètres physiques}=\text{\#paramètres total}-\text{\#générateurs brisés}.\label{eq:ComptPara}
\end{equation}
Ici, le nombre de paramètres total est 54 (secteur de Yukawa). Par
ailleurs, une transformation sous $G_{F}=U(3)^{5}$ implique 5 matrices
unitaires $3\times3$, chacune d'entre elles nécessitant 9 paramètres
réels pour la décrire\footnote{En effet, une matrice U de U(n) possède $n^{2}$ paramètres complexes,
soit $2n^{2}$ paramètres réels. En traduisant la condition d'unitarité
$U^{\dagger}U=\mathbb{I}$, on obtient $n$ contraintes réelles correspondant
aux coefficients diagonaux et $\frac{n(n-1)}{2}$ contraintes complexes
correspondant à la partie triangulaire supérieure (ou inférieure)
stricte, soit $2\times\frac{n(n-1)}{2}=n(n-1)$ contraintes réelles.
En tout, on a $n+n(n-1)=n^{2}$contraintes réelles. Par conséquent,
le nombre de coefficients indépendants nécessaires pour décrire une
matrice de U(n) est $2n^{2}-n^{2}=n^{2}$coefficients réels.}, soit un total de $5\times9=45$ paramètres réels. Or, tous ne sont
pas brisés et il reste quatre paramètres réels (générateurs) correspondant
aux quatre U(1)s résiduels. Le nombre de générateurs brisés est donc
$45-4=41$. Par conséquent, $\text{\#paramètres physiques}=54-41=13.$

Bien entendu, cette technique de comptage basée sur la symétrie de
saveur ne nous renseigne pas sur la nature de ces paramètres. Pour
les identifier, penchons nous de plus près sur les couplages de Yukawa.

Après brisure de symétrie électrofaible (EWSSB\footnote{Electroweak Spontaneous Symmetry Breaking.}),
le champ de Higgs acquiert une valeur moyenne dans le vide non nulle
(VEV\footnote{Vacuum Expectation Value.}), $<0\mid H\mid0>=v$, et
on l'exprime alors dans la jauge unitaire comme $H=\left(\begin{array}{cc}
0 & v+h\end{array}\right)^{T}$, où $h$ est le boson de Higgs. Les termes de Yukawa s'identifient
alors à des termes de masses pour les fermions:
\begin{equation}
\mathcal{L}_{Yukawa}=-v(\bar{u}_{R}^{I}Y_{u}^{IJ}u_{L}^{J}+\bar{d}_{R}^{I}Y_{d}^{IJ}d_{L}^{J}+\bar{e}_{R}^{I}Y_{e}^{IJ}e_{L}^{J})(1+\frac{h}{v})+h.c.\label{eq:Yukawa}
\end{equation}
Les matrices de couplages $(Y_{u,d,e})$ ne sont à priori pas diagonales.
Afin d'obtenir les états propres de masses des fermions, on réalise
une transformation bi-unitaire\footnote{Cette procédure appelée \emph{décomposition en valeurs singulières},
est toujours possible. $\forall M\in M_{n}(\mathbb{C}),\hspace{1em}\exists U,V\in U(n)\hspace{1em}\text{telles que}\hspace{1em}VMU^{\dagger}=D\hspace{1em}\text{avec D diagonale}$.
Les coefficients diagonaux sont des réels positifs appelés valeurs
singulières.}:
\begin{equation}
\exists V_{R}^{u,d,e},V_{L}^{u,d,e}\in U(3)\hspace{1em}\text{telles que}\hspace{1em}vV_{R}^{u,d,e}Y_{u,d,e}V_{L}^{u,d,e}=\mathbb{M}_{u,d,e}\label{eq:SVD1}
\end{equation}
On réinjecte dans le lagrangien de Yukawa:
\begin{equation}
\mathcal{L}_{Yukawa}=-(\bar{u}_{R}(V_{R}^{u})^{\dagger}\mathbb{M}_{u}(V_{L}^{u})^{\dagger}u_{L}+\bar{d}_{R}(V_{R}^{d})^{\dagger}\mathbb{M}_{d}(V_{L}^{d})^{\dagger}d_{L}+\bar{e}_{R}(V_{R}^{e})^{\dagger}\mathbb{M}_{e}(V_{L}^{e})^{\dagger}e_{L})(1+\frac{h}{v})+h.c.
\end{equation}
On peut alors définir les états propres de masses à partir des états
propres d'interactions. Le passage de l'un à l'autre se fait par l'application
d'une transformation unitaire, autrement dit, en utilisant la symétrie
de saveur. Cependant, la symétrie de jauge nous impose d'appliquer
la même transformation unitaire aux deux composantes du doublet de
quarks $Q^{I}=\left(\begin{array}{c}
u_{L}^{I}\\
d_{L}^{I}
\end{array}\right)$. On choisit d'appliquer $(V_{L}^{u})^{\dagger}$ au doublet. En redéfinissant
les champs de la manière suivante:
\begin{equation}
\begin{cases}
(V_{L}^{u})^{\dagger}u_{L}= & u_{L}^{mass}\\
(V_{L}^{u})^{\dagger}d_{L}= & d_{L}^{\prime}
\end{cases},\begin{cases}
\bar{u}_{R}(V_{R}^{u})^{\dagger}= & \bar{u}_{R}^{mass}\\
\bar{d}_{R}(V_{R}^{d})^{\dagger}= & \bar{d}_{R}^{mass}
\end{cases},\begin{cases}
(V_{L}^{e})^{\dagger}e_{L}= & e_{L}^{mass}\\
\bar{e}_{R}(V_{R}^{e})^{\dagger}= & \bar{e}_{R}^{mass}
\end{cases},
\end{equation}
On obtient:
\begin{equation}
\mathcal{L}_{Yukawa}=-(\bar{u}_{R}^{mass}\mathbb{M}_{u}u_{L}^{mass}+\bar{d}_{R}^{mass}\mathbb{M}_{d}V_{CKM}^{\dagger}d_{L}^{\prime}+\bar{e}_{R}^{mass}\mathbb{M}_{e}e_{L}^{mass})(1+\frac{h}{v})+h.c,
\end{equation}
où la rotation encore nécessaire pour atteindre les états propres
de masses des quarks gauches de type down $(d_{L},s_{L}\text{ et }b_{L})$
est la matrice de \noun{cabibbo-Kobayashi-Maskawa (CKM)} \cite{CKM}:
\begin{equation}
V_{L}^{u\dagger}V_{L}^{d}\equiv V_{CKM}=\left(\begin{array}{ccc}
V_{ud} & V_{us} & V_{ub}\\
V_{cd} & V_{cs} & V_{cb}\\
V_{td} & V_{ts} & V_{tb}
\end{array}\right).
\end{equation}
La matrice CKM étant unitaire, il faut $3^{2}=9$ paramètres réels
pour la décrire qu'on peut choisir comme étant 3 nombres réels et
6 phases\footnote{En effet, une matrice de $U(n)$ peut être décrites par $\frac{n(n-1)}{2}$
nombres réels et $\frac{n(n+1)}{2}$ phases.}. Or, nous avons vu qu'il y avait 13 paramètres libres dans ce secteur.
Parmi ces 13 paramètres, 9 correspondent aux masses des fermions (pas
de masses pour les neutrinos dans le MS) et donc il en reste 4. Par
conséquent, parmi les 9 paramètres, seuls 4 sont physiques. En fait,
en redéfinissant les champs de quarks, on peut absorber 5 phases dans
ces derniers. Plusieurs paramétrisations sont possibles, par exemple
on peut paramétrer la matrice CKM par 3 angles de mélanges ($\theta_{12},\theta_{13}$
et $\theta_{23}$ qui sont des nombres réels) et une phase $(\delta)$:

Si on note $c_{ij}=\cos\theta_{ij}$ et $s_{ij}=\sin\theta_{ij}$,
on peut exprimer la matrice CKM sous la forme:
\begin{equation}
V_{CKM}=\left(\begin{array}{ccc}
c_{12}c_{13} & s_{12}c_{13} & s_{13}e^{-i\delta}\\
-s_{12}c_{23}-c_{12}s_{23}s_{13}e^{i\delta} & c_{12}c_{23}-s_{12}s_{23}s_{13}e^{i\delta} & s_{23}c_{13}\\
s_{12}s_{23}-c_{12}c_{23}s_{13}e^{i\delta} & -c_{12}s_{23}-s_{12}c_{23}s_{13}e^{i\delta} & c_{23}c_{13}
\end{array}\right).
\end{equation}
Les valeurs actuelles mesurées des coefficients en modules (à partir
de processus à courants chargés) sont \cite{CKMFitter}:

\begin{equation}
\mid V_{CKM}\mid=\left(\begin{array}{ccc}
0.97427 & 0.22534 & 0.00351\\
0.2252 & 0.97344 & 0.0412\\
0.00867 & 0.0404 & 0.999146
\end{array}\right)
\end{equation}
 Notons que la phase $\delta$ est l'unique source de violation-$\mathcal{CP}$
faible dans le MS. 

Expérimentalement, on a $s_{13}\ll s_{23}\ll s_{12}\ll1$, si on définit
le paramètre $\lambda=s_{12}=\frac{\mid V_{us}\mid}{\sqrt{\mid V_{ud}\mid^{2}+\mid V_{us}\mid^{2}}}\ll1$
et qu'on effectue un développement limité en $\lambda$, on aboutit
à la paramétrisation dite de \noun{Wolfenstein }\cite{Wolfenstein}\noun{:
\begin{equation}
V_{CKM}=\left(\begin{array}{ccc}
1-\tfrac{\lambda^{2}}{2} & \lambda & A\lambda^{3}(\rho-i\eta)\\
-\lambda & 1-\tfrac{\lambda^{2}}{2} & A\lambda^{2}\\
A\lambda^{3}(1-\rho-i\eta) & -A\lambda^{2} & 1
\end{array}\right)+\mathcal{O}(\lambda^{4}),
\end{equation}
}où les valeurs actuelles mesurées sont \cite{CKMFitter} $\lambda=0.22548_{-0.00034}^{+0.00068}$,
$A=0.810_{-0.024}^{+0.018}$, $\bar{\rho}\equiv\rho(1-\tfrac{\lambda^{2}}{2})=0.145_{-0.007}^{+0.013}$
et $\bar{\eta}\equiv\eta(1-\tfrac{\lambda^{2}}{2})=0.343_{-0.012}^{+0.011}$.\noun{
}Remarquons que l'unique source de violation-$\mathcal{CP}$ faible,
à savoir, la partie imaginaire $i\eta$ n'intervient qu'à l'ordre
3 en $\lambda$ (très supprimée).

\subsection{Processus avec changements de saveur à courants chargés et à courants
neutres}

Les états propres de masses ne sont pas alignés avec les états propres
d'interactions. Comme nous l'avons vu, pour atteindre les états propres
de masses il faut briser la symétrie de jauge $SU(2)_{L}$ en tournant
les quarks downs $d_{L}\to V_{CKM}d_{L}$ tout en laissant $u_{L}$
fixe. Ceci affecte directement les courants chargés de quarks:
\begin{equation}
\mathcal{L}_{CC}=\frac{g}{\sqrt{2}}\sum_{I}W_{\mu}^{+}(\bar{\nu}_{L}^{I}\gamma^{\mu}e_{L}^{I}+\bar{u}_{L}^{I}\gamma^{\mu}V_{CKM}^{IJ}d_{L}^{J})+h.c.
\end{equation}
Les courants leptoniques sont diagonaux dans l'espace des saveurs
et c'est pourquoi les nombres leptoniques de saveurs sont conservés
(symétries accidentelles). Les processus violant la saveur dans le
secteur leptonique (LFV) sont strictement interdits dans le MS. Par
ailleurs, les courants neutres sont également diagonaux. Par exemple,
pour le couplage des quarks de type down avec le boson Z et le photon,
on a:
\begin{equation}
\mathcal{L}_{NC}\ni\frac{g}{2\cos\theta_{W}}Z_{\mu}\bar{d}_{L}^{I}\gamma^{\mu}\left(2T_{3}-2\sin^{2}\theta_{W}Q\right)d_{L}^{I}+eA_{\mu}\bar{d}_{L}^{I}\gamma^{\mu}Qd_{L}^{I},
\end{equation}
pour les états propres d'interactions. Le passage aux états propres
de masses $d_{L}\to V_{CKM}d_{L}$ laisse le lagrangien invariant
du fait de l'unitarité de la matrice CKM.

\begin{figure}
\includegraphics[width=1\textwidth]{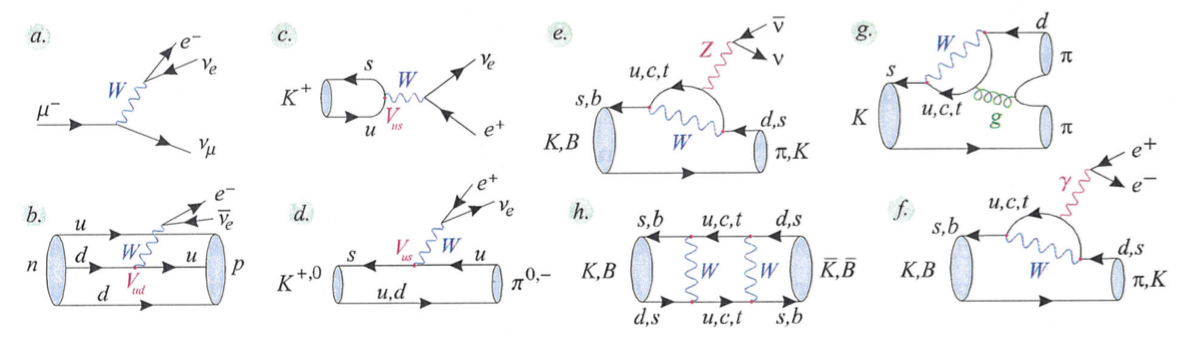}

\caption{(a - d) Processus à courants chargés. (e - h) FCNC. (e, f, g) sont
respectivement les pingouins avec Z, photon et gluon. (h) représente
un diagramme en boîte (W box). Source: \cite{HDRChristopher}.}
\label{FCNCs}
\end{figure}

A l'arbre, les courants chargés sont notre seul accès à $V_{CKM}$
et par conséquent à la violation-$\mathcal{CP}$. Expérimentalement,
les courants chargés sont sondés à basse énergie à partir de processus
avec changement de saveur (voir figure \ref{FCNCs}). Aux ordres supérieurs,
les changements de saveurs par courant neutre (FCNC\footnote{Flavor Changing Neutral Current.})
sont possibles mais défavorisés et ils permettent de tester le MS.

Considérons les diagrammes de la figure \ref{FCNCs}, appelés \emph{pingouins}
et \emph{boîtes.}

\paragraph*{Pingouin avec boson Z (diagramme e)}

Par exemple, si on prend la transition $s\to dZ$ et on somme sur
tous les quarks up intermédiaires possibles $(u,c,t)$, l'amplitude
du processus est donnée par:
\begin{equation}
\mathcal{M}=V_{ud}V_{us}^{\ast}f(m_{u}/M_{W})+V_{cd}V_{cs}^{\ast}f(m_{c}/M_{W})+V_{td}V_{ts}^{\ast}f(m_{t}/M_{W}),
\end{equation}
où la fonction d'\noun{Inami-Lim }$f(m_{q}/M_{W})$ contient la dépendance
de l'intégrale de boucle en la masse de la particule virtuelle \cite{Inami}.
Si les masses des quarks étaient égales, alors $f(m_{q}/M_{W})$ serait
constante et on peut donc la factoriser dans l'expression de l'amplitude
et on obtiendrait $\mathcal{M}=0$ du fait de l'unitarité de la matrice
CKM $V_{ud}V_{us}^{\ast}+V_{cd}V_{cs}^{\ast}+V_{td}V_{ts}^{\ast}=0$.
C'est ce que l'on appelle le \emph{mécanisme GIM }\noun{(Glashow-Iliopoulos-Maiani)}
\cite{GIM}. Le fait que les FCNCs soient possibles est une conséquence
directe des différences de masses des quarks. 

On peut montrer que l'amplitude peut se réécrire comme:

\begin{equation}
\mathcal{M}\sim G_{F}\times\frac{e}{4\pi^{2}\sin\theta_{W}}\times\bar{s}_{L}\gamma_{\mu}d_{L}\times Z^{\mu}\times\sum_{q}m_{q}^{2}V_{qs}^{\ast}V_{qd}F_{Z}(m_{q}^{2}/M_{W}^{2}).\label{eq:ZbosonP}
\end{equation}
On parle dans ce cas de violation quadratique (du fait du $m_{q}^{2}$)
du mécanisme GIM.

\paragraph*{Pingouin avec photon (diagramme f)}

L'amplitude du processus est donnée par:
\begin{equation}
\mathcal{M}\sim G_{F}\times\frac{e}{4\pi^{2}}\times m_{s}\times\bar{s}_{R}\sigma_{\mu\nu}d_{L}\times F^{\mu\nu}\times\sum_{q}V_{qs}^{\ast}V_{qd}F_{Z}^{\prime}(m_{q}^{2}/M_{W}^{2}).\label{eq:PhotonP}
\end{equation}
Notons qu'un basculement de chiralité (chirality flip) est nécessaire
car seule la partie gauche du quark est sensible à l'interaction faible.

\paragraph*{Diagrammes en boîtes }

Ce cas concerne les oscillations dans les systèmes de mésons $B-\bar{B}$
ou $K-\bar{K}$ (figure \ref{FCNCs} (h)). Les interactions effectives
correspondantes présentent une violation quadratique de GIM. En ne
retenant que la contribution du quark top (majoritaire), on obtient:
\begin{equation}
\mathcal{M}=\frac{G_{F}^{2}m_{t}^{2}}{4\pi^{2}}\times(V_{ts}V_{td}^{\ast})^{2}\times B_{WW}(m_{t}^{2}/M_{W}^{2})\times(\bar{s}_{L}\gamma_{\mu}d_{L})(\bar{s}_{L}\gamma_{\mu}d_{L}).\label{eq:Box}
\end{equation}

\section{Succès expérimentaux et problèmes}

Le MS est à ce jour la théorie la plus précise dont on dispose pour
décrire les particules élémentaires et leurs interactions. Le MS possède
19 paramètres libres dont la valeur n'est pas fixée par des principes
premiers mais doit être déterminée expérimentalement. Dès lors que
ces paramètres libres sont mesurés, le MS permet de faire toutes sortes
de prédictions sur des processus et de les confronter aux expériences.
Aucune déviation majeure n'a été observée et confirmée jusqu'à présent.
Il est un véritable succès d'un point de vue expérimental, comme l'atteste
la découverte du boson de Higgs en 2012 par la collaboration ATLAS
puis par CMS au LHC \cite{HiggsAtlas,HiggsCMS} pour ne citer que
lui. Cependant, plusieurs éléments nous amènent à penser que ce n'est
pas la fin de l'histoire. En effet, un certain nombre de points ne
sont pas expliqués par le MS et donc nous pensons que le MS est incomplet,
il doit s'inscrire dans une théorie plus large dont il est la manifestation
à basse énergie. Ces points sont: \newline
\begin{itemize}[label=\textbullet]
\item \textbf{Les masses des neutrinos:} En effet, dans le MS, les neutrinos
n'ont pas de masses. Or, on sait depuis la découverte des oscillations
de neutrinos \cite{neutrino1,neutrino2,neutrino3} que ces derniers
sont massifs. Ce point est le sujet traité dans le Chapitre \ref{Ch2}.
\item \textbf{La gravité:} Le MS n'inclut pas l'interaction gravitationnelle.
A l'échelle de Planck $M_{Planck}=10^{19}GeV$, échelle où les effets
de gravité quantique ne peuvent plus être négligés, nous savons que
le MS n'est plus valable. Au-delà, il faut de la nouvelle physique
et notamment une théorie consistante de gravité quantique \cite{QuantumG}.
\item \textbf{La matière noire et l'énergie noire:} L'existence de la matière
noire est aujourd'hui établie \cite{DM}. Le MS ne décrit pas la matière
noire dont serait composé plus de 27\% de l'univers \cite{DMPropPlanck}.
Il ne rend pas compte non plus de l'accélération de l'expansion de
l'univers.
\item \textbf{Baryogénèse: }Le MS n'explique pas l'asymétrie matière/antimatière
de notre univers. Il semble dominé par la matière \cite{BAU} alors
que matière et antimatière ont été créées en quantités égales dans
le plasma primordial. De plus, on sait que pour pouvoir expliquer
cette asymétrie, il faut que les conditions dites de Sakharov soient
remplies \cite{Sakharov}. Or, deux de ces conditions sont les violations
de la symétrie $\mathcal{CP}$ et du nombre baryonique $\mathcal{B}$,
et il n'y a clairement pas assez de ces violations dans le MS pour
résoudre ce problème (voir Chapitre \ref{Ch2}). \newline
\end{itemize}
De plus, mis à part ces pièces manquantes, il existe d'autres problèmes
dans le MS du côté de ses paramètres libres. Tout d'abord, l'argument
\og esthétique \fg{} consiste à dire que le nombre de paramètres
libres (19) est trop important. De plus, les valeurs de ces paramètres
ainsi que certains faits appellent à une compréhension plus profonde.
Citons quelques uns de ces faits: \newline
\begin{itemize}[label=\textbullet]
\item \textbf{L'unification des constantes de couplages:} Les constantes
de couplages (à priori indépendantes) dépendent de l'énergie et elles
se rejoignent autour de $10^{15}GeV$ \cite{GUT1,GUT2}. De ce fait,
il est tentant de penser que les trois groupes de jauges du MS proviennent
d'une brisure spontanée de symétrie d'un groupe de jauge plus large
les englobant.
\item \textbf{Le problème de hiérarchie: }Pourquoi le boson de Higgs est
beaucoup plus léger que la masse de Planck alors qu'on s'attendrait
à ce que les grandes corrections quantiques fassent que $M_{Higgs}^{2}$
devienne énorme, comparable à l'échelle à laquelle une nouvelle physique
apparaît, à moins qu'il n'y ait eu une incroyable annulation par ajustement
fin (fine-tuning) entre les corrections quadratiques radiatives et
la masse nue.
\item \textbf{La réplication de saveur:} Le nombre de générations dans le
MS n'est dicté par aucun principe sous-jacent. Pourquoi la nature
a-t-elle bégayé ? Il y a 3 familles identiques à l'exception des masses
qui diffèrent. Du fait des masses différentes, ce secteur est responsable
de la majorité des paramètres libres du MS. 
\item \textbf{Le problème de la violation-$\mathcal{CP}$ forte: }Les mesures
effectuées sur les EDM des neutrons \cite{Pospelov,Cheng} contraignent
le terme effectif $\Theta$ de QCD à être proche de zéro (voir Chapitre
\ref{Ch2}). Théoriquement, ce paramètre peut prendre toutes les valeurs
dans l'intervalle $[0,2\pi]$ et il n'existe pas de raison pour qu'il
soit faible ou nul. 
\item \textbf{Les symétries accidentelles: }Dans le MS, les nombres leptoniques
et baryoniques sont conservés. Or, historiquement, les symétries globales
se sont avérées être très peu fiables et rien n'indique que ces symétries
sont absolues dans la Nature. \newline
\end{itemize}
Tous ces points alimentent notre motivation pour rechercher de la
physique au-delà du MS. Dans ce manuscrit, nous nous intéressons particulièrement
au secteur de la saveur du MS (dans la Section \ref{Ch1Sec5} ainsi
qu'aux Chapitres \ref{Ch2} et \ref{Ch3}) qui soulève à lui seul
un grand nombre de questionnements. Par exemple, pourquoi y a-t-il
3 familles ? Pourquoi les masses des fermions présentent une aussi
forte hiérarchie et pourquoi les transitions inter-familles sont-elles
si défavorisées ? Pourquoi l'interaction faible viole-t-elle la symétrie-$\mathcal{CP}$
et pourquoi est-elle gérée par un seul paramètre ? Pourquoi les saveurs
des leptons chargés sont-elles si bien conservées ? Pourquoi les neutrinos
ont-ils une masse quasi-nulle ? Sont-ils des particules de Dirac ou
de Majorana ?

\chapter{Polymorphie de la violation-$\mathcal{CP}$\label{Ch2}}

\minitoc

Parmi les pièces manquantes du MS, on y trouve la baryogénèse électrofaible,
ce mécanisme qui serait à l’origine de l’asymétrie matière-antimatière
dans l’univers. D’après le modèle du big bang, matière et antimatière
ont été créés en quantités égales, alors qu’aujourd’hui, il suffit
de regarder autour de soi pour se rendre compte que la matière semble
avoir pris le pas. Le MS n’inclut pas un tel mécanisme. En 1967, un
physicien russe du nom d’Andrei Sakharov a établi trois conditions
nécessaires pour que la baryogénèse puisse avoir lieu \cite{Sakharov}.
Une de ces conditions stipule qu’il aurait existé des interactions
violant les symétries $\mathcal{C}$ et $\mathcal{CP}$. La violation-$\mathcal{CP}$
existe déjà dans le MS. En effet, l’interaction faible peut violer
$\mathcal{CP}$ (déjà observé en 1964 dans les systèmes de Kaons par
exemple \cite{CroninFitch1964}) et certains arguments théoriques
forts nous pousse à croire que l’interaction forte devrait pouvoir
également violer $\mathcal{CP}$, mais aucun processus de la sorte
n’a encore été observé. En somme, il s’avère que la violation-$\mathcal{CP}$
présente dans le MS n’est pas suffisante pour expliquer l’asymétrie
matière-antimatière. Par conséquent, un des défis de la recherche
de physique au-delà du modèle standard est de trouver des sources
de violation-$\mathcal{CP}$ supplémentaires afin d’atteindre la quantité
requise pour la baryogénèse.

\section{Sources de violation-$\mathcal{CP}$ dans le MS}

\subsection{Secteur faible : les courants chargés de quarks}

Généralement en théorie des champs, la violation-$\mathcal{CP}$ émerge
par la présence d'interactions dans un lagrangien ayant des constantes
de couplages complexes dont les phases ne peuvent être éliminées avec
une redéfinition des champs. De façon schématique, on a
\begin{equation}
\mathcal{L}=\sum_{i}a_{i}\mathcal{O}_{i}+h.c,
\end{equation}
avec $(\mathcal{CP})\mathcal{O}_{i}(\mathcal{CP})^{\dagger}=\mathcal{O}_{i}^{\dagger}$. 

Dans le MS, la violation-$\mathcal{CP}$ se manifeste à travers les
couplages de Yukawa complexes. En effet, le lagrangien de Yukawa est
\begin{equation}
-\mathcal{L}_{Yukawa}=Y_{ij}\overline{\psi_{L,i}}H\psi_{R,j}+Y_{ij}^{\ast}\overline{\psi_{R,j}}H^{\dagger}\psi_{L,i},
\end{equation}
où $i$ et $j$ sont les indices de saveurs, $\psi$ les champs fermioniques,
$H$ le champ de Higgs et $Y_{ij}$ une matrice $3\times3$ de constantes
de couplages.

L'action de la symétrie $\mathcal{CP}$ sur les formes bilinéaires
covariantes présentes dans le lagrangien de Yukawa est $\mathcal{CP}(\overline{\psi_{L,i}}H\psi_{R,j})=\overline{\psi_{R,j}}H^{\dagger}\psi_{L,i}$
et donc
\begin{equation}
\begin{aligned}\mathcal{CP}(-\mathcal{L}_{Yukawa}) & =\mathcal{CP}(Y_{ij}\overline{\psi_{L,i}}H\psi_{R,j})+\mathcal{CP}(Y_{ij}^{\ast}\overline{\psi_{R,j}}H^{\dagger}\psi_{L,i})\\
 & =Y_{ij}\overline{\psi_{R,j}}H^{\dagger}\psi_{L,i}+Y_{ij}^{\ast}\overline{\psi_{L,i}}H\psi_{R,j}.
\end{aligned}
\end{equation}
Par conséquent, le lagrangien de Yukawa est invariant sous $\mathcal{CP}$
si et seulement si $Y_{ij}=Y_{ij}^{\ast}$, c'est à dire si $Y_{ij}$
est réelle.

De même, si on se penche sur les courants chargés dans la base des
états propres de masses des quarks,
\begin{equation}
\mathcal{L}_{CC}=\frac{g}{\sqrt{2}}\overline{u_{L,i}}V_{ij}\gamma^{\mu}W_{\mu}^{-}d_{L,i}+\frac{g}{\sqrt{2}}\overline{d_{L,i}}V_{ij}^{\ast}\gamma^{\mu}W_{\mu}^{+}u_{L,i}
\end{equation}
ainsi que l'image sous $\mathcal{CP}$ de ce lagrangien
\begin{equation}
\mathcal{CP}(\mathcal{L}_{CC})=\frac{g}{\sqrt{2}}\overline{d_{L,i}}V_{ij}\gamma^{\mu}W_{\mu}^{+}u_{L,i}+\frac{g}{\sqrt{2}}\overline{u_{L,i}}V_{ij}^{\ast}\gamma^{\mu}W_{\mu}^{-}d_{L,i},
\end{equation}
on s'aperçoit que si le lagrangien des courants chargés de quarks
est invariant sous $\mathcal{CP}$ alors on a nécessairement $V_{ij}=V_{ij}^{\ast}$,
c'est à dire une matrice CKM réelle. La proposition contraposée nous
apprend alors que si la matrice CKM n'est pas réelle, c'est à dire
complexe, alors le lagrangien des courants chargés n'est pas invariant
sous $\mathcal{CP}$. D'où le fait que la nature complexe de la matrice
CKM, à travers sa phase irréductible bien connue, soit l'origine de
la violation-$\mathcal{CP}$ dans le MS.

Pour pouvoir avoir de la violation-$\mathcal{CP}$, c'est à dire une
matrice CKM complexe avec au moins une phase irréductible, Kobayashi
et Maskawa ont montré qu'il fallait au moins trois familles de particules.
Historiquement, Cronin et Fitch avaient déjà observé en 1964 de la
violation-$\mathcal{CP}$ dans les systèmes de Kaons \cite{CroninFitch1964}.
Suite à ce résultat, Kobayashi et Maskawa ont alors suggéré en 1973
la possibilité qu'il y ait une troisième famille de particules (alors
que la deuxième famille n'était pas encore établie!), ce qui expliquerait
dans le cadre du MS la violation-$\mathcal{CP}$ observée par Cronin
et Fitch 9 ans auparavant. Un an plus tard, la deuxième famille a
été complétée puis en 1977 et 1994 les quarks bottom et top ont été
respectivement découverts, ce qui a complété la troisième génération
de particules et donné raison à Kobayashi et Maskawa. Cette prédiction
faite en 1973 leur a valu le prix Nobel de physique en 2008.

\begin{figure}[h]
\begin{center}
\includegraphics[scale=0.7]{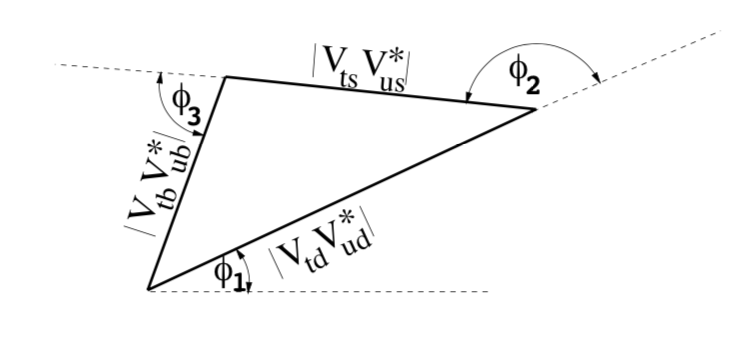}
\caption{Le triangle d'unitarité $V_{td}V_{ud}^{\ast}+V_{ts}V_{us}^{\ast}+V_{tb}V_{ub}^{\ast}=0$ avec $V_{td}V_{ud}^{\ast}=\left|V_{td}V_{ud}^{\ast}\right|e^{i\phi_{1}},\hspace{1em}V_{ts}V_{us}^{\ast}=\left|V_{ts}V_{us}^{\ast}\right|e^{i\phi_{2}},\hspace{1em}V_{tb}V_{ub}^{\ast}=\left|V_{tb}V_{ub}^{\ast}\right|e^{i\phi_{3}}$}
\label{UnitarityTriangle}
\end{center}
\end{figure}

Avec trois générations dans le MS, la matrice CKM est donnée par
\begin{equation}
V_{CKM}=\left(\begin{array}{ccc}
V_{ud} & V_{us} & V_{ub}\\
V_{cd} & V_{cs} & V_{cb}\\
V_{td} & V_{ts} & V_{tb}
\end{array}\right).
\end{equation}
Cette matrice est unitaire donc on a $V_{CKM}^{\dagger}V_{CKM}=V_{CKM}V_{CKM}^{\dagger}=\mathbb{I}_{3}$.
En traduisant $V_{CKM}V_{CKM}^{\dagger}=\mathbb{I}_{3}$ sur les coefficients
de la matrice, on aboutit aux relations d'unitarités (sur la diagonale)
\begin{equation}
\begin{aligned}V_{ud}V_{ud}^{\ast}+V_{us}V_{us}^{\ast}+V_{ub}V_{ub}^{\ast} & =1\\
V_{cd}V_{cd}^{\ast}+V_{cs}V_{cs}^{\ast}+V_{cb}V_{cb}^{\ast} & =1\\
V_{td}V_{td}^{\ast}+V_{ts}V_{ts}^{\ast}+V_{tb}V_{tb}^{\ast} & =1
\end{aligned}
\end{equation}
ainsi qu'aux conditions d'orthogonalités (hors diagonale)
\begin{equation}
\begin{aligned}V_{ud}V_{cd}^{\ast}+V_{us}V_{cs}^{\ast}+V_{ub}V_{cb}^{\ast} & =0\\
V_{ud}V_{td}^{\ast}+V_{us}V_{ts}^{\ast}+V_{ub}V_{tb}^{\ast} & =0\\
V_{cd}V_{ud}^{\ast}+V_{cs}V_{us}^{\ast}+V_{cb}V_{ub}^{\ast} & =0\\
V_{cd}V_{td}^{\ast}+V_{cs}V_{ts}^{\ast}+V_{cb}V_{tb}^{\ast} & =0\\
V_{td}V_{ud}^{\ast}+V_{ts}V_{us}^{\ast}+V_{tb}V_{ub}^{\ast} & =0\\
V_{td}V_{cd}^{\ast}+V_{ts}V_{cs}^{\ast}+V_{tb}V_{cb}^{\ast} & =0.
\end{aligned}
\end{equation}
Seules trois relations d'orthogonalités sont indépendantes dans la
mesure où il y a trois identités ainsi que leurs identités conjugués.
De plus, on peut obtenir d'autres relations d'orthogonalités dont
trois sont indépendantes à l'aide de la condition d'unitarité inverse
$V_{CKM}^{\dagger}V_{CKM}=\mathbb{I}_{3}$
\begin{equation}
\begin{aligned}V_{ud}^{\ast}V_{us}+V_{cd}^{\ast}V_{cs}+V_{td}^{\ast}V_{ts} & =0\\
V_{ud}^{\ast}V_{ub}+V_{cd}^{\ast}V_{cb}+V_{td}^{\ast}V_{tb} & =0\\
V_{us}^{\ast}V_{ud}+V_{cs}^{\ast}V_{cd}+V_{ts}^{\ast}V_{td} & =0\\
V_{us}^{\ast}V_{ub}+V_{cs}^{\ast}V_{cb}+V_{ts}^{\ast}V_{tb} & =0\\
V_{ub}^{\ast}V_{ud}+V_{cb}^{\ast}V_{cd}+V_{tb}^{\ast}V_{td} & =0\\
V_{ub}^{\ast}V_{us}+V_{cb}^{\ast}V_{cs}+V_{tb}^{\ast}V_{ts} & =0.
\end{aligned}
\end{equation}
Ces relations d'orthogonalités sont des sommes nulles de nombres complexes.
Comme chaque nombre complexe représente un vecteur, une somme nulle
de trois vecteurs signifie que les vecteurs forment un triangle (relation
de Chasles). Il est d'usage de représenter ces relations par des triangles
dans le plan complexe appelés \emph{triangle d'unitarité}. A titre
d'illustration, un des six triangles d'unitarités indépendants correspondant
à la relation $V_{td}V_{ud}^{\ast}+V_{ts}V_{us}^{\ast}+V_{tb}V_{ub}^{\ast}=0$
est représenté en figure \ref{UnitarityTriangle}.

Du fait de l'unitarité, tous les triangles ont la même aire et celle-ci
donne une mesure invariante de la quantité de violation-$\mathcal{CP}$.
L'aire de tous ces triangles est proportionnelle à $Im(V_{ud}V_{td}^{\ast}V_{tb}V_{ub}^{\ast})$.
Notons qu'une dégénérescence des masses de deux quarks de types up
ou deux quarks de types down impliquerait une annulation de la violation-$\mathcal{CP}$
dans la mesure où la phase complexe peut être éliminée par redéfinition
des phases des champs de quarks. Il a été montré \cite{Jarlskog85}
que l'invariant de Jarlskog est une bonne mesure de la violation-$\mathcal{CP}$
\begin{equation}
J=Det\left(\left[\mathbb{M}_{u},\mathbb{M}_{d}\right]\right),
\end{equation}
où $\mathbb{M}_{u}$ et $\mathbb{M}_{d}$ sont les matrices de masses
respectivement des quarks de types up et de types down. On a alors:
\begin{equation}
J=2iIm(V_{ud}V_{td}^{\ast}V_{tb}V_{ub}^{\ast})\times(m_{u}-m_{c})(m_{u}-m_{t})(m_{c}-m_{t})(m_{d}-m_{s})(m_{d}-m_{b})(m_{s}-m_{b}).
\end{equation}

En présence de trois générations, la matrice CKM peut être paramétrée
de plusieurs façons différentes par trois angles et une phase complexe.
Une paramétrisation possible adoptée dans le PDG \cite{PDG} consiste
à écrire la matrice CKM sous la forme d'un produit de trois rotations
et une matrice de phase. En définissant, 
\begin{equation}
\begin{aligned}U_{12}=\left(\begin{array}{ccc}
c_{12} & s_{12} & 0\\
-s_{12} & c_{12} & 0\\
0 & 0 & 1
\end{array}\right), & U_{13}=\left(\begin{array}{ccc}
c_{13} & 0 & s_{13}\\
0 & 1 & 0\\
-s_{13} & 0 & c_{13}
\end{array}\right)\\
U_{23}=\left(\begin{array}{ccc}
1 & 0 & 0\\
0 & c_{23} & s_{23}\\
0 & -s_{23} & c_{23}
\end{array}\right), & U_{\delta}=\left(\begin{array}{ccc}
1 & 0 & 0\\
0 & 1 & 0\\
0 & 0 & e^{-i\delta_{13}}
\end{array}\right),
\end{aligned}
\end{equation}
avec $s_{ij}=\sin(\theta_{ij})$ et $c_{ij}=\cos(\theta_{ij})$. La
paramétrisation standard est alors obtenue par
\begin{equation}
V_{CKM}=U_{23}U_{\delta}^{\dagger}U_{13}U_{\delta}U_{12}.
\end{equation}

\subsection{Secteur fort : le terme $\theta$ de QCD et le problème de $\mathcal{CP}$
fort}

Théoriquement, la violation-$\mathcal{CP}$ dans le secteur de l'interaction
forte est autorisée à travers un couplage d'origine non-perturbative
formé du tenseur de gluons $G^{\mu\nu}$ et de son tenseur dual $\tilde{G}^{\mu\nu}=\tfrac{1}{2}\epsilon_{\mu\nu\rho\sigma}G^{\rho\sigma}$
et invariant de jauge sous $SU(3)_{c}$ appelé \emph{terme theta de
QCD
\begin{equation}
\mathcal{L}_{\bar{\theta}}=-\frac{\alpha_{S}}{16\pi^{2}}\bar{\theta}G_{a}^{\mu\nu}\tilde{G}_{\mu\nu}^{a},
\end{equation}
}où $\alpha_{s}$ est la constante de couplage de l'interaction forte.
Le paramètre libre $\bar{\theta}$, représentant un angle, peut prendre
toutes les valeurs possibles dans l'intervalle $[0,2\pi]$.

Le terme $\theta$ contribue directement aux moments dipolaires électriques
(EDMs) des nucléons. Dans le cas du neutron, la contribution du terme
$\theta$ à l'EDM est
\begin{equation}
d_{n}^{\bar{\theta}}\approx-10^{-16}\bar{\theta}e\cdot cm.
\end{equation}
A partir des contraintes expérimentales sur les EDMs de systèmes hadroniques,
on peut obtenir un majorant de l'ensemble des valeurs possibles de
$\bar{\theta}$. En supposant que cette interaction est la seule source
de violation-CP et en négligeant les incertitudes sur les facteurs
de formes hadroniques et nucléaires, la borne supérieure sur le paramètre
$\bar{\theta}$ est
\begin{equation}
\bar{\theta}\lessapprox10^{-10}.
\end{equation}
Comprendre l'extrême petitesse de cette valeur par rapport à ce qui
est attendu constitue \emph{\og le problème de $\mathcal{CP}$ fort} \fg{},
un des problèmes les plus sérieux du MS.

Plusieurs solutions à ce problème ont été proposées avec une classe
de solutions se démarquant du reste: les solutions axioniques. La
solution axionique originale a été proposée en 1977 par Peccei et
Quinn \cite{PecceiQuinn} et revient à postuler l'existence d'un nouveau
champ, l'axion $a(x)$, se couplant aux gluons de la manière suivante
\begin{equation}
\mathcal{L}_{a}=\frac{1}{2}\partial^{\mu}a\partial_{\mu}a-V(a)-\frac{a(x)}{f_{a}}\frac{\alpha_{s}}{8\pi}G_{a}^{\mu\nu}\tilde{G}_{\mu\nu}^{a}.
\end{equation}
Le premier terme est le terme cinétique, le second est l'énergie potentiel
de l'axion, le troisième terme est le couplage axion-gluon dépendant
de la constante de désintégration de l'axion $f_{a}$ (analogue à
la constante de désintégration du pion $f_{\pi}$). Le champ axionique
acquière une VEV minimisant le potentiel de l'axion $V(a)$, ce qui
a pour effet d'induire un décalage de la valeur du paramètre $\bar{\theta}$
de $\bar{\theta}\rightarrow\bar{\theta}+\tfrac{\langle a\rangle}{f_{a}}$
conduisant à une annulation du terme theta. Ce mécanisme expliquerait
donc pourquoi la valeur du paramètre $\bar{\theta}$ est si petite.

\section{Au-delà du MS: masse des neutrinos et violation-$\mathcal{CP}$ leptonique}

\subsection{Phénomène d’oscillation dans le vide}

Lorsqu'il se propage librement dans le vide, un neutrino peut changer
spontanément d'une saveur donnée à une autre. Ce phénomène d'origine
purement quantique est appelé oscillation des neutrinos et a permis
d'établir que les neutrinos avaient une masse et que les leptons se
mélangeaient tout comme les quarks, contrairement à ce qui est supposé
dans le MS. Nous allons voir en détail dans cette partie le raisonnement
qui a permis d'arriver à cette conclusion. 

Historiquement, l'idée d'oscillation des neutrinos a été proposée
la première fois par Pontecorvo en 1957 \cite{Pontecorvo}, en analogie
avec les oscillations de Kaons dans les systèmes de mésons $K$. A
l'époque, seul un type de neutrino était connu, le neutrino électronique
$\nu_{e}$, donc Pontecorvo suggéra l'oscillation de ce dernier avec
son antiparticule $\nu_{e}\leftrightarrow\bar{\nu_{e}}$. Cinq ans
plus tard, Lederman, Schwartz et Steinberger ont observés le neutrino
muonique $\nu_{\mu}$ au Brookhaven National Laboratory, expérience
qui leur a valu le prix Nobel en 1988. D'autres types de neutrinos
pouvaient donc exister. Peu après cette découverte, Maki, Nakagawa
et Sakata ont suggéré la possibilité d'avoir des transitions entre
les différents types de neutrinos qu'ils appelaient \og transmutations
virtuelles \fg{}. Aujourd'hui, trois saveurs différentes de neutrinos
ont été observés correspondant aux trois générations du MS.

Penchons nous de plus près sur la théorie des oscillations des neutrinos
dans le vide. En réalité, ce phénomène est une manifestation directe
du mélange des leptons. En effet, les états propres d'interactions,
c'est à dire ceux qui interagissent par interaction faible et donc
observés par nos détecteurs, sont différents des états propres de
masses, c'est à dire les états qui se propagent librement avec une
masse bien définie. Les états propres d'interactions $\mathcal{B}=(|\nu_{e}\rangle,|\nu_{\mu}\rangle,|\nu_{\tau}\rangle)$
forment une base de l'espace des saveurs de dimension 3 tout comme
les états propres de masses $\mathcal{B}^{\prime}=(|\nu_{1}\rangle,|\nu_{2}\rangle,|\nu_{3}\rangle)$.
La matrice de passage qui permet de passer d'une base à l'autre $\mathcal{B}\rightarrow\mathcal{B}^{\prime}$
est appelée matrice de mélange leptonique ou encore matrice PMNS (du
nom de ses inventeurs Pontecorvo-Maki-Nakagawa-Sakata). Par conséquent,
n'importe quel état propre de masse $|\nu_{i}\rangle$ peut se décomposer
dans la base propre de saveur $\mathcal{B}$, c'est à dire s'exprimer
comme une combinaison linéaire des états propres de saveurs. Formellement,
on a
\begin{equation}
|\nu_{i}\rangle=\sum_{\alpha\in\{e,\mu,\tau\}}U_{\alpha i}|\nu_{\alpha}\rangle\hspace{1em}i\in\{1,2,3\},
\end{equation}
où $U$ est la matrice de mélange PMNS. La fraction de saveur $\alpha$
dans l'état propre de masse $|\nu_{i}\rangle$ est donnée par $\left|U_{\alpha i}\right|^{2}$.
La matrice de mélange étant unitaire, elle est inversible $(U^{-1}=U^{\dagger})$
et la relation précédente peut être inversée afin d'exprimer un état
propre d'interaction $|\nu_{\alpha}\rangle$ comme combinaison linéaire
des états propres de masses, c'est à dire dans la base $\mathcal{B}^{\prime}$
\begin{equation}
|\nu_{\alpha}\rangle=\sum_{i\in\{1,2,3\}}U_{\alpha i}^{\text{\ensuremath{\star}}}|\nu_{i}\rangle\hspace{1em}\alpha\in\{e,\mu,\tau\}.\label{eq:NuEigenstateSuperposition}
\end{equation}

\begin{figure}[h]
\begin{center}
\includegraphics[scale=0.7]{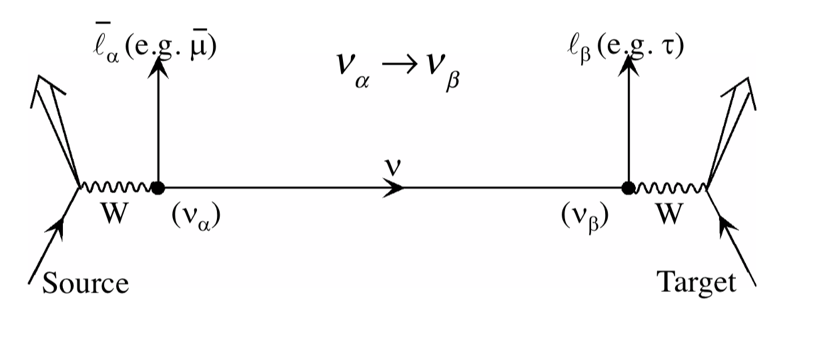}
\caption{Expérience typique d'oscillation de neutrinos. Source: Boris Kayser, Neutrino physics \cite{BKayserNu}}
\label{ExpOscNu}
\end{center}
\end{figure}

Une expérience typique d'oscillation de neutrinos est schématisée
en figure (\ref{ExpOscNu}). D'abord, une source radioactive produit
un neutrino de saveur $\alpha$ ainsi que le lepton chargé associé
$\overline{\ell_{\alpha}}$. Puis, le neutrino $\nu_{\alpha}$ se
propage librement sur une distance $L$ jusqu'à un détecteur. A son
arrivée, il interagit avec une cible et produit un lepton chargé $\ell_{\beta}$
de saveur $\beta$, ce qui signifie que le neutrino associé est de
saveur $\beta$, $\nu_{\beta}$. Il s'ensuit alors qu'au cours de
son voyage entre la source et le détecteur, le neutrino en question
s'est transformé, transmuté, métamorphosé en un neutrino de saveur
différente. Ce phénomène de changement de saveur $\nu_{\text{\ensuremath{\alpha}}}\to\nu_{\beta}$
est un effet quantique par essence lié à la superposition des états
quantiques, autrement dit à la non-unicité de la base hilbertienne
des états.

\begin{figure}[h]
\begin{center}
\includegraphics[scale=0.5]{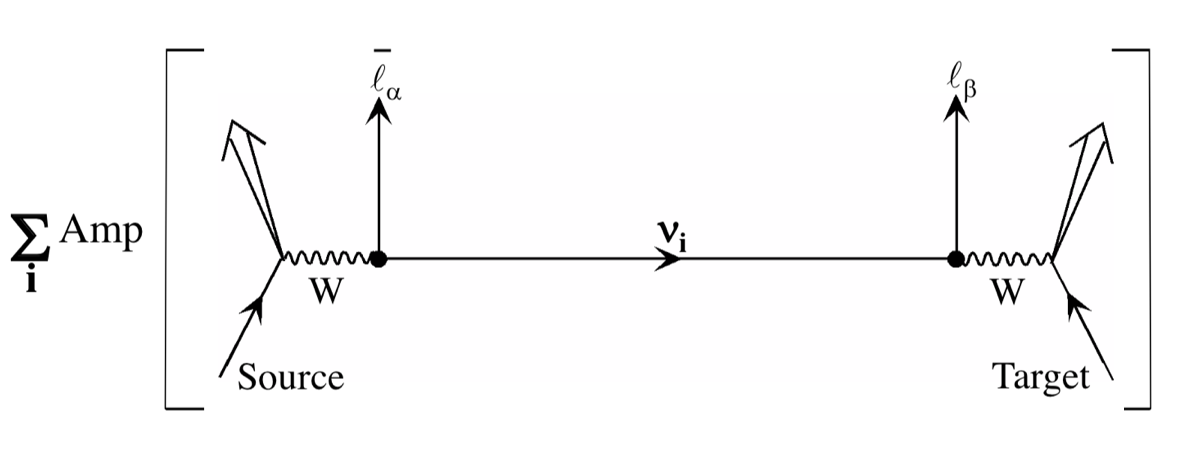}
\caption{Superposition cohérente des différentes contributions des états propres de masses $|\nu_{i}\rangle_{i=1,2,3}$. Source: Boris Kayser, Neutrino physics \cite{BKayserNu}}
\label{ExpOscNu2}
\end{center}
\end{figure}

Calculons la probabilité de changement de saveur, c'est à dire la
probabilité de transition $\nu_{\text{\ensuremath{\alpha}}}\to\nu_{\beta}$.
Comme nous l'avons vu plus haut (\ref{eq:NuEigenstateSuperposition}),
un neutrino de saveur $\alpha$ est une superposition cohérente d'états
propres de masses $|\nu_{i}\rangle$. Rappelons que les états propres
de masses sont ceux qui se propagent de la source au détecteur dans
la figure (\ref{ExpOscNu}). Chacun des trois états propres de masses
$|\nu_{1}\rangle,|\nu_{2}\rangle$ et $|\nu_{3}\rangle$ est susceptible
de s'y trouver. Ainsi, nous devons ajouter chacune de leurs contributions
de façon cohérente (voir figure (\ref{ExpOscNu2})). 

Pour un $|\nu_{i}\rangle$ donné, la transition $\nu_{\text{\ensuremath{\alpha}}}\rightarrow\nu_{\text{i}}\to\nu_{\beta}$
peut être décomposé en trois évènements consécutifs. Dans un premier
temps, il y a une transition du neutrino initial de saveur $\alpha$
produit par la source vers l'état propre de masse $|\nu_{i}\rangle$.
La probabilité de cet évènement est simplement $U_{\alpha i}^{\star}$.
Ensuite, il y a la propagation de l'état propre de masse $|\nu_{i}\rangle$
produit, d'énergie bien définie $E$, sur une distance $L$, de la
source à la cible. La probabilité de cet évènement est notée $Prop(\nu_{i},E,L)$.
Dans un dernier temps, l'état propre de masse $|\nu_{i}\rangle$ subit
une transition en interagissant avec le détecteur vers un état propre
d'interaction $|\nu_{\beta}\rangle$. La probabilité de cette transition
est $U_{\beta i}$\footnote{Le hamiltonien d'interaction couplant le neutrino, le lepton chargé
et le boson W étant hermitien, si l'amplitude de la transition $W\rightarrow\overline{\ell_{\alpha}}\nu_{i}$
est $U_{\alpha i}^{\star}$, alors l'amplitude de la transition $\nu_{i}\to\text{\ensuremath{\ell}}_{\beta}W$
est $U_{\beta i}$}. Puisque l'évènement dont nous souhaitons calculer la probabilité
est l'intersection de ces trois évènements indépendants prix consécutivement,
il s'ensuit que la contribution d'un $|\nu_{i}\rangle$ à l'amplitude
totale sera le produit des probabilités de chacun des trois évènements.
En sommant les contributions de chaque état propre de masse de façon
cohérente, on obtient l'amplitude totale de transition $\nu_{\text{\ensuremath{\alpha}}}\to\nu_{\beta}$
\begin{equation}
\mathcal{A}(\nu_{\text{\ensuremath{\alpha}}}\to\nu_{\beta})=\sum_{i\in\{1,2,3\}}U_{\alpha i}^{\star}Prop(\nu_{i},E,L)U_{\beta i}.\label{eq:AmpOsc}
\end{equation}

Il reste à déterminer $Prop(\nu_{i},E,L)$, la probabilité que $|\nu_{i}\rangle$,
d'énergie $E$, se propage de la source au détecteur (distance $L$).
Afin d'établir cette amplitude, plaçons nous dans le référentiel propre
de $|\nu_{i}\rangle$, où le temps propre est noté $\tau_{i}$. Dans
ce référentiel, le vecteur d'état $|\nu_{i}(\tau_{i})\rangle$ satisfait
à l'équation de Schrödinger
\begin{equation}
i\frac{\partial}{\partial\tau_{i}}|\nu_{i}(\tau_{i})\rangle=m_{i}|\nu_{i}(\tau_{i})\rangle,
\end{equation}
avec $m_{i}$ la masse au repos. La solution de cette équation est
donnée par
\begin{equation}
|\nu_{i}(\tau_{i})\rangle=e^{-im_{i}\tau_{i}}|\nu_{i}(0)\rangle.
\end{equation}
Par conséquent, la probabilité que $|\nu_{i}\rangle$ se propage pendant
un temps $\tau_{i}$ dans son référentiel propre est donnée simplement
par l'amplitude $\langle\nu_{i}(0)|\nu_{i}(\tau_{i})\rangle=e^{-im_{i}\tau_{i}}$.
Par suite, $Prop(\nu_{i},E,L)$ est égal à cette même amplitude, avec
$\tau_{i}$ le temps propre mis par le neutrino pour faire une distance
$L$, de la source jusqu'au détecteur.

Nous avons déterminé $Prop(\nu_{i},E,L)$ dans le référentiel propre
du neutrino, mais afin qu'il soit utile aux expériences, il est préférable
de l'exprimer dans le référentiel du laboratoire, car ce sont les
distances et les temps dans ce référentiel que nous mesurons. Les
variables dans le référentiel du laboratoire définies par l'expérience
sont la distance source-détecteur $L$, le temps mis pour faire cette
distance $t$ ainsi que l'énergie $E_{i}$ et l'impulsion $p_{i}$
de l'état propre de masse $|\nu_{i}\rangle$. Le produit de Minkowski
entre le quadri-vecteur énergie-impulsion et le quadri-vecteur position
$P\cdot X$ étant invariant de Lorentz, en l'exprimant dans le référentiel
propre (où $P=(m_{i},\overrightarrow{0})$ et $X=(\tau_{i},\overrightarrow{x})$)
et dans le référentiel du laboratoire (où $P=(E_{i},p_{i},0,0)$ et
$X=(t,L,0,0)$), on exprime la phase dans $Prop(\nu_{i},E,L)=e^{-im_{i}\tau_{i}}$
en fonction des variables du laboratoire
\begin{equation}
m_{i}\tau_{i}=E_{i}t-p_{i}L.
\end{equation}

L'impulsion $p_{i}$ d'un état propre de masse $|\nu_{i}\rangle$,
d'énergie $E$ et de masse $m_{i}$ est donnée par
\begin{equation}
p_{i}=\sqrt{E^{2}-m_{i}^{2}}=E-\frac{m_{i}^{2}}{2E}+\mathcal{O}\left(\frac{m_{i}^{4}}{E^{3}}\right),
\end{equation}
où l'on a supposé que $m_{i}^{2}\ll E^{2}$ dans la dernière égalité
et effectué un développement limité, ce qui est raisonnable comme
hypothèse pour des neutrinos compte tenu de la petitesse de leurs
masses devant leurs énergies cinétiques. En combinant les deux équations
précédentes, on réécrit la phase du propagateur comme suit
\begin{equation}
m_{i}\tau_{i}\approx E(t-L)+\frac{m_{i}^{2}}{2E}L.
\end{equation}
La phase $E(t-L)$ ne joue aucun rôle dans les interférences des états
propres de masses dans la mesure où elle est commune aux trois états
cohérents $|\nu_{i}\rangle$. En définitive, l'expression du propagateur
recherchée est donc
\begin{equation}
Prop(\nu_{i},E,L)=e^{-im_{i}^{2}\frac{L}{2E}}.
\end{equation}

En injectant cette expression dans l'équation (\ref{eq:AmpOsc}),
on obtient l'amplitude de transition d'un neutrino de saveur $\alpha$
en un neutrino de saveur $\beta$ pendant sa propagation dans le vide
avec une énergie $E$ et sur une distance $L$
\begin{equation}
\mathcal{A}(\nu_{\alpha}\to\nu_{\text{\ensuremath{\beta}}})=\sum_{i\in\{1,2,3\}}U_{\alpha i}^{\star}e^{-im_{i}^{2}\frac{L}{2E}}U_{\beta i}.
\end{equation}
Finalement, la probabilité étant donnée par le module de l'amplitude
élevé au carré, on a
\begin{equation}
\begin{aligned}P(\nu_{\alpha}\to\nu_{\text{\ensuremath{\beta}}}) & =\left|\mathcal{A}(\nu_{\alpha}\to\nu_{\text{\ensuremath{\beta}}})\right|^{2}\\
 & =\delta_{\alpha\beta}-4\sum_{i>j}Re\left(U_{\text{\ensuremath{\alpha}i}}^{\star}U_{\beta i}U_{\alpha j}U_{\text{\ensuremath{\beta}j}}^{\star}\right)\sin^{2}\left(\Delta m_{ij}^{2}\frac{L}{4E}\right)\\
 & +2\sum_{i>j}Im\left(U_{\text{\ensuremath{\alpha}i}}^{\star}U_{\beta i}U_{\alpha j}U_{\text{\ensuremath{\beta}j}}^{\star}\right)\sin\left(\Delta m_{ij}^{2}\frac{L}{2E}\right),
\end{aligned}
\label{eq:ProbaOscNu}
\end{equation}
où $\Delta m_{ij}^{2}\equiv m_{i}^{2}-m_{j}^{2}$. Cette formule est
valable pour des transitions de neutrinos. Bien évidemment, le phénomène
d'oscillation existe aussi pour les antineutrinos et la probabilité
de transition pour ces derniers peut être obtenue à partir de celle
des neutrinos en invoquant l'invariance $\mathcal{CPT}$. En effet,
la transition $\overline{\nu_{\alpha}}\to\overline{\nu_{\beta}}$
est l'image par la symétrie discrète $\mathcal{CPT}$ de $\nu_{\beta}\to\nu_{\alpha}$.
En vertu du théorème $\mathcal{CPT}$, on a
\begin{equation}
P(\overline{\nu_{\alpha}}\to\overline{\nu_{\beta}})=P(\nu_{\beta}\to\nu_{\alpha}).
\end{equation}
De plus, l'équation (\ref{eq:ProbaOscNu}) montre que la probabilité
de la transition $\nu_{\beta}\to\nu_{\alpha}$ calculée avec une matrice
de mélange $U$ est la même que celle de la transition inverse $\nu_{\alpha}\to\nu_{\beta}$
calculée avec une matrice de mélange $U^{\star}$. Autrement dit,
la probabilité reste inchangée si on inverse le processus, à condition
que la matrice $U$soit remplacée par sa conjuguée. D'où,
\begin{equation}
P(\nu_{\beta}\to\nu_{\alpha};U)=P(\nu_{\alpha}\to\nu_{\beta};U^{\star}).
\end{equation}
A l'aide des deux équations précédentes, on établit alors la probabilité
d'oscillation pour des antineutrinos
\begin{equation}
\begin{aligned}P(\overline{\nu_{\alpha}}\to\overline{\nu_{\text{\ensuremath{\beta}}}}) & =\delta_{\alpha\beta}-4\sum_{i>j}Re\left(U_{\text{\ensuremath{\alpha}i}}^{\star}U_{\beta i}U_{\alpha j}U_{\text{\ensuremath{\beta}j}}^{\star}\right)\sin^{2}\left(\Delta m_{ij}^{2}\frac{L}{4E}\right)\\
 & -2\sum_{i>j}Im\left(U_{\text{\ensuremath{\alpha}i}}^{\star}U_{\beta i}U_{\alpha j}U_{\text{\ensuremath{\beta}j}}^{\star}\right)\sin\left(\Delta m_{ij}^{2}\frac{L}{2E}\right).
\end{aligned}
\label{eq:ProbaOscAntiNu}
\end{equation}
Quelques commentaires s'imposent par rapport aux formules de probabilité
d'oscillation de neutrinos et d'antineutrinos. 

Tout d'abord, remarquons que le processus $\overline{\nu_{\alpha}}\to\overline{\nu_{\text{\ensuremath{\beta}}}}$
est l'image par la symétrie discrète $\mathcal{CP}$ de la transition
$\nu_{\alpha}\to\nu_{\beta}$. Ainsi, une probabilité différente pour
ces deux processus signifierait que la symétrie $\mathcal{CP}$ n'est
plus respectée lors de ces oscillations. A l'aide des formules établies
(\ref{eq:ProbaOscNu}) et (\ref{eq:ProbaOscAntiNu}), on voit que
seul le terme impliquant la partie imaginaire diffère entre les neutrinos
et leurs antiparticules. Par conséquent, dès lors que la matrice de
mélange $U$ est complexe, on a en général $P(\overline{\nu_{\alpha}}\to\overline{\nu_{\text{\ensuremath{\beta}}}})\neq P(\nu_{\alpha}\to\nu_{\text{\ensuremath{\beta}}})$
et donc une violation de la symétrie $\mathcal{CP}$. Jusque-là, la
violation-$\mathcal{CP}$ avait été observée uniquement dans le secteur
des quarks et donc le phénomène d'oscillation des neutrinos constituait
une première indication de violation-$\mathcal{CP}$ leptonique.

D'autre part, si les neutrinos sont de masses nulles alors $\Delta m_{ij}^{2}=0$
et donc la probabilité de changement de saveur serait nulle $P(\overset{(\_\_)}{\nu_{\alpha}}\to\overset{(\_\_)}{\nu_{\beta}})=0$.
Il s'ensuit que d'après la proposition contraposée\footnote{En logique, si $A\Rightarrow B$, alors $non(B)\Rightarrow non(A)$.}
de la précédente, si la probabilité d'oscillation est non-nulle $P(\overset{(\_\_)}{\nu_{\alpha}}\to\overset{(\_\_)}{\nu_{\beta}})\neq0$,
alors les neutrinos ne sont pas de masses nulles. Il en est de même
avec des masses dégénérés. Si les neutrinos avaient des masses égales
alors $\Delta m_{ij}^{2}=0$ et donc la probabilité d'oscillation
serait nulle. Par la contraposée, si la probabilité est non-nulle,
alors les neutrinos ne sont pas dégénérés. En somme, l'observation
d'un changement de saveur de neutrinos ou d'antineutrinos impliquerait
immédiatement que les neutrinos sont massifs. C'est d'ailleurs de
cette façon que cela a été établit.

\begin{figure}[h]
\begin{center}
\includegraphics[scale=0.7]{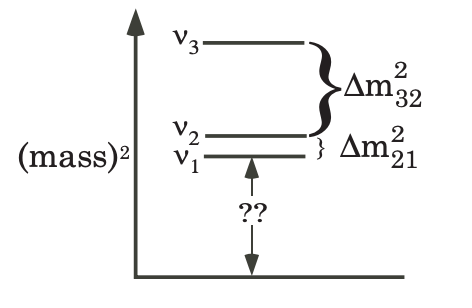}
\caption{Un spectre possible de masses-carrées des neutrinos. Les expériences d'oscillations ne peuvent pas déterminer où commence le spectre. Source: Boris Kayser, Neutrino physics \cite{BKayserNu}}
\label{NuSpectrum}
\end{center}
\end{figure}

En outre, notons que les probabilités d'oscillations de neutrinos
(équations (\ref{eq:ProbaOscNu}) et (\ref{eq:ProbaOscAntiNu})) ne
dépendent que de la différence des carrés des masses $\Delta m_{ij}^{2}$
et non de la valeur absolue de ces dernières. Cela signifie que la
seule information sur les masses accessible lors d'un changement de
saveur est le paramètre $\Delta m_{ij}^{2}=m_{i}^{2}-m_{j}^{2}$.
C'est pourquoi aucune expérience d'oscillation de neutrinos ne pourra
déterminer la masse absolue des différents état propres de masses.
Pour le dire autrement, les expériences d'oscillations peuvent établir
les écarts respectifs entre les différentes valeurs de masses, ce
qui fixe une forme, mais ne peuvent pas déterminer l'endroit où commence
cette forme (voir figure \ref{NuSpectrum}).

Pour finir cette partie sur l'oscillation des neutrinos, considérons
le cas particulier où seules deux saveurs existent. Supposons qu'on
a deux états propres de masses $\nu_{1}$ et $\nu_{2}$ ainsi que
deux états propres d'interactions $\nu_{e}$ et $\nu_{\mu}$. Dans
ces conditions, un seul paramètre de différence de masses-carrées
peut être défini $\Delta m^{2}\equiv m_{2}^{2}-m_{1}^{2}$ et la matrice
de mélange prend la forme d'une simple matrice de rotation en deux
dimensions
\begin{equation}
U=\left(\begin{array}{cc}
\cos\theta & \sin\theta\\
-\sin\theta & \cos\theta
\end{array}\right).
\end{equation}
L'angle $\theta$ est appelé angle de mélange. En utilisant cette
matrice de mélange, la probabilité d'oscillation (\ref{eq:ProbaOscNu})
devient
\begin{equation}
P(\nu_{\alpha}\to\nu_{\beta})=P(\overline{\nu_{\alpha}}\to\overline{\nu_{\beta}})=\sin^{2}2\theta\sin^{2}\left(\Delta m^{2}\frac{L}{4E}\right).
\end{equation}
Même si nous savons déjà qu'au moins trois saveurs de neutrinos existent,
ce cas reste une très bonne approximation pour plusieurs expériences
d'oscillations. En effet, pour une longueur $L$ et une énergie $E$
fixées, les expériences sont sensibles à une plage donnée de $\Delta m^{2}$,
et comme les deux paramètres de masses-carrées solaire et atmosphérique
sont très différents, ils sont en quelque sorte découplés et on peut
alors en étudier un à la fois dans le cadre de cette approximation
à deux saveurs.

\subsection{Faits expérimentaux}

En général, les expériences d'oscillations de neutrinos peuvent détecter
des changements de saveurs de deux manières différentes. Une première
façon consiste à observer dans un flux de neutrinos initialement de
saveur donnée, l'apparition de neutrinos de saveur différente de l'originale.
Ce type d'expérience est appelé \og expérience d'apparition \fg{}.
L'autre façon consiste à commencer avec un flux connu de neutrinos
d'une saveur donnée et d'observer la disparition d'une partie de ce
flux, c'est ce que l'on appelle une \og expérience de disparition \fg{}.

Une catégorie d'expériences de neutrinos consiste à utiliser des réacteurs
de fissions nucléaires comme sources de neutrinos. En effet, les neutrinos
dit de réacteurs sont en fait des antineutrinos électroniques qui
sont émis par désintégration-$\beta$ de noyaux instables dans les
réacteurs. Plus de 80\% des réacteurs commerciaux sont des réacteurs
à eau légère et fonctionnent par fissions des isotopes de l'uranium
et du plutonium suivants: $^{235}U,{}^{238}U,{}^{239}Pu$ et $^{241}Pu$.
A chaque fission, environ 6 antineutrinos électroniques d'énergies
comprises entre $0-10MeV$ sont émis ainsi qu'une énergie libérée
d'environ $200MeV$. En moyenne, les réacteurs nucléaires produisent
environ $2\times10^{20}$ $\overline{\nu_{e}}$ par $GW_{th}\cdot s$
\footnote{GigaWatts thermique-seconde} avec des énergies moyennes
autour de $4MeV$.

De la découverte de l'antineutrino par Reines et Cowan \cite{ReinesCowan1953,ReinesCowan1957,ReinesCowan1960}
jusqu'aux expériences d'oscillations en passant par l'étude du moment
magnétique du neutrino, les expériences de réacteurs ont joué un rôle
majeur dans l'histoire de la physique des neutrinos.

L'observation de changement de saveur des neutrinos solaires dans
une ligne de base d'environ 100 km avec un grand angle de mélange
et un paramètre $\Delta m^{2}\sim7.1\times10^{-5}eV^{2}$ confirme
que les (anti)neutrinos oscillent dans le vide. Le détecteur d'antineutrinos
à scintillation liquide Kamioka (KamLAND) localisé dans le laboratoire
sous-terrain de Kamioka au Japon a été spécialement conçu pour mesurer
le flux d'antineutrinos $\overline{\nu_{e}}$ de réacteurs. En 2003,
KamLAND a mis en évidence pour la première fois une disparition d'antineutrinos
de réacteurs, suite à une exposition de 162 kt-yr\footnote{kilotons-year}.
Durant cette exposition, le détecteur a enregistré 54 évènements de
plus de 2.6$MeV$, ce qui est moins que les $86.8\pm5.6$ évènements
attendus. Le déficit observé dans le flux d'antineutrinos pouvait
être expliqué par le phénomène d'oscillation de (anti)neutrinos.

Récemment, KamLAND a publié \cite{Kamland} une mesure encore plus
précise du flux de $\overline{\nu_{e}}$ de réacteurs et du spectre
d'énergie après des données collectées pendant une exposition de 766.3
t-yr\footnote{tons-year.}. Les résultats sont sans appel et indiquent
clairement l'oscillation des antineutrinos de réacteurs. Le spectre
d'énergie mesuré par KamLAND est en accord avec l'existence d'oscillations
à 99.6\% C.L (niveau de confiance). La probabilité de survie du neutrino
de départ dépend de son énergie $E_{\nu}$ de la manière suivante
$P_{ee}=1-\sin^{2}2\theta\sin^{2}\left(\Delta m^{2}\frac{L}{E_{\nu}}\right)$
(évènement complémentaire de l'oscillation) et les distorsions spectrales
constituent une signature caractéristique des oscillations.

\begin{figure}
\begin{center}
\includegraphics[scale=0.7]{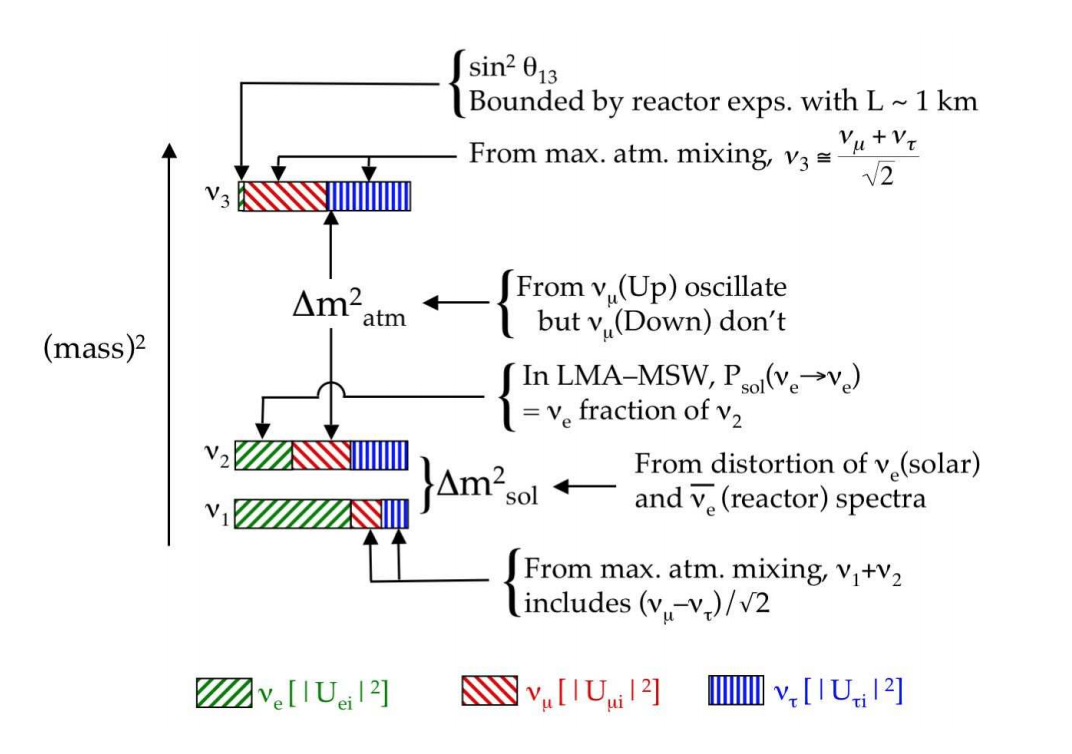}
\caption{Un spectre de masse-carrée pour trois saveurs de neutrinos en accord avec les données expérimentales. Les différentes couleurs représentent les fractions des différentes saveurs dans chaque état propre de masse. Source: Boris Kayser, Neutrino physics \cite{BKayserNu}}
\label{NuMass2Spectrum}
\end{center}
\end{figure}

\begin{figure}
\begin{center}
\includegraphics[height=3.5in,width=3in]{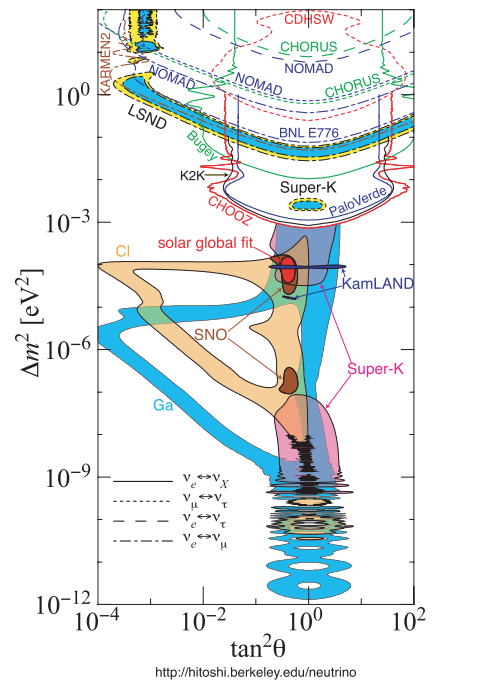}
\caption{Limites actuelles sur les paramètres d'oscillation de neutrinos établies par les expériences atmosphériques, solaires, de réacteurs et d'accélérateurs.}
\label{CurrentLimitsOScParas}
\end{center}
\end{figure}

Faisons le point sur ce qui a été établi avec les différentes expériences
d'oscillations. Les données d'oscillations de neutrinos proviennent
d'une variété d'expériences solaires (Homestake \cite{Homestake},
SAGE \cite{SAGE}, GALLEX/GNO \cite{Galex}, Super-K \cite{SuperK},
SNO \cite{SNO} and Borexino \cite{Borexino}), atmosphériques \cite{SuperK},
de réacteurs (Double-Chooz \cite{DoobleChooz}, Daya-Bay \cite{DayaBay}
et RENO \cite{RENO}) et d'accélérateurs (MINOS \cite{MINOS} et T2K
\cite{T2K}) \cite{Nakamura2010}. Les expériences de réacteurs, solaires
et atmosphériques ont permis de contraindre les paramètres d'oscillations.
Les limites posées sont résumées en figure \ref{CurrentLimitsOScParas}.
De plus, l'analyse des données des différentes expériences d'oscillations
un peu partout sur la planète ont permis de déterminer le spectre
des masses-carrées des neutrinos représenté en figure \ref{NuMass2Spectrum}
.

\begin{figure}
\begin{center}
\includegraphics[scale=0.7]{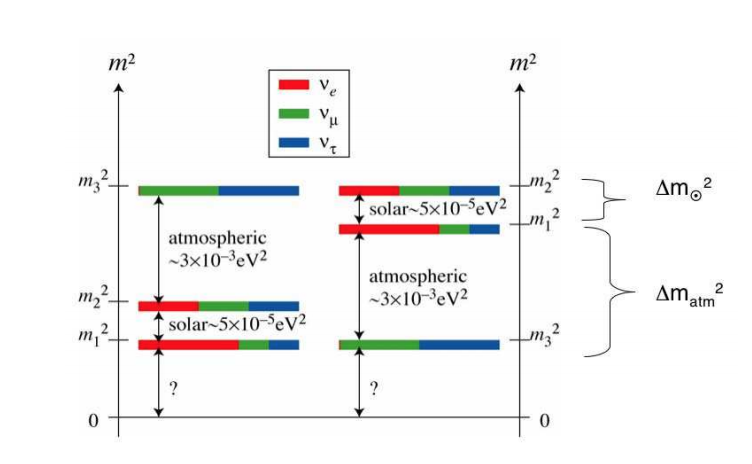}
\caption{Spectres de masses possibles dans le cas de trois neutrinos. Hiérarchie normale ou inversée des masses. Source: \cite{Heeger2004}.}
\label{NuMassSpectrumNOIO}
\end{center}
\end{figure}

En guise de résumé, les expériences d'oscillations ont confirmé le
phénomène de changement de saveur postulé théoriquement ainsi que
le mélange des neutrinos (massifs). Les différentes mesures et analyse
des données ont fait émerger deux échelles de masses décrivant les
oscillations, à savoir $\Delta m_{atm}^{2}\sim2.0\times10^{-3}eV^{2}$
et $\Delta m_{sol}^{2}\sim7.1\times10^{-5}eV^{2}$. Ces paramètres
sont qualifiés de \og solaires \fg{} et \og d'atmosphériques \fg{}
en raison du rôle important qu'ils jouent respectivement pour les
neutrinos solaires (dont l'origine se trouve dans les réactions nucléaires
au sein du soleil) et les neutrinos atmosphériques (qui proviennent
des désintégrations de pions et de kaons formés dans la haute atmosphère
suite aux interactions de rayons cosmiques). Comme expliqué dans la
partie précédente, ces valeurs nous permettent de connaitre l'écart
relatifs des masses mais pas la masse absolue. Il reste donc une ambiguïté
à ce propos que certaines expériences récentes se chargent de lever
(Mainz \cite{Mainz}, PSI \cite{PSI}, ALEPH \cite{Aleph}, KATRIN
\cite{Katrin}). Compte tenu de ces mesures, deux spectres de masses
restent possibles et sont représentés en figure \ref{NuMassSpectrumNOIO}.
En outre, l'angle de mélange associé à la transition solaire est grand
et celui associé à la transition atmosphérique est presque maximal
(45\textdegree ).

Maintenant que nous savons que les neutrinos ont une masse, il est
donc nécessaire d'étendre le MS de sorte à les inclure. C'est l'objet
de la partie suivante.

\subsection{Modèles de masses des neutrinos et brisure de la symétrie de saveur}

\subsubsection{Masse de Dirac}

Les termes de Yukawa couplant la partie gauche et la partie droite
des fermions, afin d'écrire un terme de Yukawa pour les neutrinos,
nous sommes amenés à introduire un neutrino de chiralité droite dans
notre modèle. Nous étendons alors le nombre de particules présentes
dans le MS en y ajoutant 3 neutrinos droits correspondant aux 3 saveurs
($\nu_{eR},\nu_{\mu R},\nu_{\tau R}$). Ces neutrinos ne ressentent
aucune des interactions fondamentales (à l'exception de la gravité).
En termes plus techniques, ils transforment comme des singlets sous
les différents groupes de jauges et donc ils appartiennent à la représentation
triviale du groupe de jauge du MS, $N=\nu_{R}^{\dagger}\sim(1,1)_{0}$.
De plus, la symétrie quarks-leptons est rétablie dans la mesure où
tous les fermions possèdent maintenant une composante gauche appartenant
à un doublet faible et une composante droite transformant comme un
singlet faible. Le secteur de jauge est toujours invariant sous l'action
du groupe de saveur étendu $G_{F}^{\prime}=G_{F}\times U(3)_{N}$:
\begin{equation}
G_{F}^{\prime}=U(3)^{6}=U(3)_{Q}\times U(3)_{U}\times U(3)_{D}\times U(3)_{L}\times U(3)_{E}\times U(3)_{N}.
\end{equation}
L'action de ce groupe sur le nouveau champ $N$, comme pour les autres
champs de chiralité droite, est définie par $N\overset{G_{F}^{\prime}}{\rightarrow}Ng_{N}$
avec $g_{N}\in U(3)_{N}$. Comme nous l'avons vu précédemment, cette
symétrie est brisée par les couplages de Yukawa. Ayant introduit les
neutrinos droits, on peut donc écrire un terme pour les neutrinos:

\begin{equation}
\mathcal{L}_{Yukawa}=-U^{I}Y_{u}^{IJ}Q^{J}H^{\dagger C}-D^{I}Y_{d}^{IJ}Q^{J}H^{\dagger}-E^{I}Y_{e}^{IJ}L^{J}H^{\dagger}-N^{I}Y_{\nu}^{IJ}L^{J}H^{\dagger C}+h.c.\label{Yukawa_N}
\end{equation}
 Comptons le nombre de paramètres réels libres du modèle. 

Dans l'espace des saveurs $Y_{u,d,e,\nu}\in M_{3}(\mathbb{C})$ (matrices
$3\times3$ à coefficients complexes). Par conséquent, chacune des
matrices contient 9 paramètres complexes, soit $9\times2=18$ paramètres
réels. Au total, le secteur de Yukawa introduit $18\times4=72$ paramètres
réels. Cependant, comme dans la partie précédente, ils ne sont pas
tous physiques. La symétrie $G_{F}^{\prime}$ n'est pas entièrement
brisée et il reste des symétries résiduelles. En revanche, nous n'avons
plus conservation des nombres leptoniques de saveurs du fait de l'ajout
du neutrino droit qui permet des processus avec changement de saveur
dans le secteur des leptons comme nous le verrons plus tard. Néanmoins,
nous avons conservation du nombre leptonique total $\mathcal{L}=\mathcal{L}_{e}+\mathcal{L}_{\mu}+\mathcal{L}_{\nu}$
et du nombre baryonique $\mathcal{B}$. 
\begin{equation}
G_{F}^{\prime}=U(3)^{6}\rightarrow U(1)_{\mathcal{B}}\times U(1)_{\mathcal{L}}.
\end{equation}
En appliquant la même technique de comptage que dans la Section \ref{Ch1Sec5},
nous pouvons déterminer le nombre de paramètres physiques, voir formule
(\ref{eq:ComptPara}). Ici, le nombre de paramètres total est 72 pour
le secteur de Yukawa. Comme une transformation sous $G_{F}^{\prime}=U(3)^{6}$
implique 6 matrices unitaires $3\times3$, chacune d'entre elles nécessitant
9 paramètres réels pour la décrire, soit un total de $6\times9=54$
paramètres réels. Or, il reste deux paramètres réels (générateurs
non brisés) correspondant aux $U(1)_{\mathcal{B}}$ et $U(1)_{\mathcal{L}}$
résiduels. Le nombre de générateurs brisés est donc $54-2=52$. Par
conséquent, $\text{\#paramètres physique}=72-52=20$. 

Après la brisure spontanée de symétrie électrofaible, le champ de
Higgs acquiert une VEV non nulle, $<0\mid H\mid0>=v$ et il s'exprime
alors dans la jauge unitaire comme $H=\left(\begin{array}{cc}
0 & v+h\end{array}\right)^{T}$ avec $h$ le boson de Higgs. Les termes de Yukawas donnent alors
les termes de masse des fermions:
\begin{equation}
\mathcal{L}_{Yukawa}=-v(\bar{u}_{R}^{I}Y_{u}^{IJ}u_{L}^{J}+\bar{d}_{R}^{I}Y_{d}^{IJ}d_{L}^{J}+\bar{e}_{R}^{I}Y_{e}^{IJ}e_{L}^{J}+\bar{\nu}_{R}^{I}Y_{\nu}^{IJ}\nu_{L}^{J})(1+\frac{h}{v})+h.c.
\end{equation}
 Comme les $Y_{u,d,e,\nu}$ ne sont pas diagonaux, on réalise une
transformation bi-unitaire pour les diagonaliser et obtenir les états
propres de masse des fermions,
\begin{equation}
\exists V_{R}^{u,d,e,\nu},V_{L}^{u,d,e,\nu}\in U(3)\hspace{1em}\text{telles que}\hspace{1em}vV_{R}^{u,d,e,\nu}Y_{u,d,e,\nu}V_{L}^{u,d,e,\nu}=\mathbb{M}_{u,d,e,\nu}
\end{equation}
En réinjectant dans le lagrangien de Yukawa, on obtient:
\begin{equation}
\begin{aligned}\mathcal{L}_{Yukawa}= & -(\bar{u}_{R}(V_{R}^{u})^{\dagger}\mathbb{M}_{u}(V_{L}^{u})^{\dagger}u_{L}+\bar{d}_{R}(V_{R}^{d})^{\dagger}\mathbb{M}_{d}(V_{L}^{d})^{\dagger}d_{L}+\bar{e}_{R}(V_{R}^{e})^{\dagger}\mathbb{M}_{e}(V_{L}^{e})^{\dagger}e_{L}\\
 & +\bar{\nu}_{R}(V_{R}^{\nu})^{\dagger}\mathbb{M}_{\nu}(V_{L}^{\nu})^{\dagger}\nu_{L})(1+\frac{h}{v})+h.c.
\end{aligned}
\end{equation}
En utilisant la symétrie de saveur, on applique une transformation
unitaire sur les états propres d'interaction pour récupérer les états
propres de masse. Par ailleurs, la symétrie de jauge nous impose d'appliquer
les mêmes transformations unitaires sur les deux composantes du doublet
de quarks $Q=\left(\begin{array}{c}
u_{L}\\
d_{L}
\end{array}\right)$ d'une part et du doublet de leptons $L=\left(\begin{array}{c}
\nu_{L}\\
e_{L}
\end{array}\right)$ d'autre part. On choisit d'appliquer $(V_{L}^{u})^{\dagger}$ à $Q$
et $(V_{L}^{e})^{\dagger}$ à $L$. En redéfinissant les champs de
la manière suivante:

\begin{equation}
\begin{aligned}\begin{cases}
(V_{L}^{u})^{\dagger}u_{L}= & u_{L}^{mass}\\
(V_{L}^{u})^{\dagger}d_{L}= & d_{L}^{\prime}
\end{cases}, & \begin{cases}
\bar{u}_{R}(V_{R}^{u})^{\dagger}= & \bar{u}_{R}^{mass}\\
\bar{d}_{R}(V_{R}^{d})^{\dagger}= & \bar{d}_{R}^{mass}
\end{cases}\\
\begin{cases}
(V_{L}^{\nu})^{\dagger}e_{L}= & e_{L}^{mass}\\
\bar{e}_{R}(V_{R}^{e})^{\dagger}= & \bar{e}_{R}^{mass}
\end{cases}, & \begin{cases}
(V_{L}^{\nu})^{\dagger}\nu_{L}= & \nu_{L}^{\prime}\\
\bar{\nu}_{R}(V_{R}^{\nu})^{\dagger}= & \bar{\nu}_{R}^{mass}
\end{cases}
\end{aligned}
,
\end{equation}
on obtient:
\begin{equation}
\begin{aligned}\mathcal{L}_{Yukawa}= & -(\bar{u}_{R}^{mass}\mathbb{M}_{u}u_{L}^{mass}+\bar{d}_{R}^{mass}\mathbb{M}_{d}V_{CKM}^{\dagger}d_{L}^{\prime}+\bar{e}_{R}^{mass}\mathbb{M}_{e}e_{L}^{mass}\\
 & +\bar{\nu}_{R}^{mass}\mathbb{M}_{\nu}U_{PMNS}^{\dagger}\nu_{L}^{\prime})(1+\frac{h}{v})+h.c.
\end{aligned}
\end{equation}
De la même façon que pour le secteur des quarks, la non correspondance
des deux transformations à appliquer à chacune des composantes du
doublet $L$ pour atteindre leurs états propres de masse ($V_{L}^{e}\neq V_{L}^{\nu}$)
définie la matrice de mélange des leptons appelée matrice de \noun{Pontecorvo-Maki-Nakagawa-Sakata}
(PMNS): 
\begin{equation}
V_{L}^{e\dagger}V_{L}^{\nu}\equiv U_{PMNS}=\left(\begin{array}{ccc}
V_{e1} & V_{e2} & V_{e3}\\
V_{\mu1} & V_{\mu2} & V_{\mu3}\\
V_{\tau1} & V_{\tau2} & V_{\tau3}
\end{array}\right).
\end{equation}
Motivé par l'expérience, on choisit usuellement de mettre les leptons
chargés dans leurs états propres de masse, car ils sont plus facile
à détecter. Dans ce cas, $U_{PMNS}$ lie les états propres de jauge
des neutrinos à leurs états propres de masse:
\begin{equation}
\underset{\text{états propres d'intéraction}}{\underbrace{\left(\begin{array}{c}
\nu_{e}\\
\nu_{\mu}\\
\nu_{\tau}
\end{array}\right)}}=U_{PMNS}\underset{\text{états propres de masse}}{\underbrace{\left(\begin{array}{c}
\nu_{1}\\
\nu_{2}\\
\nu_{3}
\end{array}\right)}}.
\end{equation}
Comme la matrice CKM, la matrice PMNS est unitaire, il faut donc $3^{2}=9$
paramètres réels pour la décrire qu'on peut choisir comme étant 3
nombres réels et 6 phases. Or, nous avons déjà les masses des 12 fermions
(incluant les neutrinos) élémentaires auxquelles on ajoute 4 paramètres
pour la matrice CKM (3 angles et 1 phase), ce qui fait un total de
16 paramètres physiques libres du modèle. Sachant qu'il y en a 20
en tout, seuls $20-16=4$ paramètres sont physiques dans la matrice
$U_{PMNS}$. Par exemple on peut paramétriser la matrice PMNS par
3 angles de mélange ($\theta_{12},\theta_{13}$ et $\theta_{23}$
qui sont des nombres réels) et une phase de violation de $\mathcal{CP}$
$(\delta)$ . 

Si on note $c_{ij}=\cos\theta_{ij}$ et $s_{ij}=\sin\theta_{ij}$,
on peut exprimer la matrice PMNS sous la forme:
\begin{equation}
U_{PMNS}=\left(\begin{array}{ccc}
c_{12}c_{13} & s_{12}c_{13} & s_{13}e^{-i\delta}\\
-s_{12}c_{23}-c_{12}s_{23}s_{13}e^{i\delta} & c_{12}c_{23}-s_{12}s_{23}s_{13}e^{i\delta} & s_{23}c_{13}\\
s_{12}s_{23}-c_{12}c_{23}s_{13}e^{i\delta} & -c_{12}s_{23}-s_{12}c_{23}s_{13}e^{i\delta} & c_{23}c_{13}
\end{array}\right).
\end{equation}
Les valeurs actuelles des coefficients en modules, mesurées à partir
des oscillations de neutrinos sont \cite{PMNSFit,PMNSFit2}:

\begin{equation}
\mid U_{PMNS}\mid=\left(\begin{array}{ccc}
0.82\pm0.01 & 0.54\pm0.02 & 0.15\pm0.03\\
0.35\pm0.06 & 0.70\pm0.06 & 0.62\pm0.06\\
0.44\pm0.06 & 0.45\pm0.06 & 0.77\pm0.06
\end{array}\right),
\end{equation}
ainsi que les meilleurs ajustements des angles de mélange et de la
phase: $\theta_{12}[\text{\textdegree]}=33.36_{-0.78}^{+0.81},\theta_{23}[\text{\textdegree]}=40.0{}_{-1.5}^{+2.1},\theta_{13}[\text{\textdegree]}=8.66_{-0.46}^{+0.44}$
et $\delta[\text{\textdegree]}=300{}_{-138}^{+66}$. 

Dès lors que nous considérons un secteur de neutrinos non trivial,
les transitions LFV deviennent possibles. En effet, comme les états
propres de masse et d'interaction des neutrinos ne sont plus alignés,
le passage de l'un à l'autre nécessite une rotation $\nu_{L}\to U_{PMNS}\nu_{L}$
et les courants chargés sont affectés de la manière suivante:
\begin{equation}
\mathcal{L}_{CC}=\frac{g}{\sqrt{2}}\sum_{I}W_{\mu}^{+}(\bar{\nu}_{L}^{J}U_{PMNS}^{\ast JI}\gamma^{\mu}e_{L}^{I}+\bar{u}_{L}^{I}\gamma^{\mu}V_{CKM}^{IJ}d_{L}^{J})+h.c.
\end{equation}
Le terme leptonique n'est plus diagonal dans l'espace des saveurs
et les mélanges de saveurs deviennent possibles.

\subsubsection{Masse de Majorana et mécanismes de Seesaw}

On rappelle que le spineur conjugué de charge est défini par $\psi^{C}=C(\bar{\psi})^{T}$,
où C est la matrice de conjugaison de charge $C=i\gamma^{2}\gamma^{0}$.
Les champs libres et leurs conjugués de charge doivent satisfaire
la même équation de Dirac avec la même masse:$(i\cancel{\partial}-m)\psi=0$
et $(i\cancel{\partial}-m)\psi^{C}=0$. De plus, on a $(\psi_{L})^{C}=(\psi^{C})_{R}$
et $(\psi_{R})^{C}=(\psi^{C})_{L}$, autrement dit le conjugué de
charge d'un champ chiral gauche est un champ chiral droit et inversement.
Par ailleurs, sous une transformation de Lorentz $\psi^{C}$ se transforme
comme $\psi$. Par conséquent, la forme bilinéaire $\bar{\psi}\psi^{C}$
est un invariant de Lorentz aussi acceptable que $\bar{\psi}\psi$
pour jouer le rôle de terme de masse d'un champ fermionique. Cependant,
si $\psi$ possède une charge scalaire conservée alors le lagrangien
est invariant sous $U(1)$. Le terme $\bar{\psi}\psi$ est invariant
sous $U(1)$ $(\psi\to\psi^{\prime}=e^{i\theta}\psi)$. Par contre,
le terme $\bar{\psi}\psi^{C}$ ne l'est pas $\bar{\psi}^{\prime}\psi^{\prime C}=e^{-2i\theta}\bar{\psi}\psi^{C}$.
De ce fait, comme il n'y a pas de symétrie sous $U(1)$, il n'y a
pas de charge et donc $\psi$ doit être totalement neutre. Le neutrino
est donc un bon candidat pour avoir un terme de masse construit à
partir de cette forme bilinéaire $-\frac{1}{2}\bar{\psi}M\psi^{C}$.
La masse présente dans ce terme n'a à priori rien à voir avec la masse
de Dirac et est appelée \emph{masse de Majorana}. 

On introduit une masse de Majorana pour le neutrino:
\begin{equation}
\mathcal{L}=\mathcal{L}_{SM}-\frac{1}{2}N^{I}M^{IJ}N^{J}-N^{I}(Y_{\nu})^{IJ}L^{J}H+h.c.
\end{equation}
Une des conséquences notable d'un tel terme de masse est que le nombre
leptonique total n'est plus conservé. Cela nous montre à quel point
sa conservation est accidentelle dans le MS. Celle ci disparait dès
lors qu'on étend son contenu en particules de la façon la plus simple
possible. Néanmoins, le point essentiel sans doute est le fait que
la masse de Majorana n'est pas liée à l'échelle électrofaible et donc
n'a rien à voir avec le mécanisme de Higgs. Conséquemment, elle peut
prendre n'importe quelle valeur et en particulier des valeurs (très)
grande par rapport à l'échelle électrofaible. Supposons que les neutrinos
droits $N^{I=1,2,3}$ soient très lourds et voyons ce qu'il se passe
dans ce cas:

On peut toujours prendre $M=diag(M_{1},M_{2},M_{3})$ sans perte de
généralité. On résout l'équation d'Euler-Lagrange pour les neutrinos
droits:
\begin{equation}
\frac{\partial\mathcal{L}}{\partial N^{I}}=M^{IJ}N^{J}+(Y_{\nu})^{IJ}L^{J}H=0,
\end{equation}
ce qui nous donne,
\begin{equation}
N^{K}=-(M^{-1}Y_{\nu})^{KJ}L^{J}H,
\end{equation}
En injectant la solution trouvée dans le lagrangien, on obtient \emph{l'opérateur
de Weinberg de dimension-5}, qui après SSB donne\emph{:
\begin{equation}
\mathcal{L}=\mathcal{L}_{SM}+\frac{1}{2}(L^{I}H)(Y_{\nu}^{T}M^{-1}Y_{\nu})^{IJ}(L^{J}H)+h.c.\overset{SSB}{\longrightarrow}\frac{v^{2}}{2}(Y_{\nu}^{T}M^{-1}Y_{\nu})^{IJ}\nu_{L}^{I}\nu_{L}^{J}+h.c,
\end{equation}
}qui n'est rien d'autre qu'un terme de masse de Majorana pour les
neutrinos gauches. On voit bien que si les masses des neutrinos droits
sont assez grandes, celles des neutrinos gauches seront très petites
(et inversement). C'est ce que l'on appelle le\emph{ mécanisme de
seesaw} (ou le \emph{mécanisme de la bascule).} Il est à ce jour considéré
comme le mécanisme le plus séduisant pour expliquer les faibles masses
des neutrinos sans avoir à ajuster finement les paramètres libres
du lagrangien. Cependant, pour expliquer les valeurs mesurées des
masses des neutrinos gauches, il faut que $M_{1,2,3}=10^{9}-10^{15}GeV$,
échelle relativement proche de l'échelle GUT\footnote{Grand Unified Theories.}
(échelle de grande unification). Il est donc tout à fait envisageable
que les neutrinos droits acquièrent leurs masses directement\emph{
}à cette échelle\footnote{C'est en fait le cas dans les théories $SO(10)$ GUT.}. 

A basse énergie, les rotations nécessaires pour passer des états propres
de jauge aux états propres de masse des leptons, issues respectivement
de la décomposition en valeurs singulières de $Y_{e}$ pour les leptons
chargés (que nous avons déjà vue dans la Section \ref{Ch1Sec5}) et
de la diagonalisation de la matrice symétrique\footnote{En effet, $Y_{\nu}^{T}M^{-1}Y_{\nu}$ est symétrique donc diagonalisable
orthogonalement.} $\Upsilon_{\nu}\equiv Y_{\nu}^{T}M^{-1}Y_{\nu}$ pour les neutrinos
sont:
\begin{equation}
vV_{R}^{e}Y_{e}V_{L}^{e}=\mathbb{M}_{e},\hspace{1em}v^{2}V_{L}^{\nu}{}^{T}\Upsilon_{\nu}V_{L}^{\nu}=\mathbb{M}_{\nu}.
\end{equation}
Comme précédemment, on ne peut pas atteindre simultanément (avec une
même transformation) les états propre de masse des deux composantes
du doublet de lepton $L=\left(\begin{array}{c}
\nu_{L}\\
e_{L}
\end{array}\right)$. On choisit de tourner $L$ de $V_{L}^{e}$ et on atteint la base
dans laquelle:
\begin{equation}
vY_{e}=\mathbb{M}_{e},\hspace{1em}v^{2}\Upsilon_{\nu}=V_{L}^{eT}V_{L}^{\nu\ast}\mathbb{M}_{\nu}V_{L}^{\nu\dagger}V_{L}^{e}\equiv U_{PMNS}^{\ast}\mathbb{M}_{\nu}U_{PMNS}^{\dagger},\label{eq:SpurionsMajorana}
\end{equation}
où $U_{PMNS}\equiv V_{L}^{e\dagger}V_{L}^{\nu}$.

Faisons un point sur les phases. Le terme de masse de Majorana pour
l'état propre de masse $\nu_{L}$ est donné par:
\begin{equation}
\mathcal{L}_{M}=\frac{1}{2}\sum_{I}(m_{I}\nu_{L}^{I}\nu_{L}^{I}+m_{I}^{\ast}\nu_{L}^{I\dagger}\nu_{L}^{I\dagger}).
\end{equation}
Pour une saveur donné I, l'équation du mouvement du neutrino est donné
par $(\partial^{2}+\lvert m_{I}\rvert^{2})\nu_{L}^{I}=0$ et donc
M et $\mathbb{M}_{\nu}$ ne doivent pas forcément être réels. Par
convention, ces phases ne sont pas laissées dans $\mathbb{M}_{\nu}$
mais transférées dans la matrice PMNS. En effet, grâce à la présence
de $U_{PMNS}^{\ast}$ et de $U_{PMNS}^{\dagger}$ dans l'équation
(\ref{eq:SpurionsMajorana}), on peut transférer la phase de chaque
coefficient diagonal de $\mathbb{M}_{\nu}$ dans $U_{PMNS}$ dans
la mesure où une phase globale n'est pas pertinente. Ces phases appelées
\emph{phases de Majorana} sont codées dans la redéfinition:
\begin{equation}
U_{PMNS}\rightarrow U_{PMNS}\cdot diag(1,e^{i\alpha_{M}},e^{i\beta_{M}}).
\end{equation}

\section{Un processus violant CP : le Moment Dipolaire Électrique}

\subsection{Définitions et violations de symétries}

Dans une distribution de charges quelconque, dès lors que les barycentres
des charges positives et négatives ne coïncident pas, il existe une
quantité traduisant cette séparation de charges appelée \emph{moment
dipolaire électrique} (EDM) qui est alors non nulle. L'EDM $\overrightarrow{d}$
d'un système est nécessairement de même direction que son moment cinétique
moyen $\bar{h}\langle J\rangle$, car étant la seule direction privilégiée.
Classiquement, L'EDM est défini relativement à son centre de masse
($\overrightarrow{r}=\overrightarrow{0}$) de la manière suivante
\begin{equation}
\overrightarrow{d}=\varint\overrightarrow{r}\rho_{Q}d^{3}r=d\frac{\langle\overrightarrow{J}\rangle}{J},
\end{equation}
avec $\rho_{Q}$ la distribution de charge volumique. De façon analogue,
on définit le moment dipolaire magnétique
\begin{equation}
\overrightarrow{\mu}=\frac{1}{2}\varint\overrightarrow{r}\wedge\overrightarrow{J_{Q}}d^{3}r=\mu\frac{\langle\overrightarrow{J}\rangle}{J},
\end{equation}
où $\overrightarrow{J_{Q}}$ est la densité de courant électrique.

En mécanique quantique non-relativiste, une interaction est décrite
par un opérateur hamiltonien (ou lagrangien de façon équivalente).
Ainsi, l'interaction entre un champ électromagnétique et un système
quantique de moment magnétique $\mu$ et d'EDM $D$ est dictée par
le hamiltonien suivant
\begin{equation}
H=-(\overrightarrow{\mu}\cdot\overrightarrow{B}+\overrightarrow{D}\cdot\overrightarrow{E})=-\left(\mu\frac{\overrightarrow{J}\cdot\overrightarrow{B}}{J}+D\frac{\overrightarrow{J}\cdot\overrightarrow{E}}{J}\right),\label{eq:EDMHint}
\end{equation}
avec $\overrightarrow{J}$ l'opérateur moment cinétique. Le champ
magnétique $\overrightarrow{B}$ et l'opérateur moment cinétique $\overrightarrow{J}$
sont tous les deux symétriques sous $\mathcal{P}$ mais antisymétriques
sous $\mathcal{T}$. Quant au champ électrique $\overrightarrow{E}$,
il est antisymétrique sous $\mathcal{P}$ mais symétriques sous $\mathcal{T}$.
Ainsi, le second terme en $\overrightarrow{J}\cdot\overrightarrow{E}$
est alors antisymétrique à la fois sous $\mathcal{P}$ et $\mathcal{T}$.
Or, en vertu du théorème $\mathcal{CPT}$, une violation de $\mathcal{T}$
implique une violation de $\mathcal{CP}$. Donc le terme EDM viole
la symétrie $\mathcal{CP}$.

\begin{figure}[h]
\begin{center}
\includegraphics[scale=0.7]{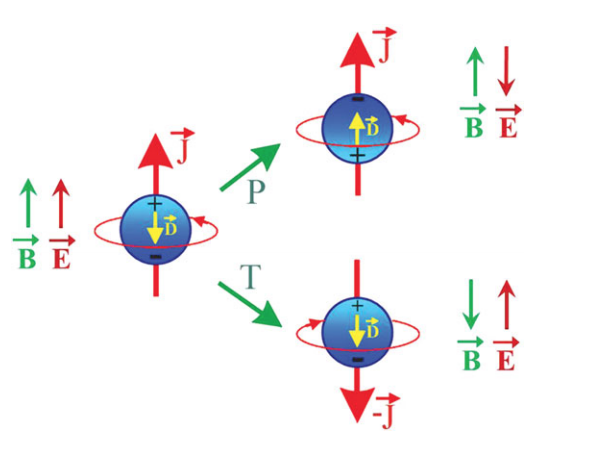}
\caption{Transformations des différentes quantités mises en jeu sous l'action des symétries $\mathcal{P}$ et $\mathcal{T}$. Source: Klaus Jungmann, Searching for electric dipole moments \cite{JungmannEDMsSearches}}
\label{EDM_CPV}
\end{center}
\end{figure}

En théorie quantique des champs, la densité lagrangienne d'interaction
correspondante est
\begin{equation}
\mathcal{L}_{dipoles}=-\frac{\mu}{2}\overline{\psi}\sigma^{\mu\nu}F_{\mu\nu}\psi-i\frac{d}{2}\overline{\psi}\sigma^{\mu\nu}\gamma^{5}F_{\mu\nu}\psi,\label{eq:Ldipoles}
\end{equation}
où $\psi$ est le champ fermionique et $F_{\mu\nu}=\partial_{\mu}A_{\nu}-\partial_{\nu}A_{\mu}$
le tenseur électromagnétique (contenant $A_{\mu}$ le champ vectoriel
photonique). Remarquons qu'à la différence du premier terme, le second
terme décrivant l'interaction de l'EDM contient la matrice de Dirac
$\gamma^{5}$, qui indique une violation de la symétrie $\mathcal{P}$,
ainsi que le nombre imaginaire $i$, dont la présence traduit une
violation de $\mathcal{T}$ et constitue donc d'après l'invariance
$\mathcal{CPT}$, un signal direct de violation $\mathcal{CP}$.

\subsection{La quête d'un signal EDM non nul}

\paragraph{Principe de mesure\newline}

L'interaction d'une particule de spin 1/2 avec un champ électromagnétique
$(\overrightarrow{E},\mathbf{\overrightarrow{B}})$ est donnée par
le hamiltonien suivant
\begin{equation}
H=-\mu\overrightarrow{\sigma}\cdot\overrightarrow{B}-d\overrightarrow{\sigma}\cdot\overrightarrow{E},
\end{equation}
où $\mu$ et $d$ sont respectivement les moments magnétiques et électriques
de la particule et $\overrightarrow{\sigma}=(\sigma^{1},\sigma^{2},\sigma^{3})$
les matrices de Pauli.

Lorsque les champs électrique $\overrightarrow{E}$ et magnétique
$\overrightarrow{B}$ sont statiques et alignés, le spin de la particule
précesse autour de cette direction d'alignement des champs. La fréquence
de précession est la fréquence de Larmor $\frac{\omega_{L}}{2\pi}$
donnée par
\begin{equation}
\bar{h}\omega_{L}=-2\mu B-2dE.
\end{equation}

Afin de mesurer un signal EDM, l'approche usuelle consiste à appliquer
un champ électrique aussi intense que possible. Dans ces conditions,
la fréquence de précession électrique est quand même très petite devant
la fréquence de précession magnétique. Par exemple, pour un champ
magnétique de $1\mu T$, la fréquence de précession magnétique est
$\left|f_{L}\right|=\left|\frac{\omega_{L}}{2\pi}\right|=29Hz$ alors
que pour un EDM de $10^{-26}e\cdot cm$ dans un champ électrique intense
de $20kV/cm$, la fréquence de précession électrique est de seulement
$10^{-7}Hz$.

Les expériences sont alors accordées pour mesurer précisément la fréquence
de Larmor dans les deux situations où les champs $\overrightarrow{E}$
et $\overrightarrow{B}$ sont parallèles ou antiparallèles. L'EDM
est alors obtenu par
\begin{equation}
d=-\frac{\bar{h}\Delta\omega_{L}}{4E},
\end{equation}
avec $\Delta\omega_{L}$ la différence des fréquences dans les deux
situations. La fréquence de précession est mesurée en utilisant la
méthode de Ramsey des champs oscillants séparés \cite{Ramsey}.

\paragraph{Précision\newline}

La limite statistique atteignable de précision dans une expérience
d'EDM est donnée par
\begin{equation}
\delta d=\frac{\bar{h}}{P\epsilon\sqrt{N\tau T}E},
\end{equation}
où $P$ est la polarisation, $\epsilon$ l'efficacité (qui peut varier
grandement en fonction des expériences et des systèmes), $T$ le temps
de mesure, $N$ le nombre de particules dans le volume observé, $\tau$
le temps de cohérence des spins et $E$ le champ électrique appliqué.
Pour des paramètres typiques d'expériences réalistes $P\approx1,\epsilon\approx0.1,N\approx10^{6},\tau\approx1s,E=10^{5}V/cm$,
on obtient une précision de $\delta d\approx10^{-27}e\cdot cm$ pour
un jour de mesure ($T\approx10^{5}s$). Généralement, la limitation
statistique n'est pas l'obstacle principal alors que les effets systématiques
sont bien plus limitants et les contrôler nécessite un effort considérable
dans les expériences de haute précision. Le lecteur désireux d'en
savoir plus à propos des différentes expériences cherchant un EDM
permanent pourra se référer à la revue \cite{Kirch}.

\paragraph{Systèmes utilisés pour chercher des EDMs\newline}

\begin{figure}[h]
\begin{center}
\includegraphics[scale=0.7]{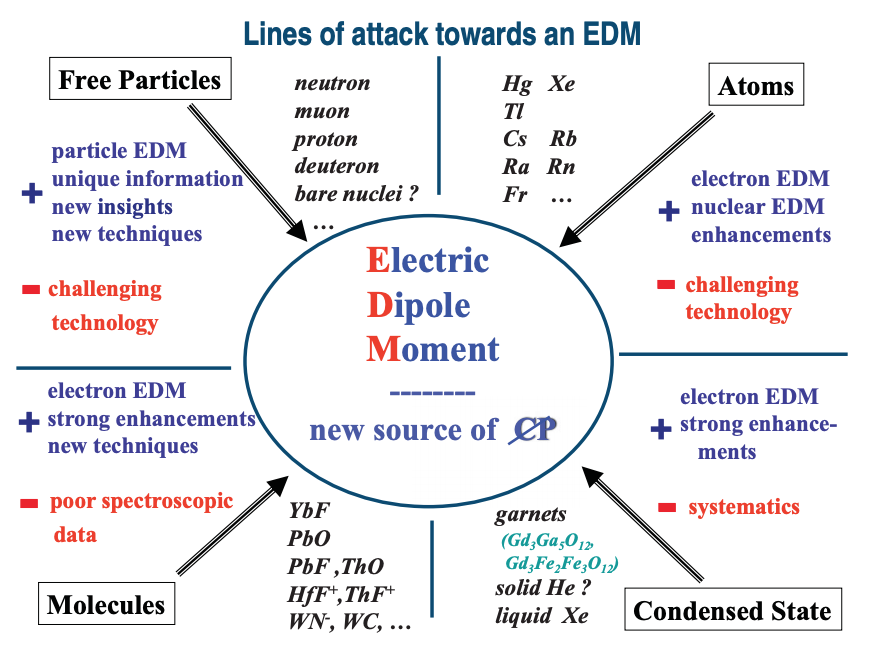}
\caption{Les différentes lignes d'attaques expérimentales dans la recherche d'EDMs. Source: Klaus Jungmann, Searching for electric dipole moments \cite{JungmannEDMsSearches}}
\label{SystemsEDMSearches}
\end{center}
\end{figure}

Plusieurs systèmes différents sont utilisés afin de mesurer un EDM.
On distingue quatre grandes catégories de systèmes des plus petits
aux plus grands:
\begin{itemize}[label=\textbullet]
\item Les particules élémentaires et nucléons (l'électron, le muon, le
tau, le neutron ou le proton...).
\item Les atomes et les ions (mercure, xenon, thallium, cesium, radon, francium,
radium, ...).
\item Les molécules (fluorure de thallium (TIF), fluorure d'ytterbium (YbF),
oxyde de plomb (PbO), ion fluorure de thorium $ThF^{+}$, ...)
\item De la matière condensée (matériaux ferroélectriques, Xenon liquide,
...)
\end{itemize}
Les systèmes utilisés par les expérimentateurs ainsi que leurs avantages
et inconvénients sont résumés dans la figure \ref{SystemsEDMSearches}.

\subsection{Contraintes établies et impacts sur la nouvelle physique}

Plusieurs expériences de précisions impliquant différents systèmes
sont réalisées dans le monde afin de mesurer un signal EDM non nul.
Le fait de mesurer un EDM de fermion non nul pourrait révéler de nouvelles
sources de violations-$\mathcal{CP}$ au-delà du MS. La précision
des expériences dédiées à cette quête a augmentée considérablement,
fruit de plusieurs décennies de développement. Cependant, les résultats
sont toujours compatibles avec une valeur d’EDM nulle, et ce pour
plusieurs systèmes différents (pour une review récente voir \cite{JungmannEDMsSearches}),
tous compatibles avec zéro. Il n’en est pas moins que ces expériences
ont un potentiel de découverte fort dans la mesure où elles sont sensibles
à certains effets de nouvelle physique siégeant à des échelles d’énergies
au-delà de ce qui est accessible aux collisionneurs.

Les différentes mesures d'EDMs ont permis d'établir des bornes supérieures
de l'ensemble des valeurs (absolues) possibles de ces derniers. Nous
présentons ici (figure \ref{EDMLimits}) plusieurs limites actuelles
sur les EDMs de particules ainsi que les facteurs d'améliorations
des précisions expérimentales nécessaires afin d'atteindre les prédictions
du MS. Concernant les EDMs des électrons, des neutrons et des muons,
plusieurs modèles de physique au-delà du modèle standard prédisent
des valeurs atteignables par les expériences actuelles en cours ou
proposées pour un futur proche.

\begin{table}[H]
\makebox[\textwidth][c]{%
\begin{tabular}{|c|c|c|c|}
\hline 
Particule & Limite à 95\% de confiance{[}$e\cdot cm${]} & Référence & facteur d'écart avec la prédiction du MS\tabularnewline
\hline 
\hline 
e & $9,7\times10^{-29}$ & \cite{expEDMe} & $10^{11}$\tabularnewline
\hline 
$\mu$ & $1,8\times10^{-19}$ & \cite{Bennett} & $10^{8}$\tabularnewline
\hline 
$\tau$ & $3,9\times10^{-17}$ & \cite{BelleEDM} & $10^{7}$\tabularnewline
\hline 
n & $3,6\times10^{-26}$ & \cite{neutronEDM} & $10^{4}$\tabularnewline
\hline 
p & $0,54\times10^{-23}$ & \cite{protonEDM} & $10^{6}$\tabularnewline
\hline 
$\nu_{e,\mu}$ & $2\times10^{-21}$ & \cite{tauEDM} & pas d'EDM dans le MS\footnote{En effet, seuls les neutrinos massifs ont un EDM non nul. Or, dans
le MS ils sont de masses nulles.}\tabularnewline
\hline 
$\nu_{\tau}$ & $5,2\times10^{-17}$ & \cite{tauNuEDM} & pas d'EDM dans le MS\tabularnewline
\hline 
\end{tabular}}

\caption{Tableau présentant les limites actuelles à 95\% de confiance sur les
EDMs de certaines particules ainsi que le facteur d'écart par rapport
aux prédictions du MS. }

\label{EDMLimits}
\end{table}

\paragraph{Impact sur la nouvelle physique\newline}

De façon générale, outre les deux sources de violation-$\mathcal{CP}$
déjà présentes dans le MS, à savoir la phase complexe de la matrice
CKM dans le secteur faible ou encore le terme $\bar{\theta}$ dans
le secteur fort, les nouvelles sources de violation-$\mathcal{CP}$
au-delà du MS sont encapsulées dans un lagrangien effectif $\mathcal{L}_{BSM}$.
Ainsi, le lagrangien associé à la violation-$\mathcal{CP}$ (CPV)
peut s'écrire
\begin{equation}
\mathcal{L}_{CPV}=\mathcal{L}_{CKM}+\mathcal{L}_{\bar{\theta}}+\mathcal{L}_{BSM}.
\end{equation}
Cette approche de théorie effective permet de connecter les résultats
expérimentaux aux théories plus fondamentales à haute énergie de façon
indépendante des modèles. En effet, l'effet des processus à haute
énergie ($E\ll\Lambda$) est codé dans les coefficients de Wilson
associés aux opérateurs effectifs non-renormalisables impliquant uniquement
les champs du MS. Dans ce formalisme, on effectue un développement
limité en $\tfrac{v}{\Lambda}$ où $v$ est la VEV du champs de Higgs.
Par conséquent, les amplitudes correspondantes sont en $\left(\frac{v}{\Lambda}\right)^{d-4}$,
où $d$ est la dimension en énergie de l'opérateur associé.

La nouvelle physique entre en jeu à partir des opérateurs de dimensions
6
\begin{equation}
\mathcal{L}_{BSM}\rightarrow\mathcal{L}_{CPV}^{eff}=\sum_{k,d}\alpha_{k}^{(d)}\left(\frac{1}{\Lambda}\right)^{d-4}\mathcal{O}_{k}^{(d)},
\end{equation}
où les $\alpha_{k}^{(d)}$ sont les coefficients de Wilson associés
aux opérateurs $\mathcal{O}_{k}^{(d)}$ invariants sous $SU(2)\times U(1)$
et ne contenant que des champs du MS. L'indice $k$ indexe les différent
opérateurs de dimension $d$. Notons que si nous travaillons seulement
avec la première génération du MS, il est suffisant de ne considérer
que les opérateurs de dimension 6 en tronquant la série ci-dessus.
Le travail du phénoménologue consiste alors à transcrire les nouvelles
mesures expérimentales en contraintes sur les coefficients de Wilson
associés aux opérateurs effectifs de physique au-delà du MS considérés
dans $\mathcal{L}_{CPV}^{eff}$.

Comme de nouvelles interactions violant $\mathcal{CP}$ (possiblement
médiées par des particules inconnues) sont requises afin d’expliquer
l’asymétrie matière-antimatière dans l’univers, un EDM peut être induit
par les effets virtuels de ces particules inconnues. Dans cette approche
de théorie effective, indépendante des modèles de nouvelle physique
considérés, les limites obtenues expérimentalement peuvent être traduites
en contraintes sur les modèles violant $\mathcal{CP}$ au-delà du
MS à l’échelle du TeV \cite{Pospelov,EngelRamseyKolck2013} ou même
à l’échelle du PeV \cite{McKeenPospelovRitz2013,AHZ2013}.

De plus, les EDMs sont très sensibles pour sonder la violation-$\mathcal{CP}$
dans le secteur de Higgs \cite{ChienCirigliano2016,ChienLi2018}.
Par conséquent, les expériences de précisions mesurant des EDMs ont
un grand potentiel de découverte et il est important de mesurer les
EDMs de différents systèmes (électron, muon, neutron, proton, atomes,
molécules, etc...) car elles constituent des informations complémentaires
\cite{Pospelov,EngelRamseyKolck2013,McKeenPospelovRitz2013,AHZ2013,ChienCirigliano2016,ChienLi2018,JungmannEDMsSearches,ChuppRamsey2015}.

\chapter{Estimation d’observables à l’aide de la symétrie de saveur\label{Ch3}}

Cette étude est basée sur l'article \cite{ST-EDM} dont quelques résultats
importants sont exposés dans \cite{TouatiMorEW18}.

\minitoc

\section{Introduction}

Dans le MS, il y a deux sources de violation-$\mathcal{CP}$. La première
est intimement liée à la physique de la saveur, elle provient des
couplages de Yukawa des quarks et est encodée dans la matrice CKM.
Pour être ressentie dans des observables, elle nécessite des transitions
de saveurs réelles ou virtuelles, et a été largement scrutée expérimentalement
dans les systèmes de mésons étranges $K$ et de mésons beaux $B$.
Quant à la seconde source, elle est plus particulière. Elle peut également
être encodée dans les couplages de Yukawa des quarks mais est intrinsèquement
insensible à la saveur et reçoit une contribution de la dynamique
de QCD. Elle est censée induire des effets conséquents à travers des
processus sans changement de saveur violant-$\mathcal{CP}$ comme
un EDM pour le neutron. Cependant, cela n'est guère confirmé expérimentalement
posant l'une des énigmes les plus sérieuses du MS.

Ces deux types de phases violant-$\mathcal{CP}$ peuvent être généralisée
au-delà du MS. Le premier type est sensible aux transitions de saveurs
et provient des mécanismes générant les masses des quarks et des leptons.
Par exemple, dans un contexte supersymétrique, les masses des squarks
et des sleptons apporteraient ce type de phases. Dès lors que la violation
minimale de saveur\footnote{Minimal Flavor Violation (MFV).} \cite{DambrosioGIS02}
est imposée, l'impact de ces nouvelles phases de saveurs sur les observables
est limitée, et ce même si l'échelle de nouvelle physique est de l'ordre
du TeV. D'autre part, le second type de phases est insensible à la
saveur et peut provenir d'un secteur scalaire étendu contenant des
paramètres complexes ou encore de la dynamique non-perturbative des
champs de jauges. Dans un cadre supersymétrique, les masses des gauginos
génèrent aussi des phases de ce type. Etant déjà problématiques dans
le MS, ces phases posent un défi pour tout modèle de nouvelle physique
voulant les accommoder dans la mesure où aucune dynamique spécifique
n'est présente pour les asservir.

Le but de cette étude est d'analyser l'impact des phases savoureuses\footnote{Flavored phases. Ce sont les phases sensibles aux transitions de saveurs
réelles ou virtuelles.} sur des observables non-savoureuses\footnote{Flavor-blind, c'est à dire insensible à la saveur.}
que sont les EDMs des quarks et des leptons. Pour procéder, on considère
le lagrangien effectif où l'on paramétrise les opérateurs magnétiques
de la façon suivante
\begin{equation}
\mathcal{L}_{eff}=e\frac{c_{u}}{\Lambda^{2}}(\bar{U}\mathbf{Y}_{u}\mathbf{X}_{u}\sigma_{\mu\nu}Q)H^{\dagger}F^{\mu\nu}+e\frac{c_{d}}{\Lambda^{2}}(\bar{D}\mathbf{Y}_{d}\mathbf{X}_{d}\sigma_{\mu\nu}Q)H^{\dagger}F^{\mu\nu}+e\frac{c_{e}}{\Lambda^{2}}(\bar{E}\mathbf{Y}_{e}\mathbf{X}_{e}\sigma_{\mu\nu}L)H^{\dagger}F^{\mu\nu}+...\;,\label{eq:EMO}
\end{equation}
où $Q,L,U,D,E$ sont respectivement les doublets faibles de quarks
et de leptons ainsi que les singlets faibles de quarks de type up,
quarks de type down et leptons chargés. Il s'agit alors d'étudier
les phases provenant des structures de saveurs $\mathbf{Y}_{u,d,e}\mathbf{X}_{u,d,e}$,
qui sont des matrices $3\times3$ dans l'espace des saveurs. Cela
inclut non seulement toutes les phases savoureuses mais encore les
phases non-savoureuses émergeant des couplages de quarks et de leptons
au(x) boson(s) de Higgs ou plus généralement celles qui peuvent être
absorbées dans ces couplages, comme par exemple la phase de violation-$\mathcal{CP}$
forte dans le secteur des quarks ou la phase globale de Majorana dans
le secteur des leptons. Au contraire, les phases non-savoureuses ne
provenant pas des couplages de quarks et de leptons sont nécessairement
encodées dans les coefficients de Wilson $c_{u,d,e}$, considérés
comme réels dans la suite.

Dans le MS, les structures de saveurs $\mathbf{X}_{u,d,e}$ sont des
polynômes en $\mathbf{Y}_{u,d,e}$ étant donné que ce sont les seules
structures de saveurs disponibles. A la manière de MFV \cite{DambrosioGIS02},
qui est exact dans le MS, la forme de ces polynômes peut être établie
directement en utilisant la symétrie de saveur et en traitant les
matrices de couplages de Yukawa brisant cette symétrie comme des spurions.
On a alors la partie imaginaire du coefficient 1-1 de la matrice $\mathbf{X}_{e}$,
$\operatorname{Im}\mathbf{X}_{e}^{11}$, contrôlant l'intensité de
l'EDM de l'électron, qui sera proportionnelle au déterminant de Jarlskog
$\det[\mathbf{Y}_{u}^{\dagger}\mathbf{Y}_{u},\mathbf{Y}_{d}^{\dagger}\mathbf{Y}_{d}]$,
car c'est le seul invariant de saveur violant-$\mathcal{CP}$ qui
peut être construit dans le MS \cite{Jarlskog85}. Cependant, il s'avère
que ce ne soit pas une mesure fiable de la violation-$\mathcal{CP}$
en raison du fait que $\operatorname{Im}\mathbf{X}_{d}^{11}$, contrôlant
l'EDM du quark down, est plus grand d'au moins dix ordres de grandeurs.
La structure $\operatorname{Im}\mathbf{X}_{d}^{11}$ n'est pas proportionnelle
au déterminant de Jarlskog mais plutôt au commutateur $\mathbf{X}_{d}=\mathbf{[Y}_{u}^{\dagger}\mathbf{Y}_{u}\;,\;\mathbf{Y}_{u}^{\dagger}\mathbf{Y}_{u}\mathbf{Y}_{d}^{\dagger}\mathbf{Y}_{d}\mathbf{Y}_{u}^{\dagger}\mathbf{Y}_{u}]$.
L'obtention de ce commutateur à partir de la symétrie de saveur ainsi
que ses propriétés seront explorées dans la troisième section de ce
chapitre (Section \ref{Ch3Sec3}).

Notons que la structure commutateur n'est pas nouvelle en soit \cite{Shabalin1982sg}.
De plus, des calculs exacts de l'EDM des quarks ont même été menés
\cite{Khriplovich1985jr,CzarneckiK}. Néanmoins, son obtention avec
les invariants violant-$\mathcal{CP}$ en utilisant uniquement la
symétrie de saveur n'a jamais été présentée auparavant. En outre,
cela permettra de mettre en place la méthode en vue de la quatrième
section où la masse des neutrinos sera introduite. Les structures
de saveurs des neutrinos additionnelles permettent de nouvelles manières
de générer des parties imaginaires pour $\mathbf{X}_{u,d,e}$. Là
aussi, les invariants de saveurs ont été largement étudiés (voir par
exemple \cite{Branco11} et les références s'y trouvant), mais l'analyse
systématique des commutateurs (non-invariants) associés ne l'a pas
été. En ce qui concerne les contributions CKM aux EDMs des quarks
et des leptons, nous allons montrer que les invariants ne sont pas
adéquats pour estimer l'ordre de grandeur des EDMs des leptons, car
les commutateurs non-invariants sont en général bien plus grands.

Avant d'entrer dans le vif du sujet, nous allons commencer par mettre
en place la méthode d'estimation utilisée, basée sur la symétrie de
saveur en la calibrant d'abord sur des processus FCNCs de quarks comme
ceux vus dans la Section \ref{Ch1Sec5}. Puis, la contribution CKM
à l'EDM est analysée dans la troisième section. De plus, nous montrons
également comment étendre la méthode pour prendre en compte le terme
theta de violation-$\mathcal{CP}$ forte afin d'estimer les EDMs de
quarks et de leptons qu'il induit. Dans la quatrième section, les
masses des neutrinos entrent en jeu. Les structures de saveurs invariantes
et non-invariantes ajustant les EDMs sont construites séparément dans
le cas de masses de Dirac, de masses de Majorana ou encore dans le
cas des trois mécanismes de seesaw les plus simples. De plus, nous
nous penchons également sur les invariants violant le nombre leptonique
pouvant exister dans le cas de Majorana et d'estimer leur possible
impact sur les EDMs.

\section{Méthode d'estimation: Comment exploiter la symétrie de saveur du
MS?\label{FlavorSymmetryMethod}}

Le secteur de jauge du MS est invariant sous l'action du groupe de
symétrie globale suivant \cite{ChivukulaG87}
\begin{equation}
G_{F}=U(3)^{5}=U(3)_{Q}\otimes U(3)_{U}\otimes U(3)_{D}\otimes U(3)_{L}\otimes U(3)_{E}\;,
\end{equation}
appelée \emph{symétrie de saveur}. L'action de ce groupe est définie
telle que les doublets gauches et les singlets droits se transforment
comme la représentation $\mathbf{3}$ de leurs $U(3)$ respectifs,
i.e., $X\rightarrow g_{X}X\ensuremath{,}g_{X}\in U(3)_{X}$ pour $X=Q,L,U,D,E$.
Cependant, cette symétrie n'est pas exacte dans le MS. En effet, elle
est explicitement brisée par les couplages de Yukawa,
\begin{equation}
\mathcal{L}_{\text{Yukawa}}=-\bar{U}\mathbf{Y}_{u}QH^{\dagger C}-\bar{D}\mathbf{Y}_{d}QH^{\dagger}-\bar{E}\mathbf{Y}_{e}LH^{\dagger}+h.c.\;.\label{eq:YukawaGauge0}
\end{equation}
Ces interactions brisent $G_{F}$ puisqu'ils mélangent différents
types de fermions.

Pour mettre en place le cadre pour l'étude ultérieure des EDMs, le
but de cette section est de montrer comment la symétrie de saveur
$G_{F}$ peut être utilisée pour établir immédiatement et analyser
les structures de saveurs des FCNCs. Premièrement, si on applique
une transformation de $G_{F}$ au lagrangien de Yukawa, on obtient
\begin{equation}
\mathcal{L}_{Yukawa}=-(\bar{U}g_{U}^{\dagger})Y_{u}(g_{Q}Q)H^{\dagger C}-(\bar{D}g_{D}^{\dagger})Y_{d}(g_{Q}Q)H^{\dagger}-(\bar{E}g_{E}^{\dagger})Y_{e}(g_{L}L)H^{\dagger}+h.c.,
\end{equation}
où $g_{X}\in U(3)_{X}$ avec $X\in\{Q,U,D,L,E\}$. On voit qu'on peut
formellement restaurer la symétrie $G_{F}$ si les matrices de couplages
$Y_{u,d,e}$ sont promues au rang de champs qui se transforment d'une
certaine manière sous $G_{F}$, c'est à dire au rang de \emph{spurions}\footnote{Un spurion est un champ statique issue de la promotion d'un paramètre
de brisure de symétrie au rang de champ.}. Pour que le lagrangien du MS soit invariant sous $G_{F}$, il suffit
que les spurions transforment ainsi
\begin{equation}
\begin{aligned}\mathbf{Y}_{u} & \sim\left(\mathbf{\bar{3}},\mathbf{3},\mathbf{1},\mathbf{1},\mathbf{1}\right)_{G_{F}}:\mathbf{Y}_{u}\overset{G_{F}}{\rightarrow}g_{U}\mathbf{Y}_{u}g_{Q}^{\dagger}\;,\;\\
\mathbf{Y}_{d} & \sim\left(\mathbf{\bar{3}},\mathbf{1},\mathbf{3},\mathbf{1},\mathbf{1}\right)_{G_{F}}:\mathbf{Y}_{d}\overset{G_{F}}{\rightarrow}g_{D}\mathbf{Y}_{d}g_{Q}^{\dagger}\;,\\
\mathbf{Y}_{e} & \sim\left(\mathbf{1},\mathbf{1},\mathbf{1},\mathbf{\bar{3}},\mathbf{3}\right)_{G_{F}}:\mathbf{Y}_{e}\overset{G_{F}}{\rightarrow}g_{E}\mathbf{Y}_{e}g_{L}^{\dagger}\;.
\end{aligned}
\end{equation}
Cette manipulation purement formelle va s'avérer très fructueuse.
En effet, dès lors que le lagrangien du MS devient invariant sous
$G_{F}$, il doit en aller de même pour l'amplitude de n'importe quel
processus. En insérant des spurions de Yukawa de manière $G_{F}$-invariante
dans l'amplitude du processus, on peut établir sa structure de saveur
sans aucuns calculs. De plus, dès lors que nous souhaiterons mettre
des chiffres dessus afin d'effectuer des prédictions quantitatives,
nous gèlerons à nouveau les spurions à leurs valeurs physiques. Pour
les identifier, considérons les couplages de Yukawa après la brisure
spontanée de la symétrie électrofaible (SSB),
\begin{equation}
\mathcal{L}_{\text{Yukawa}}=-v\left(\bar{u}_{R}\mathbf{Y}_{u}u_{L}+\bar{d}_{R}\mathbf{Y}_{d}d_{L}+\bar{e}_{R}\mathbf{Y}_{e}e_{L}\right)\left(1+\frac{h}{v}\right)+h.c.\;.
\end{equation}
A priori, aucun de ces couplages n'est diagonal dans l'espace des
saveurs. De la même manière que dans le Chapitre \ref{Ch1}, Section
\ref{Ch1Sec5}, on effectue une décomposition en valeurs singulière
(équation (\ref{eq:SVD1}))
\begin{equation}
\exists V_{R}^{u,d,e},V_{L}^{u,d,e}\in U(3)\hspace{1em}\text{telles que}\hspace{1em}vV_{R}^{u,d,e}\mathbf{Y}_{u,d,e}V_{L}^{u,d,e}=\mathbb{M}_{u,d,e},\label{eq:SVD}
\end{equation}
avec les matrices de masses $\mathbb{M}_{u,d,e}$ diagonales. Ainsi,
une transformation de $G_{F}$ invariante de jauge avec $g_{U,D,E}=V_{R}^{u,d,e}$,
$g_{L}=V_{L}^{e}$ et par exemple $g_{Q}=V_{L}^{d}$ donne
\begin{equation}
v\mathbf{Y}_{u}\overset{gel\acute{e}}{\longrightarrow}\mathbb{M}_{u}V_{CKM},\hspace{1em}v\mathbf{Y}_{d}\overset{gel\acute{e}}{\longrightarrow}\mathbb{M}_{d},\hspace{1em}v\mathbf{Y}_{e}\overset{gel\acute{e}}{\longrightarrow}\mathbb{M}_{e},\label{G=0000E8leSpurions}
\end{equation}
où $V_{CKM}\equiv V_{L}^{u}V_{L}^{d\dagger}$. Ce sont les valeurs
physiques des spurions dans la base d'interaction dans laquelle tous
les quarks sauf ceux de type up $u_{L}$ sont états propres de masses
(la base où tous sauf les quarks de type down $d_{L}$ sont états
propres de masses transférerait $V_{CKM}$ dans $\mathbf{Y}_{d}$).
Dans ces conditions, il est immédiat de prévoir que certains processus
sont très supprimés par rapport à d'autres en raison de la forte hiérarchie
des valeurs numériques dans $\mathbb{M}_{u,d,e}$ et $V_{CKM}$. De
plus, on voit aussi directement que les FCNCs leptoniques ne sont
pas autorisés étant donné que si on insère le spurion $\mathbf{Y}_{e}$
dans une amplitude avec des champs leptoniques externes, il sera gelé
à sa valeur diagonale $\mathbb{M}_{e}/v$ et ne pourra donc pas induire
de transitions de saveurs (hors-diagonale).

\begin{figure}[t]
\centering
\includegraphics[width=0.95\textwidth]{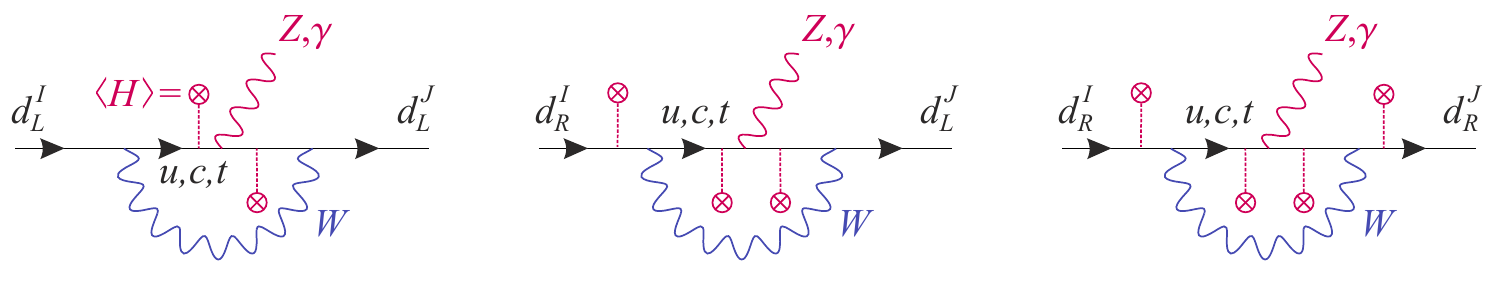}
\caption{Les diagrammes pingouins avec $Z$ et $\gamma$ contribuant à l'opérateur FCNC dans l'équation (\ref{eq:OpsFCNC}). Les insertions de spurions de Yukawa, représentées par les tadpoles avec une croix, permettent de briser la symétrie de saveur, voir équation (\ref{eq:MFVFCNC}).}
\label{FigFCNC}
\end{figure}

Voyons comment cette technique s'applique dans un cas pratique. Afin
d'utiliser la symétrie $G_{F}$ du secteur de jauge, nous nous placerons
avant brisure spontanée de symétrie électrofaible. Tous les couplages
renormalisables de dimension-4 en énergie sont déjà contenus dans
le lagrangien du MS et n'induisent pas de FCNC. En considérant les
opérateurs de dimension-6, on construit le lagrangien effectif suivant
pour estimer quatre processus FCNCs pris comme exemples d'applications
de la méthode
\begin{equation}
\mathcal{L}_{\text{eff}}=\frac{a_{1}}{\Lambda^{2}}(\bar{Q}\gamma_{\nu}Q)D_{\mu}F^{\mu\nu}+\frac{a_{2}}{\Lambda^{2}}(\bar{D}\gamma_{\nu}D)D_{\mu}F^{\mu\nu}+\frac{a_{3}}{\Lambda^{2}}(\bar{Q}\gamma_{\mu}Q)H^{\dagger}D^{\mu}H+\frac{a_{4}}{\Lambda^{2}}(\bar{D}\sigma_{\mu\nu}Q)H^{\dagger}F^{\mu\nu}+...\label{eq:OpsFCNC}
\end{equation}
Tous ces opérateurs sont générés dans le MS par des processus à boucles
(voir figure \ref{FigFCNC}), donc on pose $\Lambda^{2}\sim M_{W}^{2}/g^{2}\sim G_{F}^{-1}$.
Les coefficients de Wilson sont des matrices $3\times3$ dans l'espace
des saveurs, avec par exemple $a_{1}\bar{Q}\gamma_{\nu}Q\equiv a_{1}^{IJ}\bar{Q}^{I}\gamma_{\nu}Q^{J}$.
Comme au niveau fondamental les seuls couplages savoureux sont les
matrices de couplages de Yukawa, ces coefficients de Wilson sont fonctions
de $\mathbf{Y}_{u,d}$. En supposant que les dépendances sont polynomiales,
la symétrie de saveur $G_{F}$ impose les structures suivantes
\begin{equation}
\begin{aligned}a_{1},a_{3} & =\mathbf{1}\oplus\mathbf{Y}_{d}^{\dagger}\mathbf{Y}_{d}\oplus\mathbf{Y}_{u}^{\dagger}\mathbf{Y}_{u}\oplus...\;,\\
a_{2} & =\mathbf{1\oplus Y}_{d}\mathbf{Y}_{d}^{\dagger}\oplus\mathbf{Y}_{d}\mathbf{Y}_{u}^{\dagger}\mathbf{Y}_{u}\mathbf{Y}_{d}^{\dagger}\oplus...\;,\\
a_{4} & =\mathbf{Y}_{d}(\mathbf{1}\oplus\mathbf{Y}_{d}^{\dagger}\mathbf{Y}_{d}\oplus\mathbf{Y}_{u}^{\dagger}\mathbf{Y}_{u}\oplus...)\;,
\end{aligned}
\label{eq:MFVFCNC}
\end{equation}
où le signe $\oplus$ sert à indiquer que les coefficients de $\mathcal{O}(1)$
ne sont pas nécessairement les mêmes dans les différents termes du
développement. Une fois que les spurions ont été correctement introduit,
pour effectuer des prédictions quantitatives, nous pouvons les geler
à leurs valeurs physiques selon (\ref{G=0000E8leSpurions}). Remarquons
que la structure $Y_{u}^{\dagger}Y_{u}$ est celle qui va induire
un changement de saveur étant donné qu'elle est non diagonale. De
plus, si les masses des quarks de type up étaient toutes égales ($m_{u}=m_{c}=m_{t}=m$),
on aurait $\mathbb{M}_{u}\equiv\left(\begin{array}{ccc}
m_{u} & 0 & 0\\
0 & m_{c} & 0\\
0 & 0 & m_{t}
\end{array}\right)=m\mathbb{I}$ et par conséquent, $vY_{u}\overset{gel\acute{e}}{\longrightarrow}mV_{CKM}$,
$vY_{u}^{\dagger}\overset{gel\acute{e}}{\longrightarrow}mV_{CKM}^{\dagger}$
et donc du fait de l'unitarité de la matrice CKM $v^{2}Y_{u}^{\dagger}Y_{u}\overset{gel\acute{e}}{\longrightarrow}m^{2}\mathbb{I}$
qui est diagonal. Autrement dit, si les masses des quarks up étaient
égales, nous n'aurions pas pu induire un changement de saveur. De
ce fait, une différence de masse entre les quarks est nécessaire pour
avoir un changement de saveur. Le terme $Y_{u}^{\dagger}Y_{u}$ induit
une violation quadratique de GIM du fait de la grande masse du top.
Dans ce cas, $vY_{u}\overset{gel\acute{e}}{\longrightarrow}\mathbb{M}_{u}V_{CKM}$,
$vY_{u}^{\dagger}\overset{gel\acute{e}}{\longrightarrow}V_{CKM}^{\dagger}\mathbb{M}_{u}^{\dagger}$
et donc $v^{2}Y_{u}^{\dagger}Y_{u}\overset{gel\acute{e}}{\longrightarrow}V_{CKM}^{\dagger}\mathbb{M}_{u}^{\dagger}\mathbb{M}_{u}V_{CKM}$.
Comme $\mathbb{M}_{u}$ est diagonale à coefficients réels, elle est
donc auto-adjointe $\mathbb{M}_{u}^{\dagger}=\mathbb{M}_{u}$ et de
ce fait $v^{2}Y_{u}^{\dagger}Y_{u}\overset{gel\acute{e}}{\longrightarrow}V_{CKM}^{\dagger}\mathbb{M}_{u}^{2}V_{CKM}$
qui est alors non diagonal. On a alors
\begin{equation}
v^{2}(\mathbf{Y}_{u}^{\dagger}\mathbf{Y}_{u})^{IJ}=\sum_{q=u,c,t}m_{q}^{2}V_{qd^{I}}^{\ast}V_{qd^{J}}\approx m_{t}^{2}V_{td^{I}}^{\ast}V_{td^{J}}\;,
\end{equation}
et les coefficients de Wilson sont prédits
\begin{equation}
\begin{aligned}a_{1,3}^{I\neq J} & \rightarrow\alpha_{1,2}\frac{m_{t}^{2}}{v^{2}}V_{tI}^{\dagger}V_{tJ}\;,\\
a_{2}^{I\neq J} & \rightarrow\alpha_{3}\frac{m_{d^{I}}m_{d^{J}}}{v^{2}}\frac{m_{t}^{2}}{v^{2}}V_{tI}^{\dagger}V_{tJ}\;,\\
a_{4}^{I\neq J} & \rightarrow\alpha_{4}\frac{m_{d^{I}}}{v}\frac{m_{t}^{2}}{v^{2}}V_{tI}^{\dagger}V_{tJ}\;,
\end{aligned}
\end{equation}
avec $\alpha_{i}$ des nombres réels de $\mathcal{O}(1)$. Ceci montre
qu'en utilisant seulement la symétrie de saveur, nous sommes en mesure
de prévoir correctement non seulement les amplitudes CKM des transitions
FCNC mais aussi les basculements de chiralité (chirality flips). Dans
le cas ci-dessus, les opérateurs impliquant des quarks de chiralité
droites nécessitent des basculements de chiralité car le boson $W$
ne se couple qu'aux fermions gauches. Cela signifie en particulier
que $a_{2}\ll a_{1}$ car $m_{d,s,b}\ll v$, et par conséquent l'opérateur
correspondant peut être négligé.

En assemblant les différentes pièces, l'amplitude du pingouin avec
photon virtuel, induit par l'opérateur $(\bar{Q}\gamma_{\nu}Q)D_{\mu}F^{\mu\nu}$
est donnée par:
\begin{equation}
\mathcal{M}\left(\bar{d}^{I}d^{J}\rightarrow\gamma^{\ast}\left(q\right)\right)\sim\alpha_{1}\times G_{F}\times\frac{e}{4\pi^{2}}\times\bar{d}_{L}^{I}\gamma_{\mu}d_{L}^{J}\times(q^{\mu}q^{\nu}-q^{2}g^{\mu\nu})\times A_{\nu}\times\sum_{q}\frac{m_{q}^{2}}{v^{2}}V_{qd^{I}}^{\ast}V_{qd^{J}}\;.
\end{equation}
L'amplitude du pingouin avec boson $Z$, correspondant à l'opérateur
$(\bar{Q}\gamma_{\mu}Q)H^{\dagger}D^{\mu}H$, est
\begin{equation}
\mathcal{M}\left(\bar{d}^{I}d^{J}\rightarrow Z\left(q\right)\right)\sim\alpha_{3}\times G_{F}\times\frac{e}{4\pi^{2}\sin\theta_{W}}\times\bar{d}_{L}^{I}\gamma_{\mu}d_{L}^{J}\times v^{2}\times Z^{\mu}\times\sum_{q}\frac{m_{q}^{2}}{v^{2}}V_{qd^{I}}^{\ast}V_{qd^{J}}\;.\label{eq:PengZ}
\end{equation}
Le facteur $v^{2}$ d'amélioration par rapport au cas du pingouin
avec photon provient de $H^{\dagger}D^{\mu}H\overset{\text{SSB}}{\rightarrow}iv^{2}(g/2\cos\theta_{W})Z^{\mu}$.
Il découle de la violation de l'identité de Ward du boson $Z$ qui
permet de troquer le projecteur $q^{\mu}q^{\nu}-q^{2}g^{\mu\nu}$
contre le paramètre de brisure de $SU(2)_{L}$. Finalement, le dernier
opérateur est appelé \emph{opérateur magnétique de pingouin avec photon
(magnetic photon penguin operator)} et jouera un rôle particulier
dans cette étude. Après SSB, l'amplitude est
\begin{equation}
\mathcal{M}\left(\bar{d}^{I}d^{J}\rightarrow\gamma\left(q\right)\right)\sim\alpha_{4}\times G_{F}\times\frac{e}{4\pi^{2}}\times m_{d^{I}}\times\bar{d}_{R}^{I}\sigma_{\mu\nu}d_{L}^{J}\times F^{\mu\nu}\times\sum_{q}\frac{m_{q}^{2}}{v^{2}}V_{qd^{I}}^{\ast}V_{qd^{J}}\;.
\end{equation}

Il est intéressant de comparer ces estimations simples obtenues en
utilisant seulement la symétrie de saveur aux calculs à boucles exacts
dans le MS (voir par exemple \cite{Buras98} pour une revue). Hormis
certains facteurs numériques non-essentiels, la différence principale
se trouve dans les termes brisant GIM, où la brisure quadratique simple
est remplacée par une fonction de boucle dépendante du processus
\begin{equation}
\sum_{q}\frac{m_{q}^{2}}{v^{2}}V_{qd^{I}}^{\ast}V_{qd^{J}}\rightarrow\sum_{q}V_{qd^{I}}^{\ast}V_{qd^{J}}F(m_{q}^{2}/M_{W}^{2})\;.
\end{equation}
La fonction $F(m_{q}^{2}/M_{W}^{2})$ produit une brisure de GIM quadratique
pour le pingouin avec $Z$ mais pas pour les pingouins avec $\gamma$
\cite{Inami}. En particulier, $\bar{d}^{I}d^{J}\rightarrow\gamma^{\ast}\left(q\right)$
est seulement logarithmique, se comportant asymptotiquement comme
$\log m_{q}/M_{W}$ aux limites $m_{q}\rightarrow0$ et $m_{q}\rightarrow\infty$.
Cette différence est attendue lors de l'utilisation de $G_{F}$ seulement.
En effet, nous sommes forcés à travailler dans la phase non brisée
du MS où les fermions sont de masses nulles. Les insertions de spurions
sont comprises comme des insertions de tadpoles de Higgs qui se réduisent
après SSB à des insertions de masses. Bien que cela ne pose pas de
problème pour prévoir la structure de saveur, certains effets dynamiques
sont perdus dans un tel traitement perturbatif des masses des fermions.
En particulier, lorsque l'amplitude sans masse est dangereuse dans
l'infrarouge, la brisure de GIM quadratique s'adoucit en une brisure
logarithmique seulement. Pour notre propos, cela n'a pas de conséquence
mais doit être gardé en tête.

\section{Comment prédire l'EDM généré par la phase CKM?\label{Ch3Sec3}}

Le formalisme de la symétrie de saveur peut également être utilisé
pour des observables conservant la saveur. Par exemple, considérons
les opérateurs magnétiques diagonaux dans l'espace des saveurs de
l'équation (\ref{eq:EMO}). Après SSB, leur structure générale est\footnote{Rappelons que $2\sigma^{\mu\nu}\gamma_{5}=i\varepsilon^{\mu\nu\alpha\beta}\sigma_{\alpha\beta}$
et $\varepsilon^{\mu\nu\alpha\beta}F_{\alpha\beta}\equiv2\tilde{F}^{\mu\nu}$.}
\begin{equation}
\begin{aligned}\mathcal{H}_{{\rm eff}} & =c\,\bar{\psi}_{L}\sigma^{\mu\nu}\psi_{R}F_{\mu\nu}+c^{\ast}\,\bar{\psi}_{R}\sigma^{\mu\nu}\psi_{L}F_{\mu\nu}\\
 & =\left(\operatorname{Re}c\right)\bar{\psi}\sigma^{\mu\nu}\psi F_{\mu\nu}+i\left(\operatorname{Im}c\right)\bar{\psi}\sigma^{\mu\nu}\gamma_{5}\psi F_{\mu\nu}\equiv e\frac{a}{4m}\bar{\psi}\sigma^{\mu\nu}\psi F_{\mu\nu}+i\frac{d}{2}\bar{\psi}\sigma^{\mu\nu}\gamma_{5}\psi F_{\mu\nu}\;,
\end{aligned}
\end{equation}
ce qui définit le \emph{moment dipolaire électrique }$d$ (violant
$\mathcal{CP}$) et le \emph{moment magnétique anormal} $a=(g-2)/2$
(conservant $\mathcal{CP}$) de la particule $\psi$ (pour des revues
récentes, voir par exemple \cite{Pospelov,Raidal,Miller})). Les EDM
sont très supprimés dans le MS alors que la seconde observable est
mesurée avec une grande précision pour les leptons légers \cite{Pospelov,Raidal,Miller}.
En utilisant le formalisme de la symétrie de saveur, au-delà du fait
que $d_{\psi}\sim m_{\psi}$ et $a\sim m_{\psi}^{2}$ à partir de
la structure gauche-droite des opérateurs magnétiques de l'équation
(\ref{eq:EMO}), on peut aussi prévoir à partir de quel ordre la phase
de la matrice CKM peut générer un EDM pour un quark ou un lepton dans
le MS, ce que nous allons maintenant décrire en détails.

\subsection{EDMs de leptons générés par la phase CKM\label{SecCKMe}}

Pour commencer, considérons les EDMs des leptons obtenus à partir
de l'opérateur effectif $E\mathbf{Y}_{e}\mathbf{X}_{e}\sigma_{\mu\nu}LH^{\dagger}F^{\mu\nu}$
dans l'équation (\ref{eq:EMO}), où $\mathbf{X}_{e}$ représente une
chaîne de spurions insérée de manière $G_{F}$-invariante et $\Lambda\approx M_{W}$
dans le MS. Comme il n'y a pas de changement de saveur ici, l'opérateur
doit rester diagonal. Comme $\mathbf{Y}_{e}$ l'est déjà, seuls les
éléments diagonaux de $\mathbf{X}_{e}$ doivent intervenir. Ainsi,
$\mathbf{X}_{e}$ doit contenir $\mathbf{Y}_{u}$ et $\mathbf{Y}_{d}$
et l'invariance sous $G_{F}$ contraint la structure de la chaîne
de spurions à être la matrice identité multipliée par une trace complexe
(dans l'espace des saveurs) d'une chaîne de facteurs $\mathbf{Y}_{u}^{\dagger}\mathbf{Y}_{u}$
et $\mathbf{Y}_{d}^{\dagger}\mathbf{Y}_{d}$. En raison du fait que
ces facteurs soient hermitiens, la plus simple trace complexe de ce
type possible ne contient pas moins de douze spurions de Yukawa
\begin{equation}
\begin{aligned}\mathbf{X}_{e} & =\mathbf{\mathbb{I}}_{3}\times\langle(\mathbf{Y}_{d}^{\dagger}\mathbf{Y}_{d}\mathbf{)}^{2}\mathbf{Y}_{u}^{\dagger}\mathbf{Y}_{u}\mathbf{Y}_{d}^{\dagger}\mathbf{Y}_{d}(\mathbf{Y}_{u}^{\dagger}\mathbf{Y}_{u})^{2}-(\mathbf{Y}_{u}^{\dagger}\mathbf{Y}_{u})^{2}\mathbf{Y}_{d}^{\dagger}\mathbf{Y}_{d}\mathbf{Y}_{u}^{\dagger}\mathbf{Y}_{u}(\mathbf{Y}_{d}^{\dagger}\mathbf{Y}_{d}\mathbf{)}^{2}\rangle\\
 & =2i\mathbf{\mathbb{I}}_{3}\operatorname{Im}\langle(\mathbf{Y}_{d}^{\dagger}\mathbf{Y}_{d}\mathbf{)}^{2}\mathbf{Y}_{u}^{\dagger}\mathbf{Y}_{u}\mathbf{Y}_{d}^{\dagger}\mathbf{Y}_{d}(\mathbf{Y}_{u}^{\dagger}\mathbf{Y}_{u})^{2}\rangle=\mathbf{\mathbb{I}}_{3}\times\det[\mathbf{Y}_{u}^{\dagger}\mathbf{Y}_{u},\mathbf{Y}_{d}^{\dagger}\mathbf{Y}_{d}]\equiv2i\mathbf{\mathbb{I}}_{3}J_{\mathcal{CP}}\;.
\end{aligned}
\label{eq:CPtrace}
\end{equation}
Notons que le signe moins dans la première égalité est nécessaire
pour éviter des réductions vers des structures plus simples en utilisant
les identités de Cayley-Hamilton (CH) \cite{MercolliS09}, voir Annexe
\ref{AnnexeA}. La dernière égalité découle également du théorème
de Cayley-Hamilton\footnote{Pour le voir, il suffit d'injecter $\mathbf{X}=[\mathbf{Y}_{u}^{\dagger}\mathbf{Y}_{u},\mathbf{Y}_{d}^{\dagger}\mathbf{Y}_{d}]$
dans l'équation (\ref{eq:A30}) de l'Annexe \ref{AnnexeA}, qui se
simplifie grandement grâce au fait que $\langle\lbrack\mathbf{Y}_{u}^{\dagger}\mathbf{Y}_{u},\mathbf{Y}_{d}^{\dagger}\mathbf{Y}_{d}]\rangle=0$.
Ainsi, $\det[\mathbf{Y}_{u}^{\dagger}\mathbf{Y}_{u},\mathbf{Y}_{d}^{\dagger}\mathbf{Y}_{d}]$
est non-nul seulement s'il y a une phase violant $\mathcal{CP}$ dans
$\mathbf{Y}_{u}^{\dagger}\mathbf{Y}_{u}$ et/ou $\mathbf{Y}_{d}^{\dagger}\mathbf{Y}_{d}$.}. Cette quantité se réduit alors au très supprimé \emph{invariant
de Jarlskog} \cite{Jarlskog85}:
\begin{equation}
J_{\mathcal{CP}}=\mathcal{J}_{\mathcal{CP}}\times\prod_{\substack{i>j=d,s,b\\
i>j=u,c,t
}
}\frac{m_{i}^{2}-m_{j}^{2}}{v^{2}}\approx\mathcal{J}_{\mathcal{CP}}\times\frac{m_{b}^{4}m_{s}^{2}m_{c}^{2}}{v^{8}}\approx10^{-22}\;,\label{eq:JarlQuark}
\end{equation}
qui dans la paramétrisation standard de la matrice CKM \cite{PDG}
donne
\begin{equation}
\mathcal{J}_{\mathcal{CP}}=\frac{1}{4}\sin(2\theta_{12})\sin(2\theta_{23})\cos^{2}(\theta_{13})\sin(\theta_{13})\sin(\delta_{13})\approx3\times10^{-5}\;.\label{eq:JarlAngles}
\end{equation}
Remarquons que $J_{\mathcal{CP}}$ s'annule dès que deux quarks de
type up ou bien deux quarks de type down sont dégénérés en masses,
propriété rappelant que dans ce cas la phase de violation-$\mathcal{CP}$
est réductible, c'est à dire qu'elle peut être éliminée par des rotations
des champs.

\begin{figure}[t]
\centering
\includegraphics[width=0.95\textwidth]{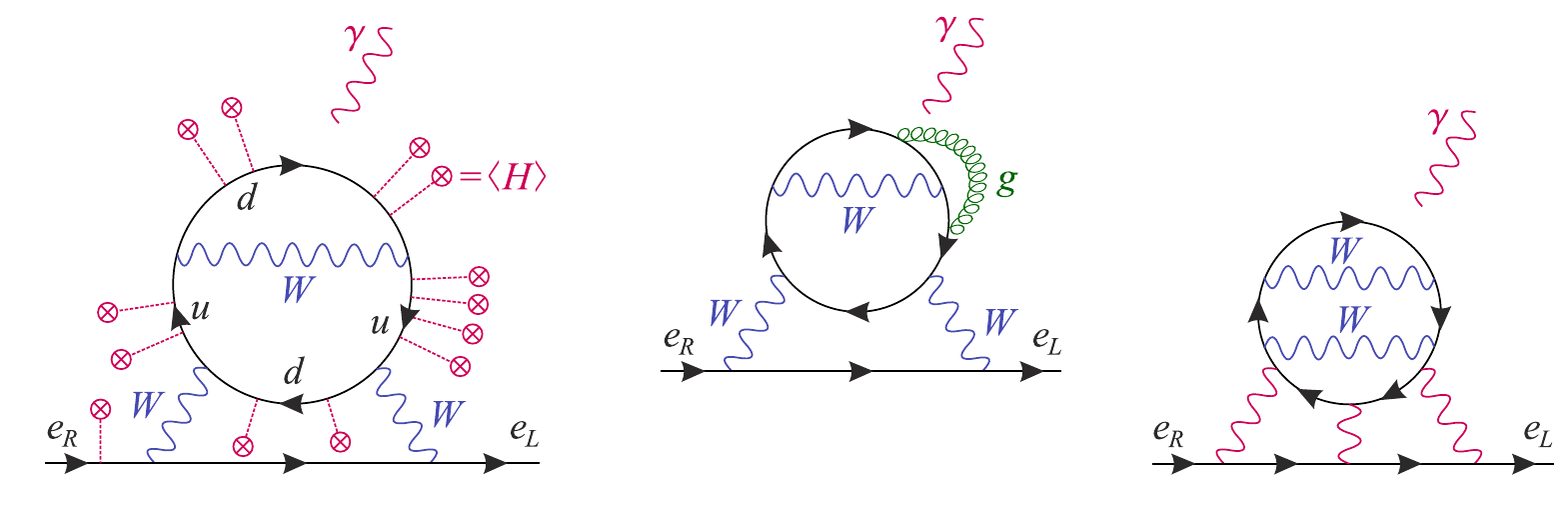}
\caption{Les EDMs de leptons induits par la matrice CKM dans le MS. La contribution électrofaible à trois boucles s'annule car la symétrie de la fonction de boucle est incompatible avec l'antisymétrie de l'invariant de Jarlskog et des corrections de QED ou de QCD sont nécessaires afin d'induire des EDMs de leptons non-nuls.}
\label{FigCKMe}
\end{figure}

Cette structure de saveur nous montre que les EDMs de leptons induits
par la matrice CKM nécessite au moins trois boucles étant donné qu'une
boucle fermée de quarks avec quatre vertex impliquant un boson $W$
est requise (voir figure \ref{FigCKMe}). Si on voit les facteurs
$\mathbf{Y}_{u}$ et $\mathbf{Y}_{d}$ comme des insertions de masses
le long d'une boucle de quark, ce qui n'est pas apparent dans ces
expressions est qu'une boucle supplémentaire de QCD est en fait requise
\cite{Donoghue}. En effet, la dynamique du MS \cite{EDM2loop1,EDM2loop2,EDM2loop3}
est telle que la fonction de boucle est la même pour les insertions
$(\mathbf{Y}_{d}^{\dagger}\mathbf{Y}_{d}\mathbf{)}^{2}\mathbf{Y}_{u}^{\dagger}\mathbf{Y}_{u}\mathbf{Y}_{d}^{\dagger}\mathbf{Y}_{d}(\mathbf{Y}_{u}^{\dagger}\mathbf{Y}_{u})^{2}$
et $(\mathbf{Y}_{d}^{\dagger}\mathbf{Y}_{d}\mathbf{)}^{2}(\mathbf{Y}_{u}^{\dagger}\mathbf{Y}_{u})^{2}\mathbf{Y}_{d}^{\dagger}\mathbf{Y}_{d}\mathbf{Y}_{u}^{\dagger}\mathbf{Y}_{u}$
(voir équation (\ref{eq:CPtrace})). Or, comme leur somme est réductible
avec les identités CH, $\mathcal{CP}$ est conservée et le processus
à trois boucles ne contribue pas aux EDMs de leptons. Pour briser
la symétrie de l'amplitude et générer une combinaison antisymétrique
d'insertions de masses, au moins une boucle supplémentaire est requise,
par exemple une correction QED ou QCD (voir figure \ref{FigCKMe}).
On arrive alors à l'estimation approximative (voir \cite{PospelovRCKM}):
\begin{equation}
d_{e}\sim e\frac{m_{e}}{M_{W}^{2}}\left(\frac{g^{2}}{16\pi^{2}}\right)^{3}\frac{\alpha_{S}}{4\pi}\tilde{J}_{\mathcal{CP}}\approx10^{-48}\;e\cdot cm\;,\label{eq:eEDMCKM}
\end{equation}
à comparer à la limite actuelle $|d_{e}|\,<8.7\cdot10^{-29}\;e\cdot cm\;(90\%)$
\cite{expEDMe}. Pour cette estimation, un facteur d'amélioration
$\tilde{J}_{\mathcal{CP}}=J_{\mathcal{CP}}(v/M_{W})^{12}\approx10^{5}$
est inclus. En effet, tandis que $J_{\mathcal{CP}}$ est défini à
partir des couplages de Yukawa et implique des rapports de masses
de quarks par la VEV électrofaible $v\approx174\,\ensuremath{GeV}$,
on s'attendrait plutôt à des rapports de masses de quarks dans la
boucle par la masse du $W$ dans l'approche diagrammatique.

La même boucle de quark contrôle l'EDM du boson $W$ (il suffit de
couper les deux propagateurs de $W$ du bas dans la figure \ref{FigCKMe})
ainsi que ceux des leptons plus lourds. Modulo des effets dynamiques
liés aux différentes échelles de ces processus, on s'attend à ce que
la relation bien connue
\begin{equation}
\frac{d_{e}}{m_{e}}=\frac{d_{\mu}}{m_{\mu}}=\frac{d_{\tau}}{m_{\tau}}
\end{equation}
soit vérifiée et donc à ce que $d_{\mu}$ et $d_{\tau}$ soient respectivement
200 et 4000 fois plus grands que $d_{e}$. Bien que ces estimations
ne soient pas très précises, elles se trouvent toutes bien au-delà
des sensibilités expérimentales dans la mesure où les limites actuelles
sont $|d_{\mu}|<1.9\cdot10^{-19}\;e\cdot cm\;(95\%)$ \cite{Bennett}
et $d_{\tau}\in\lbrack-2.2,4.5]\cdot10^{-17}\;e\cdot cm\;(95\%)$
\cite{BelleEDM}. Ces dernières sont bien plus petites que celles
pour $d_{e}$ qui exploite le champ électrique très intense présent
dans la molécule de $ThO$. En revanche, la limite sur $d_{\mu}$
a été obtenue par la mesure précise du $(g-2)_{\mu}$ et celle sur
$d_{\tau}$ par l'étude du vertex $\gamma\tau^{+}\tau^{-}$ en utilisant
le processus $e^{+}e^{-}\rightarrow\tau^{+}\tau^{-}$ à Belle.

A ce stade, adressons un mot d'avertissement à propos des dépendances
de masses. Comme pour le pingouin avec photon $sd\rightarrow\gamma^{\ast}$
discuté précédemment, ou encore similairement le pingouin avec gluon
$sd\rightarrow g^{\ast}$, l'approximation d'insertion de masse inhérente
à la technique des spurions est incapable de capturer les dépendances
de masses logarithmiques. Bien qu'un calcul explicite de cette amplitude
à quatre boucles n'a pas encore été fait, de telles dépendances ont
été trouvées pour l'opérateur similaire à trois gluons violant-$\mathcal{CP}$
induit par CKM, $f^{abc}\tilde{G}_{\mu\nu}^{a}G^{b,\nu\rho}G_{\rho\mu}^{c}$
\cite{GGG}. En particulier, des facteurs de quark lourd comme les
deux facteurs de suppressions $m_{b}^{2}/v^{2}$ dans $\mathbf{X}_{e}$
sont remplacés par des logarithmes de rapports de masses de quarks
et de boson $W$. C'est pourquoi quelques ordres de grandeurs supplémentaires
d'amélioration sont implicites pour des estimations comme dans l'équation
(\ref{eq:eEDMCKM}).

\subsection{EDMs de quarks générés par la phase CKM\label{SecCKMq}}

En ce qui concerne les EDMs de quarks, leurs générations peuvent paraître
plus simples à première vue en raison du fait que les quarks sont
directement sensibles à la phase complexe de la matrice CKM. Pourtant,
la technique des spurions montre que ce n'est pas le cas en pratique.
Considérons les interactions $U\mathbf{Y}_{u}\mathbf{X}_{u}\sigma_{\mu\nu}QH^{\dagger}F^{\mu\nu}$
et $D\mathbf{Y}_{d}\mathbf{X}_{d}\sigma_{\mu\nu}QH^{\dagger}F^{\mu\nu}$
dans le lagrangien effectif (\ref{eq:EMO}), avec $\mathbf{X}_{u,d}$
des chaînes de spurions et posons $\Lambda\approx M_{W}$. Examinons
dans un premier temps l'EDM du quark down, la composante $d_{L}$
du doublet $Q$ doit être un état propre de masse, donc nous devons
utiliser la base d'interaction de l'équation (\ref{G=0000E8leSpurions})
où $\mathbf{Y}_{d}$ est diagonale. Dans cette base, $\mathbf{X}_{d}$
doit être une chaîne de spurions construite à partir de $Y_{u}^{\dagger}Y_{u}$
et de $Y_{d}^{\dagger}Y_{d}$. Afin de générer un EDM pour le quark
down, il faut et il suffit que la partie imaginaire du coefficient
1-1 de la matrice $\mathbf{X}_{d},$ $\mathbf{X}_{d}^{11}$, soit
non nulle. Cependant, avec d'une part $\mathbf{Y}_{d}^{\dagger}\mathbf{Y}_{d}$
réelle et diagonale dans cette base et d'autre part $\mathbf{Y}_{u}^{\dagger}\mathbf{Y}_{u}$
hermitienne, cela requiert de nouveau une longue chaîne de spurions.
Il en est de même pour l'EDM du quark up, en travaillant dans la base
dans laquelle $\mathbf{Y}_{u}$ est diagonale.

Afin d'identifier la chaîne de spurions la plus simple, considérons
les développements en séries de spurions complètement génériques.
Les combinaisons $\mathbf{X}_{u}$ et $\mathbf{X}_{d}$ se transforment
comme des octets sous $SU(3)_{Q}$. En toute généralité, de tels octets
peuvent être paramétrisés comme des séries infinies de produits de
puissances des matrices hermitiennes $\mathbf{A}_{u,d}\equiv\mathbf{Y}_{u,d}^{\dagger}\mathbf{Y}_{u,d}$
\cite{ColangeloNS08}
\begin{equation}
\mathbf{X}_{octet}=\sum_{i,j,k,...=0,1,2,...}z_{ijk...}\mathbf{A}_{u}^{i}\mathbf{A}_{d}^{j}\mathbf{A}_{u}^{k}...\;,\label{eq:Generalz}
\end{equation}
avec $z_{ijk...}$ des coefficients appropriés. Comme notre but est
de quantifier l'impact de la phase CKM sur l'EDM, ces coefficients
$z_{ijk...}$ sont pris réels. Ainsi, cette série peut être partiellement
re-sommée à l'aide des identités CH, ce qui permet d'exprimer les
puissances supérieures de n'importe quelle matrice en fonctions de
ses puissances inférieures, de traces et de déterminants\footnote{Les problèmes liés à la convergence de cette série ont été traitées
dans \cite{QLMFV}, et ne devraient pas affecter l'identification
de la structure de saveur dominante violant-$\mathcal{CP}$.}. Par exemple, le terme $\mathbf{A}_{u}^{3}$ peut être absorbé dans
les redéfinitions des coefficients $z$, $z_{1}$ et $z_{2}$ en utilisant
l'équation (\ref{CH2}). Après réduction complète de la série, l'opérateur
octet $\mathbf{X}_{octet}$ ne contient plus que $17$ termes:
\begin{equation}
\begin{aligned}\mathbf{X}_{octet} & =x_{1}\mathbf{1}+x_{2}\mathbf{A}_{u}+x_{3}\mathbf{A}_{d}+x_{4}\mathbf{A}_{u}^{2}+x_{5}\mathbf{A}_{d}^{2}+x_{6}\{\mathbf{A}_{u},\mathbf{A}_{d}\}+x_{7}i[\mathbf{A}_{u},\mathbf{A}_{d}]+x_{8}\mathbf{A}_{u}\mathbf{A}_{d}\mathbf{A}_{u}\\
 & \;\;\;\;+x_{9}i[\mathbf{A}_{d},\mathbf{A}_{u}^{2}]+x_{10}\mathbf{A}_{d}\mathbf{A}_{u}\mathbf{A}_{d}+x_{11}i[\mathbf{A}_{u},\mathbf{A}_{d}^{2}]+x_{12}\mathbf{A}_{d}\mathbf{A}_{u}^{2}\mathbf{A}_{d}+x_{13}i[\mathbf{A}_{u}^{2},\mathbf{A}_{d}^{2}]\\
 & \;\;\;\;+x_{14}i[\mathbf{A}_{u},\mathbf{A}_{u}\mathbf{\mathbf{A}}_{d}\mathbf{A}_{u}]+x_{15}i[\mathbf{A}_{d},\mathbf{A}_{d}\mathbf{A}_{u}\mathbf{A}_{d}]\\
 & \;\;\;\;+x_{16}i[\mathbf{A}_{u},\mathbf{A}_{u}\mathbf{\mathbf{A}}_{d}^{2}\mathbf{A}_{u}]+x_{17}i[\mathbf{A}_{d},\mathbf{A}_{d}\mathbf{\mathbf{A}}_{u}^{2}\mathbf{A}_{d}]\;.
\end{aligned}
\label{eq:Octet}
\end{equation}
La seule réduction non-triviale est celle du terme $\mathbf{A}_{d}^{2}\mathbf{\mathbf{A}}_{u}\mathbf{A}_{d}\mathbf{\mathbf{A}}_{u}^{2}$,
qui peut être établie en injectant $\mathbf{X}=[\mathbf{A}_{u},\mathbf{A}_{d}]$
dans l'équation (\ref{CH2}). De plus, nous avons utilisé l'hermiticité
de $\mathbf{A}_{u,d}$ pour exprimer $\mathbf{X}_{octet}$ entièrement
en termes de combinaisons de spurions hermitiennes indépendantes \cite{MercolliS09}.

Notons qu'il est crucial de n'utiliser que les identités CH pour cette
réduction et non simplement une projection de $\mathbf{X}_{octet}$
sur un ensemble de neuf termes formant une base des matrices complexes
$3\times3$. D'abord, la réduction CH ne génère jamais des grands
coefficients numériques dans la mesure où les traces vérifient $\langle\mathbf{A}_{u,d}\rangle\lesssim\mathcal{O}(1)$.
Ensuite, si les $z_{ijk...}$ sont réels, les $x_{i}$ peuvent alors
développer une partie imaginaire proportionnelle à l'invariant de
Jarlskog dans l'équation (\ref{eq:CPtrace}). Ainsi, cela assure par
exemple que $\mathbf{X}^{11}$ est soit proportionnel à $\mathcal{J}_{\mathcal{CP}}$,
si par exemple $x_{1}=\xi_{1}+i\xi_{2}\mathcal{J}_{\mathcal{CP}}$
avec $\xi_{1,2}$ réels et $\xi_{1,2}\lesssim\mathcal{O}(1)$, soit
induit directement par une chaîne de spurions non-triviale.

Les chaînes les plus simples ayant une partie imaginaire intrinsèque
dans la base d'interaction (\ref{G=0000E8leSpurions}) sont celles
qui sont associées aux coefficients $x_{14}$ et $x_{16}$. N'importe
quelle chaîne plus longue ayant une partie imaginaire intrinsèque
peut être réduite à ces deux dernières, ou bien à $\mathcal{J}_{\mathcal{CP}}$,
et sera supprimée par des facteurs de traces du type $\langle\mathbf{Y}_{d}^{\dagger}\mathbf{Y}_{d}\mathbf{\rangle}$
ainsi que $\langle\mathbf{Y}_{u}^{\dagger}\mathbf{Y}_{u}\rangle$.
Concernant les EDMs des quarks de type down, le terme dominant est
alors\footnote{Cette structure a déjà été identifiée dans la littérature, voir par
exemple \cite{Romanino}. Cependant, son obtention par l'usage systématique
des identités CH n'a jamais été présentée auparavant.}
\begin{equation}
\mathbf{X}_{d}=\mathbf{[Y}_{u}^{\dagger}\mathbf{Y}_{u}\;,\;\mathbf{Y}_{u}^{\dagger}\mathbf{Y}_{u}\mathbf{Y}_{d}^{\dagger}\mathbf{Y}_{d}\mathbf{Y}_{u}^{\dagger}\mathbf{Y}_{u}]\;.\label{eq:ddchain}
\end{equation}
L'antisymétrie (signe moins) est nécessaire car la somme des deux
termes (l'anticommutateur) est hermitienne et possède donc que des
coefficients réels sur la diagonale (au-delà du fait qu'elle soit
réductible via les identités CH). En revanche, grâce au commutateur,
$\mathbf{X}_{d}$ possède des coefficients imaginaires purs sur la
diagonale, par exemple:
\begin{equation}
\mathbf{X}_{d}^{11}=-2i\mathcal{J}_{\mathcal{CP}}\times\frac{m_{b}^{2}-m_{s}^{2}}{v^{2}}\prod_{i>j=u,c,t}\frac{m_{i}^{2}-m_{j}^{2}}{v^{2}}\approx10^{-12}\;,
\end{equation}
à comparer avec $J_{\mathcal{CP}}\approx10^{-22}$ dans l'équation
(\ref{eq:JarlQuark}). Une telle dépendance en différences de masses
de quarks aux carrées a déjà été relevée dans \cite{Shabalin1982sg}.
Elle émerge ici en tant que la plus simple chaîne de spurions antihermitienne
ayant des éléments diagonaux non-nuls.

La prédiction pour les EDMs des quarks de type up $d_{u}$ est similaire.
En effet, $\mathbf{X}_{u}$ est obtenu à partir de $\mathbf{X}_{d}$
en interchangeant $\mathbf{Y}_{d}\leftrightarrow\mathbf{Y}_{u}$ dans
l'équation (\ref{eq:ddchain}) et en travaillant dans la base d'interaction
où $\mathbf{Y}_{u}$ est diagonale. Les termes associés à $x_{15}$
et $x_{17}$ ont des tailles comparables étant donné que $m_{t}\approx v$,
\begin{equation}
\mathbf{X}_{u}=a_{1}\mathbf{[Y}_{d}^{\dagger}\mathbf{Y}_{d}\;,\;\mathbf{Y}_{d}^{\dagger}\mathbf{Y}_{d}\mathbf{Y}_{u}^{\dagger}\mathbf{Y}_{u}\mathbf{Y}_{d}^{\dagger}\mathbf{Y}_{d}]+a_{2}\mathbf{[Y}_{d}^{\dagger}\mathbf{Y}_{d}\;,\;\mathbf{Y}_{d}^{\dagger}\mathbf{Y}_{d}(\mathbf{Y}_{u}^{\dagger}\mathbf{Y}_{u})^{2}\mathbf{Y}_{d}^{\dagger}\mathbf{Y}_{d}]\;,
\end{equation}
avec $a_{1,2}$ des coefficients réels de $\mathcal{O}(1)$. On a
alors:
\begin{equation}
\mathbf{X}_{u}^{11}=2i\mathcal{J}_{\mathcal{CP}}\times\left(a_{1}\frac{m_{t}^{2}-m_{c}^{2}}{v^{2}}+a_{2}\frac{m_{t}^{4}-m_{c}^{4}}{v^{4}}\right)\prod_{i>j=d,s,b}\frac{m_{i}^{2}-m_{j}^{2}}{v^{2}}\approx10^{-17}\;.
\end{equation}
En raison des facteurs supplémentaires de masses de quarks de type
down, celui-ci est complètement négligeable devant $\mathbf{X}_{d}^{11}$.

\begin{figure}[t]
\centering
\includegraphics[width=0.65\textwidth]{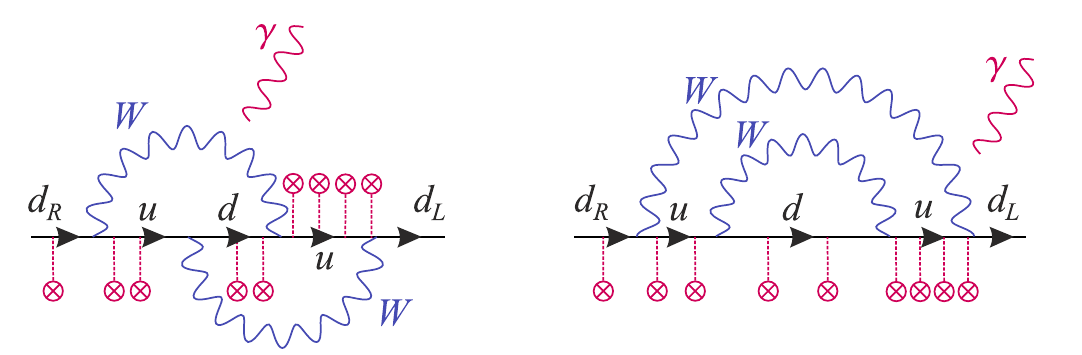}
\caption{Les diagrammes rainbow électrofaibles générant un EDM pour le quark down dans le MS. Pour chaque type, un exemple d'insertion de spurions de Yukawa est représenté. La contribution finale nécessite l'antisymétrisation des insertions sur les deux lignes de quarks de type up.}
\label{FigCKMq}
\end{figure}

Les diagrammes rainbow électrofaibles derrière de tels processus partagent
plusieurs propriétés communes avec ceux générant l'EDM de l'électron
$d_{e}$. Deux propagateurs de boson $W$ sont nécessaires (voir figure
\ref{FigCKMq}), en plus d'une correction gluonique supplémentaire
permettant de briser la symétrie de l'amplitude de boucle sous les
permutations des insertions de masses \cite{Khriplovich1985jr,CzarneckiK}.
L'ordre dominant contient trois boucles et a la forme suivante
\begin{equation}
d_{d}\sim e\frac{m_{d}}{M_{W}^{2}}\left(\frac{g^{2}}{16\pi^{2}}\right)^{2}\frac{\alpha_{S}}{4\pi}\frac{m_{b}^{2}m_{c}^{2}m_{t}^{4}}{v^{8}}\mathcal{J}_{\mathcal{CP}}\approx10^{-36}-10^{-39}\,e\cdot cm\;,\label{eq:CKMrainbow}
\end{equation}
dépendant de quel facteur $v$ ou $M_{W}$ apparaît dans le dénominateur.
De plus, là aussi l'approximation de l'insertion de masse ne reproduit
pas parfaitement le calcul explicite réalisé dans \cite{CzarneckiK},
où par exemple le facteur $m_{b}^{2}$ s'adoucit en une brisure logarithmique
de GIM, améliorant l'estimation de quelques ordres de grandeurs.

On pourrait s'étonner de la non annulation de cette expression lorsque
des quarks de type down sont dégénérés. Pour le comprendre, il faut
réaliser qu'une base spécifique est implicitement choisie pour ces
quarks en forçant les quarks de type down externes à être sur leurs
couches de masses (on-shell). Si on imagine que $m_{d}=m_{s}$, alors
les quarks de type down sur couches de masses de la première et de
la deuxième génération peuvent être des combinaisons linéaires de
ces derniers. En posant
\begin{equation}
\left(\begin{array}{c}
d_{1}^{\prime}\\
d_{2}^{\prime}
\end{array}\right)=\left(\begin{array}{cc}
\cos\theta_{ds} & \sin\theta_{ds}\\
-\sin\theta_{ds} & \cos\theta_{ds}
\end{array}\right)\left(\begin{array}{c}
d\\
s
\end{array}\right)\;,
\end{equation}
en sorte que
\begin{equation}
\mathbf{X}_{d}^{d_{1}d_{1}}=\cos^{2}\theta\mathbf{X}_{d}^{11}+\sin^{2}\theta\mathbf{X}_{d}^{22}+\cos\theta\sin\theta(\mathbf{X}_{d}^{12}+\mathbf{X}_{d}^{21})\;,
\end{equation}
et de même pour $\mathbf{X}_{d}^{d_{2}d_{2}}$, on trouve que $\mathbf{X}_{d}^{d_{2}d_{2}}=\mathbf{X}_{d}^{d_{1}d_{1}}=0$
quand $\theta_{ds}=\theta_{12}$ et $m_{d}=m_{s}$. Ceci montre que
la conservation de $\mathcal{CP}$ est retrouvée dans la limite dégénérée
comme il se doit.

Ainsi, la contribution de courte distance du MS à l'EDM du neutron
$d_{n}\approx(4d_{d}-d_{u})/3$ est prévue au plus à un ou deux ordres
de grandeurs au dessus de $10^{-36}\,e\cdot cm$. Ceci doit être comparé
aux contributions de longue distance qui pourrait amplifier la contribution
de MS jusqu'à $d_{n}\approx10^{-32}\,e\cdot cm$ \cite{EDMLD}, et
jusqu'à la limite actuelle se trouvant à $|d_{n}|<2.9\times10^{-26}\;e\cdot cm\;(90\%)$
\cite{ExpEDMn}. Pour finir, notons que
\begin{equation}
\mathbf{X}_{d}^{11}:\mathbf{X}_{d}^{22}:\mathbf{X}_{d}^{33}=m_{b}^{2}-m_{s}^{2}:m_{d}^{2}-m_{b}^{2}:m_{s}^{2}-m_{d}^{2}\;,
\end{equation}
comme attendu de $\langle\mathbf{X}_{d}\rangle=0$. Comme $m_{b}\gg m_{s,d}$,
cela signifie aussi que $\mathbf{X}_{d}^{11}\approx\mathbf{X}_{d}^{22}\gg\mathbf{X}_{d}^{33}$.
A l'échelle de l'EDM du quark, ces relations impliquent les règles
de sommes suivantes
\begin{equation}
\frac{d_{d}}{m_{d}}+\frac{d_{s}}{m_{s}}+\frac{d_{b}}{m_{b}}=0\;,\;\;\frac{d_{d}}{m_{d}}\approx-\frac{d_{s}}{m_{s}}\;,\;\;\frac{d_{d}}{d_{b}}\approx\frac{m_{b}^{2}}{m_{s}^{2}}\;,\label{eq:SumRule}
\end{equation}
modulo des effets dynamiques au-delà de notre contrôle. Des relations
analogues sont également vérifiées pour les quarks de type up.

A ce stade, on peut alors assembler les informations obtenues pour
l'EDM de l'électron et l'EDM des quarks pour obtenir
\begin{equation}
\begin{aligned}\frac{\operatorname{Im}\mathbf{X}_{d}^{11}}{\operatorname{Im}\mathbf{X}_{e}^{11}} & =\frac{-v^{4}}{(m_{b}^{2}-m_{d}^{2})(m_{s}^{2}-m_{d}^{2})}\rightarrow\left(\frac{g^{2}}{16\pi^{2}}\right)\frac{|d_{d}|}{m_{d}}\approx\frac{M_{W}^{4}}{m_{b}^{2}m_{s}^{2}}\frac{|d_{e}|}{m_{e}}\;,\\
\frac{\operatorname{Im}\mathbf{X}_{u}^{11}}{\operatorname{Im}\mathbf{X}_{e}^{11}} & =v^{4}\frac{a_{1}+a_{2}(m_{t}^{2}+m_{c}^{2})/v^{2}}{(m_{t}^{2}-m_{u}^{2})(m_{c}^{2}-m_{u}^{2})}\rightarrow\left(\frac{g^{2}}{16\pi^{2}}\right)\frac{|d_{u}|}{m_{u}}\approx\frac{M_{W}^{2}}{m_{c}^{2}}\frac{|d_{e}|}{m_{e}}\;.
\end{aligned}
\end{equation}
Numériquement, on s'attend alors à ce que les contributions CKM aux
EDMs de fermions dans le MS vérifient $|d_{d}|\approx10^{12}\times|d_{e}|$
et $|d_{u}|\approx10^{7}\times|d_{e}|$, modulo des effets dynamiques
comme toujours.

\subsection{EDMs de quarks et de leptons générés par la phase forte du MS}

Jusque-là, les combinaisons de spurions $G_{F}$-invariantes l'étaient
sous la symétrie complète $U(3)^{5}$. Cependant, ceci n'est pas cohérent
avec le MS dans la mesure où trois des cinq $U(1)s$ sont anomaux.
Dans ces conditions, seule l'invariance sous $SU(3)^{5}$ devrait
être imposée \cite{MFVRPV}. Ce changement se répercute sur la procédure
précédente de deux façons. Premièrement, il y a des nouveaux invariants
impliquant des tenseurs de Levi-Civita de $SU(3)^{5}$. Etant donnée
les propriétés de symétrie des spurions de Yukawa, ces nouveaux invariants
peuvent tous être décomposés en invariants de $U(3)^{5}$ et de puissances
de $\det\mathbf{Y}_{u}$ et/ou $\det\mathbf{Y}_{d}$. Deuxièmement,
les valeurs physiques des matrices de couplages de Yukawa dans l'équation
(\ref{G=0000E8leSpurions}) doivent inclure en général des phases
supplémentaires violant $\mathcal{CP}$ car la symétrie $SU(3)^{5}$
n'est pas suffisante pour faire que toutes les masses des fermions
soient réelles.

\subsubsection{Anomalie axiale forte}

Pour comprendre les implications sur les EDMs des quarks et des leptons,
rappelons d'abord comment ces anomalies se manifestent dans le MS.
La décomposition en valeurs singulières (SVD) de l'équation (\ref{eq:SVD})
implique les transformations de $U(1)$\footnote{Notre convention consiste à décomposer une transformation de $G_{F}$,
$g_{X}\in U(3)_{X}$ comme $g_{X}=\exp(i\alpha_{X}T^{0})\exp(i\alpha_{X}^{a}T^{a})$,
avec $T^{a}\ensuremath{,}a=1,...,8$ les générateurs de $SU(3)_{X}$
et $T^{0}=\mathbf{1}$ est le générateur de $U(1)_{X}$. Puis, en
utilisant l'identité $\det(\exp A)=\exp\langle A\rangle$ et avec
$\langle T^{0}\rangle=3$, la phase peut être extraite par $\arg\det g_{X}=3\alpha_{X}$.} $3\alpha_{Q,L}=\arg\det V_{L}^{d,e\dagger}$ et $3\alpha_{U,D,E}=\arg\det V_{R}^{u,d,e}$.
Ces phases ne sont pas fixées car les différentes matrices unitaires
de la SVD ne sont définies qu'à des phases relatives près. Cependant,
ces dernières doivent satisfaire
\begin{equation}
\arg\det\mathbf{Y}_{u}=3(\alpha_{Q}+\alpha_{U})\;,\;\arg\det\mathbf{Y}_{d}=3(\alpha_{Q}+\alpha_{D})\;,\;\arg\det\mathbf{Y}_{e}=3(\alpha_{L}+\alpha_{E})\;,\label{eq:SVDphase}
\end{equation}
pour que les masses des fermions soient réelles. En même temps, ces
transformations de $U(1)$ étant anomales, elles induisent un décalage
de la constante de couplage de l'opérateur $G_{\mu\nu}^{a}\tilde{G}^{a,\mu\nu}$
comme suit
\begin{equation}
\theta_{C}\rightarrow\theta_{C}^{eff}=\theta_{C}-3\left(2\alpha_{Q}+\alpha_{U}+\alpha_{D}\right)=\theta_{C}-\arg\det\mathbf{Y}_{u}-\arg\det\mathbf{Y}_{d}\;,\label{eq:thetashifts}
\end{equation}
où $\theta_{C}$ provient de la structure du vide de QCD. En pratique,
on a alors une phase supplémentaire de violation-$\mathcal{CP}$ dans
le MS, $\theta_{C}^{eff}$, qu'on peut interpréter librement comme
un couplage de $G_{\mu\nu}^{a}\tilde{G}^{a,\mu\nu}$ ou bien comme
des masses de quarks complexes. Notons qu'au-delà du couplage $G_{\mu\nu}^{a}\tilde{G}^{a,\mu\nu}$,
la dynamique d'instanton génère aussi l'interaction suivante \cite{tHooft76}
\begin{equation}
\mathcal{H}_{eff}^{axial}\sim\frac{g^{axial}}{\Lambda^{14}}(\varepsilon^{IJK}Q^{I}Q^{J}Q^{K})^{2}(\varepsilon^{IJK}U^{\dagger I}U^{\dagger J}U^{\dagger K})(\varepsilon^{IJK}D^{\dagger I}D^{\dagger J}D^{\dagger K})+h.c.\;,
\end{equation}
où $Q,U,D$ dénotent ici des spineurs de Weyl, et les contractions
de Lorentz, $SU(2)_{L}$ et $SU(3)_{C}$ sont considérés implicitement.
Cette interaction est invariante sous $SU(3)^{5}$ mais brise explicitement
$U(1)^{5}$ car faire que les masses des quarks soient réelles décale
la phase de ce couplage de
\begin{equation}
g^{axial}\rightarrow g^{axial}\exp3i(2\alpha_{Q}+\alpha_{U}+\alpha_{D})\;,
\end{equation}
exactement comme la phase forte $\theta_{C}$ dans l'équation (\ref{eq:thetashifts}).
Indirectement, $\arg\det\mathbf{Y}_{u}+\arg\det\mathbf{Y}_{d}$ est
en principe accessible si on connait $\arg(g^{axial})$ dans une base
donnée. La seule situation où le MS n'impliquerait pas un nouveau
paramètre libre serait quand les interactions anomales sont alignées,
i.e., $\arg(g^{axial})=\theta_{C}$ avant SSB. Dans ce cas, faire
que les masses des quarks soient réelles décale $\theta_{C}\rightarrow\theta_{C}^{eff}$
et $\arg(g^{axial})=\theta_{C}\rightarrow\arg(g^{axial})=\theta_{C}^{eff}$.

\subsubsection{Spurions impliquant la phase forte et EDMs}

Quelque soit la paramétrisation choisie pour $\theta_{C}^{eff}$,
cette phase peut induire des EDMs à travers des effets non-perturbatifs
de QCD à basse énergie. Pour spécifier, considérons les opérateurs
magnétiques effectifs de l'équation (\ref{eq:EMO}), avec des coefficients
de Wilson $c_{u,d,e}$ réels et des combinaisons de spurions $\mathbf{Y}_{u,d,e}\mathbf{X}_{u,d,e}$
sensibles à $\theta_{C}^{eff}$. Comme ce paramètre provient de la
QCD et qu'il contient potentiellement des grandes contributions de
quarks légers, les effets non-locaux à longue distance sont dominants
et $\Lambda$ devrait être fixé à l'échelle hadronique typique. En
conséquence, cela repousse le formalisme effectif au-delà de ces limites,
mais procédons quand même.

La difficulté principale est d'établir la forme des insertions de
spurions. Comme nous souhaitons utiliser la symétrie de saveur $G_{F}$
et ses termes de brisure explicite, nous devrions transférer tout
le $\theta_{eff}^{C}$ sur les masses de quarks afin de se débarrasser
du couplage $G_{\mu\nu}^{a}\tilde{G}^{a,\mu\nu}$. Ceci peut être
réalisé en modifiant les valeurs physiques de fonds des spurions de
la manière suivante (dans la base d'interaction où les quarks de type
up sont états propres de masses)
\begin{equation}
\begin{aligned}v\mathbf{Y}_{u}\overset{gel\acute{e}}{\rightarrow}e^{i\kappa\theta_{C}^{eff}\mathbb{M}_{u}^{-1}}\mathbb{M}_{u} & =\mathbb{M}_{u}+i\mathbb{I}_{3}\kappa\theta_{C}^{eff}+...\;,\\
v\mathbf{Y}_{d}\overset{gel\acute{e}}{\rightarrow}e^{i\kappa\theta_{C}^{eff}\mathbb{M}_{d}^{-1}}\mathbb{M}_{d}V_{CKM}^{\dagger} & =(\mathbb{M}_{d}+i\mathbb{I}_{3}\kappa\theta_{C}^{eff}+...\;)V_{CKM}^{\dagger}\;\;,
\end{aligned}
\label{eq:ThetaBack}
\end{equation}
avec
\begin{equation}
\kappa^{-1}\equiv\langle\mathbb{M}_{u}^{-1}+\mathbb{M}_{d}^{-1}\rangle=\frac{1}{m_{u}}+\frac{1}{m_{c}}+\frac{1}{m_{t}}+\frac{1}{m_{d}}+\frac{1}{m_{s}}+\frac{1}{m_{b}}\;,\label{eq:Kappa}
\end{equation}
tandis qu'on garde $v\mathbf{Y}_{e}\overset{gel\acute{e}}{\rightarrow}\mathbb{M}_{e}$.
Une infinité d'autres choix de transformations de $U(1)_{Q}\otimes U(1)_{U}\otimes U(1)_{D}$
pourraient remplacer le couplage $G_{\mu\nu}^{a}\tilde{G}^{a,\mu\nu}$,
mais ce choix spécifique possède les propriétés souhaitables suivantes:
\begin{itemize}[label=\textbullet]
\item Ces valeurs de fonds des spurions rendent compte correctement de
l'ensemble du terme $\theta_{C}^{eff}$, comme cela peut être vérifié
en effectuant les rotations anomales inverses pour revenir à la base
où $v\mathbf{Y}_{u}\overset{gel\acute{e}}{\rightarrow}\mathbb{M}_{u}$
et $v\mathbf{Y}_{d}\overset{gel\acute{e}}{\rightarrow}\mathbb{M}_{d}V_{CKM}^{\dagger}$,
\begin{equation}
\arg\det\mathbf{Y}_{u}+\arg\det\mathbf{Y}_{d}=\arg\det\exp(i\kappa\theta_{C}^{eff}\mathbf{m}_{u}^{-1})+\arg\det\exp(i\kappa\theta_{C}^{eff}\mathbf{m}_{d}^{-1})=\theta_{C}^{eff}\;.
\end{equation}
\item Si $g^{axial}$ est réel dans la base (\ref{eq:ThetaBack}), revenir
dans la base où $v\mathbf{Y}_{u}\rightarrow\mathbb{M}_{u}$ et $v\mathbf{Y}_{d}\rightarrow\mathbb{M}_{d}V_{CKM}^{\dagger}$
aligne automatiquement la phase de la transition de quarks induites
par instanton avec $G_{\mu\nu}^{a}\tilde{G}^{a,\mu\nu}$ étant donnée
que $g^{axial}\rightarrow g^{axial}\exp i\theta_{C}^{eff}$.
\item A partir de la base (\ref{eq:ThetaBack}), des masses de quarks réelles
sont obtenues en agissant seulement sur les champs de chiralité droite
et le couplage $SU(2)_{L}$ anomal $W_{\mu\nu}^{i}\tilde{W}^{i,\mu\nu}$
n'est pas affecté (cela sera discuté dans la dernière section).
\item Cette forme assure que les EDMs de quarks et de leptons induits par
$\theta_{C}^{eff}$ sont contrôlés par $\kappa$ (défini dans l'équation
(\ref{eq:Kappa})), ce qui garantit que la contribution de $\theta_{C}^{eff}$
disparaît dès lors qu'une seule masse de quark est nulle. De plus,
cela reproduit le facteur usuel $m_{u}m_{d}m_{s}/(m_{u}m_{d}+m_{u}m_{s}+m_{d}m_{s})$
quand $m_{c,b,t}\rightarrow\infty$ et assure donc la stabilité du
vide brisant la symétrie chirale (voir par exemple \cite{Cheng88}).
\item L'impact de $\theta_{C}^{eff}$ est rendu non-savoureux même s'il
est introduit à travers les couplages de saveurs grâce à des facteurs
appropriés se compensant $\mathbb{M}_{u}^{-1}$ et $\mathbb{M}_{d}^{-1}$.
A cet égard, notons que $V_{CKM}$ pourrait être inclus dans l'exponentielle
sans affecter les propriétés de la paramétrisation.
\end{itemize}
Pour estimer l'EDM du quark en utilisant la symétrie $G_{F}=SU(5)^{5}$,
il suffit de poser $\mathbf{X}_{u}=\mathbf{X}_{d}=\mathbb{I}_{3}$
dans l'équation (\ref{eq:EMO}) dans la mesure où les $\mathbf{Y}_{u,d}$
sont directement sensibles à $\theta_{C}^{eff}$. On a alors:
\begin{equation}
d_{u,d}\sim e\frac{1}{\Lambda_{had}^{2}}\kappa\theta_{C}^{eff}\approx\theta_{eff}\times10^{-16}\;e\cdot cm\;,
\end{equation}
avec $\Lambda_{had}\approx300$ MeV. Cela est similaire aux estimations
naïves basées sur des arguments dimensionnels et implique que $\theta_{C}^{eff}\lesssim10^{-10}$
puisque $|d_{n}|<2.9\times10^{-26}\;e\cdot cm\;(90\%)$ \cite{ExpEDMn}.
Au niveau de la symétrie $SU(3)^{5}$ et de ses termes de brisures,
il n'y a pas moyen d'en apprendre plus car la dynamique hadronique
à longue distance complexe est hors d'atteinte.

Quant aux EDMs de leptons, la chaîne de spurions la plus simple est
$\mathbf{X}_{e}=\mathbb{I}_{3}\det\mathbf{Y}_{u,d}$, qui développe
une partie imaginaire comme suit
\begin{equation}
\operatorname{Im}\det\mathbf{Y}_{u}\rightarrow\det(\mathbb{M}_{u}/v)\operatorname{Im}\det\exp(i\kappa\theta_{C}^{eff}\mathbb{M}_{u}^{-1})\approx i\kappa\theta_{C}^{eff}\det(\mathbb{M}_{u}/v)\langle\mathbb{M}_{u}^{-1}\rangle\approx\theta_{C}^{eff}\times10^{-7}\;.
\end{equation}
Une expression similaire est vérifiée pour $\operatorname{Im}\det\mathbf{Y}_{d}\approx\theta_{C}^{eff}\times10^{-10}$.
Les facteurs de masses de quarks apportent une forte suppression mais
sont inévitables pour inclure de façon cohérente $\theta_{C}^{eff}$
dans les valeurs de fonds des spurions de Yukawa. De plus, ils ne
peuvent pas être représentés comme des insertions de masses le long
d'une boucle de quark, qui sont nécessairement invariant sous $U(3)^{5}$.
Ainsi, la dépendance de la boucle de quark sur $\det\mathbf{Y}_{u,d}$
doit provenir plutôt d'effets de QCD non-perturbatifs. Comme nous
ne pouvons pas les estimer ici, le mieux qu'on puisse faire est d'établir
une limite supérieure de l'EDM de l'électron en attachant la boucle
de quark au courant leptonique soit avec trois photons soit avec deux
bosons faibles,
\begin{equation}
\begin{aligned}d_{e} & \lesssim e\frac{m_{e}}{\Lambda_{had}^{2}}\left(\frac{e^{2}}{16\pi^{2}}\right)^{3}\operatorname{Im}\det\mathbf{Y}_{u}\approx\theta_{C}^{eff}\times10^{-34}\;e\cdot cm\;,\\
d_{e} & \lesssim e\frac{m_{e}}{M_{W}^{2}}\left(\frac{g^{2}}{16\pi^{2}}\right)^{2}\operatorname{Im}\det\mathbf{Y}_{u}\approx\theta_{C}^{eff}\times10^{-32}\;e\cdot cm\;,
\end{aligned}
\end{equation}
où $\Lambda_{had}\approx300$ MeV représente l'échelle hadronique
typique. Ces contributions aux EDMs de leptons sont certainement hors
d'atteinte vu que $\theta_{C}^{eff}\lesssim10^{-10}$. Il est quand
même intéressant de noter que ces contributions peuvent être plus
grandes que celles de la phase CKM dans l'équation (\ref{eq:eEDMCKM}).

\subsubsection{Contributions d'interaction faible à la phase forte}

\begin{figure}[t]
\centering
\includegraphics[width=0.65\textwidth]{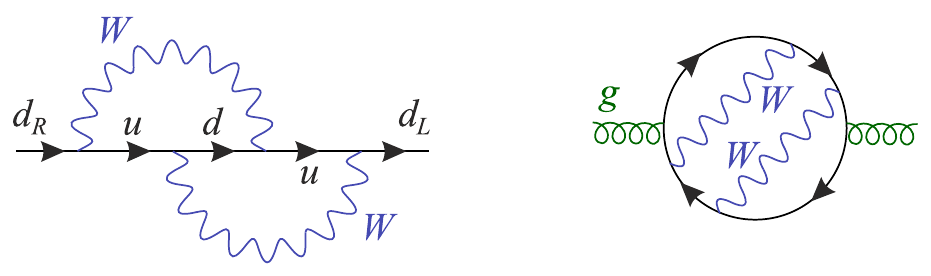}
\caption{Contributions électrofaibles dominantes à $\theta_{eff}^{C}$, provenant soit de la renormalisation des masses de quarks complexes soit de la propagation de gluon.}
\label{FigCKMth}
\end{figure}

Les combinaisons violant-$\mathcal{CP}$ de spurions établies dans
les sections \ref{SecCKMe} et \ref{SecCKMq} contrôlent également
les contributions d'interaction faible à la phase forte, voir figure
\ref{FigCKMth}. Pour spécifier, la correction violant-$\mathcal{CP}$
à la propagation du gluon provient de la boucle de quark, équation
(\ref{eq:CPtrace}), tandis que celles aux masses des quarks de type
down est proportionnelle à la combinaison (\ref{eq:ddchain}). Concernant
l'EDM, ces expressions prédisent correctement l'ordre de couplage
faible, mais ne sont pas suffisantes pour rendre compte des corrections
QCD requises pour briser la symétrie des insertions de masses. Il
a été montré dans \cite{Khriplovich1985jr} que la correction à la
propagation du gluon nécessitait une boucle QCD supplémentaire. Ainsi,
\begin{equation}
\Delta\theta_{eff}^{gluon}\sim\left(\frac{g^{2}}{4\pi^{2}}\right)^{2}\frac{\alpha_{S}}{\pi}\tilde{J}_{\mathcal{CP}}\approx10^{-23}\;,
\end{equation}
avec $\tilde{J}_{\mathcal{CP}}=J_{\mathcal{CP}}(v/M_{W})^{12}\approx10^{5}$.
Ceci est à comparer au calcul dans \cite{Khriplovich1985jr}, dans
lequel plusieurs facteurs de masses de quarks sont remplacés par des
logarithmes de rapports de masses de quarks, de sorte que $\Delta\theta_{eff}^{gluon}\sim10^{-19}$.

A partir de l'équation (\ref{eq:ddchain}), la correction violant-$\mathcal{CP}$
à la masse du quark $d$ doit être contrôlée par
\begin{equation}
\Delta\theta_{eff}^{d\ quark}\sim\mathbf{[Y}_{u}^{\dagger}\mathbf{Y}_{u}\;,\;\mathbf{Y}_{u}^{\dagger}\mathbf{Y}_{u}\mathbf{Y}_{d}^{\dagger}\mathbf{Y}_{d}\mathbf{Y}_{u}^{\dagger}\mathbf{Y}_{u}]^{11}\;.
\end{equation}
Le décalage total de la phase forte étant sommé sur les saveurs des
quarks induit une contribution qui s'annule exactement
\begin{equation}
\Delta\theta_{eff}^{d\ quark}+\Delta\theta_{eff}^{s\ quark}+\Delta\theta_{eff}^{b\ quark}=0\;,\label{eq:SumRuleTheta}
\end{equation}
et il en est de même pour les quarks de type up. Ceci n'est rien d'autre
que la règle de somme (\ref{eq:SumRule}) provenant de $\langle\mathbf{X}_{d}\rangle=0$.
En fait, en l'absence d'effets de QCD et au niveau de l'insertion
de masse, la contribution dominante à $\Delta\theta_{eff}^{all\,quarks}$
doit nécessairement provenir de
\begin{equation}
\Delta\theta_{eff}^{d\ quark}\sim((\mathbf{Y}_{d}^{\dagger}\mathbf{Y}_{d}\mathbf{)}^{2}\mathbf{Y}_{u}^{\dagger}\mathbf{Y}_{u}\mathbf{Y}_{d}^{\dagger}\mathbf{Y}_{d}(\mathbf{Y}_{u}^{\dagger}\mathbf{Y}_{u})^{2}-(\mathbf{Y}_{u}^{\dagger}\mathbf{Y}_{u})^{2}\mathbf{Y}_{d}^{\dagger}\mathbf{Y}_{d}\mathbf{Y}_{u}^{\dagger}\mathbf{Y}_{u}(\mathbf{Y}_{d}^{\dagger}\mathbf{Y}_{d}\mathbf{)}^{2})^{11}\;,
\end{equation}
et donc la somme sur les trois saveurs $d,s,b$ donne $\Delta\theta_{eff}^{quarks}\sim J_{\mathcal{CP}}\sim\Delta\theta_{eff}^{gluon}$.
Cette observation a été faite dans \cite{Khriplovich1993pf} dans
le contexte de l'étude de la contribution électrofaible divergente
dominante à $\theta^{eff}$.

Dans le monde réel, la règle de somme pourrait être secouée par des
corrections QCD dans la mesure où ces dernières adoucissent la brisure
de GIM quadratique en une dépendance logarithmique des masses des
quarks. Dans \cite{Ellis1978hq}, il y est trouvé que la contribution
dominante émerge à l'ordre $\mathcal{O}(\alpha_{S}^{3})$, donc nous
construisons la tentative d'estimation
\begin{equation}
\Delta\theta_{eff}^{quarks}\sim\left(\frac{g^{2}}{4\pi^{2}}\right)^{2}\left(\frac{\alpha_{S}}{\pi}\right)^{3}\max_{i}|\mathbf{X}_{d}^{ii}|\approx\left(\frac{g^{2}}{4\pi^{2}}\right)^{2}\left(\frac{\alpha_{S}}{\pi}\right)^{3}\frac{m_{b}^{2}m_{c}^{2}m_{t}^{4}}{M_{W}^{8}}\mathcal{J}_{\mathcal{CP}}\approx10^{-16}\;.
\end{equation}
Un calcul exact à cet ordre n'a pas encore été fait. A ce stade, on
doit également mentionner l'évaluation de \cite{Gerard2012ud}, où
les contributions à longue distance sont estimées en faisant correspondre
(matching) le processus $\eta^{(\prime)}\rightarrow\pi\pi$ induit
par $\theta^{eff}$ et celui obtenu au second ordre d'interaction
faible, avec comme résultat $\Delta\theta_{eff}^{LD}\approx10^{-17}$.

\subsection{Impact de la nouvelle physique sur les EDMs des quarks et des leptons
sous MFV}

Dans cette partie, on suppose l'existence de nouvelle physique (NP)
en imposant l'hypothèse MFV, de sorte que la totalité du secteur de
la saveur reste contrôlé par les matrices de couplages de Yukawa seulement.
Si on suppose de plus que le reste de la dynamique de NP conserve
$\mathcal{CP}$, les mêmes combinaisons de spurions que celles dans
le MS sont pertinentes pour décrire toute la violation-$\mathcal{CP}$
diagonale dans l'espace des saveurs.

Bien qu'identiques analytiquement, trois effets altèrent les estimations
numériques. Premièrement, la dynamique de NP peut être bien moins
restrictive que le MS et ces combinaisons de spurions peuvent a priori
provenir de diagrammes plus simples. Deuxièmement, les valeurs des
matrices de couplages de Yukawa peuvent être différentes si plus d'un
multiplet de Higgs est présent. Par exemple, considérons le THDM de
type II, dans lequel l'équation (\ref{G=0000E8leSpurions}) devient
\begin{equation}
v_{u}\mathbf{Y}_{u}\overset{gel\acute{e}}{\rightarrow}\mathbb{M}_{u}V_{CKM},\;\;v_{d}\mathbf{Y}_{d}\overset{gel\acute{e}}{\rightarrow}\mathbb{M}_{d},\;\;v_{d}\mathbf{Y}_{e}\overset{gel\acute{e}}{\rightarrow}\mathbb{M}_{e}\;,
\end{equation}
avec $\tan\beta=v_{u}/v_{d}$ et $v_{u,d}=\langle H_{u,d}^{0}\rangle$
les VEVs des deux scalaires neutres. Lorsque $\tan\beta$ est grand,
$\mathbf{Y}_{d}$ devient aussi grand que $\mathbf{Y}_{u}$, et
\begin{equation}
J_{\mathcal{CP}}^{NP}\approx10^{-12}\times\left(\frac{\tan\beta}{50}\right)^{6}\;\;,\;\;\;[\mathbf{X}_{d}^{NP}]^{11}\approx10^{-9}\times\left(\frac{\tan\beta}{50}\right)^{2}\;\;,\;\;\;[\mathbf{X}_{u}^{NP}]^{11}\approx10^{-7}\times\left(\frac{\tan\beta}{50}\right)^{6}\,\;.
\end{equation}
Troisièmement, les échelles apparaissant dans l'équation (\ref{eq:EMO})
doivent être au dessus de l'échelle électrofaible et est posée à $1$
TeV.

En combinant ces trois effets numériques et en supposant que les EDMs
sont déjà induits à une boucle, les prédictions pour les contributions
savoureuses sont
\begin{equation}
\begin{aligned}d_{e} & \sim e\frac{m_{e}}{\Lambda^{2}}\left(\frac{g^{2}}{16\pi^{2}}\right)J_{\mathcal{CP}}^{NP}\approx10^{-37}\times\left(\frac{1\,TeV}{\Lambda}\right)^{2}\times\left(\frac{\tan\beta}{50}\right)^{6}\;e\cdot cm\;,\\
d_{d} & \sim e\frac{m_{d}}{\Lambda^{2}}\left(\frac{g^{2}}{16\pi^{2}}\right)[\mathbf{X}_{d}^{NP}]^{11}\approx10^{-33}\times\left(\frac{1\,TeV}{\Lambda}\right)^{2}\times\left(\frac{\tan\beta}{50}\right)^{2}\,\;e\cdot cm\;,\\
d_{u} & \sim e\frac{m_{u}}{\Lambda^{2}}\left(\frac{g^{2}}{16\pi^{2}}\right)[\mathbf{X}_{u}^{NP}]^{11}\approx10^{-32}\times\left(\frac{1\,TeV}{\Lambda}\right)^{2}\times\left(\frac{\tan\beta}{50}\right)^{6}\,\;e\cdot cm\;.
\end{aligned}
\label{eq:DirectTheta}
\end{equation}
Ceci correspond par exemple à la situation dans le MSSM pour les contributions
provenant des phases violant-$\mathcal{CP}$ présentes dans les couplages
savoureux de squarks (avec MFV est imposée, voir \cite{MercolliS09}).
Notons que lorsque $\tan\beta$ augmente, ces contributions évoluent
en
\begin{equation}
\frac{|d_{d}|}{m_{d}}\approx10^{3}\times\left(\frac{50}{\tan\beta}\right)^{4}\times\frac{|d_{e}|}{m_{e}}\;,\;\;\frac{|d_{u}|}{m_{u}}\sim10^{5}\times\frac{|d_{e}|}{m_{e}}\;.
\end{equation}
Etant donné les limites actuelles, $d_{e}$ et $d_{n}$ sont similairement
sensibles à la phase de violation-$\mathcal{CP}$ à grand $\tan\beta$,
et clairement aucuns des deux n'est accessible par les expériences
actuelles et en vue dans un futur proche.

Au-delà de ces contributions directes à l'EDM, des boucles similaires
décalent $\theta_{eff}^{C}$. La différence principale est que ces
contributions ne se découplent pas dans la mesure où elles peuvent
provenir directement des self-énergies du gluon ou du quark. Pour
spécifier,
\begin{equation}
\Delta\theta_{eff}^{gluon}\sim\left(\frac{g^{2}}{4\pi^{2}}\right)J_{\mathcal{CP}}^{NP}\approx10^{-14}\times\left(\frac{\tan\beta}{50}\right)^{6}\;,\label{eq:GluonShiftNP}
\end{equation}
ce qui est plus contraignant que les contributions directes de l'équation
(\ref{eq:DirectTheta}), bien qu'encore compatible avec $\theta_{eff}^{C}\lesssim10^{-10}$.
Notons que si un mécanisme est introduit pour résoudre le problème
de $\mathcal{CP}$ fort du MS en forçant $\theta_{eff}^{C}=0$, comme
par exemple en introduisant un axion, alors ce même mécanisme tuerait
aussi $\Delta\theta_{eff}^{gluon}$, et les équations (\ref{eq:DirectTheta})
redeviennent la contribution dominante aux EDMs.

A partir des self-énergies des quarks, le décalage en $\theta_{eff}^{C}\lesssim10^{-10}$
peut être estimé
\begin{equation}
\begin{aligned}\Delta\theta_{eff}^{d-quarks} & \sim\left(\frac{g^{2}}{4\pi^{2}}\right)[\mathbf{X}_{d}^{NP}]^{11}\approx10^{-11}\times\left(\frac{\tan\beta}{50}\right)^{2}\;,\\
\Delta\theta_{eff}^{u-quarks} & \sim\left(\frac{g^{2}}{4\pi^{2}}\right)[\mathbf{X}_{u}^{NP}]^{11}\approx10^{-10}\times\left(\frac{\tan\beta}{50}\right)^{6}\;.
\end{aligned}
\end{equation}
Ces contributions pousseraient $\theta_{eff}^{C}\lesssim10^{-10}$
très proche de sa limite actuelle obtenue par l'EDM du neutron. Cependant,
on doit encore sommer sur les trois saveurs. A ce stade, des grandes
annulations peuvent être attendues. Premièrement, les insertions de
spurions ne proviennent pas forcement des insertions de masses des
quarks. Par exemple, dans un contexte supersymétrique, ils pourraient
provenir directement des termes de brisure douce (soft-breaking) de
squarks sur lesquels on impose MFV. Alternativement, en partant de
conditions aux limites universelles, ils émergeraient de l'évolution
du groupe de renormalisation jusqu'à l'échelle basse. Deuxièmement,
la séparation dynamique des contributions de chaque saveur ne serait
probablement pas aussi effective que dans le MS. Dans le MSSM avec
MFV, les squarks d'un type donné peuvent être quasi dégénérés. Pour
ces raisons, on s'attendrait à ce que la règle de somme (\ref{eq:SumRuleTheta})
soit vérifiée, du moins dans une bonne approximation, et donc que
$\Delta\theta_{eff}^{quarks}\approx\Delta\theta_{eff}^{gluon}$ apparaissant
dans l'équation (\ref{eq:GluonShiftNP}) et qui est hors d'atteinte.

\section{Estimations des EDMs en présence des masses des neutrinos\label{SecPMNS}}

De façon générale, pour prendre en compte les masses des neutrinos,
la dynamique du MS doit être supplémentée par des nouvelles interactions
savoureuses. Le contenu minimal de spurions utilisé jusqu'ici doit
alors être étendu en y incluant des spurions liés aux neutrinos. De
plus, ce contenu en spurions dépend du mécanisme adopté pour la génération
des masses des neutrinos. Ainsi, en repartant des trois opérateurs
magnétiques de l'équation (\ref{eq:EMO}), le but de cette section
est d'analyser la paramétrisation de $\mathbf{X}_{u,d,e}$ en présence
des nouveaux spurions résultants des scénarios les plus simples de
générations de masse des neutrinos.

Quelques propriétés générales peuvent être immédiatement identifiées.
D'abord, les contributions aux EDMs des quarks de type up (down) à
partir du premier (second) opérateur nécessitent que $\mathbf{X}_{u(d)}$
soit complexe dans la base où $\mathbf{Y}_{u(d)}$ est diagonale et
réelle. Comme le groupe de saveur des quarks et des leptons reste
factorisé dans tous les scénarios considérés ici, $\mathbf{X}_{u,d,e}$
doit être la matrice identité multipliée par une trace invariante
(sous le groupe de saveur) sur les spurions leptoniques. Les EDMs
de quarks émergent alors seulement quand ces traces sont complexes,
c'est à dire quand
\begin{equation}
\mathbf{X}_{u,d}=\mathbb{I}_{3}\times J_{\mathcal{CP}}\;\rightarrow d_{u,d}\sim e\frac{c_{u,d}}{\Lambda^{2}}m_{u,d}\times\operatorname{Im}J_{\mathcal{CP}}\;,\label{eq:LquarkEDM}
\end{equation}
avec $\operatorname{Im}J_{\mathcal{CP}}\neq0$. De telles traces invariantes
de saveur violant-$\mathcal{CP}$ ont déjà été largement étudiées
dans la littérature pour plusieurs modèles de masses de neutrinos
(voir en particulier \cite{Branco11}), mais seront quand même incluses
dans la suite par souci de complétude. Parce que la phase de violation-$\mathcal{CP}$
provient d'un invariant de saveur et avec $c_{u,d}$ des combinaisons
non-savoureuses de constantes de couplages de jauges et de facteurs
de boucles, on s'attend à ce que les relations suivantes
\begin{equation}
\frac{d_{u}}{m_{u}}=\frac{d_{c}}{m_{c}}=\frac{d_{t}}{m_{t}}=\frac{d_{d}}{m_{d}}=\frac{d_{s}}{m_{s}}=\frac{d_{b}}{m_{b}}\;,
\end{equation}
soient vérifiées, modulo des dépendances sous-dominantes en les masses
des particules dans les boucles.

D'autre part, concernant l'opérateur magnétique leptonique, $\mathbf{X}_{e}$
doit être une chaîne de spurions leptoniques se transformant comme
un octet sous $SU(3)_{L}$. Dans la base où $\mathbf{Y}_{e}$ est
diagonale réelle, cela induit alors des processus avec changement
de saveurs leptoniques\footnote{Lepton Flavor Violating processes, LFV.}
$\ell^{I}\rightarrow\ell^{J}\gamma$ dès que la chaîne de spurions
n'est pas diagonale, $\mathbf{X}_{e}^{IJ}\neq0$, avec un taux
\begin{equation}
\Gamma\left(\ell^{I}\rightarrow\ell^{J}\gamma\right)=\frac{\alpha m_{\ell^{I}}^{5}c_{e}^{2}}{8\Lambda^{4}}\times|\mathbf{X}_{e}^{IJ}|^{2}\;,\label{eq:LFVemo}
\end{equation}
et des EDMs de leptons $d_{\ell^{I}}$ dès que ses coefficients diagonaux
sont complexes, $\operatorname{Im}\mathbf{X}_{e}^{II}\neq0$
\begin{equation}
d_{e}\sim e\frac{c_{e}}{\Lambda^{2}}m_{e}\times\operatorname{Im}\mathbf{X}_{e}^{11}\;.\label{eq:LleptonEDM}
\end{equation}
Typiquement, la contribution dominante à $d_{\ell}$ provient d'une
chaîne de spurions de trace nulle $\langle\mathbf{X}_{e}\rangle=0$,
ce qui implique que la règle de somme suivante soit vérifiée
\begin{equation}
\frac{d_{e}}{m_{e}}+\frac{d_{\mu}}{m_{\mu}}+\frac{d_{\tau}}{m_{\tau}}=0\;.\label{eq:SRlept}
\end{equation}

\subsection{Masses de Dirac pour les neutrinos}

Dans ce scénario, les masses des neutrinos sont introduites dans le
MS en ajoutant trois neutrinos de chiralités droites afin de pourvoir
écrire une interaction de Yukawa supplémentaire
\begin{equation}
\mathcal{L}_{\text{Yukawa}}=-\bar{U}\mathbf{Y}_{u}QH^{\dagger C}-\bar{D}\mathbf{Y}_{d}QH^{\dagger}-\bar{E}\mathbf{Y}_{e}LH^{\dagger}-\bar{N}\mathbf{Y}_{\nu}LH^{\dagger C}+h.c.\;.
\end{equation}
Ces neutrinos droits possèdent des nombres quantiques de jauges triviaux
$N\sim(\mathbf{1},\mathbf{1})_{0}$ sous $SU(3)_{C}\otimes SU(2)_{L}\otimes U(1)_{Y}$.
En présence de la nouvelle structure de saveur $\mathbf{Y}_{\nu}$,
il n'est plus possible de se débarrasser de tous les mélanges de saveurs
dans le secteur des leptons. Les SVD de $\mathbf{Y}_{e}$ et $\mathbf{Y}_{\nu}$
sont respectivement $vV_{R}^{e}\mathbf{Y}_{e}V_{L}^{e}=\mathbf{m}_{e}$
et $vV_{R}^{\nu}\mathbf{Y}_{\nu}V_{L}^{\nu}=\mathbf{m}_{\nu}$. La
non-concordance des rotations gauches définit la matrice PMNS \cite{PMNS}
\begin{equation}
U_{PMNS}^{\mathrm{Dirac}}\equiv V_{L}^{e\dagger}V_{L}^{\nu}\;.
\end{equation}

Avec cela, les valeurs de fonds des spurions, dans la base d'états
propres de masses des leptons chargés, sont données par
\begin{equation}
v\mathbf{Y}_{e}\overset{gel\acute{e}}{\rightarrow}\mathbb{M}_{e}\;,\;v\mathbf{Y}_{\nu}\overset{gel\acute{e}}{\rightarrow}\mathbb{M}_{\nu}U_{PMNS}^{\mathrm{Dirac}\dagger}\;.\label{eq:DiracFreeze}
\end{equation}
Les valeurs des différents paramètres libres, extraites des données
d'oscillations de neutrinos, sont prises des meilleurs ajustements
de\cite{NeutrinoData}:
\begin{equation}
\begin{aligned}\Delta m_{21}^{2} & =\Delta m_{\odot}^{2}=7.5_{-0.17}^{+0.19}\times10^{-5}\,\text{eV}^{2},\;|\Delta m_{31}^{2}|=\Delta m_{atm}^{2}=2.524_{-0.040}^{+0.039}\times10^{-3}\,\text{eV}^{2}\;,\\
\theta_{12} & =\theta_{\odot}=(33.56_{-0.75}^{+0.77}){}^{\circ},\;\theta_{23}=\theta_{atm}=(41.6_{-1.2}^{+1.5}){}^{\circ},\;\theta_{13}=(8.46\pm0.15){}^{\circ}\;,
\end{aligned}
\end{equation}
pour une hiérarchie de masse normale, ce que nous supposerons dans
la suite de cette étude. A partir de maintenant, les prédictions pour
les FCNCs leptoniques ou encore les contributions de la phase PMNS
aux EDMs établissent un parallèle strict avec ce qui a été effectué
dans la section \ref{FlavorSymmetryMethod}. Pour mettre en place
le cadre pour les sections suivantes, établissons les explicitement
quand même.

\begin{figure}[t]
\centering
\includegraphics[width=0.95\textwidth]{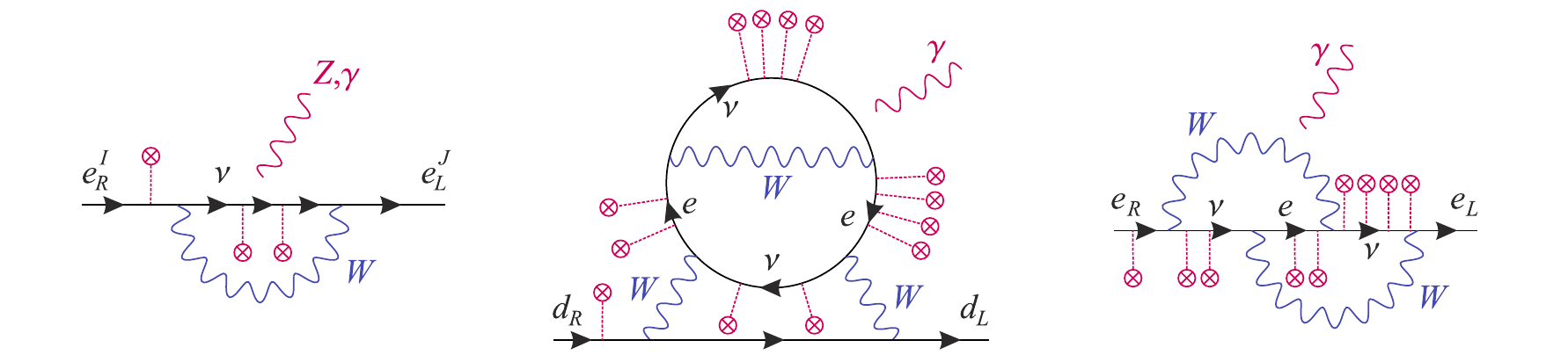}
\caption{Exemples d'insertions de spurions dominantes pour les transitions LFV $\ell^{I}\rightarrow\ell^{J}\gamma$, EDMs de quarks et de leptons, lorsque les masses des neutrinos sont purement de type Dirac.}
\label{FigDirac}
\end{figure}

\paragraph{Lepton flavor violation:}

Dans le MS étendu d'un terme de masse de Dirac pour les neutrinos,
le processus $\mu\rightarrow e\gamma$ résulte d'échanges virtuels
de $W$ (voir figure \ref{FigDirac}) et peut être paramétrisée par
l'opérateur magnétique effectif de dimension-6 de l'équation (\ref{eq:EMO})
en posant
\begin{equation}
\frac{c_{e}}{\Lambda^{2}}=\frac{G_{F}}{16\pi^{2}\sqrt{2}}\;\;,\;\;\mathbf{X}_{e}^{\mathrm{Dirac}}=\mathbf{1}\oplus\mathbf{Y}_{\nu}^{\dagger}\mathbf{Y}_{\nu}\oplus\mathbf{Y}_{e}^{\dagger}\mathbf{Y}_{e}+...\;,
\end{equation}
où $\oplus$ sert à rappeler qu'en principe, différents coefficients
de $\mathcal{O}(1)$ apparaissent devant chaque terme. Dans la base
propre de masses des leptons chargés, la contribution dominante vient
de $\mathbf{Y}_{\nu}^{\dagger}\mathbf{Y}_{\nu}$. En gelant les spurions
comme dans l'équation (\ref{eq:DiracFreeze}) et sous l'approximation
tribimaximale, les prédictions des taux de désintégrations sont les
suivantes
\begin{equation}
\begin{aligned}\mathcal{B}\left(\mu\rightarrow e\gamma\right) & \approx\lambda_{\mu}\left|\Delta m_{\odot}^{2}/3v^{2}\right|^{2}\approx10^{-58}\;,\;\;\;[\mathcal{B}^{\exp}<5.7\times10^{-13}\;\text{\cite{ExpMEG}}]\;,\\
\mathcal{B}\left(\tau\rightarrow e\gamma\right) & \approx\lambda_{\tau}\left|\Delta m_{\odot}^{2}/3v^{2}\right|^{2}\approx10^{-59}\;,\;\;\;[\mathcal{B}^{\exp}<3.3\times10^{-8}\;\text{\cite{ExpTLG}}]\;,\\
\mathcal{B}\left(\tau\rightarrow\mu\gamma\right) & \approx\lambda_{\tau}\left|\Delta m_{atm}^{2}/2v^{2}\right|^{2}\approx10^{-49}\;,\;[\mathcal{B}^{\exp}<4.4\times10^{-8}\;\text{\cite{ExpTLG}}]\;,
\end{aligned}
\label{eq:LFVrates}
\end{equation}
où l'on a posé $\lambda_{\ell}=\tau_{\ell}G_{F}^{2}\alpha m_{\ell}^{5}/2048\pi^{4}$.
Ces valeurs sont extrêmement petites, bien au-delà des sensibilités
expérimentales. Notons qu'en raison du mécanisme de GIM, ce sont les
différences de masses des particules dans la boucle électrofaible
qui importent.

\paragraph{EDMs de quarks:}

Ils sont induits par les traces invariantes de saveurs sur les spurions
leptoniques, voir figure \ref{FigDirac}. En complète analogie avec
les contributions CKM aux EDMs de leptons, on peut directement écrire
\begin{equation}
\begin{aligned}J_{\mathcal{CP}}^{\mathrm{Dirac}} & \equiv\frac{1}{2i}\det[\mathbf{Y}_{e}^{\dagger}\mathbf{Y}_{e},\mathbf{Y}_{\nu}^{\dagger}\mathbf{Y}_{\nu}]=\frac{1}{2}\operatorname{Im}\langle(\mathbf{Y}_{\nu}^{\dagger}\mathbf{Y}_{\nu})^{2}\mathbf{Y}_{e}^{\dagger}\mathbf{Y}_{e}\mathbf{Y}_{\nu}^{\dagger}\mathbf{Y}_{\nu}(\mathbf{Y}_{e}^{\dagger}\mathbf{Y}_{e})^{2}\rangle\\
 & =\mathcal{J}_{\mathcal{CP}}^{\mathrm{Dirac}}\prod_{\substack{i>j=e,\mu,\tau\\
i>j=\nu_{1},\nu_{2},\nu_{3}
}
}\frac{m_{i}^{2}-m_{j}^{2}}{v^{2}}\approx\mathcal{J}_{\mathcal{CP}}^{\mathrm{Dirac}}\frac{m_{\tau}^{4}m_{\mu}^{2}(\Delta m_{atm}^{2})^{2}\Delta m_{\odot}^{2}}{v^{12}}\approx10^{-93}\sin(\delta_{13})\;,
\end{aligned}
\label{eq:JCPdirac}
\end{equation}
avec $\mathcal{J}_{\mathcal{CP}}^{\mathrm{Dirac}}$ donné par la même
expression que pour l'invariant de Jarlskog de l'équation (\ref{eq:JarlAngles}).
Numériquement, $\mathcal{J}_{\mathcal{CP}}^{\mathrm{Dirac}}\approx0.033(2)\times\sin(\delta_{13})$
n'est pas très loin de sa valeur maximale $1/6\sqrt{3}\approx0.096$
lorsque $\delta_{13}$ est $\mathcal{O}(1)$. Néanmoins, $J_{\mathcal{CP}}^{\mathrm{Dirac}}$
est lourdement supprimé par la dépendance en $\mathcal{O}(m_{\nu}^{6})$
et ne fait pas le poids face aux contributions CKM à $d_{u,d}$.

\paragraph{EDMs de leptons:}

La combinaison $\mathbf{X}_{e}$ dans l'équation (\ref{eq:EMO}) doit
être une chaîne de spurions avec des coefficients diagonaux complexes.
Ce cas est très similaire à celui des quarks étant donné que les identités
CH permettent de construire l'équivalent de la base dans l'équation
(\ref{eq:Octet}), avec $\mathbf{Y}_{\nu}^{\dagger}\mathbf{Y}_{\nu}$
et $\mathbf{Y}_{e}^{\dagger}\mathbf{Y}_{e}$ au lieu de $\mathbf{Y}_{u}^{\dagger}\mathbf{Y}_{u}$
et $\mathbf{Y}_{d}^{\dagger}\mathbf{Y}_{d}$. La chaîne non-hermitienne
la plus simple est donc
\begin{equation}
\mathbf{X}_{e}^{\mathrm{Dirac}}=[\mathbf{Y}_{\nu}^{\dagger}\mathbf{Y}_{\nu}\;,\;\mathbf{Y}_{\nu}^{\dagger}\mathbf{Y}_{\nu}\mathbf{Y}_{e}^{\dagger}\mathbf{Y}_{e}\mathbf{Y}_{\nu}^{\dagger}\mathbf{Y}_{\nu}]\;,\label{eq:YnEDM}
\end{equation}
et correspond aux processus rainbow faibles du second ordre représentés
en figure \ref{FigDirac}. Pour l'EDM de l'électron, on obtient
\begin{equation}
(\mathbf{X}_{e}^{\mathrm{Dirac}})^{11}=2i\mathcal{J}_{\mathcal{CP}}^{\mathrm{Dirac}}\;\frac{m_{\tau}^{2}-m_{\mu}^{2}}{v^{2}}\prod_{i>j=\nu_{1},\nu_{2},\nu_{3}}\frac{m_{i}^{2}-m_{j}^{2}}{v^{2}}\approx\mathcal{J}_{\mathcal{CP}}^{\mathrm{Dirac}}\;\frac{m_{\tau}^{2}(\Delta m_{atm}^{2})^{2}\Delta m_{\odot}^{2}}{v^{8}}\approx10^{-82}\sin(\delta_{13})\;,\label{eq:DiracX11}
\end{equation}
ce qui se traduit en $d_{e}\lesssim10^{-107}\ensuremath{~}e\cdot cm$.
Cette valeur est à peine plus grande que la contribution proportionnelle
à $J_{\mathcal{CP}}^{\mathrm{Dirac}}$ dans l'équation (\ref{eq:JCPdirac})
et très petite devant la contribution CKM de l'équation (\ref{eq:eEDMCKM}).
De plus, de la même manière que pour les quarks, cet invariant ne
peut pas être obtenu à partir de diagrammes à deux boucles et le prix
à payer pour une boucle supplémentaire est une correction électromagnétique.
Finalement, la règle de somme (\ref{eq:SRlept}) est vérifiée puisque
$\langle\mathbf{X}_{e}^{\mathrm{Dirac}}\rangle=0$. En fait, on a
même $d_{e}/m_{e}\approx-d_{\mu}/m_{\mu}$ car $d_{\tau}/m_{\tau}$
est proportionnel à $m_{\mu}^{2}-m_{e}^{2}$ au lieu de $m_{\tau}^{2}-m_{\mu,e}^{2}\approx m_{\tau}^{2}$
pour $d_{e,\mu}$.

Pour des neutrinos de Dirac, il y a aussi la possibilité d'induire
des EDMs de neutrinos avec l'opérateur $\bar{N}^{I}(\mathbf{Y}_{\nu}\mathbf{X}_{\nu})^{IJ}\sigma_{\mu\nu}L^{J}F^{\mu\nu}H$.
La chaîne de spurions $\mathbf{X}_{\nu}^{\mathrm{Dirac}}$ est obtenue
à partir de $\mathbf{X}_{e}^{\mathrm{Dirac}}$ dans l'équation (\ref{eq:YnEDM})
en interchangeant $\mathbf{Y}_{\nu}\leftrightarrow\mathbf{Y}_{e}$
et en se plaçant dans la base d'interaction où les neutrinos sont
états propres de masses. Il est alors fortement amplifié par les facteurs
de masses, avec par exemple
\begin{equation}
\frac{d_{\nu_{1}}}{d_{e}}=-\frac{m_{\nu_{1}}(m_{\tau}^{2}-m_{e}^{2})(m_{\mu}^{2}-m_{e}^{2})}{m_{e}\Delta m_{atm}^{2}\Delta m_{\odot}^{2}}\approx10^{36}\;,
\end{equation}
pour $m_{\nu_{1}}\approx1$ eV, mais reste quand même totalement hors
d'atteinte expérimentalement \cite{Abel1999yz}.

\subsection{Masses de Majorana pour les neutrinos}

Au lieu d'introduire des neutrinos droits, les neutrinos gauches peuvent
être directement dotés d'un terme de masse invariant de jauge mais
violant le nombre leptonique de la manière suivante:
\begin{equation}
\mathcal{L}_{\text{Yukawa}}=-\bar{U}\mathbf{Y}_{u}QH^{\dagger C}-\bar{D}\mathbf{Y}_{d}QH^{\dagger}-\bar{E}\mathbf{Y}_{e}LH^{\dagger}-\frac{1}{2v}(L^{I}H)\left(\mathbf{\Upsilon}_{\nu}\right)^{IJ}(L^{J}H)+h.c.\;.
\end{equation}
L'interaction non-renormalisable de dimension-5, appelée \emph{opérateur
de Weinberg} \cite{BLWeinberg}, se réduit à un terme de masse de
Majorana $v\left(\mathbf{\Upsilon}_{\nu}\right)^{IJ}\nu_{L}^{I}\nu_{L}^{J}$
lorsque le champ de Higgs acquiert sa VEV. Comme dans le cas de Dirac,
il n'y a alors que deux spurions élémentaires à basse énergie. Afin
de fixer leurs valeurs de fonds, notons dans un premier temps que
les rotations unitaires nécessaires pour passer de la base d'interaction
à la base des états propres de masses sont $vV_{R}^{e}\mathbf{Y}_{e}V_{L}^{e}=\mathbb{M}_{e}$
et $vV_{L}^{\nu T}\mathbf{\Upsilon}_{\nu}V_{L}^{\nu}=\mathbb{M}_{\nu}$
où $\mathbb{M}_{\nu}=\operatorname{diag}(m_{\nu1},m_{\nu2},m_{\nu3})$
sont les masses (réelles) des neutrinos. Pour le spurion de neutrinos,
seule une matrice $V_{L}^{\nu}$ apparaît car $\mathbf{\Upsilon}_{\nu}$
est une matrice symétrique dans l'espace des saveurs et de ce fait
la SVD devient une diagonalisation orthogonale. En choisissant d'appliquer
la rotation sur le doublet de leptons $V_{L}^{e}$, on peut atteindre
la base d'interaction où
\begin{equation}
v\mathbf{Y}_{e}\overset{gel\acute{e}}{\rightarrow}\mathbb{M}_{e},\;\;\;v\mathbf{\Upsilon}_{\nu}\overset{gel\acute{e}}{\rightarrow}V_{L}^{eT}V_{L}^{\nu\ast}\mathbb{M}_{\nu}V_{L}^{\nu\dagger}V_{L}^{e}\equiv U_{PMNS}^{\ast}\mathbb{M}_{\nu}U_{PMNS}^{\dagger}\;,\label{eq:PMNSmaj2}
\end{equation}
où $U_{PMNS}\equiv V_{L}^{e\dagger}V_{L}^{\nu}$ est reliée à la matrice
PMNS vu dans le cas de Dirac comme ceci
\begin{equation}
U_{PMNS}=U_{PMNS}^{\mathrm{Dirac}}\cdot\operatorname{diag}(1,e^{i\alpha_{M}},e^{i\beta_{M}})\;.\label{eq:PMNS2}
\end{equation}
Contrairement au cas de Dirac, ces phases sont irréductibles et ne
peuvent donc pas être être éliminées par des rotations essentiellement
car le nombre leptonique n'est plus conservé. Une des phases additionnelles
est conventionnellement éliminée en tant que phase globale non pertinente,
tandis que les deux autres sont appelées \emph{phases de Majorana}.

\paragraph{Processus LFV:}

\begin{figure}[t]
\centering
\includegraphics[width=0.95\textwidth]{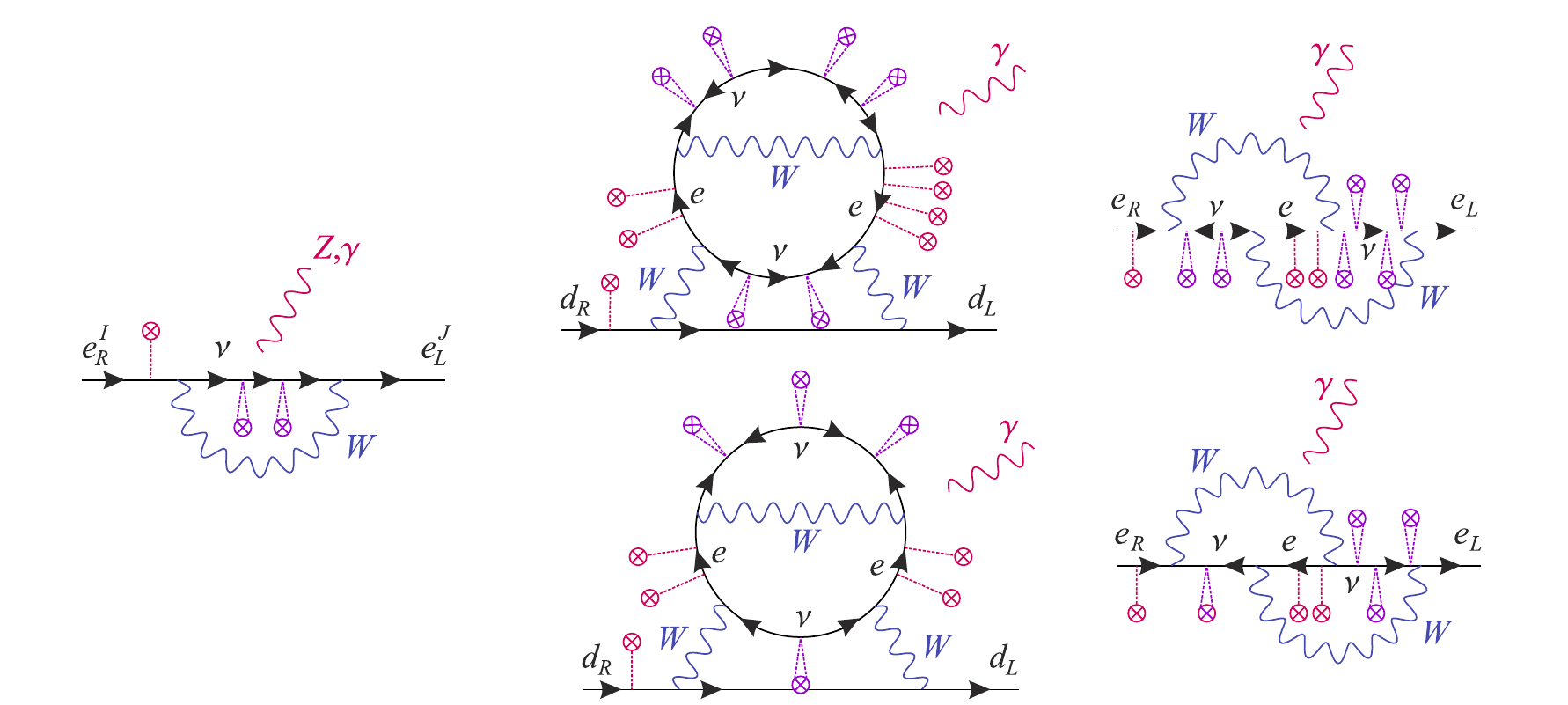}
\caption{Exemples d'insertions de spurions pour les transitions LFV, les EDMs de quarks et de leptons dans le cas des neutrinos purement de Majorana. Les doubles tadpoles dénotent les insertions de masses de Majorana, tandis que les tadpoles simples dénotent des spurions de Yukawa de leptons chargés. Les diagrammes du haut représentent les combinaisons de type Jarlskog des équations \ref{eq:MajoJ1} et \ref{MajoX1}, alors que celles du bas montrent les combinaisons spécifiques au cas de Majorana, équations \ref{eq:MajoJ2} et \ref{MajoX2}.}
\label{FigMajo}
\end{figure}

Si les neutrinos sont purement de Majorana, les processus LFV sont
encodés dans l'opérateur de l'équation (\ref{eq:EMO}) avec $\mathbf{X}_{e}$
donnée par
\begin{equation}
\mathbf{X}_{e}^{\mathrm{Majo}}=\mathbf{\Upsilon}_{\nu}^{\dagger}\mathbf{\Upsilon}_{\nu}\;.
\end{equation}
Ceci est représenté en figure \ref{FigMajo}. Ce mécanisme produit
les mêmes amplitudes que dans le cas des neutrinos de Dirac étant
donné que
\begin{equation}
\mathbf{\Upsilon}_{\nu}^{\dagger}\mathbf{\Upsilon}_{\nu}=\frac{1}{v^{2}}U_{PMNS}\mathbb{M}_{\nu}^{2}U_{PMNS}^{\dagger}=(\mathbf{Y}_{\nu}^{\dagger}\mathbf{Y}_{\nu})^{\mathrm{Dirac}}\;,\label{eq:MDequal}
\end{equation}
et les taux en $\Delta m_{\nu}^{4}$ sont les mêmes que dans l'équation
(\ref{eq:LFVrates}).

\paragraph{EDMs de quarks:}

Le spurion $\mathbf{\Upsilon}_{\nu}$ se transforme différemment des
autres matrices de couplages de Yukawa et par conséquent ouvre plusieurs
nouvelles possibilités de contracter les spurions pour former des
invariants de saveur. Afin d'organiser le développement, notons d'abord
que grâce aux identités CH, une chaîne de spurions quelconque se transformant
comme un octet sous $SU(3)_{L}$ est nécessairement une combinaison
de seulement quatre monômes hermitien élémentaires, se transformant
aussi comme des octets de $SU(3)_{L}$
\begin{equation}
\mathbf{Y}_{e}^{\dagger}\mathbf{Y}_{e}\;,\;\;\mathbf{\Upsilon}_{\nu}^{\dagger}\mathbf{\Upsilon}_{\nu}\;,\;\;\mathbf{\Upsilon}_{\nu}^{\dagger}(\mathbf{Y}_{e}^{\dagger}\mathbf{Y}_{e})^{T}\mathbf{\Upsilon}_{\nu}\;,\;\;\mathbf{\Upsilon}_{\nu}^{\dagger}((\mathbf{Y}_{e}^{\dagger}\mathbf{Y}_{e})^{T})^{2}\mathbf{\Upsilon}_{\nu}\;.
\end{equation}
Les identités CH impliquent également que l'invariant imaginaire pur
le plus simple construit uniquement à partir de deux combinaisons
de spurions hermitiennes $\mathbf{A}$ et $\mathbf{B}$ est nécessairement
$\det[\mathbf{A},\mathbf{B}]$. Avec trois combinaisons de spurions
différentes, l'invariant complexe le plus simple est $\langle\mathbf{ABC}-\mathbf{CBA}\rangle$,
tandis qu'avec quatre combinaisons, a priori il y a beaucoup de nouveaux
invariants.

Pour spécifier, étant donné l'ensemble des combinaisons de spurions
se transformant comme des octets, l'analogue de l'invariant de Dirac
n'apporte rien de nouveau dans la mesure où l'équation (\ref{eq:MDequal})
reste vérifiée
\begin{equation}
J_{\mathcal{CP}}^{\mathrm{Majo,1}}=\frac{1}{2i}\det[\mathbf{Y}_{e}^{\dagger}\mathbf{Y}_{e},\mathbf{\Upsilon}_{\nu}^{\dagger}\mathbf{\Upsilon}_{\nu}]=J_{\mathcal{CP}}^{\mathrm{Dirac}}\;.\label{eq:MajoJ1}
\end{equation}
La sensibilité aux phases de Majorana est perdue dans $\mathbf{\Upsilon}_{\nu}^{\dagger}\mathbf{\Upsilon}_{\nu}$.
Par essais et erreurs, l'invariant le plus simple sensible à ces phases
s'avère être \cite{Branco11,Branco86}
\begin{equation}
J_{\mathcal{CP}}^{\mathrm{Majo,2}}=\frac{1}{2i}\langle\mathbf{\Upsilon}_{\nu}^{\dagger}\mathbf{\Upsilon}_{\nu}\cdot\mathbf{Y}_{e}^{\dagger}\mathbf{Y}_{e}\cdot\mathbf{\Upsilon}_{\nu}^{\dagger}(\mathbf{Y}_{e}^{\dagger}\mathbf{Y}_{e})^{T}\mathbf{\Upsilon}_{\nu}-\mathbf{\Upsilon}_{\nu}^{\dagger}(\mathbf{Y}_{e}^{\dagger}\mathbf{Y}_{e})^{T}\mathbf{\Upsilon}_{\nu}\cdot\mathbf{Y}_{e}^{\dagger}\mathbf{Y}_{e}\cdot\mathbf{\Upsilon}_{\nu}^{\dagger}\mathbf{\Upsilon}_{\nu}\rangle\;.\label{eq:MajoJ2}
\end{equation}
Bien que cette quantité purement imaginaire s'annule quand tous les
leptons chargés ou les neutrinos sont dégénérés en masses, ce n'est
pas le cas quand seulement deux leptons ou deux neutrinos le sont.
En conséquence, un produit simple de différences de masses ne peut
pas être factorisé et cet invariant n'a pas d'expression analytique
simple.

Si on cherche un invariant qui ne s'annule pas pour des neutrinos
dégénérés, on doit éviter les chaînes dans lesquelles des facteurs
$\mathbf{\Upsilon}_{\nu}^{\dagger}\mathbf{\Upsilon}_{\nu}$ ou $\mathbf{\Upsilon}_{\nu}\mathbf{\Upsilon}_{\nu}^{\dagger}=(\mathbf{\Upsilon}_{\nu}^{\dagger}\mathbf{\Upsilon}_{\nu})^{T}$
apparaissent, car on a $\mathbf{\Upsilon}_{\nu}^{\dagger}\mathbf{\Upsilon}_{\nu}=\mathbf{\Upsilon}_{\nu}\mathbf{\Upsilon}_{\nu}^{\dagger}=(m_{\nu}^{2}/v^{2})\mathbb{I}_{3}$
dans la limite dégénérée. Il s'ensuit que toutes occurrences de $\mathbf{\Upsilon}_{\nu}$
ou de $\mathbf{\Upsilon}_{\nu}^{\dagger}$ doit être encadrée par
des puissances de $\mathbf{Y}_{e}^{\dagger}\mathbf{Y}_{e}$ ou de
$(\mathbf{Y}_{e}^{\dagger}\mathbf{Y}_{e})^{T}$. Les invariants de
ce type les plus simples sont
\begin{equation}
\begin{aligned}J_{\mathcal{CP}}^{\mathrm{Majo,3}} & =\operatorname{Im}\langle(\mathbf{Y}_{e}^{\dagger}\mathbf{Y}_{e}\mathbf{)}^{2}\cdot\mathbf{\Upsilon}_{\nu}^{\dagger}(\mathbf{Y}_{e}^{\dagger}\mathbf{Y}_{e})^{T}\mathbf{\Upsilon}_{\nu}\cdot\mathbf{Y}_{e}^{\dagger}\mathbf{Y}_{e}\cdot(\mathbf{\Upsilon}_{\nu}^{\dagger}(\mathbf{Y}_{e}^{\dagger}\mathbf{Y}_{e})^{T}\mathbf{\Upsilon}_{\nu})^{2}\rangle\;,\\
J_{\mathcal{CP}}^{\mathrm{Majo,4}} & =\operatorname{Im}\langle(\mathbf{Y}_{e}^{\dagger}\mathbf{Y}_{e})^{2}\cdot\mathbf{\Upsilon}_{\nu}^{\dagger}(\mathbf{Y}_{e}^{\dagger}\mathbf{Y}_{e})^{T}\mathbf{\Upsilon}_{\nu}\cdot\mathbf{Y}_{e}^{\dagger}\mathbf{Y}_{e}\cdot\mathbf{\Upsilon}_{\nu}^{\dagger}((\mathbf{Y}_{e}^{\dagger}\mathbf{Y}_{e})^{2})^{T}\mathbf{\Upsilon}_{\nu}\rangle\;.
\end{aligned}
\label{eq:MajoJ4}
\end{equation}
L'invariant $2iJ_{\mathcal{CP}}^{\mathrm{Majo,3}}=\det[\mathbf{\Upsilon}_{\nu}^{\dagger}(\mathbf{Y}_{e}^{\dagger}\mathbf{Y}_{e})^{T}\mathbf{\Upsilon}_{\nu},\mathbf{Y}_{e}^{\dagger}\mathbf{Y}_{e}]$
a déjà été trouvé dans \cite{Branco98}, mais ce n'est pas le plus
grand car $J_{\mathcal{CP}}^{\mathrm{Majo,3}}=(m_{\nu}^{2}/v^{2})\times J_{\mathcal{CP}}^{\mathrm{Majo,4}}$
dans la limite dégénérée. Notons également que pour ces deux invariants,
le facteur $(\mathbf{Y}_{e}^{\dagger}\mathbf{Y}_{e})^{2}$ doit apparaître
au lieu de simplement $(\mathbf{Y}_{e}^{\dagger}\mathbf{Y}_{e})$
car sinon les identités CH permettraient de réordonner les termes
comme ceci $\mathbf{A}\cdot\mathbf{B}\cdot\mathbf{A}\rightarrow-\mathbf{A}^{2}\cdot\mathbf{B}-\mathbf{B}\cdot\mathbf{A}^{2}+(\text{moins de facteurs})$,
auquel cas les contractions $\mathbf{\Upsilon}_{\nu}^{\dagger}\mathbf{\Upsilon}_{\nu}$
ou $\mathbf{\Upsilon}_{\nu}\mathbf{\Upsilon}_{\nu}^{\dagger}$ apparaitraient
et les invariants s'annuleraient de nouveau dans la limite dégénérée.

\paragraph{EDMs de leptons:}

Pour chacune des traces invariantes précédentes, on peut construire
une chaîne de spurions non-hermitienne correspondante. Le raisonnement
est similaire au cas de Dirac et là aussi il faut au moins quatre
insertions de masses de neutrinos:

\begin{align}
\mathbf{X}_{e}^{\mathrm{Majo,1}}  &  =[\mathbf{\Upsilon}_{\nu}^{\dagger
}\mathbf{\Upsilon}_{\nu}\;,\;\mathbf{\Upsilon}_{\nu}^{\dagger}\mathbf{\Upsilon
}_{\nu}\mathbf{Y}_{e}^{\dagger}\mathbf{Y}_{e}\mathbf{\Upsilon}_{\nu}^{\dagger
}\mathbf{\Upsilon}_{\nu}]\;,\label{MajoX1}\\
\mathbf{X}_{e}^{\mathrm{Majo,2}}  &  =[\mathbf{\Upsilon}_{\nu}^{\dagger
}\mathbf{\Upsilon}_{\nu}\;,\;\mathbf{\Upsilon}_{\nu}^{\dagger}(\mathbf{Y}%
_{e}^{\dagger}\mathbf{Y}_{e})^{T}\mathbf{\Upsilon}_{\nu}]\;,\label{MajoX2}\\
\mathbf{X}_{e}^{\mathrm{Majo,3}}  &  =[\mathbf{\Upsilon}_{\nu}^{\dagger
}((\mathbf{Y}_{e}^{\dagger}\mathbf{Y}_{e})^{2})^{T}\mathbf{\Upsilon}_{\nu
}\;,\;\mathbf{Y}_{e}^{\dagger}\mathbf{Y}_{e}\cdot\mathbf{\Upsilon}_{\nu
}^{\dagger}(\mathbf{Y}_{e}^{\dagger}\mathbf{Y}_{e})^{T}\mathbf{\Upsilon}_{\nu
}\cdot\mathbf{Y}_{e}^{\dagger}\mathbf{Y}_{e}]\;,\\
\mathbf{X}_{e}^{\mathrm{Majo,4}}  &  =\mathbf{\Upsilon}_{\nu}^{\dagger
}((\mathbf{Y}_{e}^{\dagger}\mathbf{Y}_{e})^{2})^{T}\mathbf{\Upsilon}_{\nu
}\cdot\mathbf{Y}_{e}^{\dagger}\mathbf{Y}_{e}\cdot\mathbf{\Upsilon}_{\nu
}^{\dagger}(\mathbf{Y}_{e}^{\dagger}\mathbf{Y}_{e})^{T}\mathbf{\Upsilon}_{\nu
}\nonumber\\
&  \;\;\;\;\;\;\;\;\;-\mathbf{\Upsilon}_{\nu}^{\dagger}(\mathbf{Y}
_{e}^{\dagger}\mathbf{Y}_{e})^{T}\mathbf{\Upsilon}_{\nu}\cdot\mathbf{Y}%
_{e}^{\dagger}\mathbf{Y}_{e}\cdot\mathbf{\Upsilon}_{\nu}^{\dagger}%
((\mathbf{Y}_{e}^{\dagger}\mathbf{Y}_{e})^{2})^{T}\mathbf{\Upsilon}_{\nu}\;.
\label{MajoX4}%
\end{align}Ces structures de saveurs partagent plusieurs propriétés avec les
$J_{\mathcal{CP}}^{\mathrm{Majo,i}}$. Du fait de l'équation (\ref{eq:MDequal}),
la combinaison $\mathbf{X}_{e}^{\mathrm{Majo,1}}$ reproduit l'invariant
de Dirac de l'équation (\ref{eq:YnEDM}). Le $\mathbf{X}_{e}^{\mathrm{Majo,2}}$
est spécifique au cas de Majorana, il existe car il y a plus de deux
combinaisons de spurions se transformant comme des octets et les combinaisons
$\mathbf{\Upsilon}_{\nu}^{\dagger}\mathbf{\Upsilon}_{\nu}$ et $\mathbf{\Upsilon}_{\nu}^{\dagger}(\mathbf{Y}_{e}^{\dagger}\mathbf{Y}_{e})^{T}\mathbf{\Upsilon}_{\nu}$
sont toutes les deux non-diagonales dans la base d'interaction où
$\mathbf{Y}_{e}^{\dagger}\mathbf{Y}_{e}$ est diagonale. En outre,
comme $J_{\mathcal{CP}}^{\mathrm{Majo,2}}$, il ne dépend que de façon
quartique des masses de neutrinos, est sensible aussi bien à la phase
PMNS qu'aux phases de Majorana, a une expression analytique très compliquée,
mais s'annule dès lors que les trois leptons chargés ou les trois
neutrinos sont dégénérés en masses. D'autres structures de ce type
peuvent être construites mais elles impliquent toutes un plus grand
nombre d'insertions $\mathbf{Y}_{e}^{\dagger}\mathbf{Y}_{e}$ ou $\mathbf{\Upsilon}_{\nu}^{\dagger}\mathbf{\Upsilon}_{\nu}$
et sont donc plus supprimées.

Finalement, les deux derniers $\mathbf{X}_{e}^{\mathrm{Majo,3}}$
et $\mathbf{X}_{e}^{\mathrm{Majo,4}}$ sont les combinaisons les plus
simples survivant dans la limite dégénérée stricte des masses de neutrinos,
lorsque $\mathbf{\Upsilon}_{\nu}^{\dagger}\mathbf{\Upsilon}_{\nu}\rightarrow(m_{\nu}^{2}/v^{2})\mathbb{I}_{3}$.
Dans ce cas, notons que $(\mathbf{X}_{e}^{\mathrm{Majo,3}})^{ii}=(m_{\ell^{i}}/v)^{2}\times(\mathbf{X}_{e}^{\mathrm{Majo,4}})^{ii}$
étant donné que la chaîne de spurions $\mathbf{X}_{e}^{\mathrm{Majo,3}}$
finit ou commence par le facteur diagonal $\mathbf{Y}_{e}^{\dagger}\mathbf{Y}_{e}$.
Tout de même, la règle de somme (\ref{eq:SRlept}) reste vérifiée
dans les deux cas puisqu'on a $\langle\mathbf{X}_{e}^{\mathrm{Majo,3}}\rangle=\langle\mathbf{X}_{e}^{\mathrm{Majo,4}}\rangle=0$.

\paragraph{Estimations numériques des EDMs:\newline}

Pour estimer la taille des EDMs de quarks et de leptons, plusieurs
pièces doivent être assemblées. Premièrement, les combinaisons de
spurions sont évaluées en y injectant les valeurs de fonds dans l'équation
(\ref{eq:PMNSmaj2}). A ce stade, les expressions analytiques dans
la plupart des cas sont bien trop compliquées pour être écrites explicitement.
Néanmoins, afin d'illustrer les dépendances dans les différents paramètres,
donnons un exemple. On considère $J_{\mathcal{CP}}^{\mathrm{Majo,2}}$
en ne gardant que les termes dominants en $\mathcal{O}(m_{\tau}^{4})$
et jusqu'à l'ordre $\mathcal{O}(\sin\theta_{13})$:
\begin{equation}
\begin{aligned}\frac{1}{2i}J_{\mathcal{CP}}^{\mathrm{Majo,2}} & =\frac{m_{\tau}^{4}}{v^{8}}s_{23}^{2}\left(s_{12}^{2}c_{12}^{2}s_{23}^{2}\mu_{12}^{4}\sin(2\alpha_{M})+s_{12}^{2}c_{23}^{2}\mu_{13}^{4}\sin(2\beta_{M})+c_{12}^{2}c_{23}^{2}\mu_{32}^{4}\sin(2(\alpha_{M}-\beta_{M}))\right)\\
 & \;\;\;\;+s_{13}\frac{2m_{\tau}^{4}}{v^{8}}c_{12}c_{23}s_{12}s_{23}^{3}\mu_{12}^{4}\left(s_{12}^{2}\sin(2\alpha_{M}+\delta_{13})-c_{12}^{2}\sin(2\alpha_{M}-\delta_{13})\right)\\
 & \;\;\;\;+s_{13}\frac{2m_{\tau}^{4}}{v^{8}}c_{12}c_{23}^{3}s_{23}s_{12}\left(\mu_{32}^{4}\sin(2\alpha_{M}-2\beta_{M}+\delta_{13})-\mu_{13}^{4}\sin(2\beta_{M}-\delta_{13})\right)\;,
\end{aligned}
\label{eq:explicitJNum}
\end{equation}
où l'on a posé $\mu_{ab}^{4}=m_{\nu^{a}}m_{\nu^{b}}(m_{\nu^{a}}^{2}-m_{\nu^{b}}^{2})$
ainsi que $s_{ij}=\sin\theta_{ij}$, $c_{ij}=\cos\theta_{ij}$. Cette
expression reproduit à 5\% près la série complète exacte, sur la plage
autorisée pour l'échelle de masse de neutrino. Numériquement, le terme
en $\mathcal{O}(\sin\theta_{13})$ est sous dominant, mais est cependant
pertinent quand les phases de Majorana sont suffisamment petites pour
permettre à la phase de Dirac de contribuer significativement. Hormis
quand $\alpha_{M}-\beta_{M}$ est proche de $\pi/2$, le troisième
terme domine, tandis que la dépendance en $\delta_{13}$ provient
essentiellement des termes $\sin(2\beta_{M}-\delta_{13})$ et $\sin(2\alpha_{M}-2\beta_{M}+\delta_{13})$.
Notons finalement que l'expression analytique de $(\mathbf{X}_{e}^{\mathrm{Majo,2}})^{11}$
est similaire et possède en particulier les mêmes dépendances en masses
de neutrinos $\mu_{ab}^{4}$ (en accord avec les calculs exacts à
deux boucles \cite{CalcXeM1,CalcXeM2,CalcXeM3}), mais dépend différemment
des phases violant-$\mathcal{CP}$. Explicitement, ses termes dominants
en $\mathcal{O}(m_{\tau}^{2})$ et jusqu'à $\mathcal{O}(\sin\theta_{13})$
sont,
\begin{equation}
\begin{aligned}\frac{1}{2i}(\mathbf{X}_{e}^{\mathrm{Majo,2}})^{11} & =\frac{m_{\tau}^{2}}{v^{6}}s_{23}^{2}s_{12}^{2}c_{12}^{2}\mu_{12}^{4}\sin(2\alpha_{M})\\
 & +s_{13}\frac{m_{\tau}^{2}}{v^{6}}c_{12}c_{23}s_{12}s_{23}\mu_{12}^{4}\left(s_{12}^{2}\sin(2\alpha_{M}+\delta_{13})-c_{12}^{2}\sin(2\alpha_{M}-\delta_{13})\right)\\
 & +s_{13}\frac{m_{\tau}^{2}}{v^{6}}c_{12}c_{23}s_{23}s_{12}\left(\mu_{32}^{4}\sin(2\alpha_{M}-2\beta_{M}+\delta_{13})-\mu_{13}^{4}\sin(2\beta_{M}-\delta_{13})\right)\;.
\end{aligned}
\label{eq:explicitXNum}
\end{equation}

\begin{figure}[t]
\centering
\includegraphics[width=0.95\textwidth]{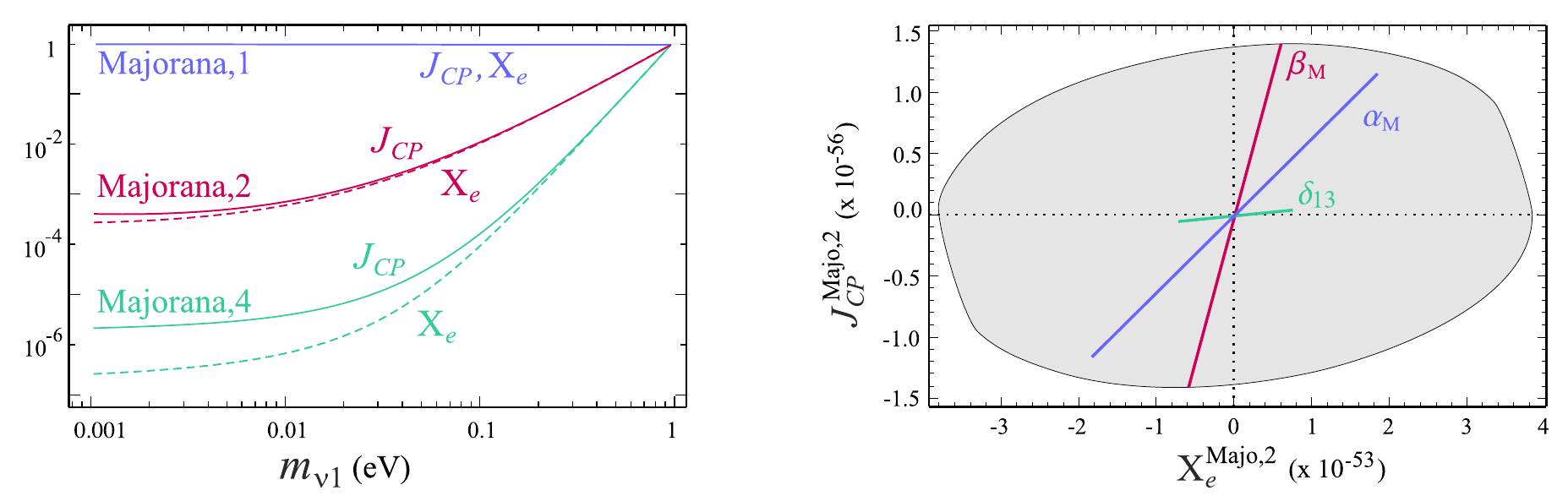}
\caption{\textbf{Gauche}: Evolutions de $J_{CP}^{\mathrm{Majo,}i}$ et $(\mathbf{X}_{e}^{\mathrm{Majo,}i})^{11}$, $i=1,2,4$, en fonction de la masse du neutrino le plus légers $m_{\nu1}$, normalisés à leurs valeurs respective pour $m_{\nu1}=1\,$eV. Le cas $i=1$ correspond à la combinaison de type Jarlskog des équations (\ref{eq:MajoJ1}) and (\ref{MajoX1}) et ne dépend que des différence de masses des neutrinos. Le cas $i=2$ représente les combinaisons plus simples des équations (\ref{eq:MajoJ2}) and (\ref{MajoX2}), et $i=4$ les combinaisons dominantes survivant dans le cas de neutrinos dégénérés des équations (\ref{eq:MajoJ4}) et (\ref{MajoX4}). \textbf{Droite}: Aire balayée par $J_{CP}^{\mathrm{Majo,2}}$ et $(\mathbf{X}_{e}^{\mathrm{Majo,2}})^{11}$ pour $m_{\nu1}=1$ eV lorsque la phase PMNS $\delta_{13}$ et les phases de Majorana $\alpha_{M},\beta_{M}$ sont autorisées à prendre n'importe quelles valeurs. Les droites montrent la corrélation stricte se réalisant quand une seule phase est non nulle. Le non-alignement de ces trois droites explique la décorrélation exhibée par l'aire grise.}
\label{FigPlotJX}
\end{figure}

En pratique, comme aucune des phases de violation-$\mathcal{CP}$
leptonique ne sont connues, on reporte dans le tableau (\ref{TableNum})
les valeurs absolues maximales atteignables lorsque $\delta_{13}$,
$\alpha_{M}$, et $\beta_{M}$ sont autorisées à prendre n'importe
quelles valeurs, c'est à dire varient sur l'intervalle $[0,2\pi]$.
La grande plage d'ordres de grandeurs balayée par les différentes
combinaisons peut être comprise par les dépendances en masses des
leptons chargés et neutrinos. Spécifiquement, le mécanisme de GIM
pour les leptons est toujours effectif et toutes les combinaisons
de spurions s'annulent dans la limite $m_{e}=m_{\mu}=m_{\tau}$. Dans
le cas plus restrictif de deux leptons chargés dégénérés, seuls $J_{\mathcal{CP}}^{\mathrm{Majo,1}}$
et $\mathbf{X}_{e}^{\mathrm{Majo,1}}$ s'annulent. D'autre part, le
mécanisme de GIM pour les neutrinos est seulement effectif pour $\mathbf{\Upsilon}_{\nu}^{\dagger}\mathbf{\Upsilon}_{\nu}\rightarrow\mathcal{O}(\Delta m_{\nu}^{2}/v^{2})$
et les masses absolues des neutrinos surviennent pour $\mathbf{\Upsilon}_{\nu}^{\dagger}(\mathbf{Y}_{e}^{\dagger}\mathbf{Y}_{e})^{T}\mathbf{\Upsilon}_{\nu}\rightarrow\mathcal{O}(m_{\nu}^{2}m_{\ell}^{2}/v^{4})$
et $\mathbf{\Upsilon}_{\nu}^{\dagger}((\mathbf{Y}_{e}^{\dagger}\mathbf{Y}_{e})^{T})^{2}\mathbf{\Upsilon}_{\nu}\rightarrow\mathcal{O}(m_{\nu}^{2}m_{\ell}^{4}/v^{6})$.
Ces comportements sont illustrés en figure \ref{FigPlotJX}.

S'il n'y a pas de nouvelle physique au-delà d'un terme de masse de
Majorana pour les neutrinos, ces structures de saveurs doivent découler
d'interactions électrofaibles, voir figure \ref{FigMajo}. L'ordre
EW auquel cela se produit peut être déterminé en comptant le nombre
de transitions de courants chargés, i.e., les contractions entre $\mathbf{\Upsilon}_{\nu}$
et $\mathbf{Y}_{e}$ ou leur transposée (adjoint) dans la chaîne de
spurions. Ceci correspond aussi aux nombres de matrices PMNS survivantes.
Pour cela, un ordre supplémentaire doit être ajouté pour l'EDM de
quarks, car la boucle de lepton doit être connectée au courant hadronique.
Les diagrammes avec trois photons contribuent aussi et sont de tailles
comparables comme
\begin{equation}
\frac{g^{2}}{4\pi}\frac{1}{M_{W}^{2}}\approx\left(\frac{e^{2}}{4\pi}\right)^{3}\frac{1}{\Lambda_{had}^{2}}\approx10^{-6}\;,
\end{equation}
avec $\Lambda_{had}$ l'échelle hadronique typique. Finalement, pour
le premier invariant $J_{\mathcal{CP}}^{\mathrm{Majo,1}}=J_{\mathcal{CP}}^{\mathrm{Dirac}}$
et sa structure rainbow associée $\mathbf{X}_{e}^{\mathrm{Majo,1}}=\mathbf{X}_{e}^{\mathrm{Dirac}}$,
au moins une boucle électrofaible additionnelle est requise afin d'obtenir
des résultats non-nuls, en analogie avec les contributions CKM des
équations (\ref{eq:eEDMCKM}) et (\ref{eq:CKMrainbow}). Par opposition,
les boucles électrofaibles dans le cas de Majorana $J_{\mathcal{CP}}^{\mathrm{Majo,2}}$
et $\mathbf{X}_{e}^{\mathrm{Majo,2}}$ ont des propriétés symétriques
différentes et aucune boucle supplémentaire n'est nécessaire \cite{CalcXeM1,CalcXeM2,CalcXeM3,XeOneGen}.
A priori, il en est de même pour les cas dégénérés, bien que ça n'ait
pas été vérifié explicitement. Quoi qu'il en soit, ce n'est pas pertinent
numériquement dans la mesure où les neutrinos étant plus légers qu'environ
$1$ eV, ils ne sont jamais assez dégénérés pour inverser la forte
hiérarchie $J_{\mathcal{CP}}^{\mathrm{Majo,3}}\ll J_{\mathcal{CP}}^{\mathrm{Majo,4}}\ll J_{\mathcal{CP}}^{\mathrm{Majo,2}}$.
Les ordres totaux EW attendus auxquels chaque combinaison de spurions
survient sont listés dans le tableau \ref{TableNum}.

\begin{table}[t] \centering
\begin{tabular}
[c]{|cccccccc|}\hline
\multicolumn{1}{|c|}{} & \multicolumn{2}{c}{$m_{\nu1}\lesssim0.01$ eV} &
\multicolumn{2}{|c}{$m_{\nu1}=1\,$eV} & \multicolumn{3}{|c|}{EDM scaling}\\
\multicolumn{1}{|c|}{} & Only $\delta_{13}$ & \multicolumn{1}{c|}{All} & Only
$\delta_{13}$ & \multicolumn{1}{c|}{All} & Flavor & Gauge &
\begin{tabular}
[c]{c}%
Prefactor\\
\lbrack$e\cdot cm$]
\end{tabular}
$\;$\\\hline
\multicolumn{1}{|c|}{$J_{\mathcal{CP}}^{\mathrm{Majo,1}}$} & $10^{-93}$ &
\multicolumn{1}{c|}{$10^{-93}$} & $10^{-93}$ & \multicolumn{1}{c|}{$10^{-93}$}
& $\frac{(\Delta m_{\nu}^{2})^{3}(\Delta m_{\ell}^{2})^{3}}{v^{12}}$ &
$\alpha_{W}^{2+1+1}$ & 10$^{-14}$\\
\multicolumn{1}{|c|}{$J_{\mathcal{CP}}^{\mathrm{Majo,2}}$} & $10^{-60}$ &
\multicolumn{1}{c|}{$10^{-59}$} & $10^{-58}$ & \multicolumn{1}{c|}{$10^{-56}$}
& $\frac{m_{\nu}^{2}\Delta m_{\nu}^{2}m_{\ell}^{4}}{v^{8}}$ & $\alpha
_{W}^{2+1}$ & 10$^{-15}$\\
\multicolumn{1}{|c|}{$J_{\mathcal{CP}}^{\mathrm{Majo,3}}$} & $10^{-107}$ &
\multicolumn{1}{c|}{$10^{-107}$} & $10^{-100}$ & \multicolumn{1}{c|}{$10^{-97}%
$} & $\frac{m_{\nu}^{6}m_{\ell}^{12}}{v^{18}}$ & $\alpha_{W}^{5+1}$ &
10$^{-12}$\\
\multicolumn{1}{|c|}{$J_{\mathcal{CP}}^{\mathrm{Majo,4}}$} & $10^{-83}$ &
\multicolumn{1}{c|}{$10^{-80}$} & $10^{-77}$ & \multicolumn{1}{c|}{$10^{-75}$}
& $\frac{m_{\nu}^{4}m_{\ell}^{12}}{v^{16}}$ & $\alpha_{W}^{4+1}$ & 10$^{-12}%
$\\
\multicolumn{1}{|c|}{$(\mathbf{X}_{e}^{\mathrm{Majo,1}})^{11}$} & $10^{-82}$ &
\multicolumn{1}{c|}{$10^{-82}$} & $10^{-82}$ & \multicolumn{1}{c|}{$10^{-82}$}
& $\frac{(\Delta m_{\nu}^{2})^{3}\Delta m_{\ell}^{2}}{v^{8}}$ & $\alpha
_{W}^{2+1}$ & 10$^{-16}$\\
\multicolumn{1}{|c|}{$(\mathbf{X}_{e}^{\mathrm{Majo,2}})^{11}$} & $10^{-56}$ &
\multicolumn{1}{c|}{$10^{-56}$} & $10^{-53}$ & \multicolumn{1}{c|}{$10^{-53}$}
& $\frac{m_{\nu}^{2}\Delta m_{\nu}^{2}m_{\ell}^{2}}{v^{6}}$ & $\alpha_{W}^{2}$
& 10$^{-17}$\\
\multicolumn{1}{|c|}{$(\mathbf{X}_{e}^{\mathrm{Majo,3}})^{11}$} & $10^{-82}$ &
\multicolumn{1}{c|}{$10^{-82}$} & $10^{-78}$ & \multicolumn{1}{c|}{$10^{-75}$}
& $\frac{m_{\nu}^{4}m_{e}^{10}}{v^{14}}$ & $\alpha_{W}^{4}$ & 10$^{-14}$\\
\multicolumn{1}{|c|}{$(\mathbf{X}_{e}^{\mathrm{Majo,4}})^{11}$} & $10^{-71}$ &
\multicolumn{1}{c|}{$10^{-71}$} & $10^{-67}$ & \multicolumn{1}{c|}{$10^{-64}$}
& $\frac{m_{\nu}^{4}m_{\ell}^{8}}{v^{12}}$ & $\alpha_{W}^{4}$ & 10$^{-15}%
$\\\hline
\end{tabular}
\caption{Estimations numériques des combinaisons de spurions construites dans les sections précédentes.
Les valeurs obtenues ne changent pas significativement lorsque la masse du neutrino le plus légers est plus petite qu'environ 0.01 eV, voir figure \ref{FigPlotJX}. Dans chaque cas, la dépendance en masses de neutrinos et de leptons chargés est indiquée. Les estimations correspondantes des EDMs de quarks et de leptons sont obtenues en multipliant le facteur de jauge, le préfacteur et la combinaison de spurions.}
\label{TableNum}
\end{table}

En tant que dernière pièce pour estimer les EDMs, les basculements
de chiralités et l'échelle globale de l'opérateur $\Lambda\approx M_{W}$
apparaissant dans les équations (\ref{eq:LquarkEDM}) et (\ref{eq:LleptonEDM})
sont combinés dans les préfacteurs reportés dans le tableau \ref{TableNum}.
De plus, on inclut aussi dans ces préfacteurs les puissances adéquates
de $v/M_{W}$ pour compenser les normalisations des spurions, vu qu'en
pratique des rapports de masses de fermions par $M_{W}$ devraient
survenir des boucles EW. Bien entendu, ces estimations d'ordres de
grandeurs sont à considérer comme très approximatifs dans la mesure
où des effets dynamiques sont négligés.

Ayant les structures de saveurs des EDMs de leptons et de quarks,
on peut étudier leurs corrélations. Cela nous renseignera sur la sensibilité
relative de ces EDMs vis-à-vis des phases de violation-$\mathcal{CP}$
sous-jacentes. Pour les structures de type Jarlskog $J_{\mathcal{CP}}^{\mathrm{Majo,1}}$
et $(\mathbf{X}_{e}^{\mathrm{Majo,1}})^{11}$, qui ne dépendent pas
de l'échelle de masse absolue des neutrinos ou des phases de Majorana,
le rapport des deux expressions (voir équations (\ref{eq:JCPdirac})
et (\ref{eq:DiracX11})) est entièrement fixé par les masses des leptons
\begin{equation}
\frac{\operatorname{Im}(\mathbf{X}_{e}^{\mathrm{Dirac}})^{11}}{2J_{\mathcal{CP}}^{\mathrm{Dirac}}}=\frac{v^{4}}{(m_{\tau}^{2}-m_{e}^{2})(m_{\mu}^{2}-m_{e}^{2})}\approx10^{11}\;.
\end{equation}
Au contraire, pour les masses de Majorana, le présence de trois sources
séparées de violation-$\mathcal{CP}$ décorrèle les EDMs de quarks
et de leptons. La figure \ref{FigPlotJX} montre le résultat d'un
scan faisant varier $\delta_{13}$, $\alpha$ et $\beta$ sur toute
la plage autorisée et $m_{\nu1}\in[0,1]$ eV. De ce graphique, il
est apparent que même si les expressions analytiques de $J_{\mathcal{CP}}^{\mathrm{Majo,2}}$
et $(\mathbf{X}_{e}^{\mathrm{Majo,2}})^{11}$ sont similaires, voir
les équations (\ref{eq:explicitJNum}) et (\ref{eq:explicitXNum}),
leurs dépendances différentes à travers les fonctions trigonométriques
ont d'importantes conséquences. Bien que cela requiert certainement
un niveau d'ajustement fin, il est même possible d'inverser la hiérarchie
et d'avoir $J_{\mathcal{CP}}^{\mathrm{Majo,2}}>(\mathbf{X}_{e}^{\mathrm{Majo,2}})^{11}$.
Cependant, augmenter le rapport $d_{u,d}/d_{e}$ de cette manière
est limité. Lorsque $(\mathbf{X}_{e}^{\mathrm{Majo,2}})^{11}\lesssim J_{\mathcal{CP}}^{\mathrm{Majo,2}}$,
la contribution dominante aux EDMs de leptons provient de $\mathbf{X}_{e}=\mathbb{I}_{3}\times J_{\mathcal{CP}}^{\mathrm{Majo,2}}$,
voir équation (\ref{eq:LquarkEDM}). Cela correspond à la situation
dans laquelle les EDMs de quarks et de leptons sont induits par la
même boucle de lepton, voir figure \ref{FigMajo}. Etant contrôlé
par le même invariant et à moins d'une annulation finement ajustée
entre les contributions rainbow et bubble à $d_{e}$, les EDMs doivent
vérifier
\begin{equation}
\frac{d_{d}}{m_{d}}\lesssim\frac{d_{e}}{m_{e}}\;.\label{eq:BoundMajo}
\end{equation}
Bien sûr, toutes ces valeurs sont bien au-delà des sensibilités prévues
mais nous allons discuter dans la prochaine partie comment amplifier
ces valeurs afin qu'elles deviennent atteignables par les expériences.

\subsection{Mécanismes de seesaw}

Les valeurs de fonds des spurions de neutrinos dans les deux cas de
Dirac et de Majorana sont extrêmement supprimés, simplement car les
neutrinos sont très légers. Il s'ensuit que les combinaisons de spurions
contrôlant les transitions LFV ou les EDMs finissent bien trop supprimées
pour les rendre accessibles expérimentalement. D'un point de vue théorique,
ces valeurs de fonds sont trop petites pour être naturelle et il est
généralement accepté que cette suppression a une origine dynamique.
Après tout, l'opérateur de Weinberg duquel le petit terme de masse
de Majorana provient n'est pas renormalisable. S'il survient à une
échelle très haute, les neutrinos gauches seraient automatiquement
légers. Il y a trois façons de réaliser ceci dynamiquement à l'arbre
\cite{Ma98}, dépendant de comment étendre de façon minimale le contenu
en particules du MS. Le mécanisme de seesaw de type I introduit des
singlets faibles de neutrinos droits lourds \cite{TypeI1,TypeI2,TypeI3,TypeI4},
celui de type III ajoute des triplets faibles de neutrinos droits
tandis que le seesaw de type II étend le secteur scalaire du MS d'un
triplet faible de champs scalaires \cite{TypeII}.

Dès lors que la suppression des masses de neutrinos est gérée dynamiquement,
on se retrouve avec des structures de saveurs bien moins supprimées.
Bien entendu, en l'absence de toute NP, le seul accès à basse énergie
à ces structures de saveurs est à travers le terme de masse de neutrinos,
les rendant à nouveau inobservables. Cependant, si on suppose que
de la NP existe pas trop loin de l'échelle EW, alors les structures
de saveurs de neutrinos non supprimées pourraient directement impacter
les transitions LFV et les EDMs. C'est l'objet de cette partie de
traiter ces scénarios en utilisant les techniques développées dans
les précédentes sections.

\subsubsection{Mécanisme de seesaw de type II}

On introduit un triplet faible de champs scalaires $\Delta_{i}\ensuremath{,}i=1,2,3$
avec une hypercharge $2$. Les couplages renormalisables autorisés
sont (voir par exemple \cite{MaTypeII} pour une description détaillée)
\begin{equation}
\begin{aligned}\mathcal{L} & =\mathcal{L}_{SM}+D_{\mu}\vec{\Delta}^{\dagger}\cdot D^{\mu}\vec{\Delta}-\vec{\Delta}^{\dagger}\vec{\Delta}M_{\Delta}^{2}-\delta V(H,\vec{\Delta})\\
 & \;\;\;\;\;\;\;+\frac{1}{2}(\bar{L}^{\mathrm{C}}\mathbf{\Upsilon}_{\Delta}\vec{\sigma}L+\lambda_{\Delta}M_{\Delta}H^{\dagger}\vec{\sigma}H^{\dagger})\cdot\vec{\Delta}+h.c.\;\;,
\end{aligned}
\end{equation}
où $\delta V(H,\vec{\Delta})$ dénote le reste du potentiel scalaire.
En intégrant le triplet $\vec{\Delta}$ donne un terme de dimension-4
ainsi que l'opérateur de Weinberg de dimension-5:
\begin{equation}
\mathcal{L}_{eff}=\mathcal{L}_{SM}+2|\lambda_{\Delta}|^{2}(H^{\dagger}H)^{2}+\frac{1}{2}(\bar{L}^{\mathrm{C}}H)\mathbf{\Upsilon}_{\Delta}\frac{\lambda_{\Delta}}{M_{\Delta}}(LH)+...
\end{equation}
La matrice de masse des neutrinos est alors linéaire en $\mathbf{\Upsilon}_{\Delta}$:
\begin{equation}
v\mathbf{Y}_{e}=\mathbb{M}_{e},\;\;\;v\mathbf{\Upsilon}_{\nu}\equiv v^{2}\mathbf{\Upsilon}_{\Delta}\frac{\lambda_{\Delta}}{M_{\Delta}}\equiv U_{PMNS}^{\ast}\mathbb{M}_{\nu}U_{PMNS}^{\dagger}\;.\label{eq:SSTypeII}
\end{equation}
Avec un mécanisme de seesaw de type II, la vraie matrice de couplage
de saveur élémentaire est $\mathbf{\Upsilon}_{\Delta}$ de l'équation
(\ref{eq:SSTypeII}), qui peut être d'ordre 1 quand $M_{\Delta}/\lambda_{\Delta}$
est assez grand. Cependant, en l'absence de NP, il n'y a pas de sensibilité
directe à $\mathbf{\Upsilon}_{\Delta}$ étant donné que tout ce qui
compte à basse énergie c'est $\mathbf{\Upsilon}_{\nu}$. Les taux
de transitions LFV sont toujours ceux dans (\ref{eq:LFVrates}).

Imaginons qu'il y ait une nouvelle dynamique à une échelle intermédiaire
$\Lambda\ll M_{\Delta}$, et que cette nouvelle dynamique soit contrôlée
par $\mathbf{\Upsilon}_{\Delta}$. La dépendance des taux LFV dans
les paramètres de mélanges des neutrinos est inchangée car $\mathbf{\Upsilon}_{\Delta}$
et $\mathbf{\Upsilon}_{\nu}$ se transforment de la même manière,
cependant ils sont globalement redimensionné par
\begin{equation}
\mathbf{X}_{e}^{\mathrm{Type\,II}}=\mathbf{\Upsilon}_{\Delta}^{\dagger}\mathbf{\Upsilon}_{\Delta}=\left(\frac{M_{\Delta}}{v\lambda_{\Delta}}\right)^{2}\mathbf{\Upsilon}_{\nu}^{\dagger}\mathbf{\Upsilon}_{\nu}\;.
\end{equation}
En injectant ceci dans l'équation (\ref{eq:LFVemo}), on peut établir
à partir de la limite expérimentale une valeur maximale pour le paramètre
d'échelle de seesaw $M_{\Delta}/v\lambda_{\Delta}$ en fonction de
l'échelle $\Lambda$:
\begin{equation}
\frac{M_{\Delta}}{v\lambda_{\Delta}}\lesssim10^{12}\times\left[\frac{\Lambda}{1\,\text{TeV}}\right]\;.
\end{equation}
Pour ceci, on suppose que les processus LFV se produisent encore à
l'ordre de boucle, i.e., $c_{e}\approx g^{2}/16\pi^{2}$ dans l'équation
(\ref{eq:LFVemo}). Fixer $c_{e}\approx1$ diminue la limite d'un
ordre de grandeur. D'autre part, la limite de perturbativité $\mathbf{\Upsilon}_{\Delta}^{IJ}\lesssim4\pi$
limite $M_{\Delta}/v\lambda_{\Delta}$ à
\begin{equation}
\frac{M_{\Delta}}{v\lambda_{\Delta}}\lesssim\frac{4\pi v}{m_{\nu}^{\max}}\;.\label{eq:PertSTII}
\end{equation}

\begin{figure}[t]
\centering
\includegraphics[width=0.95\textwidth]{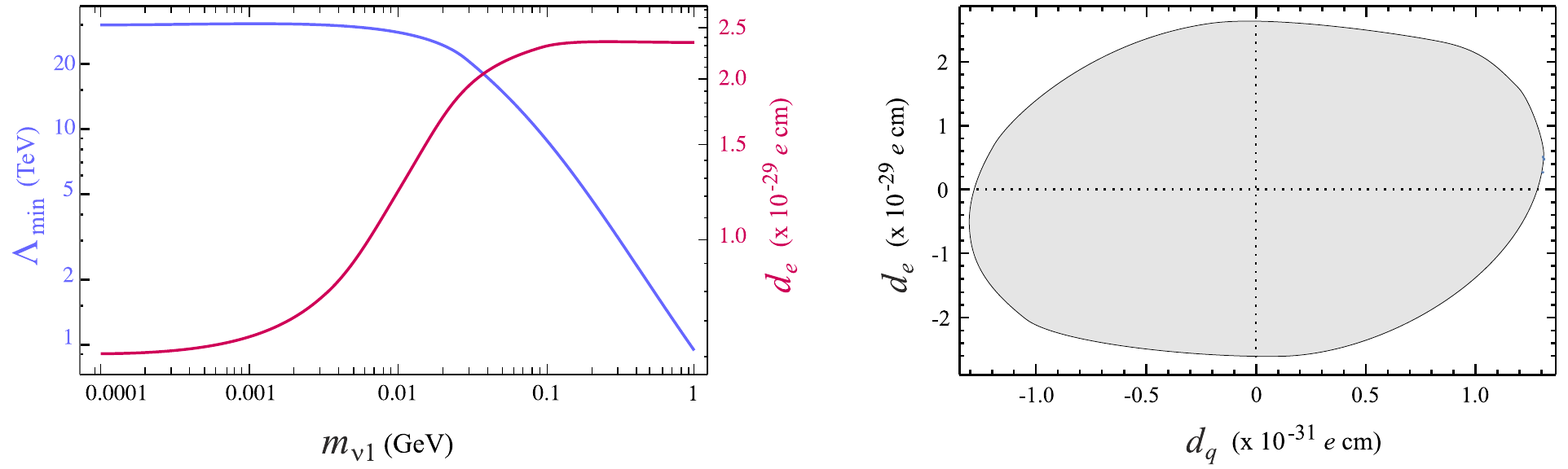}
\caption{\textbf{Gauche}: L'échelle minimale $\Lambda$ des opérateurs magnétiques, équation (\ref{eq:EMO}), et l'EDM de l'électron correspndant, tel que $\mu\rightarrow e\gamma$ sature sa limite expérimentale quand $M_{\Delta}/v \lambda_{\Delta}$ est fixée à la limite de perturbativité, équation (\ref{eq:PertSTII}), en fonction de la masse du neutrino le plus légers $m_{\nu1}$. \textbf{Droite}: Plage accessible des EDMs de quarks et de leptons dans le seesaw de type II, étant donnée la limite de perturbativité $\mathbf{\Upsilon}_{\Delta}^{IJ}\lesssim4\pi$ ainsi que la limite expérimentale sur $\mathcal{B}(\mu\rightarrow e\gamma)$.}
\label{FigEDMtII}
\end{figure}

Notre but est de vérifier à quel point les EDMs peuvent être grands
étant données les deux limites. On suppose que les opérateurs magnétiques
surviennent tous à une boucle et contiennent donc un facteur $g^{2}/16\pi^{2}$
dans les amplitudes LFV et EDMs. Sachant que $\mu\rightarrow e\gamma$
est en $\left(M_{\Delta}/v\lambda_{\Delta}\right)^{4}/\Lambda^{4}$
tandis que les EDMs sont en $\left(M_{\Delta}/v\lambda_{\Delta}\right)^{4}/\Lambda^{2}$,
notre stratégie consiste à d'abord fixer $M_{\Delta}/v\lambda_{\Delta}$
en saturant la limite de perturbativité, et puis de trouver pour cette
valeur l'échelle minimale $\Lambda_{\min}$ pour laquelle $\mu\rightarrow e\gamma$
est compatible avec sa limite expérimentale, voir figure \ref{FigEDMtII}.
La dépendance de $\mu\rightarrow e\gamma$ dans les phases violant-$\mathcal{CP}$
est faible, donc cette borne inférieure de $\Lambda$ est plutôt dure.
Avec ces deux données, $M_{\Delta}/v\lambda_{\Delta}$ et $\Lambda_{\min}$,
on maximise alors les EDMs de quarks et de leptons en scannant sur
les trois phases violant-$\mathcal{CP}$ $\delta_{13}$, $\alpha_{M}$
et $\beta_{M}$. Comme expliqué dans la partie précédente (voir figure
\ref{FigPlotJX}), les deux types d'EDMs sont décorrélés et balaye
uniformément l'aire représentée en figure \ref{FigEDMtII}. A condition
que $\mathbf{\Upsilon}_{\Delta}^{IJ}$ puisse saturer sa borne de
perturbativité, l'EDM de l'électron peut se rapprocher de sa limite
expérimentale $|d_{e}|\,<8.7\cdot10^{-29}\,e\cdot cm\;(90\%)$ \cite{expEDMe}.
A cet égard, notons que la borne de perturbativité joue le rôle crucial.
Si on impose $\mathbf{\Upsilon}_{\Delta}^{IJ}\lesssim1$ au lieu de
$\mathbf{\Upsilon}_{\Delta}^{IJ}\lesssim4\pi$, la valeur maximale
pour $M_{\Delta}/v\lambda_{\Delta}$ est réduite de $4\pi$, et il
en est de même pour $\Lambda_{\min}$ si la limite de $\mu\rightarrow e\gamma$
reste saturée, mais l'EDM de l'électron se voit réduit de $(4\pi)^{2}\approx160$.

Concernant les quarks, d'abord, la limite $\mu\rightarrow e\gamma$
implique $|J_{\mathcal{CP}}^{\mathrm{Majo,2}}|_{\max}\approx10^{-6}$.
Naïvement, cela pousserait leurs EDMs provenant des opérateurs magnétiques
au-delà de $10^{-31}$ $e\cdot cm$. Néanmoins, en même temps, $J_{\mathcal{CP}}^{\mathrm{Majo,2}}$
décale également le terme $\theta$ de
\begin{equation}
\Delta\theta_{eff}^{quarks}\sim\frac{g^{2}}{4\pi^{2}}\times J_{\mathcal{CP}}^{\mathrm{Majo,2}}\sim10^{-8}\;.
\end{equation}
Etant données les approximations grossières impliquées, on considère
ceci comme (à peine) compatible avec la borne $\theta_{C}^{eff}\lesssim10^{-10}$.
Cela montre que l'EDM du neutron induit par des phases de violation-$\mathcal{CP}$
leptoniques pourrait en principe saturer sa limite expérimentale.

Le décalage du terme $\theta$pose des contraintes strictes. Considérons
par exemple le seesaw de type II étendu pour accommoder plusieurs
doublets de Higgs. Les valeurs de fonds des spurions sont alors contrôlées
par différentes VEVs,
\begin{equation}
v_{d}\mathbf{Y}_{e}=\mathbb{M}_{e},\;\;\;v_{u}\mathbf{\Upsilon}_{\nu}\equiv v_{u}^{2}\mathbf{\Upsilon}_{\Delta}\frac{\lambda_{\Delta}}{M_{\Delta}}\equiv U_{PMNS}^{\ast}\mathbb{M}_{\nu}U_{PMNS}^{\dagger}\;.
\end{equation}
De façon cruciale, les combinaisons de spurions pertinentes pour les
EDMs ont un comportement différent dans la limite de grand $\tan\beta=v_{u}/v_{d}$
\begin{equation}
J_{\mathcal{CP}}^{\mathrm{Majo,2}}\sim\left(\tan\beta\right)^{4}\;\;,\;\;\;(\mathbf{X}_{e}^{\mathrm{Majo,2}})^{11}\sim\left(\tan\beta\right)^{2}\;,
\end{equation}
tandis que les taux LFV ne sont pas directement affectés. Si $\tan\beta$
est grand, la masse du neutrino le plus léger et/ou les phases de
violation-$\mathcal{CP}$ doivent être telles que $J_{\mathcal{CP}}^{\mathrm{Majo,2}}$
soit loin de sa valeur maximale afin de satisfaire la limite $\theta_{C}^{eff}\lesssim10^{-10}$.
A ce stade, il est possible que $d_{e}$ soit trop petit pour être
observé, mais $d_{n}$ est proche de sa limite expérimentale. Alternativement,
un tel grand décalage $\Delta\theta_{eff}^{quarks}$ serait totalement
non pertinent si le mécanisme résolvant le problème de $\mathcal{CP}$
fort du MS éliminait complètement $\theta_{C}^{eff}$. Ainsi, l'EDM
est encore entièrement induit par les opérateurs magnétiques de l'équation
(\ref{eq:EMO}). Comme $J_{\mathcal{CP}}^{\mathrm{Majo,2}}$ augmente
plus vite avec $\tan\beta$ que $(\mathbf{X}_{e}^{\mathrm{Majo,2}})^{11}$,
on peut même imaginer que la limite actuelle sur l'EDM de l'électron
est saturée par un pur $J_{\mathcal{CP}}^{\mathrm{Majo,2}}$. Etant
donné $|d_{e}|\,<8.7\cdot10^{-29}\,e\cdot cm$, ceci correspond à
\begin{equation}
|J_{\mathcal{CP}}^{\mathrm{Majo,2}}|_{\max}\approx10^{-3}\times\left[\frac{\Lambda_{\min}}{1\,\text{TeV}}\right]^{2}\;,
\end{equation}
à comparer à $|J_{\mathcal{CP}}^{\mathrm{Majo,2}}|_{\max}\approx10^{-6}$
quand $\tan\beta=1$. A ce stade, la borne (\ref{eq:BoundMajo}) implique
que les EDMs de quarks ne peuvent pas au-dessus de $d_{q}\approx10^{-27}$
$e\cdot cm$, qui représente donc la valeur maximale atteignable en
l'absence de $\theta_{C}^{eff}$.

\subsubsection{Mécanismes de seesaw de type I et III}

Le mécanisme de seesaw de type I étend le contenu en particules du
MS en y ajoutant un triplet de saveurs de neutrinos droits. Les interactions
de jauges autorisent alors à la fois un terme de masse de Dirac et
de Majorana
\begin{equation}
\mathcal{L}=\mathcal{L}_{SM}+i\bar{N}\!\not\!\partial N+\left(-\frac{1}{2}\bar{N}^{\mathrm{C}}\mathbf{M}N-\bar{N}^{\mathrm{C}}\mathbf{Y}_{\nu}LH+h.c.\right)\;.
\end{equation}
De plus, le terme de masse de Majorana n'a a priori pas de lien avec
l'échelle électrofaible et peut très bien être bien plus grand. En
supposant $\mathbf{M}=\operatorname*{diag}(M_{1},M_{2},M_{3})$ sans
perte de généralité et en intégrant le champ $N$, on obtient de nouveau
l'opérateur de Weinberg
\begin{equation}
\mathcal{L}_{eff}=\mathcal{L}_{SM}+\frac{1}{2}(\bar{L}^{\mathrm{C}}H)\mathbf{Y}_{\nu}^{T}\mathbf{M}^{-1}\mathbf{Y}_{\nu}(LH)+h.c.\;.\label{eq:SeesawI}
\end{equation}
A condition que $\mathbf{M}$ soit suffisamment grand, les masses
des neutrinos gauches seront très petits et ce même avec des couplages
de Yukawa de neutrinos de taille naturelle, $\mathbf{Y}_{\nu}\sim\mathcal{O}(\mathbf{Y}_{u,d,e})$.

Au lieu d'un triplet de saveurs de singlets faibles $N$, on pourrait
introduire des triplets de saveurs de triplets faibles $\Sigma_{i}\ensuremath{,}i=1,2,3$,
avec une hypercharge nulle. C'est le seesaw de type III. De tels champs
peuvent se coupler aux doublets faibles à travers leur courant vectoriel
comme ceci
\begin{equation}
\mathcal{L}=\mathcal{L}_{SM}+i\bar{\Sigma}_{i}\!\not\!D\Sigma_{i}+\left(-\frac{1}{2}\bar{\Sigma}_{i}^{\mathrm{C}}\mathbf{M}\Sigma_{i}-\bar{\Sigma}_{i}^{\mathrm{C}}\mathbf{Y}_{\nu}(L\sigma^{i}H)+h.c.\right)\;.
\end{equation}
Les champs $\Sigma_{i}$ étant dans la représentation adjointe, les
couplages adéquats aux bosons de jauges sont contenus dans la dérivée
covariante et un terme de masse de Majorana invariant de jauge est
autorisé. Clairement, du point de vue de la symétrie de saveur, le
contenu en spurions est identique à celui du mécanisme de seesaw de
type I. En outre, le fait d'intégrer les champs $\Sigma_{i}$ produit
exactement le même opérateur de Weinberg que dans l'équation (\ref{eq:SeesawI}).
Dans le reste de cette section, on procédera donc avec avec le seesaw
de type I mais les développements seront tout aussi applicable au
mécanisme type III.

A l'échelle de seesaw, il y a deux paramètres élémentaires brisant
la symétrie de saveur, $\mathbf{M}$ et $\mathbf{Y}_{\nu}$, qui se
transforment sous le groupe de saveur étendu $G_{F}^{\prime}\equiv G_{F}\times U(3)_{N}$
\cite{CiriglianoGIW05}. Cependant, comme $\nu_{R}$ n'est pas dynamique
à basse énergie, aucune amplitude ne se transforme de façon de non-trivialle
sous $U(3)_{N}$. Seulement les combinaisons de $\mathbf{Y}_{\nu}$
et $\mathbf{M}$ se transformant comme des singlets sous $U(3)_{N}$
sont requises. De plus, intégrer $\nu_{R}$ génère un développement
en masse-inverse et avec $\mathbf{M}\sim10^{10}-10^{13}$ GeV, seules
les combinaisons de spurions dominantes doivent être gardées:
\begin{equation}
\begin{aligned}\mathbf{Y}_{e} & \sim\left(\mathbf{\bar{3}},\mathbf{\bar{3}},\mathbf{1}\right)_{L,E,N}:\mathbf{Y}_{e}\overset{G_{F}}{\rightarrow}g_{E}^{\dagger}\mathbf{Y}_{e}g_{L}^{\dagger}\;,\;\\
\mathbf{Y}_{\nu}^{\dagger}\mathbf{Y}_{\nu} & \sim\left(\mathbf{8},\mathbf{1},\mathbf{1}\right)_{L,E,N}:\mathbf{Y}_{\nu}^{\dagger}\mathbf{Y}_{\nu}\overset{G_{F}}{\rightarrow}g_{L}\mathbf{Y}_{\nu}^{\dagger}\mathbf{Y}_{\nu}g_{L}^{\dagger}\;,\\
\mathbf{\Upsilon}_{\nu}\overset{}{\equiv}v\mathbf{Y}_{\nu}^{T}(\mathbf{M}^{-1})\mathbf{Y}_{\nu} & \sim\left(\mathbf{\bar{6}},\mathbf{1},\mathbf{1}\right)_{L,E,N}\;:\mathbf{\Upsilon}_{\nu}\overset{G_{F}}{\rightarrow}g_{L}^{\ast}\mathbf{\Upsilon}_{\nu}g_{L}^{\dagger}\;.
\end{aligned}
\label{eq:leptSpur}
\end{equation}
La matrice symétrique $\mathbf{\Upsilon}_{\nu}^{T}=\mathbf{\Upsilon}_{\nu}$
correspond au très petit terme de masse de Majorana des neutrinos
gauches. L'échelonnement $\mathbf{Y}_{e},\mathbf{Y}_{\nu}^{\dagger}\mathbf{Y}_{\nu}\gg\mathbf{\Upsilon}_{\nu}$
est stable étant donné que ces combinaisons de spurions vivent dans
des classes de trialités différentes de $SU(3)^{5}$.

Dans ce cas, il n'est pas possible de fixer sans ambiguïtés les valeurs
de fonds de $\mathbf{Y}_{\nu}^{\dagger}\mathbf{Y}_{\nu}$ à partir
des données de neutrinos disponibles. Sans perte de généralités, cette
indétermination peut être paramétrée \cite{CasasI01} en termes d'une
matrice complexe orthogonale inconnue $\mathbf{R}$ avec $v\mathbf{Y}_{\nu}=(\mathbf{M}^{1/2})\mathbf{R}(\mathbf{m}_{\nu}^{1/2})U_{PMNS}^{\dagger}$
où $U_{PMNS}^{\dagger}$ est définie par la diagonalisation de $\mathbf{Y}_{e}$
et $\mathbf{\Upsilon}_{\nu}$ et contient les phases de Majorana,
voir équation (\ref{eq:PMNS2}). Pour procéder, on suppose que les
neutrinos droits sont dégénérés, du moins en bonne approximation \cite{CiriglianoIP06}.
Ceci signifie que $\mathbf{M}$ ne brise pas complètement $U(3)_{N}$
mais laisse un sous-groupe $O(3)$ conservé, et donc trois paramètres
peuvent être éliminés. En effet, en partant de la décomposition polaire
$\mathbf{R}=\mathbf{U\,H}$ avec $\mathbf{U}=(\mathbf{R}^{\dagger}\mathbf{R})^{1/2}$
unitaire et $\mathbf{H}=\mathbf{U}^{\dagger}\mathbf{R}$ hermitienne,
et en imposant $\mathbf{R}^{T}\mathbf{R}=\mathbb{I}_{3}$, la matrice
orthogonale à six paramètres $\mathbf{R}$ se décompose en $\mathbf{R}=\mathbf{O}\mathrm{\,}\mathbf{H}$
avec $\mathbf{O}$ une matrice orthogonale réelle et $\mathbf{H}$
une matrice hermitienne. La dégénérescence $\mathbf{M}=M_{R}\mathbb{I}_{3}$
permet de se débarrasser de la matrice orthogonale avec la redéfinition
inoffensive $N\rightarrow\mathbf{O}^{T}N$, de sorte que $\mathbf{Y}_{\nu}^{\dagger}\mathbf{Y}_{\nu}$
se simplifie de la manière suivante \cite{PascoliPY03}
\begin{equation}
\mathbf{\mathbf{Y}}_{\nu}^{\dagger}\mathbf{\mathbf{Y}}_{\nu}=\frac{M_{R}}{v^{2}}U_{PMNS}(\mathbf{m}_{\nu}^{1/2})\,\mathbf{H}^{2}\mathbf{\,}(\mathbf{m}_{\nu}^{1/2})U_{PMNS}^{\dagger}\;,\label{eq:Spurion}
\end{equation}
avec la matrice $\mathbf{H}$ exprimée en termes d'une matrice antisymétrique
réelle $\mathbf{\Phi}^{IJ}=\varepsilon^{IJK}\phi_{K}$ comme suit
\cite{PascoliPY03,CiriglianoIP06}
\begin{equation}
\mathbf{H}=e^{i\mathbf{\Phi}}=\mathbf{1}+\frac{\cosh r-1}{r^{2}}i\mathbf{\Phi}\cdot i\mathbf{\Phi}+\frac{\sinh r}{r}i\mathbf{\Phi},\;\;r=\sqrt{\phi_{1}^{2}+\phi_{2}^{2}+\phi_{3}^{2}}\;.
\end{equation}
Les trois paramètres réels $\phi_{1}$, $\phi_{2}$ et $\phi_{3}$
affectent la taille des coefficients conservant-$\mathcal{CP}$ dans
$\mathbf{\mathbf{Y}}_{\nu}^{\dagger}\mathbf{\mathbf{Y}}_{\nu}$ et
induisent des parties imaginaires violant-$\mathcal{CP}$.

En l'absence de NP au-delà des neutrinos droits, les taux LFV surviendraient
seulement à l'ordre $\mathcal{O}(\mathbf{M}^{-2})$,
\begin{equation}
\mathbf{X}_{e}^{\mathrm{Type\,I}}=\mathbf{Y}_{\nu}^{\dagger}\frac{v}{\mathbf{M}^{\dagger}}\frac{v}{\mathbf{M}}\mathbf{Y}_{\nu}+\mathbf{Y}_{\nu}^{\dagger}\frac{v}{\mathbf{M}^{\dagger}}\mathbf{Y}_{\nu}^{\ast}\mathbf{Y}_{\nu}^{T}\frac{v}{\mathbf{M}}\mathbf{Y}_{\nu}+...=\frac{v^{2}}{M_{R}^{2}}\mathbf{Y}_{\nu}^{\dagger}\mathbf{Y}_{\nu}+\mathbf{\Upsilon}_{\nu}^{\dagger}\mathbf{\Upsilon}_{\nu}+....\label{eq:NoNP}
\end{equation}
Le second terme reproduit exactement le cas de Majorana pur dans l'équation
(\ref{eq:MDequal}) et mène aux taux en dépendances quartiques en
masses des neutrinos, voir équation (\ref{eq:LFVrates}). Le premier
terme donne plutôt des taux quadratiques en masses des neutrinos mais
n'est que légèrement moins supprimé (la même combinaison de spurions
contrôle d'autres opérateurs FCNC, voir par exemple \cite{Broncano02}).
La situation change si de la NP est présente à une échelle intermédiaire
$\Lambda\ll M_{R}$. Cette dynamique pourrait directement amener la
sensibilité à $\mathbf{Y}_{\nu}$, de sorte que
\begin{equation}
\mathbf{X}_{e}^{\mathrm{Type\,I}}=\mathbf{Y}_{\nu}^{\dagger}\mathbf{Y}_{\nu}\;.\label{eq:LFVext}
\end{equation}
Ce n'est pas seulement que les taux LFV sont quadratiques en masses
des neutrinos au lieu de quartiques, mais ils sont également amplifiés
par $M_{R}^{4}/\Lambda^{4}$ comparé à la situation dans l'équation
(\ref{eq:NoNP}). Ceci se produit typiquement à une boucle en supersymétrie,
où les masses des squarks et sleptons fixent l'échelle $\Lambda$
tandis que les termes de brisure douce (soft-breaking) des sleptons
apporte la dépendance en $\mathbf{Y}_{\nu}^{\dagger}\mathbf{Y}_{\nu}$.
En injectant ceci dans les taux LFV, on établit à partir de la limite
expérimentale sur $\mu\rightarrow e\gamma$,
\begin{equation}
|\mathbf{Y}_{\nu}^{\dagger}\mathbf{Y}_{\nu}|^{21}\lesssim(10^{-2}-10^{-4})\times\left[\frac{\Lambda}{1\,\text{TeV}}\right]^{2}\;,
\end{equation}
dépendant de si $c_{e}\approx g^{2}/16\pi^{2}$ ou $c_{e}\approx1$
dans l'équation (\ref{eq:LFVemo}). Ceci est très proche de la borne
de perturbativité, $|\mathbf{Y}_{\nu}^{\dagger}\mathbf{Y}_{\nu}|\lesssim4\pi$,
qui limite indirectement $M_{R}$ pour des valeurs donnés de masses
du neutrino légers et des paramètres $\phi_{i}$
\begin{equation}
\dfrac{m_{\nu}^{\max}}{1\text{\thinspace eV}}\frac{M_{R}}{10^{13}\,\text{GeV}}\lesssim12\pi e^{-2\sqrt{3}\max\phi_{i}}\;.\label{eq:UB}
\end{equation}
Du fait des dépendances exponentielles des $\phi_{i}$, l'échelle
de seesaw doit rapidement décroître lorsque les $\phi_{i}$ sont au-dessus
de l'unité.

Quant aux EDMs, les deux combinaisons de spurions non supprimées par
l'échelle de seesaw sont $\mathbf{Y}_{e}$ et $\mathbf{Y}_{\nu}^{\dagger}\mathbf{Y}_{\nu}$,
à partir desquels on peut seulement construire:
\begin{equation}
\begin{aligned}J_{\mathcal{CP}}^{\mathrm{Type\,I}} & =\frac{1}{2i}\det[\mathbf{Y}_{\nu}^{\dagger}\mathbf{Y}_{\nu},\mathbf{Y}_{e}^{\dagger}\mathbf{Y}_{e}]\;,\\
\mathbf{X}_{e}^{\mathrm{Type\,I}} & =[\mathbf{Y}_{\nu}^{\dagger}\mathbf{Y}_{\nu}\;,\;\mathbf{Y}_{\nu}^{\dagger}\mathbf{Y}_{\nu}\mathbf{Y}_{e}^{\dagger}\mathbf{Y}_{e}\mathbf{Y}_{\nu}^{\dagger}\mathbf{Y}_{\nu}]\;.
\end{aligned}
\end{equation}
D'un point de vue de symétrie, ce sont les mêmes que dans le cas des
neutrinos de Dirac, équations (\ref{eq:JCPdirac}) et (\ref{eq:YnEDM}).
Au-delà de cette similarité superficielle, la situation est différente
étant donné que $\mathbf{Y}_{\nu}^{\dagger}\mathbf{Y}_{\nu}$ possède
plus de degrés de libertés et ne dépend que linéairement des masses
de neutrinos légers. Par exemple, lorsque $\phi_{i}=0$, on trouve
$v^{2}(\mathbf{\mathbf{Y}}_{\nu}^{\dagger}\mathbf{\mathbf{Y}}_{\nu})^{\mathrm{Type\,I}}\rightarrow M_{R}U_{PMNS}\mathbf{m}_{\nu}U_{PMNS}^{\dagger}$
à comparer avec $v^{2}(\mathbf{\mathbf{Y}}_{\nu}^{\dagger}\mathbf{\mathbf{Y}}_{\nu})^{\mathrm{Dirac}}\rightarrow U_{PMNS}\mathbf{m}_{\nu}^{2}U_{PMNS}^{\dagger}$
dans le cas de Dirac. Autrement dit, $J_{\mathcal{CP}}^{\mathrm{Type\,I}}$
et $\mathbf{X}_{e}^{\mathrm{Type\,I}}$ dépendent linéairement du
produit des trois différences de masses de neutrinos dans cette limite
et sont insensibles aux phases de Majorana. En revanche, ces deux
propriétés sont perdues dès que $\phi_{i}\neq0$: ni $J_{\mathcal{CP}}^{\mathrm{Type\,I}}$
ni $\mathbf{X}_{e}^{\mathrm{Type\,I}}$ ne s'annule quand seulement
deux neutrinos sont dégénérés et tous les deux sont sensibles aux
phases de Majorana. Ce qui est préservé quand même est leur dépendance
en les masses des leptons chargés,
\begin{equation}
J_{\mathcal{CP}}^{\mathrm{Type\,I}}\sim\prod_{i>j=e,\mu,\tau}\frac{m_{i}^{2}-m_{j}^{2}}{v^{2}}\;,\;\;(\mathbf{X}_{e}^{\mathrm{Type\,I}})^{11}\sim\frac{m_{\tau}^{2}-m_{\mu}^{2}}{v^{2}}\;.
\end{equation}
Remarquablement, ces différences de masses de leptons chargés sont
multipliées par le même facteur dans les deux expressions, et ce même
quand $\phi_{i}\neq0$. Ceci signifie que contrairement au cas de
Majorana (voir figures \ref{FigPlotJX} et \ref{FigEDMtII}), le rapport
est fixé à
\begin{equation}
\frac{\operatorname{Im}(\mathbf{X}_{e}^{\mathrm{Type\,I}})^{11}}{2J_{\mathcal{CP}}^{\mathrm{Type\,I}}}=\frac{\operatorname{Im}(\mathbf{X}_{e}^{\mathrm{Dirac}})^{11}}{2J_{\mathcal{CP}}^{\mathrm{Dirac}}}=\frac{v^{4}}{(m_{\tau}^{2}-m_{e}^{2})(m_{\mu}^{2}-m_{e}^{2})}\approx10^{11}\;\rightarrow\frac{d_{q}}{m_{q}}\approx10^{-11}\times\frac{d_{e}}{m_{e}}\;.\label{eq:ScalingTI}
\end{equation}
En nette différence avec le seesaw de type II, les contributions savoureuses
aux EDMs de quarks et de leptons sont strictement corrélées dans les
mécanismes de seesaw de type I et III, et celle du type III reste
beaucoup plus petite que celle du type I. Bien sûr, des effets dynamiques
peuvent venir altérer cette stricte corrélation, par exemple à travers
des dépendances logarithmiques en les masses de leptons chargés. Néanmoins,
les ordres de grandeurs relatifs des EDMs de quarks et de leptons
devraient être bien prédits par le comportement de ces combinaisons
de spurions.

Une conséquence immédiate de la suppression de $J_{\mathcal{CP}}^{\mathrm{Type\,I}}$
se voit dans les EDMs de quarks. Les contributions magnétiques et
générées par le décalage du terme $\theta$ sont contrôlées par $J_{\mathcal{CP}}^{\mathrm{Type\,I}}$,
qui est au moins 11 ordres de grandeurs en dessous de $(\mathbf{X}_{e}^{\mathrm{Type\,I}})^{11}$.
L'EDM du neutron reste alors entièrement dominé par les contributions
CKM, quoi qu'il arrive dans le secteur leptonique. Cette conclusion
reste vraie en présence de deux doublets de Higgs, étant donné que
l'équation (\ref{eq:ScalingTI}) est modifié en
\begin{equation}
\frac{\operatorname{Im}(\mathbf{X}_{e}^{\mathrm{Type\,I}})^{11}}{2J_{\mathcal{CP}}^{\mathrm{Type\,I}}}\approx\frac{v^{4}}{(m_{\tau}^{2}-m_{e}^{2})(m_{\mu}^{2}-m_{e}^{2})}\frac{1}{(\tan\beta)^{4}}\approx10^{5}\times\left(\frac{50}{\tan\beta}\right)^{4}\;.
\end{equation}
Avec une aussi grande hiérarchie, la limite actuelle de $d_{e}$ exclut
tout signal dans $d_{q}$.

\begin{figure}[t]
\centering
\includegraphics[width=0.95\textwidth]{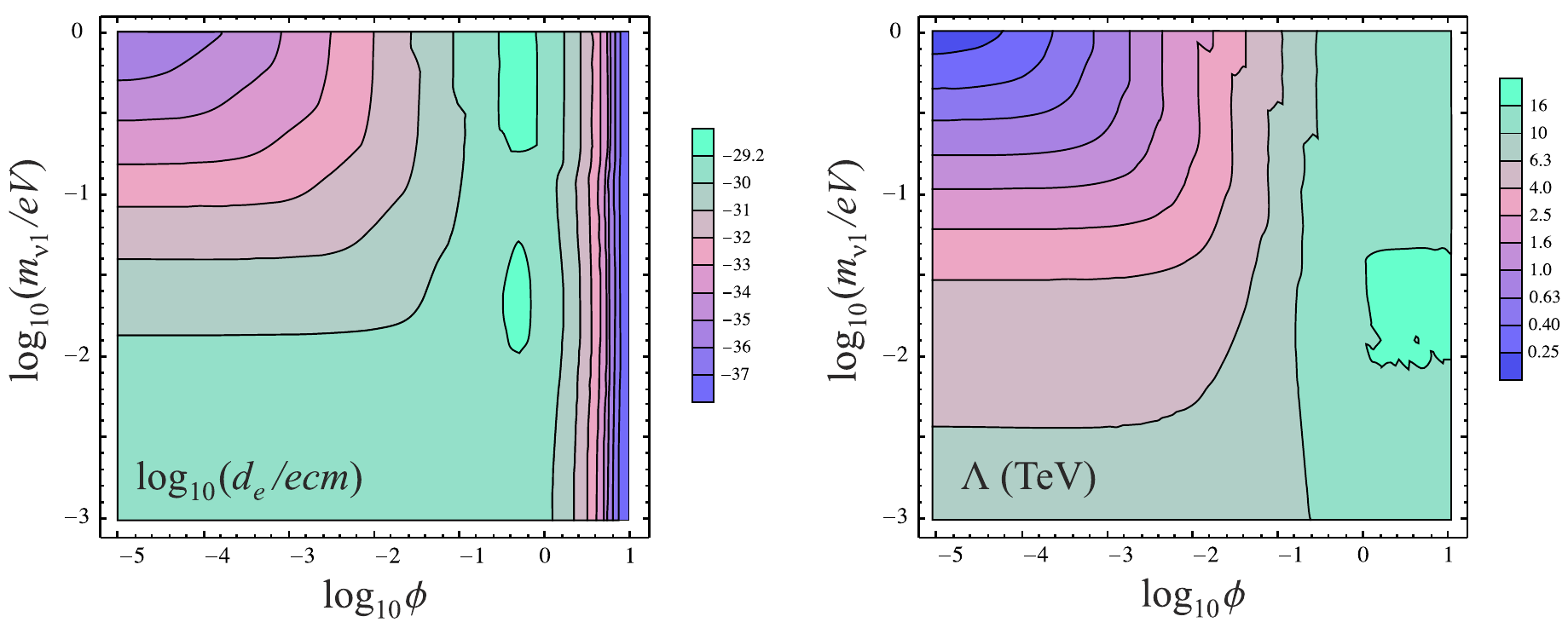}
\caption{EDM de l'électron et échelle minimale de l'opérateur effectif $\Lambda_{\min}$ à laquelle $\mu\rightarrow e\gamma$ est compatible avec sa limite expérimentale, en fonction de la masse du neutrino le plus légers et du paramètre $\phi=\phi_{1}=\phi_{2}=\phi_{3}$. Implicitement, l'échelle de seesaw est fixée pour chaque valeurs de ces paramètres en saturant la borne de perturbativité, $|\mathbf{Y}_{\nu}^{\dagger }\mathbf{Y}_{\nu}|\lesssim4\pi$, tandis que l'EDM est maximisé sur les trois phases de violation-$\mathcal{CP}$ (à ce stade des fluctuations numériques peuvent se produire, d'où les petites irrégularités au niveau des contours).}
\label{FigEDMtI}
\end{figure}

La situation pour les EDMs de leptons est différente. Afin de trouver
les valeurs maximales atteignables (avec $\tan\beta=1$), suivons
la même stratégie que pour le seesaw de type II, avec $\mu\rightarrow e\gamma$
se comportant comme $M_{R}^{2}/\Lambda^{4}$ et les EDMs comme $M_{R}^{3}/\Lambda^{2}$.
Ainsi, pour une masse donnée de neutrino le plus légers $m_{\nu1}$,
le paramètre $\phi\equiv\phi_{1}=\phi_{2}=\phi_{3}$, ainsi que les
phases violant-$\mathcal{CP}$, on ajuste $M_{R}$ pour saturer la
borne de perturbativité $|\mathbf{Y}_{\nu}^{\dagger}\mathbf{Y}_{\nu}|\lesssim4\pi$.
Alors, on détermine l'échelle minimale $\Lambda_{\min}$ pour laquelle
$\mu\rightarrow e\gamma$ est compatible avec sa limite expérimentale,
puis on calcule les EDMs en supposant qu'ils surviennent à la même
échelle. De plus, pour maintenir le parallèle avec le seesaw de type
II discuté précédemment, on inclut un facteur $g^{2}/16\pi^{2}$ pour
tous les opérateurs magnétiques. Le résultat de cette analyse est
montré en figure \ref{FigEDMtI}. Comme dans le seesaw de type II,
il est possible de ramener $d_{e}$ proche de sa limite expérimentale
à condition que la borne de perturbativité soit appliquée $|\mathbf{Y}_{\nu}^{\dagger}\mathbf{Y}_{\nu}|\lesssim4\pi$.
En demandant que $|\mathbf{Y}_{\nu}^{\dagger}\mathbf{Y}_{\nu}|\lesssim1$
réduise $M_{R}$ de $4\pi$ tout en augmentant $\Lambda_{\min}$ de
$\sqrt{4\pi}$, pour une réduction nette de tous les EDMs de deux
ordres de grandeurs, $(4\pi)^{2}\approx160$. Dans ce cas, augmenter
$\tan\beta$ est nécessaire pour ramener à nouveau $d_{e}$ au-dessus
de $10^{-30}$ $e\cdot cm$.

\subsection{Termes de masses de Majorana et invariants anomaux}

Le terme de masse de Majorana viole explicitement le nombre leptonique
qui n'est rien d'autre qu'une combinaison linéaire spécifique des
cinq $U(1)$s du groupe de saveur $U(3)^{5}$. Comme dans le secteur
des quarks, cela signifie que les invariants de $SU(3)^{5}$ devraient
être considérés. En effet, en généralisant l'équation (\ref{eq:PMNS2})
à $U_{PMNS}=U_{PMNS}^{\mathrm{Dirac}}\cdot\operatorname{diag}(e^{i\gamma_{M}},e^{i\alpha_{M}},e^{i\beta_{M}})$,
les structures de saveurs leptoniques invariantes sous $SU(3)^{5}$
mais non sous $U(3)^{5}$ impliquent nécessairement
\begin{equation}
\begin{aligned}\operatorname{Im}\det\mathbf{\Upsilon}_{\nu} & =\frac{m_{\nu1}m_{\nu2}m_{\nu3}}{v^{3}}\sin(\alpha_{M}+\beta_{M}+\gamma_{M})\;,\\
\operatorname{Im}\varepsilon^{AJL}\varepsilon^{IKM}\mathbf{\Upsilon}_{\nu}^{IJ}\mathbf{\Upsilon}_{\nu}^{KL}\mathbf{\Upsilon}_{\nu}^{MB} & =\operatorname{Im}\det\mathbf{\Upsilon}_{\nu}\varepsilon^{AJL}\varepsilon^{BJL}=2\delta^{AB}\operatorname{Im}\det\mathbf{\Upsilon}_{\nu}\;.
\end{aligned}
\end{equation}
Avec des masses de neutrinos de l'ordre de $0.1$ eV, ces structures
sont toutes les deux d'ordre $\mathcal{O}(10^{-34})$, i.e., beaucoup
plus grandes que celles invariantes sous le $U(3)^{5}$ total. De
plus, dans un seesaw de type II, on s'attendrait à ce que $\operatorname{Im}\det\mathbf{\Upsilon}_{\Delta}$
apparaisse à la place, ce qui pourrait atteindre des valeurs $\mathcal{O}(1)$.
La question que nous adressons ici consiste à savoir si ces derniers
contribuent aux observables physiques comme des EDMs ou s'ils peuvent
être éliminés par des rotations.

\subsubsection{Interactions anomales électrofaibles}

Comme première étape, les relations entre le terme de masse de Majorana
et le couplage violant-$\mathcal{CP}$ du MS $\theta_{W}W_{\mu\nu}\tilde{W}^{\mu\nu}$
doivent être identifiées. Pour spécifier, sous des transformations
de $U(1)$ de savoureuses avec les paramètres $3\alpha_{Q,L}=\arg\det V_{L}^{d,e\dagger}$
et $3\alpha_{U,D,E}=\arg\det V_{R}^{u,d,e}$, on a
\begin{equation}
\theta_{W}\rightarrow\theta_{W}^{eff}=\theta_{W}-3\left(\alpha_{L}+3\alpha_{Q}\right)\;.\label{eq:EWCP}
\end{equation}
Dans le MS, vu que $U(1)_{\mathcal{B}+\mathcal{L}}$ est anomal, il
est toujours possible de choisir $\alpha_{L}+3\alpha_{Q}$ de sorte
à fixer $\theta_{W}^{eff}=0$ dans la mesure où $\alpha_{L}$ et $\alpha_{Q}$
sont laissés libres dès que les trois conditions dans l'équation (\ref{eq:SVDphase})
sont imposées. Ceci ne fixe pas séparément $\alpha_{L}$ et $\alpha_{Q}$
car $U(1)_{\mathcal{B}-\mathcal{L}}$ reste une symétrie exacte non-anomale.
En présence d'un terme de masse de Majorana, les deux groupes $U(1)_{\mathcal{B}+\mathcal{L}}$
et $U(1)_{\mathcal{B}-\mathcal{L}}$ sont brisés explicitement et
toutes les rotations $U(1)$ sont fixées. En effet, la convention
pour les phases de Majorana détermine $\alpha_{L}$ vu qu'en plus
des trois conditions (\ref{eq:SVDphase}), il y'a maintenant
\begin{equation}
\arg\det\mathbf{\Upsilon}_{\nu}=2\arg\det U_{PMNS}^{\dagger}+2\arg\det V_{L}^{e\dagger}=-2(\alpha_{M}+\beta_{M}+\gamma_{M})+6\alpha_{L}\;.
\end{equation}
Dès que $\alpha_{L}$ est choisie pour éliminer disons $\gamma_{M}$,
$\alpha_{Q}$ doit être ajustée en conséquence pour annuler $\theta_{W}^{eff}$,
tandis que $\alpha_{U,D,E}$ sont fixées par la condition de masses
réelles pour les fermions, équation (\ref{eq:SVDphase}). La conséquence
principale de tout ceci est que la phase de l'invariant $\det\mathbf{\Upsilon}_{\nu}$
ne peut pas être physique étant donné qu'il est toujours possible
de choisir $\alpha_{L}=(\alpha_{M}+\beta_{M}+\gamma_{M})/3$, auquel
cas $\arg\det\mathbf{\Upsilon}_{\nu}=0$. Notons que ceci explique
aussi a posteriori notre choix dans l'équation (\ref{eq:ThetaBack}).
Il est obligatoire de tenir compte de $\theta_{C}^{eff}$ en agissant
uniquement sur les champs singlets faibles droits, car sinon $\alpha_{Q}$
ne serait pas libre mais dépendrait de $\theta_{C}^{eff}$. A son
tour, $\alpha_{L}$ aurait dû être fixée en termes de $\theta_{C}^{eff}$
afin d'éliminer $\theta_{W}$, et les phases de Majorana des neutrinos
dépendraient de $\theta_{C}^{eff}$.

La réalité de $\det\mathbf{\Upsilon}_{\nu}$ repose sur la possibilité
de choisir $\alpha_{L}$ même après que $\theta_{W}$ ait été éliminée.
Parce que le lagrangien total sauf le terme $\mathbf{\Upsilon}_{\nu}$
est invariant sous $U(1)_{\mathcal{B}-\mathcal{L}}$, cette symétrie
peut être utilisée pour éliminer la phase globale du terme de masse
de Majorana. A ce stade, on peut se demander ce qu'il se passerait
si d'autres interactions à part $\mathbf{\Upsilon}_{\nu}$ violaient
$\mathcal{B}$ et/ou $\mathcal{L}$. Clairement, les interactions
anomales électrofaibles $\mathcal{B}+\mathcal{L}$
\begin{equation}
\mathcal{L}_{SM}^{\mathcal{B}+\mathcal{L}}\sim g_{SM}^{\mathcal{B}+\mathcal{L}}(\varepsilon^{IJK}Q^{I}Q^{J}Q^{K})^{3}\times(\varepsilon^{IJK}L^{I}L^{J}L^{K})\;,
\end{equation}
ne sont pas affectées par la convention de phase adoptée pour les
phases de Majorana dans la mesure où elles se transforment de la façon
suivante
\begin{equation}
g_{SM}^{\mathcal{B}+\mathcal{L}}\rightarrow g_{SM}^{\mathcal{B}+\mathcal{L}}\exp3i(3\alpha_{Q}+\alpha_{L})=g_{SM}^{\mathcal{B}+\mathcal{L}}\exp3i\theta_{W}\;,
\end{equation}
lorsque $\theta_{W}^{eff}=0$. Vu que la même combinaison $3\alpha_{Q}+\alpha_{L}$
que celle dans l'équation (\ref{eq:EWCP}) apparaît, cette phase est
fixée de façon non-univoque dès que l'absence du terme $W_{\mu\nu}\tilde{W}^{\mu\nu}$
est imposée \cite{Perez2014fja}. Il en est de même pour les opérateurs
de Weinberg de dimension-6 \cite{BLWeinberg}, puisqu'ils préservent
aussi $\mathcal{B}-\mathcal{L}$.

\subsubsection{Invariants de Majorana à partir de couplages violant les nombres
baryoniques et leptoniques}

C'est seulement en présence de couplages violant $\mathcal{B}$ et/ou
$\mathcal{L}$, non-alignés ni avec la masse de Majorana ($\Delta\mathcal{L}=2n$,
$n$ entier) ni avec le couplage anomal $\mathcal{B}+\mathcal{L}$
que leurs phases ne peuvent être définies sans ambiguïtés. Pour illustrer
ce propos, considérons les deux opérateurs de dimension-9 (ici écris
en termes de spineurs de Weyl gauches) \cite{MFVBL}
\begin{equation}
\mathcal{H}_{eff}=\delta_{1}\frac{EL^{2}U^{3}}{\Lambda^{5}}+\delta_{2}\frac{U^{2}D^{4}}{\Lambda^{5}}+h.c.\;,
\end{equation}
où $\delta_{1}$ induit des transitions $\Delta\mathcal{L}=3,\Delta\mathcal{B}=1$
et $\delta_{2}$ induit des transitions $\Delta\mathcal{L}=0,\Delta\mathcal{B}=2$.
Sous les transformations de $U(1)^{5}$,
\begin{equation}
\begin{aligned}\delta_{1} & \rightarrow\delta_{1}\exp i(-\alpha_{E}+2\alpha_{L}-3\alpha_{U})\rightarrow\delta_{1}\exp i(3(\alpha_{L}+\alpha_{Q})-\frac{1}{3}\arg\det\mathbf{Y}_{e}-\arg\det\mathbf{Y}_{u})\;,\\
\delta_{2} & \rightarrow\delta_{2}\exp i(-2\alpha_{U}-4\alpha_{D})\rightarrow\delta_{2}\exp i(6\alpha_{Q}-\frac{2}{3}\arg\det\mathbf{Y}_{u}-\frac{4}{3}\arg\det\mathbf{Y}_{d})\;,
\end{aligned}
\label{eq:deltas}
\end{equation}
où nous avons imposé l'équation (\ref{eq:SVDphase}). Parce-que les
deux opérateurs dans $\mathcal{H}_{eff}$ induisent des différents
patterns de $\Delta\mathcal{B}$ et $\Delta\mathcal{L}$ que le terme
de masse de Majorana ou les couplages anomaux du MS, ils dépendent
différemment des rotations $U(1)$. Un choix donné pour $\alpha_{L}$
et $\alpha_{Q}$ peut enlever la violation-$\mathcal{CP}$ de certains
couplages, mais ces derniers ne peuvent pas être tous réels simultanément.
Ceci ressemble fortement à la façon dont la phase de violation-$\mathcal{CP}$
forte peut être transférée entre le terme $G_{a,\mu\nu}\tilde{G}^{a,\mu\nu}$
et les termes de masses des quarks, mis à part que les contenus physiques
des différents couplages sont très différents ici. En ayant différentes
charges $\mathcal{B}$ et $\mathcal{L}$, ceux-ci n'induisent pas
les mêmes types d'observables, donc c'est plutôt déroutant de pouvoir
déplacer une phase de violation-$\mathcal{CP}$ de cette manière.

\begin{figure}[t]
\centering
\includegraphics[width=0.45\textwidth]{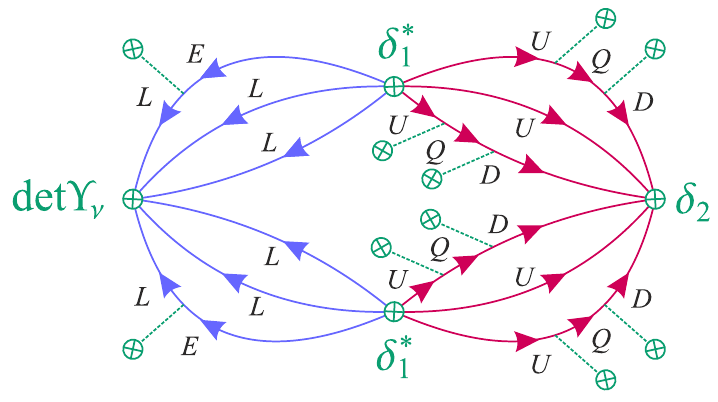}
\caption{Représentation diagrammatique de la contribution à l'EDM provenant de la combinaison des couplages effectifs dans la première  équation de (\ref{eq:CPdim9}), où les tadpoles représentent des insertions de matrices de couplages de Yukawa.}
\label{FigEffB}
\end{figure}

La clef pour résoudre ceci est de supposer que l'invariant $\det\mathbf{\Upsilon}_{\nu}$
peut seulement provenir d'une interaction portant une charge globale
$\mathcal{L}=6$. Qu'elle soit locale ou pas, cette interaction ne
contribue alors jamais directement aux EDMs. Au lieu de cela, au moins
deux autres interactions non-alignées violant $\mathcal{B}$ et/ou
$\mathcal{L}$ sont requises pour construire des combinaisons globales
$\Delta\mathcal{B}=\Delta\mathcal{L}=0$. A ce stade, seules les différences
de phases entre les couplages impliqués comptent, et celles-ci ne
dépendent pas des choix spécifiques faits pour $\alpha_{L}$ et $\alpha_{Q}$.
Par exemple, en présence des opérateurs anomaux de dimension-9, les
phases de violation-$\mathcal{CP}$ induisant potentiellement des
EDMs peuvent provenir de
\begin{equation}
\begin{aligned}\arg(\det\mathbf{\Upsilon}_{\nu}\cdot\delta_{1}^{\ast2}\cdot\delta_{2})-\frac{2}{3}\arg\det\mathbf{Y}_{e}-\frac{4}{3}\arg\det\mathbf{Y}_{u}+\frac{4}{3}\arg\det\mathbf{Y}_{d} & =-2(\alpha_{M}+\beta_{M}+\gamma_{M})\;,\\
\arg(\det\mathbf{\Upsilon}_{\nu}\cdot\delta_{1}^{\ast3}\cdot g_{SM}^{\mathcal{B}+\mathcal{L}})-\arg\det\mathbf{Y}_{e}-3\arg\det\mathbf{Y}_{u} & =-2(\alpha_{M}+\beta_{M}+\gamma_{M})\;,\\
\arg(\det\mathbf{\Upsilon}_{\nu}\cdot\delta_{2}^{3}\cdot(g_{SM}^{\mathcal{B}+\mathcal{L}})^{\ast2})+2\arg\det\mathbf{Y}_{u}+4\arg\det\mathbf{Y}_{d} & =-2(\alpha_{M}+\beta_{M}+\gamma_{M})\;,
\end{aligned}
\label{eq:CPdim9}
\end{equation}
où $\delta_{1,2}$ et $g_{SM}^{\mathcal{B}+\mathcal{L}}$ sont supposés
réels, modulo les rotations de $U(1)^{5}$ (à savoir, $\operatorname{Im}\delta_{1,2}=0$
dans le membre de droite de l'équation (\ref{eq:deltas})). Les trois
phases de Majorana originales apparaissent dans la même combinaison
pour les trois mécanismes, indépendamment des choix des $\alpha$'s
vu qu'ils s'annulent. Les différents termes $\arg\det\mathbf{Y}_{u,d,e}$
apparaissent du fait que certaines transitions de saveurs sont nécessaires
pour relier les interactions effectives, voir figure \ref{FigEffB}.
En définitive, les membres de gauches de ces équations sont invariants
sous le groupe total $U(1)^{5}$. Sous cette forme, il est alors clair
qu'un choix différent de $\alpha_{U,D,E}$ dans lequel les phases
violant-$\mathcal{CP}$ sont déplacées sur les masses des fermions
ne change pas les EDMs. Finalement, notons que dès que la présence
des interactions $\Delta\mathcal{B}$ et $\Delta\mathcal{L}$ rend
physique la combinaison de phases de Majorana, il se pourrait aussi
qu'elle soit accessible par d'autres processus. Considérons par exemple
la désintégration di-proton, induite par l'interaction $\delta_{1}^{2}$.
La phase $\mathcal{CP}$ de cette amplitude est accessible seulement
par interférence avec une autre amplitude, mais tout ce qui est disponible
est $\delta_{2}\det\mathbf{\Upsilon}_{\nu}$. La différence de phases
de ces deux amplitudes est précisément celle donnée dans l'équation
(\ref{eq:CPdim9}).

En pratique, l'existence de cette contribution supplémentaire à l'EDM
n'a pas d'impact si $\operatorname{Im}\det\mathbf{\Upsilon}_{\nu}\lesssim\mathcal{O}(10^{-34})$.
Même dans le cas le plus favorable $\mathbf{\Upsilon}_{\Delta}\approx10^{12}\times\mathbf{\Upsilon}_{\nu}$,
le facteur $\delta_{1}^{\ast2}\delta_{2}$ apporte une suppression
prohibitive en $\Lambda^{-15}$ de sorte que
\begin{equation}
d_{e}\approx e\frac{m_{e}}{M_{W}^{2}}\frac{M_{W}^{15}}{\Lambda^{15}}\operatorname{Im}(\det\mathbf{\Upsilon}_{\Delta}\cdot\delta_{1}^{\ast2}\cdot\delta_{2})\lesssim10^{-37}\times\left[\frac{1\,\text{TeV}}{\Lambda}\right]^{15}\;e\cdot cm\;,
\end{equation}
lorsque $\operatorname{Im}(\det\mathbf{\Upsilon}_{\Delta}\cdot\delta_{1}^{\ast2}\cdot\delta_{2})$
est d'ordre $\mathcal{O}(1)$. Alternativement, la limite actuelle
sur $d_{e}$ nécessite $\Lambda\gtrsim250$ GeV. Etant donné que la
désintégration du proton ou encore les oscillations neutron-antineutron
devraient pousser $\Lambda$ au-dessus du TeV, même en supposant que
MFV s'applique à ces opérateurs \cite{MFVBL}, cette contribution
est trop petite pour être vue. Bien que la situation décrite ici est
plutôt particulière, avec des opérateurs effectifs de grandes dimensions
uniquement, cette conclusion devrait être assez robuste. En tout cas,
pour rester du bon côté, il vaudrait mieux garder ce mécanisme en
tête dès lors que les neutrinos ont des termes de masses de Majorana
ainsi que certaines interactions s'avérant aussi violer $\mathcal{B}$
et/ou $\mathcal{L}$.

\newpage\null\thispagestyle{empty}\newpage

\chapter{Théorie effective des champs (EFT)\label{Ch4}}

\minitoc

\section{Introduction aux EFT}

Cette section est inspirée des revues \cite{Georgi,Manohar,Skiba}.

\subsection{De la notion de théorie effective}

De la taille d'un quark à la taille de l'univers, la physique (connue)
s'étend sur plus de 40 ordres de grandeurs et les phénomènes intéressants
mis en jeu à chaque échelle sont d'une grande diversité. Dès lors
que l'on s'intéresse à un ensemble de phénomènes ayant lieux à une
échelle particulière, il est souhaitable de pouvoir démêler cet ensemble
de phénomènes du reste de la physique afin d'être capable de le décrire
sans pour autant tout connaître. Le concept de théorie effective,
bien qu'étant toujours implicite, est une idée centrale en physique.
Lorsque le physicien tente de modéliser un système, les premières
questions auxquelles il est confronté sont : quels sont les degrés
de libertés pertinents permettant de décrire ce système ? Et quelles
sont les interactions à considérer entre ces derniers ? Les réponses
à ces questions définissent une théorie \og effective \fg{} qui
permettra de décrire un phénomène physique. Généralement, la dynamique
mise en jeu à une certaine échelle est découplée de celle ayant lieu
à une autre échelle (très) différente (basse énergie/haute énergie
ou de façon équivalente longue distance/courte distance). Il en résulte
que la physique à basse énergie peut être décrite par une théorie
effective impliquant seulement quelques degrés de libertés (pertinents
à cette échelle), en ignorant ceux qui sont en action à des échelles
d'énergies supérieures (ou courtes distances). Comme tous les degrés
de libertés et toutes les interactions ne sont pas considérées, les
résultats obtenus dans ce cadre sont approximatifs et de ce fait,
la question de la précision de ces résultats, c'est à dire de l'erreur
commise se pose naturellement. Le cadre des EFT, de par sa transcription
mathématique sous forme de développement perturbatif en un certain
paramètre d'expansion, fournit une manière de quantifier et contrôler
cette erreur en choisissant la précision de façon arbitraire à travers
l'ordre auquel est tronqué la série perturbative. Construire une théorie
effective consiste donc à trouver le cadre le plus simple (avec le
moins de degrés de libertés et d'interactions) qui capture l'essentiel
de la physique du phénomène considéré et ce avec une précision arbitraire.

Par exemple, afin de déterminer les niveaux d'énergie de l'atome d'hydrogène,
Böhr considéra comme seuls degrés de libertés le proton et l'électron
en interaction coulombienne sans se préoccuper de la sous-structure
du proton (inconnue à l'époque) ou d'autres interactions qui apporteraient
de légères corrections à ce résultat. Mais est-ce vraiment nécessaire
d'inclure toutes les contributions à une certaine quantité ? Cela
dépend de la précision souhaitée. Si le fait de prendre en compte
l'existence des quarks ne corrige les valeurs des niveaux d'énergie
trouvées que de façon négligeable et au prix de calculs lourds, il
serait plus raisonnable d'adopter une approche effective tout en quantifiant
l'erreur commise due à l'omission de la contribution des quarks. L'approche
EFT simplifie grandement les calculs, car toutes l'information contenue
dans le proton est résumée en seulement deux paramètres, sa masse
et sa charge. En faisant cela, les résultats obtenus pour les niveaux
d'énergie de l'atome d'hydrogène sont satisfaisant en première approximation.
Si on souhaite affiner la description, par exemple pour expliquer
les niveaux hyperfins de l'atome d'hydrogène, alors il faudrait prendre
en compte des paramètres supplémentaires qui sont le spin du proton
ainsi que son moment magnétique. On pourrait également obtenir des
résultats encore plus précis pour les niveaux d'énergie en considérant
que le proton n'est pas ponctuel et en ajoutant aux paramètres déjà
considérés le rayon de celui-ci.

En physique des particules, le seul paramètre pertinent est l'échelle
d'énergie ou de façon équivalente l'échelle de distance ou de temps\footnote{En effet, dans un système d'unités naturelles $\bar{h}=c=1$, on a
$[Energie]=[distance]^{-1}=[temps]^{-1}$.}. Dans une théorie quantique et relativiste, là où les particules
peuvent être créées et annihilées, le concept de théorie effective
(quantique) est particulièrement intéressant. L'idée centrale d'une
théorie des champs quantiques est qu'à des énergies petites devant
une échelle caractéristique $\Lambda$, tous les effets des degrés
de libertés plus lourds que $\Lambda$ peuvent être encodés dans de
nouvelles interactions entre les champs restés actifs en dessous de
$\Lambda$. Ce concept s'avère extrêmement utile car il consiste à
négliger les phénomènes physiques siégeant au-delà d'une certaine
énergie (ou en-deçà d'une certaine distance) et parmi ces effets qui
peuvent être ignorés se trouvent toutes les particules trop lourdes
pour être produites, même celles qu'on ne connait pas. Ainsi, en éliminant
les particules trop lourdes, la théorie se voit grandement simplifiée. 

\subsection{Principe de construction en théorie quantique des champs}

La première étape dans la construction d'une théorie effective des
champs (EFT) consiste à identifier les degrés de libertés pertinents
(généralement les champs légers) pouvant décrire le phénomène physique
auquel on s'intéresse. Considérons un exemple simple d'EFT où l'on
choisit de décrire le système uniquement avec les particules légères,
c'est à dire en omettant les champs lourds de la théorie. Généralement,
une échelle d'énergie caractéristique correspondant au seuil de production
on-shell des particules peut être définie et établit une séparation
entre champs légers et champs lourds. Bien entendu, toutes les théories
des champs sont forcément des théories effectives dans la mesure où
nous ne connaissons pas tous les états lourds jusqu'à l'échelle de
Planck, et donc nous omettons leurs effets (inconnus) de la théorie.
En ce sens, les EFTs sont un moyen puissant de paramétrer la nouvelle
physique, qui par définition est alors inconnue. Cette technique a
été appliquée de façon plus ou moins sophistiquée dans un grand nombre
de thématiques \cite{Manohar,Rothstein,Kaplan,Goldberger} et se révèle
être un outil puissant.

Dans le formalisme des intégrales de chemins, les champs lourds sont
intégrés (integrated out) au niveau de l'action et par conséquent
sont éliminés de la théorie
\begin{equation}
\varint\mathcal{D}\phi_{H}e^{i\varint\mathcal{L}(\phi_{L},\phi_{H})}=e^{i\varint\mathcal{L}_{eff}(\phi_{L})},\label{eq:IntOut}
\end{equation}
où $\phi_{L}$ et $\phi_{H}$ dénotent respectivement les champs légers
et lourds de la théorie. Comme souvent en pratique, pour des raisons
de simplicité (ou parfois même de faisabilité), les calculs sont réalisés
dans l'approche des diagrammes de Feynman plutôt que dans celle des
intégrales de chemins. Lorsqu'on intègre les particules lourdes, on
obtient une EFT entièrement définie par un nouveau lagrangien effectif
impliquant seulement les champs légers (voir équation (\ref{eq:IntOut})).
Le lagrangien effectif alors obtenu peut se décomposer en une composante
finie contenant uniquement des opérateurs de dimension en énergie
inférieure ou égale à quatre puis d'une autre composante contenant
des opérateurs de dimensions supérieures à quatre organisée en série
infinie de dimension croissante, appelée \og tour d'opérateurs de
dimensions supérieures \fg{}. Comme une longue phrase ne vaut pas
une formule, on a:
\begin{equation}
\mathcal{L}_{eff}(\phi_{L})=\mathcal{L}_{d\leq4}+\sum_{i}\frac{O_{i}}{\Lambda^{dim(O_{i})-4}},\label{eq:InfTowerOps}
\end{equation}
où $\Lambda$ est une échelle caractéristique d'énergie et $dim(O_{i})$
la dimension en énergie de l'opérateur $O_{i}$. Notons que lorsque
ce lagrangien effectif est utilisé pour des processus à basse énergie,
c'est à dire $\Lambda\ll m_{\phi_{H}}$, les termes obtenus dans le
développement (\ref{eq:InfTowerOps}) sont locaux (dans l'espace-temps). 

En pratique, la tour infinie d'opérateurs de dimensions supérieures
dans (\ref{eq:InfTowerOps}) est tronquée à un ordre arbitraire et
seuls quelques termes s'avèrent être pertinents. L'idée est que la
théorie effective doit reproduire les résultats expérimentaux avec
une précision finie. Comme la précision d'une expérience est toujours
finie, le nombre de termes considérés dans la série l'est aussi. Plus
la dimension d'un opérateur est grande et plus sa contribution aux
observables à basse énergie est petite. Ainsi, le nombre de termes
gardés dans la série sera d'autant plus grand que la précision souhaitée.
Ces termes de dimensions supérieures sont constitués d'opérateurs
non-renormalisables, or étant donné que la série infinie est tronquée,
on se retrouve seulement avec un nombre fini de paramètres libres
requis pour pouvoir faire des prédictions, ce qui est exactement la
même situation que dans les théories renormalisables. C'est pourquoi
dans cette approche, les théories non-renormalisables conviennent
tout autant que celles qui le sont.

\subsection{Comptage de puissance et correspondance à l'arbre\label{TreeLevelMatchingPart}}

Le formalisme des EFTs est basé sur un développement en série systématique
en un certain paramètre d'expansion, qui peut être par exemple le
rapport des échelles d'énergies considérées\footnote{Ces échelles d'énergies peuvent être: les masses des particules lourdes,
l'énergie à laquelle se déroule l'expérience, le transfert d'impulsion,
etc...}. Dans ce cadre, on peut garder une trace des puissances du paramètre
d'expansion (ordre de la série), car il est souhaitable que l'EFT
construite permette de prédire les ordres de grandeurs des différents
termes dans le lagrangien effectif. En comptant les puissances (power
counting) d'un terme, on peut alors déduire le comportement de ce
dernier lorsque l'énergie est modifiée. Dans les cas les plus aisés,
le comptage de puissance se réduit à une simple analyse dimensionnelle
en unités naturelles ($\bar{h}=c=1$), pour lesquelles toutes les
unités peuvent être ramenées à une certaine puissance de l'énergie
(ou de façon équivalente de la masse), qu'on appelle la dimension
en énergie (ou abusivement la dimension tout court). La dimension
d'un champ est généralement déterminée à partir de son terme cinétique
dans la densité lagrangienne.

Prenons un exemple de théorie contenant un champ fermionique $\psi$
sans masse ainsi que deux champs scalaires réels (un léger $\phi$
de masse $m$ et un lourd $\Phi$ de masse $M\gg m$) couplés par
une interaction de type Yukawa. Le lagrangien de la théorie UV (à
haute énergie) est donc:
\begin{equation}
\mathcal{L}=i\bar{\psi}\slashed\partial\psi+\frac{1}{2}(\partial_{\mu}\Phi)^{2}-\frac{M^{2}}{2}\Phi^{2}+\frac{1}{2}(\partial_{\mu}\phi)^{2}-\frac{m^{2}}{2}\phi^{2}-\lambda\bar{\psi}\psi\Phi-\eta\bar{\psi}\psi\phi.\label{eq:LagrangianUV}
\end{equation}
On souhaite construire une EFT impliquant seulement les champs légers
comme degrés de libertés, à savoir le fermion $\psi$ et le scalaire
léger $\phi$. En intégrant le scalaire lourd $\Phi$, celui-ci sera
éliminé de la théorie UV et ses effets seront encodés dans des nouvelles
interactions de dimension supérieure impliquant les champs légers.

\begin{figure}[h]
\begin{center}
\includegraphics[height=1.2315in,width=6in]{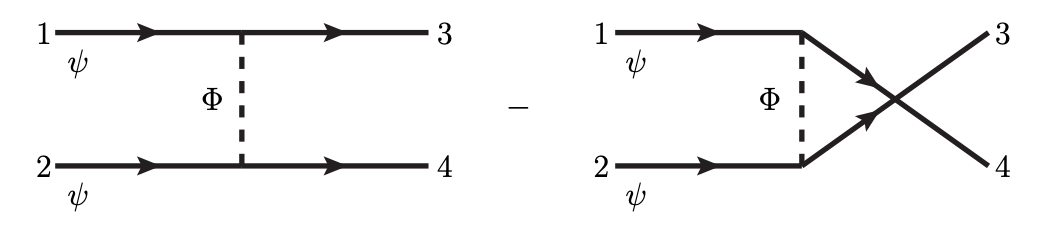}
\caption{Diagrammes à l'arbre contribuant à la diffusion $\psi\psi\rightarrow\psi\psi$ à l'ordre $\mathcal{O}(\lambda^{2})$. Source: \cite{Skiba}}
\label{TreeLevelMatching}
\end{center}
\end{figure}

Pour commencer, examinons le processus de diffusion $\psi\psi\rightarrow\psi\psi$
à l'ordre $\mathcal{O}(\lambda^{2}\eta^{0})$ dans les constantes
de couplages, en gardant les termes jusqu'à l'ordre deux en impulsions
externes. Les diagrammes à l'arbre y contribuant sont représentés
figure \ref{TreeLevelMatching}.

Intégrer des champs dans l'approche diagrammatique consiste tout d'abord
à calculer l'amplitude du processus considéré d'une part dans la théorie
complète (à haute énergie, aussi appelée théorie UV) et d'autre part
avec la théorie effective construite (à basse énergie, aussi appelée
théorie IR), puis en égalant la limite à basse énergie de l'amplitude
UV et l'amplitude EFT. Cette étape clef est appelée la correspondance
(matching), car elle fait correspondre les deux théories à basse énergie,
comme attendu pour une EFT.

Dans notre exemple, l'amplitude UV à l'ordre $\mathcal{O}(\lambda^{2})$
est donnée par les diagrammes à l'arbre sur la figure \ref{TreeLevelMatching},
et on obtient:
\begin{equation}
\mathcal{A}_{UV}=\bar{u}(p_{3})u(p_{1})\bar{u}(p_{4})u(p_{2})(-i\lambda)^{2}\frac{i}{(p_{3}-p_{1})^{2}-M^{2}}-\{3\leftrightarrow4\},
\end{equation}
où $\{3\leftrightarrow4\}$ signifie le même terme en interchangeant
$p_{3}$ et $p_{4}$. La structure tensorielle de l'amplitude (ici
spinorielle) est la même dans la théorie UV et dans l'EFT donc nous
pouvons nous concentrer uniquement sur le propagateur, dont on gardera
le développement limité jusqu'à l'ordre 2 en impulsion externes 
\begin{equation}
(-i\lambda)^{2}\frac{i}{(p_{3}-p_{1})^{2}-M^{2}}=i\frac{\lambda^{2}}{M^{2}}\frac{1}{1-\frac{(p_{3}-p_{1})^{2}}{M^{2}}}\approx i\frac{\lambda^{2}}{M^{2}}(1+\frac{(p_{3}-p_{1})^{2}}{M^{2}}+\mathcal{O}(\frac{p^{4}}{M^{4}})).
\end{equation}
Ici, le paramètre d'expansion sur lequel est construite l'EFT est
le rapport $\frac{p^{2}}{M^{2}}$ et il servira à contrôler la précision
souhaitée.

A l'ordre $\mathcal{O}(p^{0})$, l'amplitude de diffusion $\psi\psi\rightarrow\psi\psi$
peut être reproduite par le lagrangien effectif suivant:
\begin{equation}
\mathcal{L}_{p^{0},\lambda^{2}}=i\bar{\psi}\slashed\partial\psi+\frac{c}{2}\bar{\psi}\psi\bar{\psi}\psi,
\end{equation}
où l'on a omis les termes impliquant le scalaire léger car ce dernier
ne joue aucun rôle ici. L'amplitude EFT calculée en utilisant ce lagrangien
effectif est:
\begin{equation}
\mathcal{A}_{EFT}=\bar{u}(p_{3})u(p_{1})\bar{u}(p_{4})u(p_{2})(ic)-\{3\leftrightarrow4\}.
\end{equation}
En comparant les amplitudes calculées dans les deux théories, on effectue
la correspondance et on obtient $c=\frac{\lambda^{2}}{M^{2}}$.

A l'ordre suivant $\mathcal{O}(p^{2})$, le lagrangien effectif suivant
peut être construit
\begin{equation}
\mathcal{L}_{p^{2},\lambda^{2}}=i\bar{\psi}\slashed\partial\psi+\frac{\lambda^{2}}{2M^{2}}\bar{\psi}\psi\bar{\psi}\psi+d\partial_{\mu}\bar{\text{\ensuremath{\psi}}}\partial^{\mu}\psi\bar{\psi}\psi.
\end{equation}
De la même manière qu'à l'ordre précédent, il s'agit de comparer l'amplitude
UV à basse énergie et l'amplitude EFT calculée à l'aide de ce lagrangien
effectif. Au cours de cette procédure de correspondance, tout choix
judicieux d'impulsions externes (par exemple une condition on-shell)
qui pourrait simplifier les calculs peut être fait, à condition de
faire le même choix dans les deux théories pour la correspondance.
En effet, les impulsions des particules externes n'ont rien à voir
avec la dynamique à haute énergie. Ici par exemple, on peut faire
le choix utile de supposer les champs externes sur couche de masse
(on-shell), c'est à dire $p_{1,2,3,4}^{2}=0$, ce qui simplifie grandement
les expressions dans la mesure où les seules dépendances possibles
sont celles qui s'expriment sous la forme d'un produit de Minkowski
des 4-impulsions $p_{i}\cdot p_{j}$. Par conséquent, la seule partie
de l'amplitude UV à basse énergie que doit reproduire l'EFT est $-2i\frac{\lambda^{2}}{M^{2}}\frac{p_{1}\cdot p_{3}}{M^{2}}-\{3\leftrightarrow4\}$.
Le terme impliquant le coefficient $d$ dans le lagrangien effectif
$\mathcal{L}_{p^{2},\lambda^{2}}$ donne l'amplitude
\begin{equation}
\mathcal{A}_{EFT}=id(p_{1}\cdot p_{3}+p_{2}\cdot p_{4})\bar{u}(p_{3})u(p_{1})\bar{u}(p_{4})u(p_{2})-\{3\leftrightarrow4\}.
\end{equation}
En effet, les impulsions $p_{1,2}$ sont supposées entrantes et contribuent
alors à l'amplitude avec $-ip_{1,2}^{\mu}$ alors que les impulsions
$p_{3,4}$ sont sortantes et donc contribuent avec $+ip_{3,4}^{\mu}$.
La conservation de l'impulsion $p_{1}+p_{2}=p_{3}+p_{4}$ nous donne
$p_{1}\cdot p_{2}=p_{3}\cdot p_{4}$, $p_{1}\cdot p_{3}=p_{2}\cdot p_{4}$
et $p_{1}\cdot p_{4}=p_{2}\cdot p_{3}$. En somme, on effectue la
correspondance entre les deux théories et on obtient $d=-\frac{\lambda^{2}}{M^{4}}$.

Enfin, remarquons que l'opérateur dérivatif associé au coefficient
$d$ n'est pas le seul possible qu'on puisse écrire avec quatre fermions
et deux dérivées. Par exemple, on pourrait inclure également $(\partial^{2}\bar{\psi})\psi\bar{\psi}\psi$
ou encore $\partial_{\mu}\bar{\psi}\psi\bar{\psi}\partial^{\mu}\psi$.
En fait, lorsqu'on construit une EFT en général, il est crucial de
considérer tous les termes possibles à un ordre donné. Il peut s'avérer
ensuite que certains termes ne soit pas indépendants sous l'application
de l'équation du mouvement (EOM) ou par intégration partie (nous verrons
ceci plus en détail ultérieurement), dans ce cas le nombre d'opérateur
à considérer se voit réduit (comme dans l'exemple ici) mais a priori,
il faut tous les considérer dans un premier temps.

\subsection{Groupe de renormalisation}

Jusque là, nous avons intégré le scalaire lourd $\Phi$ de la théorie
UV et étudié les interactions effectives du fermion. Néanmoins, nous
avons totalement ignoré le deuxième scalaire $\phi$ présent dans
la théorie complète. Nous avions obtenu le lagrangien effectif à l'arbre
suivant
\begin{equation}
\mathcal{L}_{p^{2},\lambda^{2}}=i\bar{\psi}\slashed\partial\psi+\frac{c}{2}\bar{\psi}\psi\bar{\psi}\psi+d\partial_{\mu}\bar{\text{\ensuremath{\psi}}}\partial^{\mu}\psi\bar{\psi}\psi+\frac{1}{2}(\partial_{\mu}\phi)^{2}-\frac{m^{2}}{2}\phi^{2}-\eta\bar{\psi}\psi\phi,
\end{equation}
et en effectuant la correspondance avec la théorie UV, nous avions
calculé les coefficients $c$ et $d$.

Considérons maintenant les contributions à une boucle à l'amplitude
de diffusion $\psi\psi\rightarrow\psi\psi$, à l'ordre $\mathcal{O}(\lambda^{2}\eta^{2})$.
A une boucle, on s'attendrait à une correction par rapport à l'amplitude
à l'arbre de $\frac{\eta^{2}}{(4\pi)^{2}}$, le facteur $\tfrac{1}{(4\pi)^{2}}$
provient de l'intégration de la boucle. Cependant, cette estimation
n'est pas correcte dès lors que plusieurs échelles d'énergies sont
en jeu comme ici si on considère le deuxième scalaire. Les deux échelles
d'énergies sont les masses des deux scalaires avec $m\ll M$. En raison
du fait que les deux échelles présentes une forte hiérarchie, l'amplitude
peut contenir des grands logarithmes ($\log(\frac{M}{m}))$. En effet,
dans une EFT, on sépare généralement les contributions à l'amplitude
en deux types, ceux qui contiennent des grands logarithmes et ceux
qui n'en contiennent pas. La partie sans logarithme provient de la
correspondance alors que celle qui en contient est générée dans le
processus d'évolution des paramètres sous le groupe de renormalisation
(RG). Remarquons que lorsque nous effectuons la correspondance, nous
comparons deux théories dont les contenus en particules sont différents
vu que dans l'une des deux nous avons intégré (éliminé) des champs.
Ainsi, cette étape doit impérativement être réalisée à la même échelle
d'énergie pour les deux théories, c'est à dire avec la même échelle
de renormalisation. Cette échelle de correspondance (matching scale)
est usuellement prise comme étant la masse de la particule lourde
qui a été intégrée. Dans ces conditions, une seule échelle d'énergie
est en jeu et de ce fait nous ne générerons pas de grands logarithmes.
Etant donné que les deux théories sont les mêmes à basse énergie par
construction, les logarithmes du rapport de l'échelle de correspondance
et d'une échelle d'énergie basse seront identiques dans les deux théories. 

Afin de pouvoir changer d'échelle, nous avons besoin d'établir la
dépendance en énergie des paramètres de notre lagrangien effectif,
autrement dit: les équations du groupe de renormalisation (RGEs).
Supposons que nous voulions connaître l'amplitude à l'échelle $m$,
en négligeant le terme proportionnel à $d$ dans le lagrangien effectif
car nous nous intéresserons seulement à la partie de l'amplitude indépendante
des impulsions. Le comptage de puissance nous apprend que l'amplitude
recherchée est proportionnelle à $\frac{\lambda^{2}\eta^{2}}{16\pi^{2}M^{2}}$.
Le terme contenant deux dérivées sera proportionnel à $\tfrac{1}{M^{4}}$,
et donc ce dernier doit être supprimé d'un facteur $\tfrac{m^{2}}{M^{2}}$
par rapport au terme dominant. En fait, ce raisonnement n'est valable
que si l'on utilise un régulateur indépendant d'échelle (mass-independant
regulator), comme par exemple la régularisation dimensionnelle avec
une prescription de soustraction minimal (MS). Généralement, les autres
méthodes de régularisation (cut-off, Pauli-Villars etc...) compliquent
grandement le comptage de puissance.

Pour établir la dépendance en énergie du coefficient $c$ (RG running),
nous devons déterminer au préalable les constantes de renormalisation
perturbative ($Z$ factors). Tout d'abord, calculons l'amplitude de
la self-énergie du fermion $\mathcal{A}_{self}$, dont le diagramme
est le suivant:

\begin{figure}[h!]
\begin{center}
\includegraphics[scale=1]{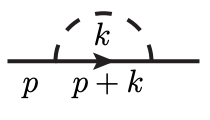}
\caption{Diagramme de Feynman de la self-énergie du fermion.}
\label{FermionSelfEnergy}
\end{center}
\end{figure}On a alors:
\begin{equation}
\begin{aligned}\mathcal{A}_{self} & =(-i\eta)^{2}\varint\frac{d^{d}k}{(2\pi)^{d}}\frac{i(\slashed k+\slashed p)}{(k+p)^{2}}\frac{i}{k^{2}-m^{2}}=\eta^{2}\varint\frac{d^{d}l}{(2\pi)^{d}}\int_{0}^{1}dx\frac{\slashed l+(1-x)\slashed p}{(l^{2}-\Delta^{2})^{2}}\\
 & =\frac{i\eta^{2}}{(4\pi)^{2}}\frac{1}{\epsilon}(\int_{0}^{1}dx(1-x)\slashed p)+\text{termes finis}=\frac{i\eta^{2}\slashed p}{2(4\pi)^{2}}\frac{1}{\epsilon}+\text{termes finis},
\end{aligned}
\label{eq:SelfDiv}
\end{equation}
où l'on a utilisé la technique des paramètres de Feynman pour combiner
les dénominateurs puis nous avons translaté l'impulsion de boucle
$l=k+xp$ pour enfin utiliser les résultats standards des intégrales
de références. Comme souvent, on effectue un développement en série
en la dimension $d=4-2\epsilon$ puis on garde seulement les termes
ayant une structure de pôle en $\tfrac{1}{\epsilon}$, car les termes
finis ne jouent pas de rôle dans le calcul des RGEs.

\begin{figure}[h!]
\begin{center}
\includegraphics[scale=0.5]{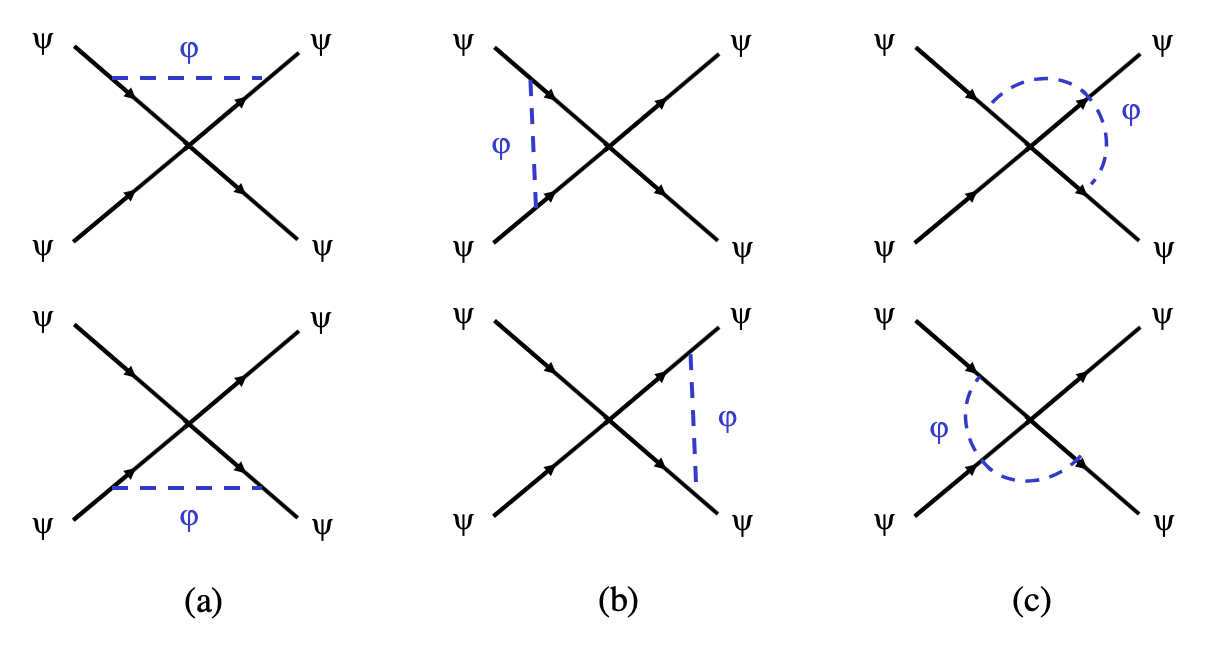}
\caption{Diagramme à une boucle contribuant à la renormalisation de l'interaction à quatre fermion. Figure modifiée à partir de \cite{Skiba}.}
\label{1LoopDiags}
\end{center}
\end{figure}

Intéressons nous maintenant aux corrections à une boucle apportées
aux vertex effectif à quatre-fermions. Les diagrammes concernés sont
représentés en figure \ref{1LoopDiags} et correspondent à toutes
les façons possibles d'apparier deux lignes externes. 

Tous ces diagrammes possèdent des divergences UV logarithmiques et
comme nous sommes uniquement intéressé par les parties divergentes,
nous pouvons négliger les impulsions externes ainsi que les masses
des particules. En ce qui concerne les deux diagrammes $(a)$ de la
figure \ref{1LoopDiags}, leur partie divergente est:
\begin{equation}
2(-i\eta)^{2}ic\varint\frac{d^{d}k}{(2\pi)^{d}}\frac{i\slashed k}{k^{2}}\frac{i\slashed k}{k^{2}}\frac{i}{k^{2}}=-2c\eta^{2}\varint\frac{d^{d}k}{(2\pi)^{d}}\frac{1}{k^{4}}=-\frac{2ic\eta^{2}}{(4\pi)^{2}}\frac{1}{\epsilon}+\text{termes finis}.\label{eq:DIVdiagsA}
\end{equation}

Quant aux diagrammes $b$, ils sont un peu plus délicats dans la mesure
où l'intégration de la boucle implique deux lignes fermioniques différentes.
La partie divergente s'écrit:
\begin{equation}
2(-i\eta)^{2}ic\varint\frac{d^{d}k}{(2\pi)^{d}}\bar{u}(p_{3})\frac{i\slashed k}{k^{2}}u(p_{1})\bar{u}(p_{4})\frac{-i\slashed k}{k^{2}}u(p_{2})\frac{i}{k^{2}}=\frac{ic\eta^{2}}{2(4\pi)^{2}}\frac{1}{\epsilon}\bar{u}(p_{3})\gamma^{\mu}u(p_{1})\bar{u}(p_{4})\gamma_{\mu}u(p_{2})+\text{termes finis},
\end{equation}
et est exactement annulée par celle des diagrammes $c$:
\begin{equation}
2(-i\eta)^{2}ic\varint\frac{d^{d}k}{(2\pi)^{d}}\bar{u}(p_{3})\frac{i\slashed k}{k^{2}}u(p_{1})\bar{u}(p_{4})\frac{i\slashed k}{k^{2}}u(p_{2})\frac{i}{k^{2}}.
\end{equation}

Pour calculer les RGEs, considérons que la partie impliquant le fermion
en négligeant le terme proportionnel à $d$. Le lagrangien original
étant exprimé en termes des champs nus (bare fields) et des paramètres
nus, on renormalise ces derniers pour avoir les champs et paramètres
renormalisés $\psi_{0}=\sqrt{Z_{\psi}}\psi$ et $c_{0}=c\mu^{2\epsilon}Z_{c}$.
Le lagrangien devient alors
\begin{equation}
\begin{aligned}\mathcal{L}_{p^{0},\lambda^{2}\eta^{2}log} & =i\bar{\psi}_{0}\slashed\partial\text{\ensuremath{\psi}}_{0}+\frac{c_{0}}{2}\bar{\psi}_{0}\psi_{0}\bar{\psi}_{0}\psi_{0}=iZ_{\psi}\bar{\psi}\slashed\partial\text{\ensuremath{\psi}}+\frac{c}{2}Z_{c}Z_{\psi}^{2}\mu^{2\epsilon}\bar{\psi}\psi\bar{\psi}\psi\\
 & =i\bar{\psi}\slashed\partial\text{\ensuremath{\psi}}+\mu^{2\epsilon}\frac{c}{2}\bar{\psi}\psi\bar{\psi}\psi+i(Z_{\psi}-1)\bar{\psi}\slashed\partial\text{\ensuremath{\psi}}+\mu^{2\epsilon}\frac{c}{2}(Z_{c}Z_{\psi}^{2}-1)\bar{\psi}\psi\bar{\psi}\psi.
\end{aligned}
\end{equation}
Les contre-termes doivent annuler les divergences obtenues dans les
équations (\ref{eq:SelfDiv}) et (\ref{eq:DIVdiagsA}). En imposant
cela, les constantes multiplicatives de renormalisation doivent satisfaire:
\begin{equation}
\begin{cases}
Z_{\psi}-1 & =-\frac{\eta^{2}}{2(4\pi)}\frac{1}{\epsilon}\\
c(Z_{c}Z_{\psi}^{2}-1) & =\frac{2c\eta^{2}}{(4\pi)^{2}}\frac{1}{\epsilon},
\end{cases}\label{eq:RenormCond}
\end{equation}
où l'on a utilisé la prescription de soustraction minimal (MS) qui
consiste à ne garder que les pôles en $\tfrac{1}{\epsilon}$. En combinant
les deux conditions précédentes (\ref{eq:RenormCond}), obtient 
\begin{equation}
Z_{c}=1+\frac{3\eta^{2}}{(4\pi)^{2}}\frac{1}{\epsilon}.
\end{equation}

La méthode usuelle pour calculer les RGEs repose sur le fait que les
quantités nus ne dépendent pas de l'échelle de renormalisation.
\begin{equation}
0=\mu\frac{d}{d\mu}c_{0}=\mu\frac{d}{d\mu}(c\mu^{2\epsilon}Z_{c})=\beta_{c}\mu^{2\epsilon}Z_{c}+2\epsilon c\mu^{2\text{\ensuremath{\epsilon}}}Z_{c}+c\mu^{2\epsilon}\mu\frac{d}{d\mu}Z_{c},
\end{equation}
avec $\beta_{c}\equiv\mu\frac{dc}{d\mu}$. De plus, on a $\mu\frac{d}{d\mu}Z_{c}=\frac{3}{(4\pi)^{2}}2\eta\beta_{\eta}\frac{1}{\epsilon}$.
En refaisant les mêmes étapes pour $\eta$, on obtient $\beta_{\eta}=-\epsilon\eta-\eta\frac{d\log Z_{\eta}}{d\log\mu}$.
Le second terme est d'ordre supérieur donc nous ne garderons que le
premier $\beta_{\eta}=-\epsilon\eta$, ce qui donne $\mu\frac{d}{d\mu}Z_{c}=-\frac{6\eta^{2}}{(4\pi)^{2}}$.
Finalement, on obtient:
\begin{equation}
\beta_{c}=\frac{6\eta^{2}}{(4\pi)^{2}}c.\label{eq:Betac}
\end{equation}

Ainsi, à l'ordre dominant, la constante de couplage $c$ à basse énergie
est donné par 
\begin{equation}
c(m)=c(M)-\frac{6\eta^{2}}{(4\pi)^{2}}c\log(\frac{M}{m})=\frac{\lambda^{2}}{M^{2}}\left(1-\frac{6\eta^{2}}{(4\pi)^{2}}\log(\frac{M}{m})\right).\label{eq:Runningc}
\end{equation}
A ce stade, les RGEs doivent être résolues afin de pouvoir re-sommer
les logarithmes. Sans le faire en détail, en résolvant l'équation
pour $\eta$, on obtient
\begin{equation}
\frac{1}{\eta^{2}(\mu_{2})}-\frac{1}{\eta^{2}(\mu_{1})}=\frac{10}{(4\pi)^{2}}\log\frac{\mu_{1}}{\mu_{2}}.\label{eq:RunningEta}
\end{equation}
En injectant l'équation (\ref{eq:RunningEta}) dans l'équation (\ref{eq:Betac})
puis en intégrant, on arrive au résultat suivant:
\begin{equation}
c(m)=c(M)\left(\frac{\eta^{2}(m)}{\eta^{2}(M)}\right)^{\frac{3}{5}}.
\end{equation}

Le fait que la fonction beta ne dépende pas de l'échelle de renormalisation
est caractéristique de l'utilisation de régulateurs indépendants d'échelle,
comme par exemple la régularisation dimensionnelle avec la prescription
MS.

\begin{figure}[h!]
\begin{center}
\includegraphics[scale=0.5]{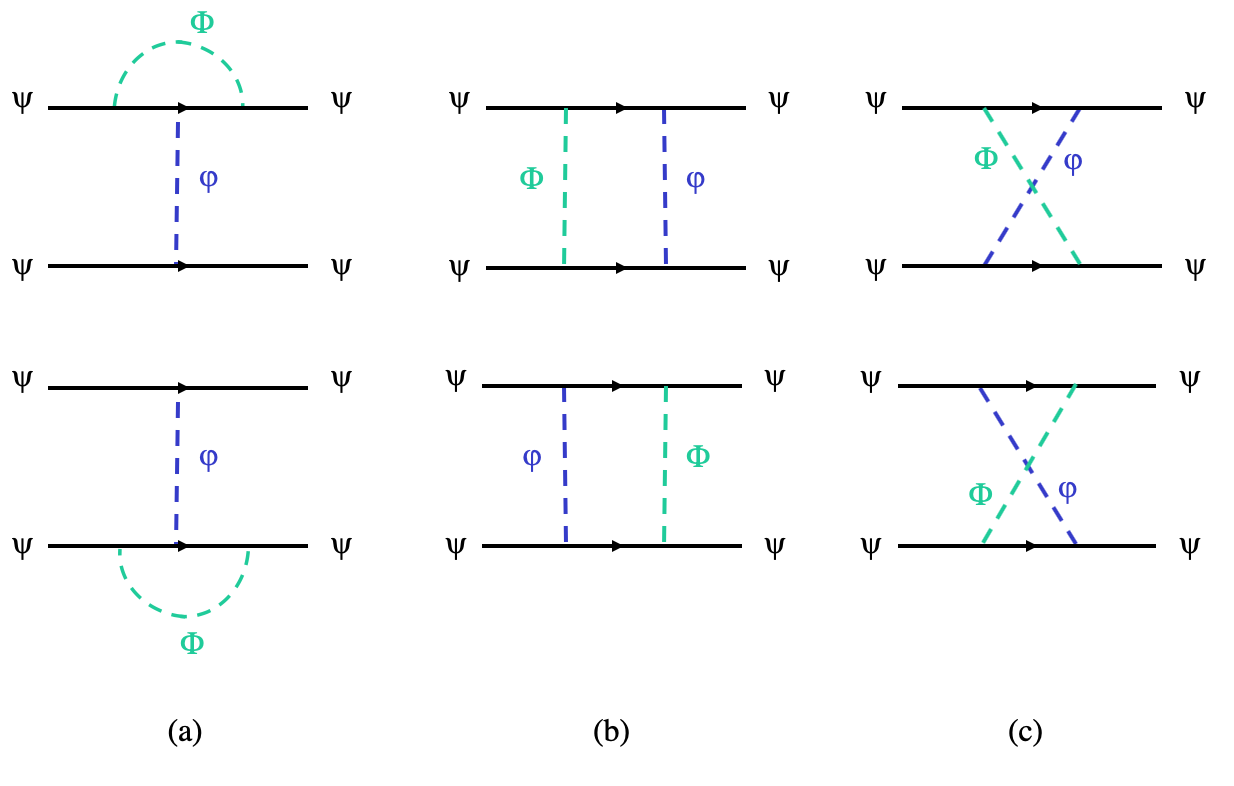}
\caption{Les diagrammes analogues dans la théorie complète de ceux de la figure $\ref{1LoopDiags}$ dans l'EFT. La ligne en pointillé bleu représente le scalaire lourd $\Phi$ tandis que celle en vert le scalaire léger $\phi$. Figure modifiée à partir de \cite{Skiba}.}
\label{DiagsUVTheory}
\end{center}
\end{figure}

Les diagrammes que l'on a dû calculer pour obtenir l'équation (\ref{eq:Runningc})
sont en bijection avec les diagrammes dans la théorie complète (représentés
en figure \ref{DiagsUVTheory}). Evidemment, les diagrammes dans l'EFT
sont plus simples à calculer que ceux de la théorie UV dans la mesure
où il y a moins de propagateurs (ceux impliquant les particules intégrées).
Dans la théorie complète, on devrait également calculer les contributions
finies des diagrammes en boîtes (box diagrams).

En outre, lorsque nous avions intégré le scalaire lourd, le seul opérateur
ne dépendant pas des impulsions générés à l'arbre est l'opérateur
à quatre fermions $\bar{\psi}\psi\bar{\psi}\psi$. En réalité, ce
n'est pas le seul opérateur à quatre fermions sans dérivées, comme
par exemple l'opérateur $\bar{\psi}\gamma^{\mu}\psi\bar{\psi}\gamma_{\mu}\psi$
qui possède la même dimension et le même contenu en particules. Supposons
que nous avions intégré un champ vectoriel massif de masse $M$ et
que notre lagrangien effectif soit:
\begin{equation}
\mathcal{L}_{p^{0},V}=i\bar{\psi}\slashed\partial\psi+\frac{c_{V}}{2}\bar{\psi}\gamma^{\mu}\psi\bar{\psi}\gamma_{\mu}\psi+\frac{1}{2}(\partial_{\mu}\phi)^{2}-\frac{m^{2}}{2}\phi^{2}-\eta\bar{\psi}\psi\phi.\label{eq:LeffCv}
\end{equation}
On pourrait alors se poser les mêmes questions concernant la diffusion
à basse énergie dans cette théorie, c'est à dire calculer les RGE
pour le coefficient $c_{V}$. Le calcul est parfaitement similaire
à la seule différence près que le vertex à quatre fermion contient
dans ce cas des matrices de Dirac $\gamma^{\mu}$. Les diagrammes
$a$ donnent une contribution divergente à l'opérateur $\bar{\psi}\gamma^{\mu}\psi\bar{\psi}\gamma_{\mu}\psi$
alors qu'en sommant les parties divergentes des diagrammes $b$ et
$c$, il en résulte un terme proportionnel à $\bar{\psi}\sigma^{\mu\nu}\psi\bar{\psi}\sigma_{\mu\nu}\psi$
plutôt qu'à l'opérateur de départ. Autrement dit, cela signifie que
sous l'évolution des RGE, les deux opérateurs se mélangent. En effet,
les deux opérateurs ont la même dimension, le même contenu en particules
ainsi que les mêmes propriétés de symétrie, c'est pourquoi les corrections
radiatives peuvent interchanger l'un en l'autre.

Conséquemment, il n'est pas cohérent à une boucle de considérer un
seul opérateur à quatre fermions dans le lagrangien effectif (\ref{eq:LeffCv})
dans la mesure où un contre-terme supplémentaire est nécessaire afin
d'absorber la divergence. A une boucle, le lagrangien effectif suivant
est suffisant
\begin{equation}
\mathcal{L}_{p^{0},VT}=i\bar{\psi}\slashed\partial\psi+\frac{c_{V}}{2}\bar{\psi}\gamma^{\mu}\psi\bar{\psi}\gamma_{\mu}\psi+\frac{c_{T}}{2}\bar{\psi}\sigma^{\mu\nu}\psi\bar{\psi}\sigma_{\mu\nu}\psi+\frac{1}{2}(\partial_{\mu}\phi)^{2}-\frac{m^{2}}{2}\phi^{2}-\eta\bar{\psi}\psi\phi.
\end{equation}

Nous ne calculerons pas en détail les fonctions beta pour les nouveaux
coefficients mais le calcul est complètement analogue à celui effectué
précédemment pour $\beta_{c}$. L'opérateur vectoriel induit une contribution
divergente absorbée par lui même et par l'opérateur tensoriel alors
que ce dernier génère une contribution divergente entièrement absorbée
par l'opérateur vectoriel. Les coefficients pour les contre-termes
sont
\begin{equation}
\begin{cases}
c_{v}(Z_{V}Z_{\psi}^{2}-1) & =\frac{\text{\ensuremath{\eta}}^{2}}{(4\pi)^{2}}(-c_{V}+6c_{T})\frac{1}{\epsilon},\\
c_{T}(Z_{T}Z_{\psi}^{2}-1) & =\frac{\text{\ensuremath{\eta}}^{2}}{(4\pi)^{2}}c_{V}\frac{1}{\epsilon},
\end{cases}
\end{equation}
avec des constantes $Z$ différentes pour les deux opérateurs. On
obtient alors les fonctions beta suivantes

\begin{equation}
\begin{cases}
\beta_{c_{V}}= & 12c_{T}\frac{\text{\ensuremath{\eta}}^{2}}{(4\pi)^{2}}\\
\beta_{c_{T}}= & 2(c_{T}+c_{V})\frac{\text{\ensuremath{\eta}}^{2}}{(4\pi)^{2}}.
\end{cases}
\end{equation}

\subsection{Correspondance à une boucle}

Jusque là, nous avons fait correspondre les deux théories à l'arbre,
c'est à dire à l'ordre le plus bas dans le développement perturbatif
en constante de couplage. La situation actuelle est que les deux théories
reproduisent les mêmes résultats à basse énergie, seulement à l'arbre.
En effet, si on augmente la précision des calculs, les deux théories
vont commencer à différer. Il est possible de les faire à nouveau
correspondre et ce avec la précision souhaitée, moyennant des calculs
de plus en plus lourds. Par exemple, afin d'améliorer la précision
de la correspondance des deux théories pour le calcul de la diffusion
$\psi\psi\rightarrow\psi\psi$, il suffit de faire la correspondance
à l'ordre suivant, c'est à dire à une boucle. Comme exemple illustratif,
étudions les termes d'ordre $\mathcal{O}(\lambda^{4})$.

\begin{figure}[h]
\begin{center}
\includegraphics[scale=0.5]{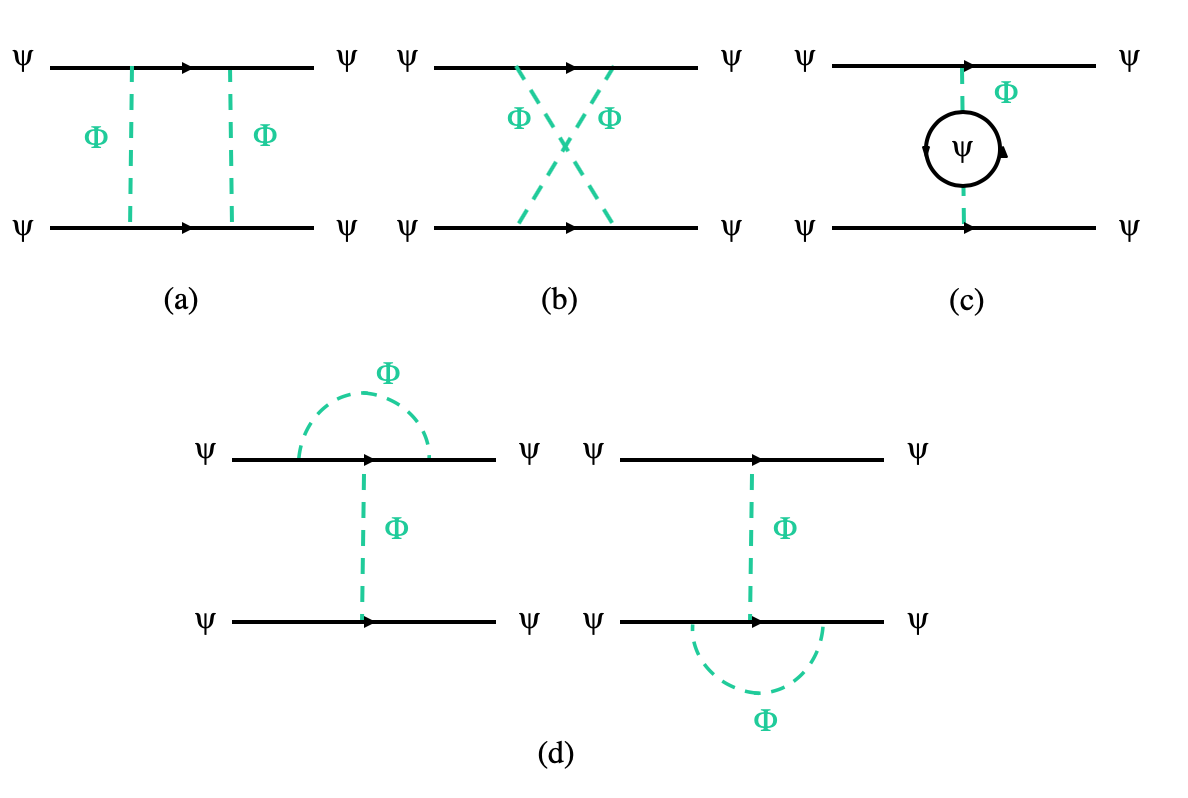}
\caption{Diagrammes dans la théorie complète à l'ordre $\mathcal{O}(\lambda^{4})$. Figure modifiée à partir de \cite{Skiba}.}
\label{DiagsUVOneLoop}
\end{center}
\end{figure}

Pour commencer, considérons le lagrangien de la théorie UV (\ref{eq:LagrangianUV})
en négligeant le champ scalaire léger pour le moment dans la mesure
où nous nous intéressons qu'au champ scalaire lourd $\Phi$. De plus,
on y ajoute une petite masse $\sigma$ au champ fermionique afin d'éviter
les problèmes de divergences infrarouges. 
\begin{equation}
\mathcal{L}=i\bar{\psi}\slashed\partial\psi-\sigma\bar{\psi}\psi+\frac{1}{2}(\partial_{\mu}\Phi)^{2}-\frac{M^{2}}{2}\Phi^{2}-\lambda\bar{\psi}\psi\Phi.
\end{equation}
Les diagrammes à une boucle contribuant à la diffusion $\psi\psi\rightarrow\psi\psi$
sont représentés en figure \ref{DiagsUVOneLoop}.

Comme précédemment, nous ne considérerons que la partie de l'amplitude
indépendante des impulsions et nous n'écriront pas les termes impliquant
des échanges de fermions externes. Dans ces conditions, pour le diagramme
$(a)$, on a
\begin{equation}
\begin{aligned}\mathcal{A}_{(a)} & =(-i\lambda)^{4}\varint\frac{d^{d}k}{(2\pi)^{d}}\bar{u}(p_{3})\frac{i(\slashed k+\sigma)}{k^{2}-\sigma^{2}}u(p_{1})\bar{u}(p_{4})i\frac{i(-\slashed k+\sigma)}{k^{2}-\sigma^{2}}u(p_{2})\frac{i^{2}}{(k^{2}-M^{2})^{2}}\\
 & =\lambda^{4}\left[-\bar{u}(p_{3})\gamma^{\alpha}u(p_{1})\bar{u}(p_{4})\gamma^{\beta}u(p_{2})\varint\frac{d^{d}k}{(2\pi)^{d}}\frac{k_{\alpha}k_{\beta}}{(k^{2}-\sigma^{2})^{2}(k^{2}-M^{2})^{2}}\right.\\
 & \left.+\bar{u}(p_{3})u(p_{1})\bar{u}(p_{4})u(p_{2})\varint\frac{d^{d}k}{(2\pi)^{d}}\frac{\sigma^{2}}{(k^{2}-\sigma^{2})^{2}(k^{2}-M^{2})^{2}}\right].
\end{aligned}
\end{equation}
Les intégrales de boucles sont évaluées à l'aide de la paramétrisation
de Feynman
\begin{equation}
\frac{1}{(k^{2}-\sigma^{2})^{2}(k^{2}-M^{2})^{2}}=6\int_{0}^{1}dx\frac{x(1-x)}{(k^{2}-xM^{2}-(1-x)\sigma^{2})^{4}}.
\end{equation}
Finalement, on obtient pour le diagramme $(a)$
\begin{equation}
\begin{aligned}\mathcal{A}_{(a)}^{UV} & =\frac{i\lambda^{4}}{(4\pi)^{2}}\left[\frac{U_{V}}{2}\int_{0}^{1}dx\frac{x(1-x)}{xM^{2}+(1-x)\sigma^{2}}+\sigma^{2}U_{S}\int_{0}^{1}dx\frac{x(1-x)}{(xM^{2}+(1-x)\sigma^{2})^{2}}\right]\\
 & =\frac{i\lambda^{4}}{(4\pi)^{2}}\left[U_{V}\left(\frac{1}{4M^{2}}+\frac{\sigma^{2}}{4M^{4}}\left(3-2\log\left(\frac{M^{2}}{\sigma^{2}}\right)\right)\right)+U_{S}\frac{\sigma^{2}}{M^{4}}\left(\log\left(\frac{M^{2}}{\sigma^{2}}\right)-2\right)\right]+...,
\end{aligned}
\end{equation}
où l'on a noté $U_{S}=\bar{u}(p_{3})u(p_{1})\bar{u}(p_{4})u(p_{2})$
et $U_{V}=\bar{u}(p_{3})\gamma^{\alpha}u(p_{1})\bar{u}(p_{4})\gamma_{\alpha}u(p_{2})$.
Les points de suspension traduisent l'omission des termes d'ordre
$\mathcal{\mathcal{O}}(\tfrac{1}{M^{6}})$ et plus. 

En ce qui concerne le diagramme $(b)$, l'amplitude est quasi identique
hormis un signe d'impulsion qui change. On a alors
\begin{equation}
\mathcal{A}_{(b)}^{UV}=\frac{i\lambda^{4}}{(4\pi)^{2}}\left[-U_{V}\left(\frac{1}{4M^{2}}+\frac{\sigma^{2}}{4M^{4}}\left(3-2\log\left(\frac{M^{2}}{\sigma^{2}}\right)\right)\right)+U_{S}\frac{\sigma^{2}}{M^{4}}\left(\log\left(\frac{M^{2}}{\sigma^{2}}\right)-2\right)\right]+....
\end{equation}

Les diagrammes $(c)$ et $(d)$ sont quant à eux divergents. Pour
le $(c)$, on a
\begin{equation}
\mathcal{A}_{(c)}^{UV}=-4\frac{i\lambda^{4}}{(4\pi)^{2}}\frac{\sigma^{2}}{M^{4}}U_{S}\left[\frac{3}{\hat{\epsilon}}+3\log\left(\frac{\mu^{2}}{\sigma^{2}}\right)+1\right]+...,
\end{equation}
avec $\tfrac{1}{\hat{\epsilon}}=\tfrac{1}{\epsilon}-\gamma+\log(4\pi)$
et $\mu$ l'échelle de régularisation. Ensuite, pour les deux diagrammes
$(d)$, on a
\begin{equation}
\mathcal{A}_{(d)}^{UV}=-2\frac{i\lambda^{4}}{(4\pi)^{2}M^{2}}U_{S}\left[\frac{1}{\hat{\epsilon}}+1+\log\left(\frac{\mu^{2}}{M^{2}}\right)+\frac{\sigma^{2}}{M^{2}}\left(2-3\log\left(\frac{M^{2}}{\sigma^{2}}\right)\right)\right]+....
\end{equation}
La somme de toutes ces contributions est alors 
\begin{equation}
\mathcal{A}_{totale}^{UV}=\frac{2i\lambda^{4}U_{S}}{(4\pi)^{2}M^{2}}\left[-\frac{1}{\hat{\epsilon}}-1-\log\left(\frac{\mu^{2}}{M^{2}}\right)+\frac{\sigma^{2}}{M^{2}}\left(-\frac{6}{\hat{\epsilon}}-6-6\log\left(\frac{\mu^{2}}{\sigma^{2}}\right)+4\log\left(\frac{M^{2}}{\sigma^{2}}\right)\right)\right].
\end{equation}

D'autre part, nous avons besoin de calculer la fonction à deux-points
du fermion (avec la partie finie) pour la renormalisation de la fonction
d'onde dans l'EFT. Le calcul est similaire à celui effectué plus haut
pour la self-énergie (\ref{eq:SelfDiv}). Pour la partie de l'amplitude
linéaire en impulsion, on trouve $i\slashed p\frac{\lambda^{2}}{2(4\pi)^{2}}\left(\frac{1}{\hat{\epsilon}}+\frac{1}{2}+\log\left(\frac{\mu^{2}}{M^{2}}\right)+...\right)$.

\begin{figure}[h]
\begin{center}
\includegraphics[scale=0.5]{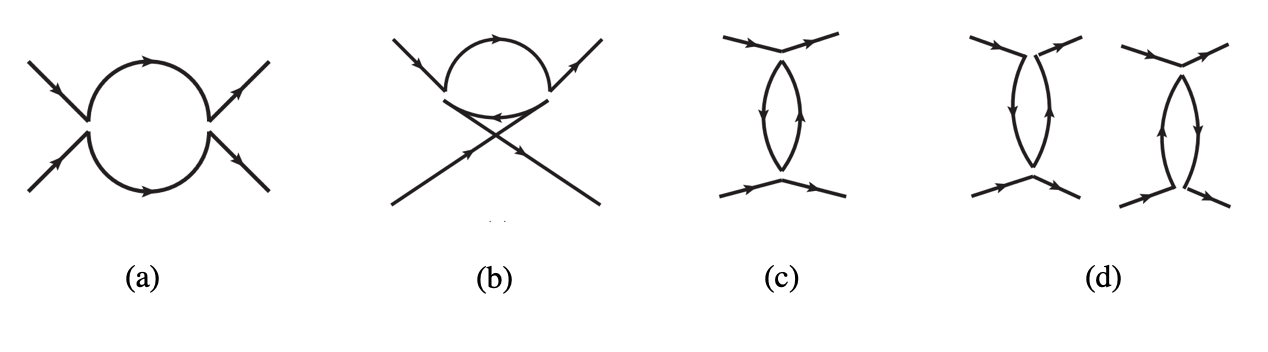}
\caption{Diagrammes dans la théorie effective à l'ordre $\mathcal{O}(c^{2})$. Les vertex à quatre fermions sont représentés de façon non ponctuelle afin de pouvoir suivre les lignes fermioniques. Figure modifiée à partir de \cite{Skiba}.}
\label{DiagsEFTOrderc2}
\end{center}
\end{figure}

Pour faire la correspondance, il faut maintenant calculer l'amplitude
à une boucle dans la théorie effective. On a vu qu'une interaction
à quatre fermions était déjà induite à l'arbre. Toujours en négligeant
le scalaire léger $\phi$, le lagrangien effectif est
\begin{equation}
\mathcal{L}_{eff}=iz\bar{\psi}\slashed\partial\psi-\sigma\bar{\psi}\psi+\frac{c}{2}\bar{\psi}\psi\bar{\psi}\psi.
\end{equation}
De plus, en faisant correspondre les théories UV et EFT à l'arbre
dans la partie \ref{TreeLevelMatchingPart}, nous avons déjà établit
précédemment que $c=\tfrac{\lambda^{2}}{M^{2}}$. L'amplitude à deux-points
est nulle dans la théorie effective. Les diagrammes à quatre-points
dans l'EFT sont représentés en figure \ref{DiagsEFTOrderc2}. Généralement,
le degré superficiel de divergence d'un diagramme donné est plus élevé
dans la théorie effective que dans la théorie complète vu qu'il y
a moins de propagateurs. En effet, le diagramme $(a)$ dans la théorie
UV (figure \ref{DiagsUVOneLoop}) était convergent alors que son équivalent
dans l'EFT ($(a)$ dans la figure \ref{DiagsEFTOrderc2}) diverge
quadratiquement.

En premier lieu, considérons les diagrammes $(a)$ et $(b)$ dont
les propagateurs fermioniques ne diffèrent que par le signe de leurs
impulsions respectives. Dans ce cas, les termes proportionnels à $U_{V}$
sont opposés dans les deux diagrammes et s'annulent donc lorsqu'on
les somme tandis que la partie proportionnelle à $U_{S}$ est la même
dans les deux diagrammes et contribue donc à la somme des amplitudes.
Ainsi, la partie de l'amplitude qui nous intéresse pour la somme des
deux diagrammes $(a)$ et $(b)$ est
\begin{equation}
\mathcal{A}_{(a+b)}^{EFT}=2\frac{ic^{2}\sigma^{2}}{(4\pi)^{2}}U_{S}\left[\frac{1}{\hat{\epsilon}}+\log\left(\frac{\mu^{2}}{\sigma^{2}}\right)\right]+....
\end{equation}

Ensuite, le diagramme $(c)$ est identique dans les deux théories,
la partie de l'amplitude indépendante des impulsions s'écrit alors
\begin{equation}
\mathcal{A}_{(c)}^{EFT}=-4\frac{ic^{2}\sigma^{2}}{(4\pi)^{2}}U_{S}\left[\frac{3}{\hat{\epsilon}}+3\log\left(\frac{\mu^{2}}{\sigma^{2}}\right)+1\right]+....
\end{equation}

En dernier lieu, en incluant le facteur $2$ pour les deux diagrammes
équivalents $(d)$, on a
\begin{equation}
\mathcal{A}_{(d)}^{EFT}=2\frac{ic^{2}\sigma^{2}}{(4\pi)^{2}}U_{S}\left[\frac{3}{\hat{\epsilon}}+3\log\left(\frac{\mu^{2}}{\sigma^{2}}\right)+1\right]+....
\end{equation}
Finalement, la somme de tous les diagrammes est
\begin{equation}
\mathcal{A}_{(a+b+c+d)}^{EFT}=-2\frac{ic^{2}\sigma^{2}}{(4\pi)^{2}}U_{S}\left[\frac{2}{\hat{\epsilon}}+2\log\left(\frac{\mu^{2}}{\sigma^{2}}\right)+1\right]+....
\end{equation}

Les structures des divergences (pôles en $\tfrac{1}{\hat{\text{\ensuremath{\epsilon}}}}$)
sont différentes dans les deux théories. De ce fait, on renormalise
les deux théories séparément en y ajoutant les contre-termes adéquats
permettant d'absorber les divergences respectives. Par suite, pour
effectuer la correspondance, on égale les amplitudes de diffusions
physiques, c'est à dire renormalisées, obtenues dans les deux théories.
Ici, nous allons utiliser la prescription $\overline{MS}$ qui consiste
à retirer les pôles en $\tfrac{1}{\hat{\text{\ensuremath{\epsilon}}}}$.
Remarquons que les coefficients de Wilson dans la théorie effective
dépendent du choix du régulateur étant donné que nous ajoutons des
contre-termes différents aux deux théories. Bien entendu, les résultats
physiques n'en dépendent pas.

On choisit comme échelle de correspondance la masse du scalaire lourd
intégré $\mu=M$ et on égale les amplitudes de diffusion à une boucle
calculées dans les deux théories: $\mathcal{A}_{(a+b+c+d)}^{EFT}=\mathcal{A}_{(a+b+c+d)}^{UV}$.
\begin{equation}
c(\mu=M)=\frac{\lambda^{2}}{M^{2}}-\frac{2\lambda^{4}}{(4\pi)^{2}M^{2}}-\frac{10\lambda^{4}\sigma^{2}}{(4\pi)^{2}M^{4}},
\end{equation}
où l'on a remplacé le coefficient de Wilson $c$ par sa valeur déterminée
à l'aide de la correspondance à l'arbre: $c=\tfrac{\text{\ensuremath{\lambda}}^{2}}{M^{2}}$.
Afin de reproduire la fonction à deux-points dans la théorie UV, absente
de la théorie effective, il est nécessaire de poser $z=1+\tfrac{\lambda^{2}}{4(4\pi)^{2}}$
(dans la prescription $\overline{MS}$). De plus, pour obtenir une
amplitude de diffusion physique, nous devons encore renormaliser le
champ fermionique à l'aide du facteur $Z$(wave function renormalization),
à savoir $\psi_{physique}=\sqrt{Z}\psi$. Cette procédure ajoute une
contribution supplémentaire au terme $\text{\ensuremath{\tfrac{\lambda^{4}}{(4\pi)^{2}M^{2}}}}$
dans l'amplitude provenant du produit du facteur $Z$ et de la contribution
à l'arbre.

\subsection{Redéfinitions des champs}

Les redéfinitions des champs ne modifient pas la matrice de diffusion
$S$. En effet, considérons la fonctionnelle génératrice
\begin{equation}
Z[J]=\varint\mathcal{D}\phi e^{i\varint d^{4}x\left[\mathcal{L}[\phi]+J\phi\right]}.\label{eq:FoncGenDepart}
\end{equation}
Les fonctions de Green
\begin{equation}
\langle0|T\{\phi(x_{1})\ldots\phi(x_{r})\}|0\rangle=\frac{\varint\mathcal{D}\phi\phi(x_{1})\ldots\phi(x_{r})e^{iS[\phi]}}{\varint\mathcal{D}\phi e^{iS[\phi]}},
\end{equation}
sont alors obtenues à l'aide de la fonctionnelle génératrice de la
façon suivante
\begin{equation}
\langle0|T\{\phi(x_{1})\ldots\phi(x_{r})\}|0\rangle=\left.\frac{1}{Z[J]}\left(\frac{-i\delta}{\delta J(x_{1})}\right)\ldots\left(\frac{-i\delta}{\delta J(x_{r})}\right)Z[J]\right|_{J=0}.
\end{equation}

Considérons la redéfinition locale $\phi(x)=F\left[\phi^{\prime}(x)\right]$,
où $F:X\longmapsto X+c_{1}\partial^{2}X+c_{2}X^{3}$. Généralement,
l'application $F$ est la somme d'un polynôme et d'une application
contenant un nombre finis d'opérateurs dérivatifs. Suite à la redéfinition
du champ $\phi$, on obtient alors un nouveau lagrangien $\mathcal{L}^{\prime}$
donné par
\begin{equation}
\mathcal{L}[\phi(x)]=\mathcal{L}\left[F\left[\phi^{\prime}(x)\right]\right]=\mathcal{L}^{\prime}\left[\phi^{\prime}(x)\right].
\end{equation}
Par suite, la nouvelle fonctionnelle génératrice exprimée en fonction
de $\phi^{\prime}$ et de $\mathcal{L}^{\prime}$ s'écrit alors
\begin{equation}
Z^{\prime}[J]=\varint\mathcal{D}\phi^{\prime}e^{i\varint d^{4}x\left[\mathcal{L}^{\prime}[\phi^{\prime}]+J\phi^{\prime}\right]}=\varint\mathcal{D}\phi e^{i\varint d^{4}x\left[\mathcal{L}^{\prime}[\phi]+J\phi\right]},
\end{equation}
où l'on a renommé la variable muette d'intégration $\phi^{\prime}\rightarrow\phi$
dans la dernière égalité. D'autre part, si on effectue le changement
de variable $\phi(x)=F\left[\phi^{\prime}(x)\right]$ dans l'intégrale
fonctionnelle de départ (\ref{eq:FoncGenDepart}), on obtient
\begin{equation}
Z[J]=\varint\mathcal{D}\phi^{\prime}\left|\frac{\delta F}{\delta\phi^{\prime}}\right|e^{i\varint d^{4}x\left[\mathcal{L}^{\prime}[\phi^{\prime}]+JF\left[\phi^{\prime}\right]\right]}.
\end{equation}
En l'absence d'anomalies et en régularisation dimensionnelle, le jacobien
$\left|\frac{\delta F}{\delta\phi^{\prime}}\right|$ vaut $1$. En
renommant la variable d'intégration muette par $\phi$, on a donc
\begin{equation}
Z[J]=\varint\mathcal{D}\phi e^{i\varint d^{4}x\left[\mathcal{L}^{\prime}[\phi]+JF\left[\phi\right]\right]}.
\end{equation}
Ainsi, $Z[J]$ permet à la fois d'obtenir les fonctions de Green de
$\phi$, calculées avec $\mathcal{L}[\phi]$ (avec l'équation (\ref{eq:FoncGenDepart}))
et les fonctions de Green de $F[\phi]$, calculées avec $\mathcal{L}^{\prime}[\phi]$.
En revanche, $Z^{\prime}[J]$ permet de calculer les fonctions de
corrélations de $\phi$ en utilisant le nouveau lagrangien $\mathcal{L}^{\prime}[\phi]$.
Les deux fonctions de corrélations sont différentes et donc les fonctions
de Green changent sous une redéfinition de champ. Par contre, la matrice
de diffusion $S$ est inchangée. En effet, $Z[J]$ permet de calculer
la matrice $S$ à l'aide du lagrangien $\mathcal{L}^{\prime}[\phi]$
et du champ interpolateur $\phi$. Le choix du champ interpolateur,
c'est à dire de la redéfinition du champ, n'a pas d'effet sur la matrice
de diffusion $S$ tant que $\langle p|F[\phi]|0\rangle\neq0$. Dans
ces conditions, la matrice $S$ est invariante sous une redéfinition
des champs.

Enfin, lorsque nous utilisons des lagrangien renormalisables, seules
les transformations linéaires sont autorisées pour redéfinir les champs
\begin{equation}
\phi_{i}^{\prime}=C_{ij}\phi_{j}.
\end{equation}
Ce genre de transformation est utilisée pour mettre le terme cinétique
sous forme canonique $\frac{1}{2}\partial_{\mu}\phi_{i}\partial^{\mu}\phi^{i}$.

Cependant, dans une EFT, du fait de la présence d'opérateurs de dimensions
supérieures à $4$, non renormalisables, il y a plus de libertés dans
les redéfinitions de champs. On effectue des redéfinitions de champs
qui préservent le comptage de puissance de la théorie effective, par
exemple la redéfinition
\begin{equation}
\phi\rightarrow\phi+\frac{1}{\Lambda^{2}}\phi^{3}+\ldots,
\end{equation}
permet de travailler ordre par ordre dans le développement en $\tfrac{1}{\Lambda}$.
Les redéfinitions de ce genre sont souvent utilisées pour mettre le
lagrangien effectif sous forme canonique. Le lagrangien EFT est alors
obtenu par correspondance (matching) avec la théorie complète, suivi
d'une redéfinition des champs. Il en résulte que les champs dans l'EFT
sont différents de ceux dans la théorie UV.

\subsection{Equations du mouvement}

Un cas particulier de redéfinition des champs est l'utilisation de
l'équation du mouvement (EOM). En effet, l'équation du mouvement classique
$E[\phi]$ est définie par:
\begin{equation}
E[\phi]\equiv\frac{\delta S}{\delta\phi}.
\end{equation}
Par exemple, prenons le cas de la théorie $\phi^{4}$
\begin{equation}
\mathcal{L}_{\phi^{4}}=\frac{1}{2}\partial_{\mu}\phi\partial^{\mu}\phi-\frac{1}{2}m^{2}\phi^{2}-\frac{1}{4!}\lambda\phi^{4},
\end{equation}
en appliquant l'équation d'Euler-Lagrange, on obtient l'équation du
mouvement suivante
\begin{equation}
E[\phi]=-\partial^{2}\phi(x)-m^{2}\phi(x)-\frac{1}{3!}\lambda\phi^{3}(x).
\end{equation}
Considérons maintenant un opérateur $\hat{O}[\phi]$ qui fait manifestement
apparaître l'équation du mouvement, c'est à dire qui est \og divisible \fg{}
par $E[\phi]$
\begin{equation}
\hat{O}[\phi]=\mathcal{F}[\phi]E[\phi]=\mathcal{F}[\phi]\frac{\delta S}{\delta\phi},\label{eq:OpEOM}
\end{equation}
et considérons la fonctionnelle génératrice
\begin{equation}
Z[J,\tilde{J}]=\varint\mathcal{D}\phi e^{i\varint d^{4}x\left[\mathcal{L}_{\phi^{4}}[\phi]+J\phi+\tilde{J}\hat{O}[\phi]\right]}.\label{eq:FonctGen}
\end{equation}
La fonction de corrélation $\langle0|T\{\phi(x_{1})...\phi(x_{n})\hat{O}[x]\}|0\rangle$
où l'on a inséré l'opérateur $\hat{O}$ contenant l'EOM est donnée
par 
\begin{equation}
\langle0|T\{\phi(x_{1})...\phi(x_{n})\hat{O}[x]\}|0\rangle=\left.\frac{1}{Z[J,\tilde{J}]}\left(-i\frac{\delta}{\delta J(x_{1})}\right)\ldots\left(-i\frac{\delta}{\delta J(x_{n})}\right)\left(-i\frac{\delta}{\delta\tilde{J}(x)}\right)Z[J,\tilde{J}]\right|_{J=\tilde{J}=0}.
\end{equation}
Effectuons le changement de variable $\phi=\phi^{\prime}-\tilde{J}F[\phi^{\prime}]$
dans la fonctionnelle génératrice (\ref{eq:FonctGen}). On obtient
alors
\begin{equation}
\begin{aligned}Z[J,\tilde{J}] & =\varint\mathcal{D}\phi^{\prime}\left|\frac{\delta\phi}{\delta\phi^{\prime}}\right|e^{i\varint d^{4}x\left[\mathcal{L}_{\phi^{4}}[\phi^{\prime}]-\left.\frac{\delta S}{\delta\phi}\right|_{\phi^{\prime}}\tilde{J}F[\phi^{\prime}]+J\phi^{\prime}-J\tilde{J}F[\phi^{\prime}]+\tilde{J}\hat{O}[\phi^{\prime}]+\mathcal{O}(\tilde{J}^{2})\right]}\\
 & =\varint\mathcal{D}\phi^{\prime}\left|\frac{\delta\phi}{\delta\phi^{\prime}}\right|e^{i\varint d^{4}x\left[\mathcal{L}_{\phi^{4}}[\phi^{\prime}]+J\phi^{\prime}-J\tilde{J}F[\phi^{\prime}]+\mathcal{O}(\tilde{J}^{2})\right]},
\end{aligned}
\end{equation}
où on a utilisé l'équation (\ref{eq:OpEOM}). Dans le cadre de la
régularisation dimensionnelle (adopté ici), le Jacobien $\left|\frac{\delta\phi(x)}{\delta\phi^{\prime}(y)}\right|$
de la transformation vaut $1$. En renommant par $\phi$ la variable
muette d'intégration fonctionnelle, on a
\begin{equation}
Z[J,\tilde{J}]=\varint\mathcal{D}\phi e^{i\varint d^{4}x\left[\mathcal{L}_{\phi^{4}}[\phi]+J\phi-J\tilde{J}F[\phi]+\mathcal{O}(\tilde{J}^{2})\right]}.
\end{equation}
Par suite, appliquons la dérivée fonctionnelle par rapport à $\tilde{J}$
puis évaluons le résultat en $\tilde{J}=0$
\begin{equation}
\varint\mathcal{D}\phi\hat{O}(x)e^{i\varint d^{4}x\left[\mathcal{L}_{\phi^{4}}[\phi]+J\phi\right]}=-\varint\mathcal{D}\phi J(x)\mathcal{F}[\phi(x)]e^{i\varint d^{4}x\left[\mathcal{L}_{\phi^{4}}[\phi]+J\phi\right]}.
\end{equation}
En dérivant plusieurs fois par rapport $J$, on obtient l'identité
de Ward qui n'est rien d'autre que l'EOM
\begin{equation}
\langle0|T\{\phi(x_{1})...\phi(x_{n})\hat{O}[x]\}|0\rangle=i\sum_{r}\delta(x-x_{r})\langle0|T\{\phi(x_{1})...\cancel{\phi(x_{r})}\ldots\phi(x_{n})\mathcal{F}\left[\phi(x_{r})\right]\}|0\rangle.\label{eq:EOMWard}
\end{equation}
L'élément de matrice de diffusion $S$ avec une insertion de l'opérateur
$\hat{O}$ s'annule
\begin{equation}
_{out}\langle q_{1},\ldots,q_{m}|\hat{O}|p_{1},\ldots,p_{n}\rangle_{in}=0,
\end{equation}
car il est obtenu en choisissant le terme ayant $m+n$ pôles dans
$\langle0|T\{\phi(x_{1})...\phi(x_{n})\hat{O}[x]\}|0\rangle$. Cependant,
le membre droit de l'équation (\ref{eq:EOMWard}) montre que l'élément
de matrice du r-ième terme ne possède pas de pôle en $p_{r}$ du fait
de la présence de la distribution $\delta$ de Dirac. En définitive,
tous les termes de la somme sont nuls et on obtient l'annulation de
l'élément de matrice $_{out}\langle q_{1},\ldots,q_{m}|\hat{O}|p_{1},\ldots,p_{n}\rangle_{in}$.
Par conséquent, les opérateurs EOM peuvent être omis étant donné qu'ils
ne contribuent pas à la matrice de diffusion $S$.

Remarquons que l'utilisation des EOMs est un cas particulier de la
redéfinition d'un champ. En effet, considérons la redéfinition suivante
\begin{equation}
\phi(x)=\phi^{\prime}(x)+\epsilon\mathcal{F}\left[\phi^{\prime}(x)\right],
\end{equation}
avec $\epsilon\ll1$. Suite à cette redéfinition du champ $\phi$,
le lagrangien se voit modifié comme suit
\begin{equation}
\mathcal{L}[\phi]=\mathcal{L}[\phi^{\prime}]+\epsilon\mathcal{F}[\phi^{\prime}]\frac{\delta S[\phi^{\prime}]}{\delta\phi^{\prime}}+\mathcal{O}(\epsilon^{2})=\mathcal{L}[\phi^{\prime}]+\epsilon\hat{O}[\phi^{\prime}]+\mathcal{O}(\epsilon^{2}).
\end{equation}
Nous avons déjà vu dans la partie précédente que la redéfinition d'un
champ ne modifiait pas la matrice $S$. C'est pourquoi un élément
de matrice donné peut tout aussi bien être calculé avec le nouveau
lagrangien qu'avec l'ancien. Ainsi, on peut toujours redéfinir un
lagrangien de sorte à lui ajouter des termes EOM vu qu'ils ne contribuent
pas. En pratique, les EOMs sont utilisées dans le but d'éliminer des
opérateurs contenant des dérivées afin de réduire la base d'opérateurs
effectifs.

Comme exemple d'utilisation des EOMs, considérons le lagrangien effectif
suivant
\begin{equation}
\mathcal{L}=\frac{1}{2}\partial_{\mu}\phi\partial^{\mu}\phi-\frac{m^{2}}{2}\phi^{2}-\frac{\lambda}{4!}\phi^{4}+\frac{c_{1}}{\Lambda^{2}}\phi^{3}\partial^{2}\phi+\frac{c_{6}}{\Lambda^{2}}\phi^{6}+\ldots,
\end{equation}
puis effectuons la redéfinition du champ suivante $\phi\rightarrow\phi+\tfrac{c_{1}}{\Lambda^{2}}\phi^{3}$,
on obtient alors un nouveau lagrangien
\begin{equation}
\begin{aligned}\mathcal{L} & =\frac{1}{2}\partial_{\mu}\phi\partial^{\mu}\phi-\frac{m^{2}}{2}\phi^{2}-\frac{\lambda}{4!}\phi^{4}+\frac{c_{1}}{\Lambda^{2}}\phi^{3}\partial^{2}\phi+\frac{c_{6}}{\Lambda^{2}}\phi^{6}\\
 & +\frac{c_{1}}{\Lambda^{2}}\phi^{3}\left[-\partial^{2}\phi-m^{2}\phi-\frac{\lambda}{3!}\phi^{3}\right]+\ldots\\
 & =\frac{1}{2}\partial_{\mu}\phi\partial^{\mu}\phi-\frac{m^{2}}{2}\phi^{2}-\left[\frac{\lambda}{4!}+\frac{c_{1}}{\Lambda^{2}}m^{2}\right]\phi^{4}+\left[\frac{c_{6}}{\Lambda^{2}}-\frac{c_{1}}{\Lambda^{2}}\frac{\lambda}{3!}\right]\phi^{6}+\ldots.
\end{aligned}
\end{equation}
Insistons sur le fait que les deux lagrangien donnent le même élément
de matrice. Dans le nouveau lagrangien, l'opérateur $\phi^{3}\partial^{2}\phi$
a été éliminé au prix d'une redéfinition des coefficients associés
aux opérateurs $\phi^{4}$ et $\phi^{6}$. De plus, le comptage de
puissance (power counting) de l'EFT est maintenu. En somme, la redéfinition
du champ $\phi$ a permis de réduire la base d'opérateurs effectifs.
Conséquemment, il est plus aisé de calculer avec le nouveau lagrangien
dans la mesure où il contient moins d'opérateurs indépendants. En
général, lors de la procédure de construction de la base d'opérateurs
effectifs, les EOMs sont utilisées pour éliminer le maximum d'opérateurs
contenant des dérivées possible, afin de réduire le nombre d'opérateurs
indépendants qui constituent la base de l'EFT.

Pour conclure cette partie, notons que les opérateurs EOM, $E_{i}$,
sont nombreux et que les RGEs ont pour effet de les mélanger entre
eux. En effet,
\begin{equation}
\mu\frac{d}{d\mu}E_{i}=\gamma_{ij}E_{j},
\end{equation}
où les $\gamma_{ij}$ peuvent dépendre de la jauge choisie. Les $E_{i}$
ne sont pas des quantités physiques observables et leurs dimensions
anomales peuvent dépendre du choix de jauge. Pour des opérateurs non-EOM\footnote{Un opérateur est dit non-EOM s'il contribue aux éléments de matrice
de diffusion.}, $O_{i}$, la dimension anomale s'écrit
\begin{equation}
\mu\frac{d}{d\mu}O_{i}=\gamma_{ij}O_{j}+\Gamma_{ik}E_{k}.
\end{equation}
Sous l'évolution des RGEs, ces derniers peuvent se mélanger avec les
opérateurs EOM $\{E_{i}\}$, en raison de leurs contributions nulles
aux éléments de matrice de diffusion $S$. Comme les opérateurs non-EOM
$\{O_{i}\}$ sont observables, il en résulte que $\gamma_{ij}$ est
indépendant du choix de jauge, alors que $\Gamma_{ik}$ peut en dépendre.

\newpage

\section{Une théorie effective pour bosons de jauges }

Cette étude est basée sur l'article \cite{QST-GBEFT}.

\subsection{Introduction: le lagrangien d'Euler-Heisenberg}

La théorie de Maxwell de l'électromagnétisme est décrite par le lagrangien
classique $\mathcal{L}_{Maxwell}=-\tfrac{1}{4}F^{\mu\nu}F_{\mu\nu}$.
En lui appliquant les équations d'Euler-Lagrange, on obtient les équations
de Maxwell dans le vide qui décrivent la dynamique des champs électromagnétiques.
Ces équations différentielles sont linéaires et donc le principe de
superposition est vérifié. Par conséquent, deux ondes électromagnétiques
dans le vide se propagent indépendamment l'une de l'autre, autrement
dit elles n'interagissent pas.

Dans la théorie de Dirac, du fait de l'existence du positron, la structure
du vide est différente et des paires électron-positron peuvent être
créées en un très court instant\footnote{En effet, le principe d'incertitude d'Heisenberg peut s'écrire avec
les variables énergie-temps, on a alors $\Delta E\Delta t\geq\tfrac{\bar{h}}{2}$.} puis annihilées aussitôt. Dans ces conditions, l'électrodynamique
devient non-linéaire (même dans le vide) et c'est ainsi que deux ondes
électromagnétique peuvent interagir indirectement par l'intermédiaire
des paires électron-positron créées dans le vide.

En 1936, bien avant le développement de la QED, H. Euler et W. Heisenberg
ont calculé les interactions non-linéaires entre photons dans un champ
électromagnétique constant induites par des paires virtuelles électron-positron
extraites du vide quantique \cite{EulerH36}. Leur démarche demeure
l'archétype d'une théorie effective et est considéré comme un exemple
canonique d'EFT.

En QED, pour des énergies petites devant la masse de l'électron $m_{e}$,
les photons peuvent interagir entre eux via une boucle fermionique.
Ces interactions sont supprimées par des puissances de l'inverse de
la masse de l'électron $(\tfrac{1}{m_{e}})^{i}$ par rapport au terme
de Maxwell et sont donc très petits. En intégrant l'électron dans
le lagrangien de QED, on récupère une tour de nouvelles interactions
photoniques, qui devraient être invariantes de Lorentz, de jauge et
sous parité $\mathcal{P}$. La première interaction non-triviale entre
photons correspond aux opérateurs de dimension 8 formant le lagrangien
d'Euler-Heisenberg
\begin{equation}
\mathcal{L}_{\text{EH}}=-\mathcal{F}+\frac{8}{45}\left(\frac{\alpha^{2}}{m_{e}^{4}}\right)\mathcal{F}^{2}+\frac{14}{45}\left(\frac{\alpha^{2}}{m_{e}^{4}}\right)\mathcal{G}^{2},\label{eq:EH}
\end{equation}
avec
\begin{equation}
\mathcal{F}=\frac{1}{4}F_{\mu\nu}F^{\mu\nu}=\frac{1}{2}({\mathbf{B}}^{2}-{\mathbf{E}}^{2}),\quad\mathcal{G}=\frac{1}{8}\epsilon^{\mu\nu\lambda\rho}F_{\mu\nu}F^{\lambda\rho}={\mathbf{E}}\cdot{\mathbf{B}}\ ,
\end{equation}
où ${\mathbf{B}}$ et ${\mathbf{E}}$ sont les champs magnétiques
et électriques, $\alpha=e^{2}/4\pi$ la constante de structure fine,
$e$la charge électrique de l'électron et $\epsilon_{\mu\nu\lambda\rho}$
le tenseur complètement antisymétrique. Le premier terme dans (\ref{eq:EH})
correspond au lagrangien classique de Maxwell et n'autorise pas l'interaction
de deux photons. Cependant, dès lors que les corrections apportées
par les deux derniers termes du lagrangien effectif d'Euler-Heisenberg
entrent en jeu, deux photons peuvent interagir à travers une boucle.
En ce sens, observer la diffusion photon-photon (light-by-light scattering)
serait une confirmation de la nature quantique de la QED (effet de
boucle).

Le but de cette étude est de généraliser les résultats d'Euler et
Heisenberg pour les photons à des bosons de jauge d'un groupe de jauge
arbitraire, dont les interactions effectives sont induites par des
boucles de champs lourds dans des représentations génériques du groupe
de jauge et de spin 0, 1/2 ou 1.

\subsection{Interactions effectives de photons}

Dans l'approche des intégrales de chemins, l'action effective est
obtenue en intégrant les champs lourds \cite{Dobado}. En général,
cela génère une infinité d'interactions effectives entre les champs
légers restants. La renormalisabilité nous permet d'absorber les couplages
effectifs de dimensions en énergie inférieures à 4 dans les paramètres
libres du lagrangien impliquant les champs légers. Les interactions
de dimensions supérieures à 4 sont finies et peuvent être organisées
dans un développement perturbatif en l'inverse de la masse du champs
lourd \cite{ApplequistC75}.

Considérons la fonctionnelle génératrice de QED
\begin{equation}
Z_{QED}\left[J^{\mu},\eta,\overline{\eta}\right]=\int DA^{\mu}D\psi D\overline{\psi}\;\exp i\int dx(\mathcal{L}_{QED}+\overline{\eta}\psi+\overline{\psi}\eta+J^{\mu}A_{\mu})\ ,
\end{equation}
avec
\begin{equation}
\mathcal{L}_{QED}=-\frac{1}{4}F_{\mu\nu}F^{\mu\nu}+\overline{\psi}(i\slashed{D}-m)\psi\ ,
\end{equation}
et $D^{\mu}$ la dérivée covariante usuelle. Pour des raisons de simplicité,
on omet le terme de fixation de jauge ainsi que les fantômes associés.
A très basse énergie, c'est à dire à une échelle d'énergie $\Lambda$
très petite devant la masse de l'électron $\Lambda\ll m_{e}$, seuls
les photons sont actifs. Pour construire la théorie effective valide
dans ce régime, le champ fermionique doit être intégré. L'intégration
se réalise aisément étant donné que la partie fermionique de l'intégrande
est gaussienne lorsque les sources $\eta,\overline{\eta}$ sont éteintes:
\begin{equation}
\begin{aligned}Z_{QED}\left[J^{\mu},0,0\right] & =\int DA^{\mu}\;\exp i\int dx\left\{ -\frac{1}{4}F_{\mu\nu}F^{\mu\nu}+J^{\mu}A_{\mu}\right\} \times\det(i\slashed{D}-m)\\
 & \equiv\int DA^{\mu}\;\exp i\int dx(\mathcal{L}_{eff}+J^{\mu}A_{\mu})\ .
\end{aligned}
\end{equation}
En exponentiant le déterminant de l'opérateur de Dirac, on obtient
le lagrangien effectif de QED suivant:
\begin{equation}
\mathcal{L}_{eff}=-\frac{1}{4}F_{\mu\nu}F^{\mu\nu}-iTr\ \ln(i\slashed{D}-m)\ .
\end{equation}
A ce stade, il s'agit de développer perturbativement $Tr\ \ln(i\slashed{D}-m)$
en $\tfrac{1}{m}$. Plusieurs méthodes existent dans la littérature.

Les techniques universelles les plus puissantes sont celles qui reposent
sur des méthodes fonctionnelles comme par exemple l'approche de Gaillard
\cite{Gaillard1985uh} et Cheyette \cite{Cheyette1987qz} dans laquelle
ils introduisent une méthode manifestement covariante de jauge basée
sur un développement en série de la dérivée covariante (Covariant
Derivative Expansion ou CDE en abrégé). Cette méthode élégante, qui
a été récemment remise au goût du jour dans \cite{EffAction}, simplifie
grandement l'évaluation du terme quadratique en champ lourd dans l'intégrale
de chemin et permet d'obtenir la théorie effective à basse énergie.

\begin{figure}[h]
\begin{center}
\includegraphics[height=1.2315in,width=3.0346in]{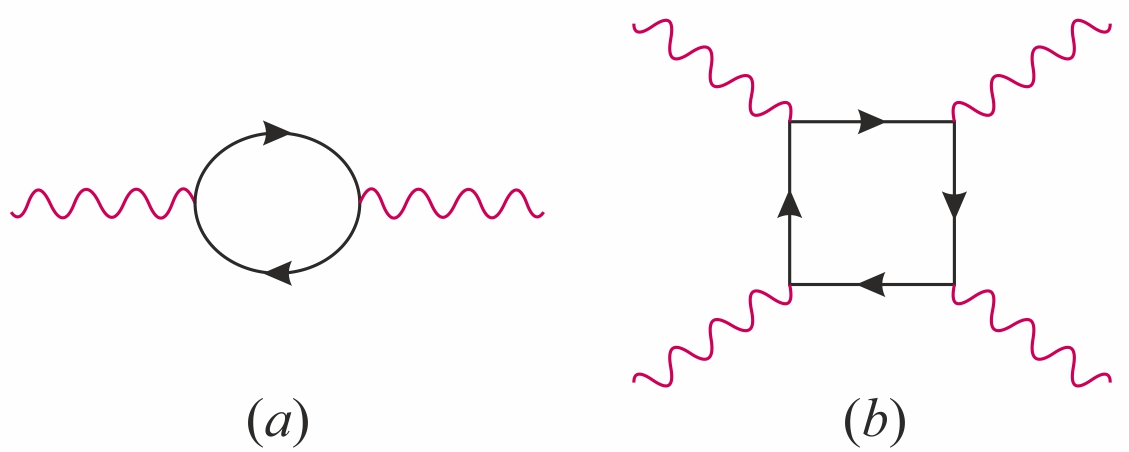}
\caption{Amplitudes 1PI à une boucle générant l'action effective de QED jusqu'aux opérateurs de dimension 8. Les six permutations des photons externes dans le diagramme ($b$) sont considérées implicitement.}
\label{EHphot}
\end{center}
\end{figure}

Cependant, ici nous souhaitons travailler dans l'approche diagrammatique
avec des champs de jauges externes. Dans ce cas, la trace est développée
de la manière suivante:
\begin{equation}
\mathcal{L}_{eff}=-\frac{1}{4}F_{\mu\nu}F^{\mu\nu}+i\sum_{n=1}^{\infty}\frac{e^{n}}{n}Tr\left(\frac{1}{i\slashed{\partial}-m}\slashed{A}\right)^{n}\ .
\end{equation}
Graphiquement, cette série est représentée par une tour (infinie)
de diagrammes 1PI\footnote{One particle irreducible.} à une boucle
contenant des opérateurs de dimensions de plus en plus grandes. Si
on tronque la série au-delà des opérateurs de dimension 8, on obtient
les diagrammes 1PI à une boucle générant l'action effective de QED
jusqu'aux opérateurs de dimension 8 (voir Fig.~\ref{EHphot}). Tous
les diagrammes avec un nombre impair de photons externes sont d'amplitude
nulle en vertu du théorème de Furry pour la QED\footnote{Ce théorème est une conséquence directe de l'invariance de la QED
sous l'action de la symétrie discrète de conjugaison de charge C.
Sous l'action de C ($e\rightarrow-e$), chaque vertex impliquant un
photon est multiplié par -1 et par conséquent un diagramme contenant
un nombre impair de photons briserait l'invariance de la QED sous
C sauf si cette amplitude est identiquement nulle.} \cite{Furry37}). L'avantage principal à exprimer l'action effective en termes de diagrammes
1PI est que des outils informatiques (déjà bien testés) automatisant
les calculs existent et permettent de calculer ces amplitudes à une
boucle. Pour ce travail, nous nous sommes appuyés sur les librairies
Mathematica \textit{FeynArts} \cite{FeynArts}, \textit{FeynCalc}
\cite{FeynCalc1,FeynCalc2} et \textit{Package X} \cite{PackageX}
(implémenté à travers \textit{FeynHelpers} \cite{FeynHelper}).

Construisons le lagrangien effectif jusqu'à l'ordre $m^{-4}$ du développement
perturbatif, c'est à dire jusqu'aux interactions de dimension 8 en
énergie. La contribution à la polarisation du vide de QED d'un fermion
de charge unité (de masse $m$ et de 4-vecteur énergie-impulsion $p^{\mu}$),
une fois développée en l'inverse de la masse du fermion s'écrit
\begin{equation}
\Pi^{\mu\nu}(p^{2})=i\frac{8e^{2}}{(4\pi)^{2}}\left(g^{\mu\nu}p^{2}-p^{\mu}p^{\nu}\right)\left\{ \frac{1}{6}D_{\varepsilon}+\frac{p^{2}}{30m^{2}}+\frac{p^{4}}{280m^{4}}+\mathcal{O}\left(p^{6}/m^{6}\right)\right\} \;,
\end{equation}
où $D_{\varepsilon}=2/\varepsilon-\gamma+\log4\pi\mu^{2}/m^{2}$.
Les interactions effectives correspondantes impliquant deux photons
sont les suivantes:
\begin{equation}
\mathfrak{L}_{eff}^{(0+2)}=-\frac{1}{4}\left\{ 1+\frac{\alpha}{3\pi}D_{\varepsilon}\right\} F_{\mu\nu}F^{\mu\nu}+\frac{\alpha}{60\pi m^{2}}F_{\mu\nu}\Box F^{\mu\nu}-\frac{\alpha}{560\pi m^{4}}F_{\mu\nu}\Box^{2}F^{\mu\nu}+\mathcal{O}(m^{-6})\;.
\end{equation}
Avec quatre photons, le lagrangien effectif est donné par les deux
couplages suivants:
\begin{equation}
\mathfrak{L}_{eff}^{(4)}=\frac{\alpha^{2}}{90m^{4}}(F_{\mu\nu}F^{\mu\nu})^{2}+\frac{7\alpha^{2}}{360m^{4}}(F_{\mu\nu}\tilde{F}^{\mu\nu})^{2}+\mathcal{O}(m^{-6})\;,
\end{equation}
où le tenseur électromagnétique dual est défini avec la convention
$\tilde{F}^{\mu\nu}=\frac{1}{2}\varepsilon^{\mu\nu\rho\sigma}F_{\rho\sigma}$,
en sorte que $(F_{\mu\nu}\tilde{F}^{\mu\nu})^{2}=2(F_{\mu\nu}F^{\mu\nu})^{2}-4F_{\mu\nu}F^{\nu\rho}F_{\rho\sigma}F^{\sigma\mu}$.
Le premier terme dans $\mathfrak{L}_{eff}^{(0+2)}$, divergent (possède
un pôle lorsque $\epsilon\rightarrow0$), traduit la renormalisation
de la fonction d'onde du photon afin d'y absorber la divergence de
la polarisation du vide de QED. Le second terme représente l'interaction
de Uehling \cite{Uehling35} et $\mathfrak{L}_{eff}^{(4)}$ est exactement
le lagrangien d'Euler-Heisenberg \cite{EulerH36} vu dans la partie
précédente.

Notons que dans la plupart des bases d'opérateurs \cite{SMEFT1,SMEFT2},
le couplage dérivatif est éliminé à l'aide de l'équation du mouvement
(EOM) de la manière suivante:
\begin{equation}
F_{\mu\nu}\Box F^{\mu\nu}=F_{\mu\nu}\partial^{\rho}\partial_{\rho}F^{\mu\nu}=F_{\mu\nu}\partial^{\rho}\partial^{\mu}F_{\rho}^{\,\,\,\nu}+F_{\mu\nu}\partial^{\rho}\partial^{\nu}F_{\,\,\,\rho}^{\mu}=-2\partial^{\mu}F_{\mu\rho}\partial_{\nu}F^{\nu\rho}=-2j_{\nu}j^{\nu}\;,
\end{equation}
où l'identité de Jacobi $\partial_{\mu}F_{\rho\nu}-\partial_{\rho}F_{\mu\nu}+\partial_{\nu}F_{\mu\rho}=0$
a été utilisée dans la première égalité, suivie d'une intégration
par partie et de l'utilisation de l'EOM $\partial_{\mu}F^{\mu\nu}=j^{\nu}$.
Cela fait sens physiquement car le potentiel de Uehling n'a un effet
que sur le couplage entre courants avec un transfert d'impulsion non
nul.

Dans la théorie effective considérée ici, tous les fermions ont été
intégrés et $\partial_{\mu}F^{\mu\nu}=0$. Ceci illustre une idée
générale du formalisme de l'action effective, à savoir que l'intégralité
de l'effet des champs lourds est codé dans les couplages effectifs
entre les champs légers à travers les corrections radiatives. Remarquons
que les champs légers ne sont jamais supposés sur couche de masse
(on-shell) et donc il est possible que certaines interactions effectives
ne contribuent jamais aux processus physiques, même si ces dernières
sont nécessaires pour encoder toute l'information sur la dynamique
sous-jacente du champ lourd.

Notons que les interactions effectives peuvent être construites a
priori en invoquant uniquement l'invariance de jauge de QED (sous
le groupe $U(1)_{em}$). En effet, la base d'opérateurs pour une dimension
en énergie fixée $D$ sera constituée de tous les couplages invariants
de jauge possibles de dimension $D$. En appliquant ce principe pour
construire les interactions effectives de photons, le lagrangien effectif
le plus général obtenu jusqu'aux opérateurs de dimension 8 est
\begin{equation}
\begin{aligned}\mathfrak{L}_{eff} & =-\frac{1}{4}\left\{ 1+\alpha_{0}\frac{e^{2}}{4!\pi^{2}}\right\} F_{\mu\nu}F^{\mu\nu}+\alpha_{2}\frac{e^{2}}{5!\pi^{2}m^{2}}\partial^{\mu}F_{\mu\nu}\partial_{\rho}F^{\rho\nu}+\alpha_{4}\frac{e^{2}}{6!\pi^{2}m^{4}}\partial^{\mu}F_{\mu\nu}\Box\partial_{\rho}F^{\rho\nu}\\
 & +\gamma_{4,1}\frac{e^{4}}{6!\pi^{2}m^{4}}(F_{\mu\nu}F^{\mu\nu})^{2}+\gamma_{4,2}\frac{e^{4}}{6!\pi^{2}m^{4}}(F_{\mu\nu}\tilde{F}^{\mu\nu})^{2}+\mathcal{O}(m^{-6})\;.
\end{aligned}
\label{eq:EffPhotons}
\end{equation}
Les couplages dérivatifs sont écrits sous une forme qui fait manifestement
apparaître l'EOM. Ceci s'avérera utile quand il s'agira de comparer
avec les résultats obtenus dans le cas non-abélien dans la prochaine
sous-section, pour lequel ce choix de base d'opérateurs est bien plus
judicieux. La nomenclature adoptée à partir d'ici et pour le reste
de cette section consiste à noter respectivement $\alpha_{i}\ensuremath{,}\beta_{i}$
et $\gamma_{i}$ les coefficients de Wilson associés aux couplages
impliquant deux, trois et quatre tenseurs électromagnétiques et à
l'ordre $\mathcal{O}(m^{-i})$. Ces coefficients, à travers leurs
valeurs spécifiques, contiennent toute l'information sur le champ
lourd. En faisant la correspondance (matching) entre cette théorie
effective de photons et la QED à basse énergie, nous avons calculé
les coefficients de Wilson induit par un fermion, un scalaire ou un
vecteur dans la boucle. Les résultats obtenus sont présentés dans
la Table ci-dessous:

\begin{table}[h] \centering 
$
\begin{tabular}[c]{cccccc}
\hline 
& $\alpha_{0}$ & $\alpha_{2}$ & $\alpha_{4}$ & $\gamma_{4,1}$ & $\gamma_{4,2}$ 
\smallskip\\
\hline
\multicolumn{1}{r}{Scalar} & $\dfrac{1}{2}D_{\varepsilon}Q^{2}$ & $-\dfrac{1}{8}Q^{2}$ & $\dfrac{3}{56}Q^{2}$ & $\dfrac{7}{32}Q^{4}$ &  $\dfrac{1}{32}Q^{4}$\rule[-0.28in]{0in}{0.4in}\\ 
\multicolumn{1}{r}{Fermion} & $2D_{\varepsilon}Q^{2}$ & $-Q^{2}$ & $\dfrac {9}{14}Q^{2}$ & $\dfrac{1}{2}Q^{4}$ & $\dfrac{7}{8}Q^{4}\rule[-0.28in]{0in}{0.4in}$\\ 
\multicolumn{1}{r}{Vector} & $-\dfrac{21D_{\varepsilon}+2}{2}Q^{2}$ & $\dfrac{37}{8}Q^{2}$ & $-\dfrac{159}{56}Q^{2}$ & $\dfrac{261}{32}Q^{4}$ & $\dfrac{243}{32}Q^{4}\rule[-0.18in]{0in}{0.4in}$\\\hline 
\end{tabular}
$
\caption{Coefficients de Wilson des opérateurs effectifs de photons induits par un scalaire, un fermion ou un vecteur de charge $Q$. Pour le cas du vecteur, la correspondance des amplitudes 1PI sur les opérateurs invariants de jauges $U(1)$ de l'équation (\ref{eq:EffPhotons}) n'est possible qu'en utilisant une jauge non-linéaire pour les vecteurs massifs, et les valeurs reportées de $\alpha_{1,2,3}$ sont spécifiques à cette jauge (($\kappa=1$ dans les équations (\ref{eq:LNLG}) et (\ref{eq:GaugeDep})).} 

\label{TableU1}

\end{table}Notons que le seul interêt de la normalisation non conventionnelle
des constantes de couplages dans le lagrangien effectif (\ref{eq:EffPhotons})
est d'obtenir des coefficients d'ordre $\mathcal{O}(1)$ (pour le
cas des fermions) afin de rendre le tableau \ref{TableU1} plus lisible.

Le calcul dans le cas d'un scalaire circulant dans la boucle est similaire
a celui des fermions et ne présente pas de difficulté supplémentaire.
Les sfermions chargés sous $U(1)_{em}$ (sleptons chargés et squarks)
du MSSM ont été utilisés comme particules représentatives pour effectuer
les calculs d'amplitudes à boucles et les faire correspondre (à basse
énergie) aux amplitudes calculées dans la théorie effective avec (\ref{eq:EffPhotons}).
En intégrant les champs scalaires lourds au niveau de l'intégrale
de chemin, on obtient la tour de diagrammes 1PI qui génèrent l'action
effective. Si on ne s'intéresse qu'aux opérateurs de dimensions inférieures
ou égales à 8 (on tronque la série au delà), les diagrammes 1PI sont
représentés dans la figure \ref{EHphotSV} ci-dessous:

\begin{figure}[h]
\begin{center}
\includegraphics[height=1.2315in]{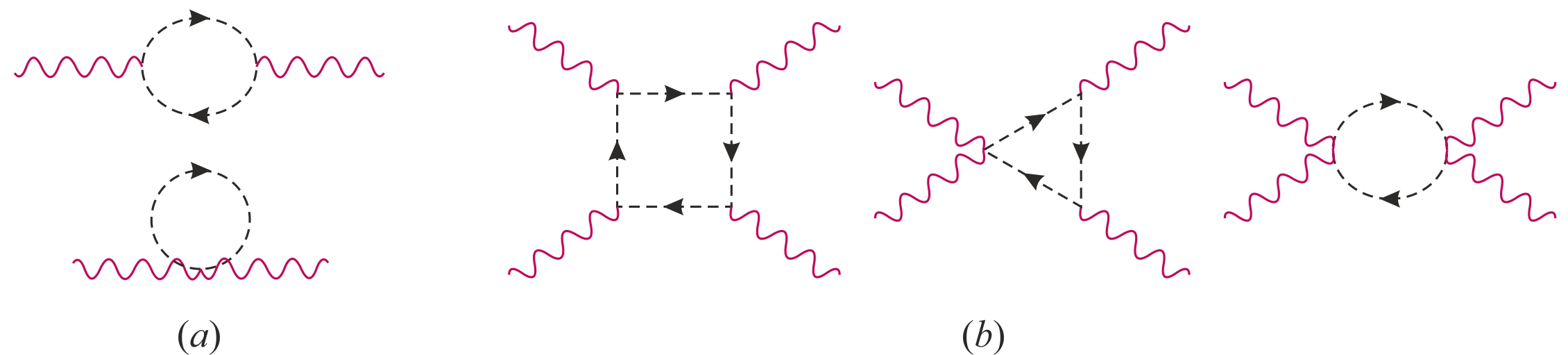}
\caption{Diagrammes 1PI à une boucle scalaire générant l'action effective de QED jusqu'aux opérateurs de dimension 8. Les permutations des photons externes sont considérées implicitement pour les diagrammes ($b$). Pour les bosons vecteurs massifs, les topologies sont les mêmes mais il faut alors inclure les diagrammes contenant les bosons de Goldstone appropriés et les fantômes.}
\label{EHphotSV}
\end{center}
\end{figure}Cependant, pour le cas des vecteurs circulant dans la boucle, les
choses sont nettement moins aisées. En effet, intégrer des vecteurs
pose problème à cause de la procédure de fixation de jauge. Afin de
comprendre où se trouve le problème précisément, considérons le MS
supplémenté du mécanisme de Higgs. Dans la jauge (usuelle) de 't Hooft-Feynman,
les identités de Ward de QED ne sont plus respectées au niveau de
l'amplitude pour des photons hors de leur couche de masse (off-shell).
Par conséquent, la correspondance entre l'amplitude à quatre photons
en QED et les couplages effectifs à l'ordre $\mathcal{O}(m^{-4})$
ne peut se faire que si les quatre photons externes sont on-shell
\cite{FanchiottiGS72}, et donc la procédure usuelle permettant de
construire l'action effective échoue ici. Le problème réside dans
la procédure de fixation de jauge. Par exemple, dans la jauge $R_{\xi}$,
on ajoute usuellement le terme suivant:
\begin{equation}
\mathfrak{L}_{gauge-fixing}^{R_{\xi},linear}=-\frac{1}{\xi}|\partial^{\mu}W_{\mu}^{+}+\xi M_{W}\phi^{+}|^{2}\ ,\label{eq:Rksi}
\end{equation}
où $\phi^{\pm}$ sont les bosons de Goldstone scalaires associés aux
$W^{\pm}$ (would-be Goldstone ou WBG en abrégé). Ceci brise explicitement
$U(1)_{QED}$. Bien que l'amplitude de la polarisation du vide de
QED reste transverse et est reproduite par les opérateurs effectifs
dans (\ref{eq:EffPhotons}), l'amplitude à quatre photons off-shell
n'est pas invariante de jauge et nécessite des opérateurs supplémentaires
dès l'ordre $\mathcal{O}(m^{-2})$ \cite{DongJZ93}. Bien entendu,
les processus physiques doivent être invariants de jauge et ceci ne
devrait pas avoir de conséquence. Néanmoins, en pratique, il y a des
réticences à ajouter des opérateurs non-invariants de jauge au lagrangien
effectif. On pourrait essayer de résoudre ce problème en se plaçant
dans la jauge unitaire, dans laquelle les couplages des $W$ aux photons
proviennent de:
\begin{equation}
\mathcal{L}_{unitary-gauge}=-\frac{1}{2}(D_{\mu}W_{\nu}^{+}-D_{\nu}W_{\nu}^{+})(D^{\mu}W^{-\nu}-D^{\nu}W^{-\mu})+ieF^{\mu\nu}W_{\mu}^{+}W_{\nu}^{-}+M_{W}^{2}W_{\mu}^{+}W^{-\mu}\ ,
\end{equation}
où $D_{\mu}W_{\nu}^{\pm}=\partial_{\mu}W_{\nu}^{\pm}\mp ieA_{\mu}W_{\nu}^{\pm}$.
Le terme de moment magnétique $F^{\mu\nu}W_{\mu}^{+}W_{\nu}^{-}$,
invariant de jauge, est donné par la symétrie $SU(2)_{L}$ sous-jacente.
Comme cela a été montré dans \cite{g2}, la présence de ce terme assure
un comportement cohérent des amplitudes de diffusion à haute énergie.
Cependant, cela n'est pas suffisant pour assurer le bon comportement
off-shell et tout comme dans le cas précédent de la jauge linéaire
$R_{\xi}$, le matching off-shell échoue encore \cite{PreucilH17}.

La solution à ce problème consiste à utiliser une jauge non-linéaire
où l'on remplace la dérivée partielle par la dérivée covariante dans
(\ref{eq:Rksi}) $\partial^{\mu}W_{\mu}^{\pm}\rightarrow D^{\mu}W_{\mu}^{\pm}=\partial^{\mu}W_{\mu}^{\pm}\pm ieA^{\mu}W_{\mu}^{\pm}$.
Cette modification est analogue à la contrainte imposée lors de la
construction de la CDE dans l'approche des intégrales de chemin \cite{EffAction}.
Dans l'approche diagrammatique considérée ici, du fait de l'utilisation
de la jauge non-linéaire \cite{Boudjema86}, l'amplitude à 4-photons
est alors invariante de jauge et ce même pour des photons externes
off-shell. Nous avons vérifié ceci explicitement en utilisant le model
FeynArts dédié \cite{FeynArtsNLG} pour le MS dans la jauge non-linéaire.
et nous avons obtenu un matching cohérent off-shell sur les opérateurs
d'Euler-Heisenberg. Les résultats des coefficients de Wilson dans
cette jauge sont dans le tableau \ref{TableU1}. Remarquons que les
trois premiers coefficients dépendent de la jauge et seuls $\gamma_{4,1}$
et $\gamma_{4,2}$ sont physiques. Afin d'examiner de plus près ces
dépendances, considérons le terme de fixation de jauge suivant \cite{BaceH75,GavelaGMS81}:
\begin{equation}
\mathfrak{L}_{gauge-fixing}^{non-linear}=-\frac{1}{\xi}|\partial^{\mu}W_{\mu}^{+}+i\kappa eA^{\mu}W_{\mu}^{+}+\xi M_{W}\phi^{+}|^{2}\ ,\label{eq:LNLG}
\end{equation}
qui permet d'interpoler entre le cas linéaire ($\kappa=0$) et non-linéaire
en jauge $U(1)$ (pour $\kappa=1$). Le développement en $\tfrac{1}{m}$
de la polarisation du vide de QED dans la jauge de 't Hooft-Feynman
($\xi=1$) donne des coefficients dépendant de $\kappa$de la manière
suivante:
\begin{equation}
\alpha_{0}=-\dfrac{12\kappa+9}{2}D_{\varepsilon}-1\ ,\ \ \alpha_{2}=\dfrac{20\kappa+17}{8}\ ,\ \ \alpha_{4}=-\dfrac{84\kappa+75}{56}\ .\label{eq:GaugeDep}
\end{equation}
Bien évidemment, ces dépendances dans la jauge ne sont pas physiques.
A très basse énergie, lorsque le photon est le seul degré de liberté
actif restant, le premier coefficient ($\alpha_{0}$) est absorbé
dans le champ de photon tout comme la constante de renormalisation
de la fonction d'onde. Les deux autres coefficients ne contribuent
pas étant donné que $\partial_{\mu}F^{\mu\nu}=0$. Si certains champs
restent actifs tels que $\partial_{\mu}F^{\mu\nu}=j^{\nu}\neq0$,
alors d'autres types de processus sont aussi présents. Dans ce cas,
l'opérateur associé à $\alpha_{2}$ doit être éliminé en faveur de
l'opérateur de dimension 6 $j_{\mu}j^{\mu}/m^{2}$, pour lequel d'autres
diagrammes sont présents. Dans le MS, même si les champs dans le courant
$j^{\mu}$ ne sont pas couplés directement aux $W^{\pm}$, ils sont
nécessairement couplés au boson $Z$. La dépendance en $\kappa$des
contributions des $W^{\pm}$ aux polarisations du vide $Z\gamma$
et $ZZ$ \cite{GRACE} doit s'annuler avec la dépendance en $\kappa$
de $\alpha_{2}$ et ainsi on obtient un coefficient invariant de jauge
et physique pour l'opérateur $j_{\mu}j^{\mu}/m^{2}$. La conclusion
de ceci est que dans le MS, le potentiel d'Uehling ne peut pas être
défini de manière cohérente en terme de l'opérateur $F_{\mu\nu}\Box F^{\mu\nu}$
et on doit utiliser à la place les opérateurs effectifs à 4-fermions.
Ceci est cohérent physiquement dans la mesure où le potentiel d'Uehling
ne fait sens qu'en présence de champs fermioniques restés actifs à
cette échelle d'énergie.

\subsection{Interactions effectives de gluons }

\begin{figure}[h]
\begin{center}
\includegraphics[height=1.2315in]{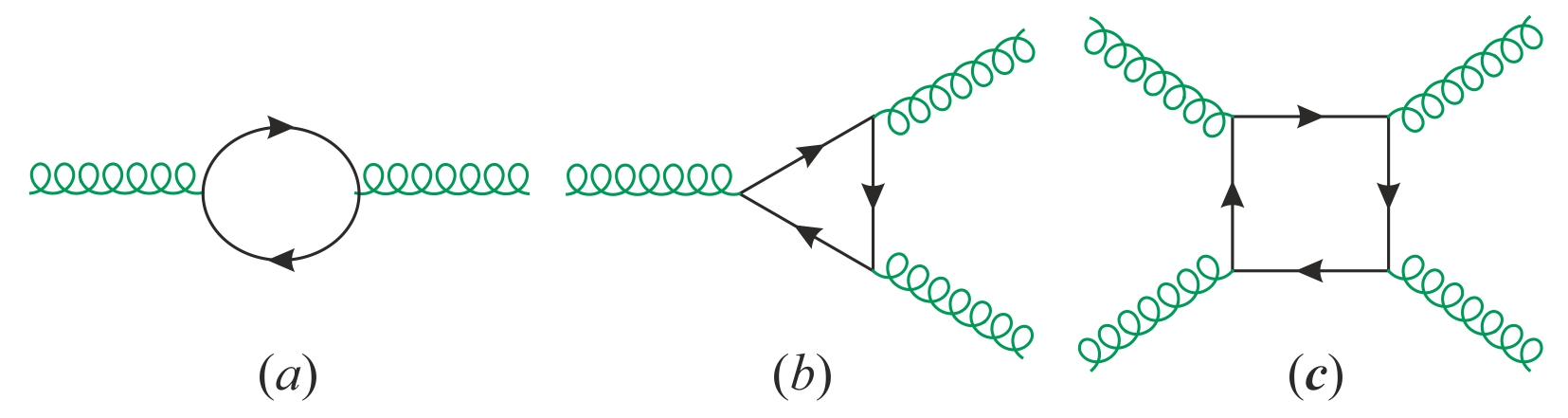}
\caption{Amplitude 1PI à une boucle fermionique générant l'action effective gluonique. Les permutations des gluons sont considérées implicitement pour les diagrammes $(b)$ et $(c)$. Comme pour le cas de la QED (figure \ref{EHphotSV}), des diagrammes supplémentaires sont considérés dans les cas d'un scalaire ou vecteur dans la boucle.}
\label{EHglue}
\end{center}
\end{figure}

Essayons de généraliser l'étude faite sur les interactions de photons
aux cas des gluons. L'action effective pour les champs de gluons est
construite de la même manière que celle des photons dans l'approche
diagrammatique. En intégrant un fermion lourd au niveau de l'intégrale
de chemin, on obtient le lagrangien effectif suivant:
\begin{equation}
\begin{aligned}\mathcal{L}_{eff} & =-\frac{1}{4}G_{\mu\nu}^{a}G^{a,\mu\nu}-iTr\ \ln(i\slashed{D}-m)\\
 & =-\frac{1}{4}G_{\mu\nu}^{a}G^{a,\mu\nu}+i\sum_{n=1}^{\infty}\frac{e^{n}}{n}Tr\left(\frac{1}{i\slashed{\partial}-m}\slashed{G}^{a}T^{a}\right)^{n}\ ,
\end{aligned}
\end{equation}
où les $T^{a}$ sont les générateurs de $SU(3)$ et la trace agit
à la fois sur les indices de Dirac et de couleur. Ce développement
en série génère la tour infinie d'amplitudes 1PI qui contribuent à
l'action effective et si on s'arrête aux opérateurs de dimension 8,
les diagrammes 1PI sont ceux représentés dans la figure \ref{EHglue}.

Contrairement au cas de la QED avec les photons, les diagrammes possédant
un nombre impair de gluons externes ne s'annulent pas. De plus, une
autre différence fondamentale avec la QED se trouve dans le caractère
non-linéaire du tenseur gluonique ($G^{\mu\nu}$). Cette non-linéarité
par nature a pour conséquence d'invalider la relation entre l'ordre
$\mathcal{O}(m^{-i})$ dominant d'une amplitude donnée dans la série
en $\tfrac{1}{m^{i}}$ et le nombre de gluons externes. Une autre
conséquence directe de la non-linéarité est que les diagrammes à 3-gluons
et 4-gluons sont divergents. Néanmoins, ces deux divergences renormalisent
le même opérateur local $G_{\mu\nu}^{a}G^{a,\mu\nu}$ et de ce fait,
elles sont cohérentes avec la divergence de la polarisation du vide
de QCD (amplitude à 2-gluons). Voyons plus en détail comment cela
se passe en pratique.

Tout d'abord, construisons la base d'opérateurs la plus générale possible
jusqu'à l'ordre $\mathcal{O}(m^{-4})$ (opérateurs de dimension 8).
Avec deux tenseurs gluoniques, la généralisation du cas de la QED
se fait aisément et on obtient le lagrangien effectif suivant:
\begin{equation}
\begin{aligned}\mathfrak{L}_{eff}^{(0+2)} & =-\frac{1}{4}\left\{ 1+\alpha_{0}\frac{g_{S}^{2}}{4!\pi^{2}}\right\} G_{\mu\nu}^{a}G^{a,\mu\nu}\\
 & +\alpha_{2}\frac{g_{S}^{2}}{5!\pi^{2}m^{2}}D^{\nu}G_{\nu\mu}^{a}D_{\rho}G^{a,\rho\mu}+\alpha_{4}\frac{g_{S}^{2}}{6!\pi^{2}m^{4}}D^{\nu}G_{\nu\mu}^{a}D^{2}D_{\rho}G^{a,\rho\mu}\;,
\end{aligned}
\end{equation}
avec $G_{\mu\nu}^{a}=\partial_{\mu}A_{\nu}^{a}-\partial_{\nu}A_{\mu}^{a}+gf^{abc}A_{\mu}^{b}A_{\nu}^{c}$
et $D_{\rho}G_{\mu\nu}^{a}=(\partial_{\rho}\delta^{ac}+gf^{abc}G_{\rho}^{b})G_{\mu\nu}^{c}$.
Il ne peut y avoir qu'un seul opérateur dérivatif par ordre $\mathcal{O}(m^{-i})$
\cite{SMEFT1,SMEFT2}. En effet, remarquons que les dérivées peuvent
être déplacées par intégrations par parties de sorte à agir sur un
des tenseurs gluoniques. Ainsi, seul un ordre précis des dérivées
covariantes est pertinent car le fait de changer l'ordre des dérivées
génère un tenseur gluonique supplémentaire étant donné qu'elles ne
commutent pas, $[D^{\rho},D^{\sigma}]G_{\mu\nu}^{a}=gf^{abc}G_{\rho\sigma}^{b}G_{\mu\nu}^{c}$.
Finalement, en combinant ceci avec l'identité de Bianchi
\begin{equation}
D_{[\mu}G_{\rho\sigma]}^{a}=D_{\mu}G_{\rho\sigma}^{a}+D_{\rho}G_{\sigma\mu}^{a}+D_{\sigma}G_{\mu\rho}^{a}=0\;,
\end{equation}
les opérateurs effectifs peuvent être réécrit sous une forme qui fait
manifestement apparaître l'EOM des gluons, $D^{\mu}G_{\mu\nu}^{a}=0$.
Insistons sur le fait que les EOMs n'ont pas été utilisées car cela
rendrait la correspondance avec la QCD à basse énergie impossible.

Considérons maintenant les interactions impliquant trois tenseurs
gluoniques. Nous avons un seul opérateur à l'ordre $\mathcal{O}(m^{-2})$
mais plusieurs à l'ordre $\mathcal{O}(m^{-4})$. Cependant, après
intégrations par parties, utilisation de l'identité de Bianchi et
en écartant les termes impliquant quatre tenseurs gluoniques ou plus,
seules deux contractions non équivalentes subsistent \cite{Simmons89}.
Là encore, nous faisons en sorte de les écrire sous une forme qui
fait apparaître l'EOM des champs de gluons:
\begin{equation}
\begin{aligned}\mathfrak{L}_{eff}^{(3)} & =\beta_{2}\frac{g_{S}^{3}}{5!\pi^{2}m^{2}}f^{abc}G_{\mu}^{a\;\nu}G_{\nu}^{b\;\rho}G_{\rho}^{c\;\mu}\\
 & +\beta_{4,1}\frac{g_{S}^{3}}{6!\pi^{2}m^{4}}f^{abc}G^{a,\mu\nu}D^{\alpha}G_{\mu\nu}^{b}D^{\beta}G_{\alpha\beta}^{c}+\beta_{4,2}\frac{g_{S}^{3}}{6!\pi^{2}m^{4}}f^{abc}G^{a,\mu\nu}D^{\alpha}G_{\alpha\mu}^{b}D^{\beta}G_{\beta\nu}^{c}\;.
\end{aligned}
\end{equation}

Avec quatre tenseurs gluoniques, les opérateurs jusqu'à l'ordre $\mathcal{O}(m^{-4})$
ne contiennent pas de dérivées covariantes. Pour atteindre un nombre
minimal d'opérateurs, on utilise la généralisation de l'identité de
QED:
\begin{equation}
G_{\mu\nu}^{a}\tilde{G}^{b,\mu\nu}G_{\rho\sigma}^{c}\tilde{G}^{d,\rho\sigma}=G_{\mu\nu}^{a}G^{c,\mu\nu}G_{\rho\sigma}^{b}G^{d,\rho\sigma}+G_{\mu\nu}^{a}G^{d,\mu\nu}G_{\rho\sigma}^{b}G^{c,\rho\sigma}-4G_{\mu\nu}^{a}G^{c,\nu\rho}G_{\rho\sigma}^{b}G^{d,\sigma\mu}\;,\label{eq:IdGGGG}
\end{equation}
et notons l'absence de contractions impliquant le tenseur complètement
symétrique $d^{abc}$ du fait de leurs réductions avec l'identité
suivante (voir Annexe \ref{AnnexeC}):
\begin{equation}
3d^{abe}d^{cde}=\delta^{ac}\delta^{bd}-\delta^{ab}\delta^{cd}+\delta^{ad}\delta^{bc}+f^{ace}f^{bde}+f^{ade}f^{bce}\;.\label{eq:IddSU3}
\end{equation}
Les contractions mixtes entre des tenseurs $f$ (antisymétriques)
et des tenseurs $d$ (symétriques) sont identiquement nulles du fait
de leurs propriétés de symétries. En tenant compte de tout cela, on
se retrouve avec six opérateurs à l'ordre $\mathcal{O}(m^{-4})$ dans
le lagrangien effectif $\mathfrak{L}_{eff}^{(4)}$:
\begin{equation}
\begin{aligned}\mathfrak{L}_{eff}^{(4)} & =\gamma_{4,1}\frac{g_{S}^{4}}{6!\pi^{2}m^{4}}G_{\mu\nu}^{a}G^{a,\mu\nu}G_{\rho\sigma}^{b}G^{b,\rho\sigma}+\gamma_{4,2}\frac{g_{S}^{4}}{6!\pi^{2}m^{4}}G_{\mu\nu}^{a}\tilde{G}^{a,\mu\nu}G_{\rho\sigma}^{b}\tilde{G}^{b,\rho\sigma}\\
 & +\gamma_{4,3}\frac{g_{S}^{4}}{6!\pi^{2}m^{4}}G_{\mu\nu}^{a}G^{b,\mu\nu}G_{\rho\sigma}^{a}G^{b,\rho\sigma}+\gamma_{4,4}\frac{g_{S}^{4}}{6!\pi^{2}m^{4}}G_{\mu\nu}^{a}\tilde{G}^{b,\mu\nu}G_{\rho\sigma}^{a}\tilde{G}^{b,\rho\sigma}\\
 & +\gamma_{4,5}\frac{g_{S}^{4}}{6!\pi^{2}m^{4}}f^{abe}f^{cde}G_{\mu\nu}^{a}G^{c,\mu\nu}G_{\rho\sigma}^{b}G^{d,\rho\sigma}+\gamma_{4,6}\frac{g_{S}^{4}}{6!\pi^{2}m^{4}}f^{abe}f^{cde}G_{\mu\nu}^{a}\tilde{G}^{c,\mu\nu}G_{\rho\sigma}^{b}\tilde{G}^{d,\rho\sigma}\;.
\end{aligned}
\label{eq:EffSUNA}
\end{equation}
Cette base correspond a celle trouvée dans \cite{Gracey17_1,Gracey17_2}
à la différence de quelques remplacements via l'identité (\ref{eq:IdGGGG}).

\begin{figure}[h]
\begin{center}
\includegraphics[height=0.9608in,width=5.4578in]{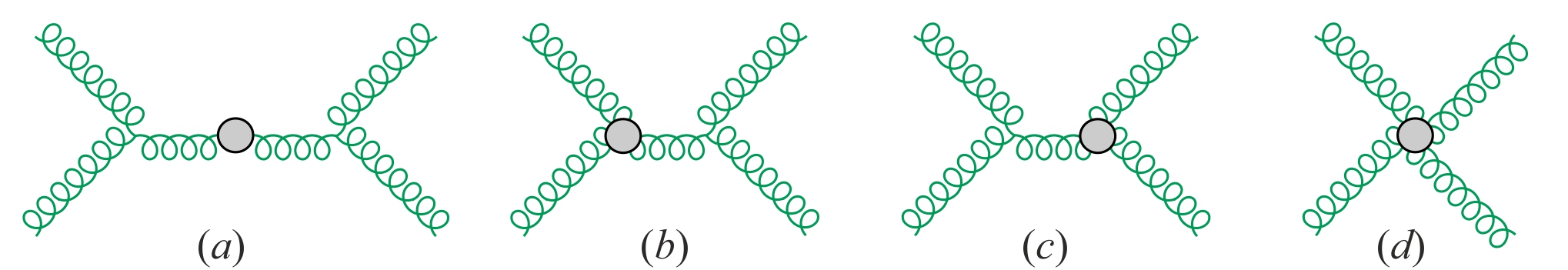}
\caption{Les quatre topologies de bases en voie $s$ pour l'amplitude de diffusion gluon-gluon. Celles pour les voies $t$ et $u$ sont considérées implicitement. Les disques gris représentent les insertions des vertex effectifs.}
\label{EHEffGlue}
\end{center}
\end{figure}

Le caractère non-abélien de la QCD rend le développement en série
de l'action effective très différent du cas de la QED. Les opérateurs
s'annulant avec l'EOM doivent être gardés car ils contribuent à plusieurs
diagrammes 1PI off-shell. Par exemple, l'opérateur $D^{\nu}G_{\nu\mu}^{a}D_{\rho}G^{a,\rho\mu}$
contribue aux processus 1PI à deux, à trois et à quatre gluons off-shell
(représentés figure \ref{EHglue}) du fait des termes non-abéliens
présents dans les tenseurs gluoniques. D'autre part, pour un processus
physique impliquant des gluons externes on-shell, ces opérateurs ne
doivent pas contribuer et la base se voit simplifiée. Vérifions cela
pour un processus simple, à savoir l'amplitude de diffusion gluon-gluon
\begin{equation}
\mathcal{A}(g(p_{1},\varepsilon_{p_{1}}^{\mu_{1}})g(p_{2},\varepsilon_{p_{2}}^{\mu_{2}})\rightarrow g(p_{3},\varepsilon_{p_{3}}^{\mu_{3}})g(p_{4},\varepsilon_{p_{4}}^{\mu_{4}}))=\varepsilon_{p_{1}}^{\mu_{1}}\varepsilon_{p_{2}}^{\mu_{2}}\varepsilon_{p_{3}}^{\mu_{3}\ast}\varepsilon_{p_{4}}^{\mu_{4}\ast}\mathcal{M}_{\mu_{1}\mu_{2}\mu_{3}\mu_{4}}\;.
\end{equation}
Cette amplitude est calculée en utilisant le lagrangien effectif jusqu'à
l'ordre $\mathcal{O}(m^{-4})$, les topologies de bases à considérer
sont représentées figure \ref{EHEffGlue}.

Outre les termes locaux à quatre points, nous devons ajouter les contributions
non-locales provenant des opérateurs à trois gluons et deux gluons
ainsi que le terme à l'arbre contenant la fonction d'onde renormalisée.
Après avoir effectué les calculs, on observe que:
\begin{itemize}[label=\textbullet]
\item La correction de la fonction d'onde est automatiquement prise en
compte en renormalisant le champ et la constante de couplage $g_{s}$.
\item Les opérateurs dans $\mathfrak{L}_{eff}^{(2)}$ contribuent à toutes
les topologies, ceux dans $\mathfrak{L}_{eff}^{(3)}$ contribuent
aux topologies ($b-d$) et ceux de $\mathfrak{L}_{eff}^{(4)}$ seulement
à la topologie ($d$).
\item Pour les opérateurs impliquant l'EOM, toutes leurs contributions s'annulent
précisément. Comme attendu, ces derniers ne jouent aucun rôle pour
des processus physiques.
\item Indépendamment pour chaque opérateur $Q_{i}$ ne contenant pas l'EOM,
la somme de toutes les contributions $\mathcal{M}_{\mu_{1}\mu_{2}\mu_{3}\mu_{4}}(Q_{i})$
satisfait bien les quatre identités de Ward $p_{k}^{\mu_{k}}\mathcal{M}_{\mu_{1}\mu_{2}\mu_{3}\mu_{4}}(Q_{i})=0\ensuremath{,}k=1,2,3,4$.
\end{itemize}
Le fait que les opérateurs EOM s'éliminent de l'amplitude physique
totale peut être compris qualitativement. Par exemple, en prenant
l'opérateur de dimension 6 $D^{\nu}G_{\nu\mu}^{a}D_{\rho}G^{a,\rho\mu}$
dans $\mathfrak{L}_{eff}^{(2)}$ et en développant les dérivées covariantes,
on obtient:
\begin{equation}
\begin{aligned}D^{\nu}G_{\nu\mu}^{a}D_{\rho}G^{a,\rho\mu} & =\partial^{\nu}G_{\nu\mu}^{a}\partial_{\rho}G^{a,\rho\mu}+gf^{abc}G^{b,\nu}G_{\nu\mu}^{c}\partial_{\rho}G^{a,\rho\mu}\\
 & +gf^{abc}\partial^{\nu}G_{\nu\mu}^{a}G_{\rho}^{b}G^{c,\rho\mu}+g^{2}f^{abc}f^{ade}G^{b,\nu}G_{\nu\mu}^{c}G_{\rho}^{d}G^{e,\rho\mu}\;.
\end{aligned}
\label{eq:DGDGcancel}
\end{equation}
En remplaçant le tenseur gluonique par son expression $G_{\mu\nu}^{a}\rightarrow\partial_{\mu}G_{\nu}^{a}-\partial_{\nu}G_{\mu}^{a}$,
les quatre termes obtenus sont précisément ceux qui entrent en jeu
dans les quatre topologies de la figure \ref{EHEffGlue}. On peut
voir que l'annulation se réalise car les pôles du propagateur de gluon
sont précisément compensés par les contributions à l'ordre dominant
provenant du vertex à trois gluons ainsi que par les dérivées dans
les trois premiers termes de l'équation (\ref{eq:DGDGcancel}). Un
raisonnement similaire est appliqué aux termes non-abélien qui s'annulent
de la même manière.

Calculons maintenant les coefficients de Wilson de ces opérateurs
effectifs induits par un fermion, un scalaire ou un vecteur dans la
représentation fondamentale de $SU(3\text{)}$. Généralement, la procédure
a suivre pour faire correspondre la théorie effective construite et
la QCD à basse énergie est la suivante. On commence par la polarisation
du vide de QCD (figure \ref{EHglue}$a$) qui nous permet de fixer
les coefficients $\alpha_{0,2,4}$. Ensuite, les amplitudes 1PI à
une boucle impliquant 3-gluons (figure \ref{EHglue}$b$) génèrent
de nouveau les opérateurs de $\mathfrak{L}_{eff}^{(2)}$ ainsi que
ceux de $\mathfrak{L}_{eff}^{(3)}$, et ceci permet de fixer $\beta_{2}\ensuremath{,}\beta_{4,1}$
et $\beta_{4,2}$. Notons que le choix de la base pour $\mathfrak{L}_{eff}^{(0+2)}$
influe sur la valeur des trois coefficients de Wilson de $\mathfrak{L}_{eff}^{(3)}$.
Finalement, les amplitudes 1PI à 4-gluons (figure \ref{EHglue}$c$)
correspondent avec les opérateurs effectifs locaux impliquant quatre
gluons présents dans $\mathfrak{L}_{eff}^{(0+2+3+4)}$ et connaissant
les coefficients trouvés dans les précédentes étapes permet de fixer
les six derniers coefficients de Wilson $\gamma_{4,i}$. Les résultats
finaux trouvés sont exposés dans le tableau \ref{TableSU3} et sont
en accord avec \cite{EffAction} pour les opérateurs de dimension-6.

\begin{table}[h] \centering 
$ 
\begin{tabular} [c]{ccccccc}\hline 
& $\alpha_{0}$ & $\alpha_{2}$ & $\alpha_{4}$ & $\beta_{2}$ & $\beta_{4,1}$ & $\beta_{4,2}$\smallskip\\\hline 
\multicolumn{1}{r}{Scalar} & $\dfrac{1}{4}D_{\varepsilon}$ & $-\dfrac{1}{16} $ & $\dfrac{3}{112}$ & $\dfrac{1}{48}$ & $-\dfrac{1}{28}$ & $0\rule[-0.18in]{0in}{0.4in}$\\ 
\multicolumn{1}{r}{Fermion} & $D_{\varepsilon}$ & $-\dfrac{1}{2}$ & $\dfrac {9}{28}$ & $-\dfrac{1}{24}$ & $\dfrac{1}{14}$ & $-\dfrac{3}{4}\rule[-0.18in]{0in}{0.4in}$\\ 
\multicolumn{1}{r}{Vector} & $-\dfrac{21D_{\varepsilon}+2}{4}$ & $\dfrac {37}{16}$ & $-\dfrac{159}{112}$ & $\dfrac{1}{16}$ & $-\dfrac{3}{28}$ & $3\rule[-0.18in]{0in}{0.4in}$\\\hline 
\multicolumn{1}{r}{} & $\gamma_{4,1}$ & $\gamma_{4,2}$ & $\gamma_{4,3}$ & $\gamma_{4,4}$ & $\gamma_{4,5}$ & $\gamma_{4,6}$\smallskip\\\hline 
\multicolumn{1}{r}{Scalar} & $\dfrac{7}{768}$ & $\dfrac{1}{768}$ & $\dfrac {7}{384}$ & $\dfrac{1}{384}$ & $\dfrac{1}{96}$ & $\dfrac{1}{672}\rule[-0.18in]{0in}{0.4in}$\\ 
\multicolumn{1}{r}{Fermion} & $\dfrac{1}{48}$ & $\dfrac{7}{192}$ & $\dfrac {1}{24}$ & $\dfrac{7}{96}$ & $\dfrac{1}{96}$ & $\dfrac{19}{672}\rule[-0.18in]{0in}{0.4in}$\\ 
\multicolumn{1}{r}{Vector} & $\dfrac{87}{256}$ & $\dfrac{81}{256}$ & $\dfrac{87}{128}$ & $\dfrac{81}{128}$ & $-\dfrac{3}{32}$ & $-\dfrac{27}{224}\rule[-0.18in]{0in}{0.4in}$\\\hline 
\end{tabular} 
$ 
\caption{Coefficients de Wilson des opérateurs effectifs gluoniques pour un scalaire, un fermion et un vecteur dans la représentation fondamentale de $SU(3)$. Ceci correspond par exemple aux contributions des squarks dans le MSSM, ou des quarks lourds dans le MS. Dans le cas des vecteurs, nous avons utilisé les leptoquarks du modèle minimal grand-unifié $SU(5)$ GUT, quantifié en utilisant une jauge non-linéaire (voir Annexe \ref{AnnexeB}).} 

\label{TableSU3} 

\end{table}

Cette procédure est directe pour les cas des fermions et des scalaires
circulants dans la boucle. Les calculs ont été effectués en utilisant
les modèles SM et MSSM implémentés dans FeynArts et en prenant des
quarks et des squarks dans la représentation fondamentale comme particules
représentatives.

Concernant le cas des vecteurs, le calcul donne bien plus de fil à
retordre. En effet, il s'agit d'abord de construire un modèle cohérent
qui autorise un champ vectoriel massif dans la représentation fondamentale
de QCD. De plus, nous savons grâce au calcul dans le cas de la QED
que travailler dans la jauge unitaire ne fonctionnera pas, et ce même
en introduisant un mécanisme approprié du type mécanisme de Higgs
pour donner la masse à ces vecteurs. Nous devons alors généraliser
la jauge non-linéaire pour qu'elle puisse préserver la symétrie de
QCD tout au long de la quantification sinon on ne peut pas faire correspondre
les amplitudes 1PI off-shell et les opérateurs effectifs invariants
de jauge. Ceci serait particulièrement dérangeant ici étant donné
que les diagrammes 1PI à 3-gluons sont interdits cinématiquement et
ne peuvent exister on-shell. Du fait du caractère non-abélien, les
amplitudes 1PI à 3-gluons (nécessairement off-shell) interviennent
dans la correspondance des diagrammes à 4-gluons qui doit donc nécessairement
se faire off-shell.

Notre stratégie consiste à utiliser le modèle minimal $SU(5)$ GUT,
brisé spontanément en $SU(3)_{C}\times SU(2)_{L}\times U(1)_{Y}$
par un Higgs scalaire dans la représentation adjointe. Au cours de
ce processus, douze des bosons de jauge de $SU(5)$ acquièrent une
masse, ce sont les leptoquarks. Ces derniers possèdent précisément
les nombres quantiques souhaités. Le doublet faible de leptoquarks
$(X,Y)$ se transforme comme les antitriplets de couleurs ($\bar{3}$
de $SU(3)_{c}$) et donc intégrer les leptoquarks génèrera les opérateurs
effectifs gluoniques. Nous ne nous intéressons pas à la seconde étape
de brisure de symétrie $SU(3)_{C}\times SU(2)_{L}\times U(1)_{Y}\rightarrow SU(3)_{C}\times U(1)_{em}$.
Dans l'Annexe \ref{AnnexeB}, nous décrivons en détail le modèle minimal
$SU(5)$ GUT dont la quantification est réalisée en jauge non-linéaire
pour les leptoquarks. En notant $H_{X}^{k}$ et $H_{Y}^{k}$ les bosons
de Goldstone scalaires associés (would-be Goldstone) aux leptoquarks
$X_{\mu}^{k}$ et $Y_{\mu}^{k}$, le point clef consiste à modifier
la jauge $R_{\xi}$ 
\begin{equation}
\mathcal{L}_{\text{GF}}=-\frac{1}{\xi}|\partial^{\mu}X_{\mu}^{k+}-i\xi M_{XY}H_{X}^{k+}|^{2}-\frac{1}{\xi}|\partial^{\mu}Y_{\mu}^{k+}-i\xi M_{XY}H_{Y}^{k+}|^{2}+....
\end{equation}
en remplaçant la dérivée partielle par la dérivée covariante de $SU(5)$
\begin{equation}
\begin{aligned}\partial^{\mu}X_{\mu}^{i+} & \rightarrow\partial^{\mu}X_{\mu}^{i+}-ig_{5}\left(-\alpha_{G}X_{\nu}^{j+}T_{ji}^{a}G_{\mu}^{a}+\frac{\alpha_{W}}{2}W_{\mu}^{3}X_{\nu}^{i+}+\frac{\alpha_{W}}{\sqrt{2}}W_{\mu}^{+}Y_{\nu}^{i+}+\alpha_{B}\sqrt{\frac{\ 5}{12}}B_{\mu}X_{\nu}^{i+}\right)\ ,\\
\partial^{\mu}Y_{\nu}^{i+} & \rightarrow\partial^{\mu}Y_{\mu}^{i\pm}-ig_{5}\left(-\alpha_{G}Y_{\nu}^{j+}T_{ji}^{a}G_{\mu}^{a}-\frac{\alpha_{W}}{2}W_{\mu}^{3}Y_{\nu}^{i+}+\frac{\alpha_{W}}{\sqrt{2}}W_{\mu}^{-}X_{\nu}^{i+}+\alpha_{B}\sqrt{\frac{\ 5}{12}}B_{\mu}Y_{\nu}^{i+}\right)\ ,
\end{aligned}
\label{eq:NLGSU5}
\end{equation}
où les $T^{\alpha}$ sont les générateurs de $SU(3)$ dans la représentation
fondamentale et $i,j,k$ les indices correspondants. Les paramètres
de jauges $\alpha_{G}\ensuremath{,}\alpha_{W}\ensuremath{,}\alpha_{B}$
servent à interpoler entre la jauge de 't Hooft Feynman ($\alpha_{G}=\alpha_{W}=\alpha_{B}=0$)
et la jauge non-linéaire ($\alpha_{G}=\alpha_{W}=\alpha_{B}=1$) où
les termes \ref{eq:NLGSU5} coïncident avec $D^{\mu}X_{\mu}^{i+}$
et $D^{\mu}Y_{\mu}^{i+}$. Dans cette limite, les symétries de jauges
du modèle standard sont préservées, exactement comme pour le cas du
MS avec $U(1)_{em}$ dans la jauge non-linéaire. D'un point de vue
technique, cette jauge possède aussi la plaisante propriété de réduire
drastiquement le nombre de diagrammes pour un processus donné \cite{GavelaGMS81}.
En effet, rappelons que l'interêt de la jauge $R_{\xi}$ usuelle est
d'éliminer les termes de mélanges comme $X_{\mu}^{k}\partial^{\mu}H_{X}^{k}$.
Cependant, dès lors que le vecteur est chargé sous une symétrie résiduelle,
ce terme est nécessairement de la forme $X_{\mu}^{k}D^{\mu}H_{X}^{k}$
puisqu'il provient du terme cinétique du Higgs scalaire qui est invariant
sous les symétries résiduelles. Dans la jauge non-linéaire, tous ces
termes sont éliminés et il n'y a donc plus de couplages $X-V_{SM}-H_{X}$.
Ainsi, toutes les boucles mélangeant un vecteur massif avec son boson
de Goldstone disparaissent, ce qui est apprécié compte tenu du grand
nombre de diagrammes.

Afin d'effectuer le calcul en pratique, nous avons là aussi utiliser
FeynArts \cite{FeynArts} mais avec un fichier-modèle $SU(5)$ personnalisé
que nous avons implémenté. Une fois le calcul effectué, les coefficients
de Wilson obtenus sont rapportés dans le tableau \ref{TableSU3}.
Quelques commentaires s'imposent:
\begin{itemize}[label=\textbullet]
\item La correspondance (matching) ne peut se faire que dans le cas ($\alpha_{G}=\alpha_{W}=\alpha_{B}=1$).
Sans cette condition, des opérateurs effectifs non-invariants de jauge
sont nécessaires. Notons qu'en partant de 207 diagrammes 1PI à 4-gluons,
une fois la condition ($\alpha_{G}=\alpha_{W}=\alpha_{B}=1$) de jauge
non-linéaire appliquée, on se retrouve avec seulement 21 boucles de
bosons de jauge, 21 boucles de Goldstone et 42 boucles de fantômes
(ghosts). Le fait d'éliminer les mélanges leptoquarks-WBG réduit le
nombre de diagrammes de plus d'un facteur 2.
\item Plusieurs propriétés découvertes dans \cite{Boudjema86} pour les
photons survivent à la généralisation non-abélienne, à savoir qu'en
jauge non-linéaire, les contributions des fantômes et des bosons de
Goldstone sont séparément invariantes de jauge. Faire correspondre
séparément les contributions des $H_{X}^{k}$ avec les opérateurs
effectifs redonne les mêmes coefficients de Wilson que dans le cas
scalaire du tableau \ref{TableSU3} alors que si on considère uniquement
la contribution des fantômes $c_{X}$ et $c_{X}^{\dagger}$, on obtient
-2 fois les coefficients dans le cas scalaire du tableau \ref{TableSU3}.
Avec la jauge non-linéaire, les fantômes se comportent exactement
comme des particules scalaires mais avec la statistique de Fermi-Dirac.
\item Comme vérification, nous avons calculé l'amplitude physique de diffusion
gluon-gluon en laissant le paramètre de jauge $\alpha_{G}$ libre.
En incluant des topologies 1PI et non-1PI avec des gluons on-shell,
la seule dépendance restante en $\alpha_{G}$ peut être absorbée dans
la redéfinition de la fonction d'onde. En d'autres termes, le développement
en série en $\tfrac{1}{m}$ de l'amplitude totale correspond avec
les opérateurs effectifs non-EOM (ne faisant pas apparaître l'EOM)
et à part pour $\alpha_{0}$, les coefficients de Wilson de ces opérateurs
sont indépendants de la jauge, comme il se doit pour un processus
physique.
\item Pour effectuer des vérifications supplémentaires, nous avons calculer
les diagrammes 1PI à deux, trois et quatre bosons de $SU(2)_{L}$
externes. Comme la symétrie $SU(2)_{L}$ n'est pas brisée et comme
les leptoquarks $(X,Y)$ forment un doublet de $SU(2)_{L}$, on peut
utiliser (quasiment) la même base d'opérateurs que pour les gluons
(moyennant des changements évidents). En effectuant cela, nous retrouvons
encore les coefficients du tableau \ref{TableSU3}.
\item Finalement, nous avons aussi calculé les opérateurs effectifs impliquant
deux et quatre bosons de jauge de $U(1)_{Y}$, et on retrouve bien
les mêmes résultats que dans le tableau \ref{TableU1} pour les contributions
des $W^{\pm}$ aux opérateurs effectifs de photons dans la jauge non-linéaire.
\end{itemize}
Pour conclure cette partie, la même remarque à propos de l'interaction
d'Uehling concluant la partie sur les photons peut se faire pour les
opérateurs gluoniques EOM. Ces derniers ne jouent pas de rôle pour
des processus où les gluons sont on-shell mais contribuent dès lors
que d'autres champs comme par exemple des quarks légers restent actifs.
Dans ce cas, il est nécessaire d'inclure aussi tous les opérateurs
effectifs impliquant les champs de quarks. Bien que les opérateurs
EOM sont invariants de jauge par construction, leurs coefficients
de Wilson ne le sont pas. Par exemple, la jauge choisie pour les champs
de leptoquarks $X_{\mu}^{k}$ et $Y_{\mu}^{k}$ a une influence sur
leurs valeurs (l'équation (\ref{eq:GaugeDep}) reste valable pour
la polarisation du vide de QCD). D'un point de vue phénoménologique,
il n'y aurait pas vraiment de sens à considérer par exemple l'opérateur
$D^{\nu}G_{\nu\mu}^{a}D_{\rho}G^{a,\rho\mu}$ sans inclure tous les
opérateurs à 4-quarks. Dans le cas de $SU(5)$, il est clair que les
boucles de $X_{\mu}^{k}$ et $Y_{\mu}^{k}$ contribuent à la fois
à l'opérateur $D^{\nu}G_{\nu\mu}^{a}D_{\rho}G^{a,\rho\mu}$ et aux
opérateurs à 4-quarks, et seules leur combinaison donnerait un résultat
physique, invariant de jauge à l'ordre $\mathcal{O}(m^{-2})$. Mentionnons
aussi le fait que le coefficient dépendant de la jauge de l'opérateur
$D^{\nu}G_{\nu\mu}^{a}D_{\rho}G^{a,\rho\mu}$ rapporté dans le tableau
\ref{TableSU3}, est en accord avec la référence \cite{EffAction},
le calcul par la méthode CDE étant réalisé dans la même jauge non-linéaire.

\subsection{Généralisation aux bosons de jauges SU(N) }

L'étude réalisée dans le cas de la QCD peut être généralisée à des
représentations arbitraires d'autres groupes de Lie. Pour cela, il
suffit de remplacer les traces de générateurs de $SU(3)$ dans la
représentation fondamentale apparaissant dans chaque diagrammes 1PI
de la partie précédente par des traces de générateurs dans une représentation
générique $\mathbf{R}$. Les notations utilisées ainsi que plusieurs
résultats de théorie des groupes utiles ici sont collectés dans l'Annexe
\ref{AnnexeC}. De plus, nous nous baserons ici sur le groupe de jauge
spécial unitaire $SU(N)$ mais les résultats s'étendent trivialement
à d'autres algèbres de Lie.

D'abord, la polarisation du vide est proportionnelle à $\operatorname*{Tr}(T_{\mathbf{R}}^{a}T_{\mathbf{R}}^{b})=I_{2}(\mathbf{R})\delta^{ab}$
avec $I_{2}(\mathbf{R})$ l'invariant de Casimir quadratique. Par
conséquent, les coefficients $\alpha_{i}$ sont simplement obtenus
en multipliant ceux du tableau \ref{TableSU3} par un facteur $I_{2}(\mathbf{R})/I_{2}(\mathbf{F})=2I_{2}(\mathbf{R})$.
De la même manière, les diagrammes à 3-bosons sont proportionnels
à:
\begin{equation}
\operatorname*{Tr}(T_{\mathbf{R}}^{a}[T_{\mathbf{R}}^{b},T_{\mathbf{R}}^{c}])=iI_{2}(\mathbf{R})f^{abc}\;.\label{eq:TrTTT}
\end{equation}
Le fait que les amplitudes à deux et à trois bosons soient proportionnelles
au même invariant $I_{2}(\mathbf{R})$ assure une bonne correspondance
avec la théorie effective. En particulier, la divergence des diagrammes
à 3-bosons est correctement prise en compte par les couplages de $\mathfrak{L}_{eff}^{(2)}$. 

En ce qui concerne les amplitudes à 4-bosons, la situation est plus
délicate. Les diagrammes 1PI à une boucle, que ce soit dans le cas
des fermions, des scalaires ou des vecteurs, sont équivalents deux
à deux par renversement de l'impulsion circulant dans la boucle. C'est
pourquoi les amplitudes totales peuvent toujours s'écrire sous la
forme
\begin{equation}
\mathcal{M}^{abcd}=C_{1}^{abcd}\mathcal{M}_{1}+C_{2}^{abcd}\mathcal{M}_{2}+C_{3}^{abcd}\mathcal{M}_{3}\;,\ \ \left\{ \begin{array}{c}
C_{1}^{abcd}=\operatorname*{Tr}(T_{\mathbf{R}}^{a}T_{\mathbf{R}}^{b}T_{\mathbf{R}}^{d}T_{\mathbf{R}}^{c})+\operatorname*{Tr}(T_{\mathbf{R}}^{a}T_{\mathbf{R}}^{c}T_{\mathbf{R}}^{d}T_{\mathbf{R}}^{b})\;,\\
C_{2}^{abcd}=\operatorname*{Tr}(T_{\mathbf{R}}^{a}T_{\mathbf{R}}^{b}T_{\mathbf{R}}^{c}T_{\mathbf{R}}^{d})+\operatorname*{Tr}(T_{\mathbf{R}}^{a}T_{\mathbf{R}}^{d}T_{\mathbf{R}}^{c}T_{\mathbf{R}}^{b})\;,\\
C_{3}^{abcd}=\operatorname*{Tr}(T_{\mathbf{R}}^{a}T_{\mathbf{R}}^{c}T_{\mathbf{R}}^{b}T_{\mathbf{R}}^{d})+\operatorname*{Tr}(T_{\mathbf{R}}^{a}T_{\mathbf{R}}^{d}T_{\mathbf{R}}^{b}T_{\mathbf{R}}^{c})\;.
\end{array}\right.\label{eq:Decomp4p}
\end{equation}
En développant $\mathcal{M}^{abcd}$ en l'inverse de la masse de la
particule (lourde) circulant dans la boucle, seules deux combinaisons
de traces indépendantes interviennent aux ordres $\mathcal{O}(m^{0})$
et $\mathcal{O}(m^{-2})$ et ces dernières peuvent être exprimées
entièrement en termes d'invariants quadratiques de la manière suivante:
\begin{equation}
\begin{aligned}D_{1}^{abcd} & =2C_{1}^{abcd}-C_{2}^{abcd}-C_{3}^{abcd}=I_{2}(\mathbf{R})(2f^{ace}f^{bde}-f^{ade}f^{bce})\;,\\
D_{2}^{abcd} & =2C_{2}^{abcd}-C_{1}^{abcd}-C_{3}^{abcd}=I_{2}(\mathbf{R})(2f^{ade}f^{bce}-f^{ace}f^{bde})\;,\\
D_{3}^{abcd} & =2C_{3}^{abcd}-C_{1}^{abcd}-C_{2}^{abcd}=I_{2}(\mathbf{R})(-f^{ade}f^{bce}-f^{ace}f^{bde})=-D_{1}^{abcd}-D_{2}^{abcd}\;,
\end{aligned}
\end{equation}
où l'on a utilisé $[T_{\mathbf{R}}^{a},T_{\mathbf{R}}^{b}]=if^{abc}T_{\mathbf{R}}^{c}$,
l'équation (\ref{eq:TrTTT}) et l'identité de Jacobi $f^{abe}f^{cde}=f^{ace}f^{bde}-f^{ade}f^{bce}$.
Par suite, la correspondance entre $\mathcal{M}^{abcd}$ et l'amplitude
à 4-bosons calculée avec les couplages effectifs $\mathfrak{L}_{eff}^{(2)}$
et $\mathfrak{L}_{eff}^{(3)}$ aux ordres $\mathcal{O}(m^{0})$ et
$\mathcal{O}(m^{-2})$.

A l'ordre $\mathcal{O}(m^{-4})$, ces mêmes combinaisons de traces
$D_{1,2,3}^{abcd}$ induisent les opérateurs impliquant les constantes
de structures (associés aux coefficients de Wilson $\gamma_{4,5}$
et $\gamma_{4,6}$). Le reste est proportionnel à la trace totalement
symmétrisée:
\begin{equation}
D_{0}^{abcd}=C_{1}^{abcd}+C_{2}^{abcd}+C_{3}^{abcd}=\frac{1}{4}S\operatorname*{Tr}(T_{\mathbf{R}}^{a}T_{\mathbf{R}}^{b}T_{\mathbf{R}}^{c}T_{\mathbf{R}}^{d})\ .\label{eq:D01}
\end{equation}
Pour une algèbre $SU(N)$ générale (voir détail en Annexe \ref{AnnexeC}),
la trace totalement symétrisée se décompose en invariants quadratiques
et quartiques. En injectant la relation (\ref{eq:GenQuartic}) dans
l'équation (\ref{eq:D01}), on obtient:
\begin{equation}
D_{0}^{abcd}=6I_{4}(\mathbf{R})d^{abcd}+6\Lambda(\mathbf{R})(\delta^{ab}\delta^{cd}+\delta^{ac}\delta^{bd}+\delta^{ad}\delta^{bc})\;,\label{eq:D02}
\end{equation}
où $d^{abcd}$ est le symbole d'ordre quatre totalement symétrique,
normalisé tel que $I_{4}(\mathbf{F})=1$ pour la représentation fondamentale
et où 

\begin{equation}
\Lambda(\mathbf{R})=\left(\frac{N(\mathbf{A})I_{2}(\mathbf{R})}{N(\mathbf{R})}-\frac{I_{2}(\mathbf{A})}{6}\right)\frac{I_{2}(\mathbf{R})}{2+N(\mathbf{A})}\ ,\label{eq:ConvLambda}
\end{equation}
avec $\mathbf{A}$ la représentation adjointe et $N(\mathbf{R})$
la dimension de le représentation $\mathbf{R}$. Le terme proportionnel
à $\Lambda(\mathbf{R})$ correspond aux opérateurs associés aux coefficients
de Wilson de $\gamma_{4,1}$ jusqu'à $\gamma_{4,4}$ alors que le
terme proportionnel à $d^{abcd}$ nécessite d'étendre le lagrangien
effectif $\mathfrak{L}_{eff}^{(4)}$ (\ref{eq:EffSUNA}) en y ajoutant
deux opérateurs supplémentaires. Ainsi, le lagrangien effectif total
est le suivant:
\begin{equation}
\begin{aligned}\mathfrak{L}_{eff}^{(4)} & =\gamma_{4,1}\frac{g_{S}^{4}}{6!\pi^{2}m^{4}}G_{\mu\nu}^{a}G^{a,\mu\nu}G_{\rho\sigma}^{b}G^{b,\rho\sigma}+\gamma_{4,2}\frac{g_{S}^{4}}{6!\pi^{2}m^{4}}G_{\mu\nu}^{a}\tilde{G}^{a,\mu\nu}G_{\rho\sigma}^{b}\tilde{G}^{b,\rho\sigma}\\
 & +\gamma_{4,3}\frac{g_{S}^{4}}{6!\pi^{2}m^{4}}G_{\mu\nu}^{a}G^{b,\mu\nu}G_{\rho\sigma}^{a}G^{b,\rho\sigma}+\gamma_{4,4}\frac{g_{S}^{4}}{6!\pi^{2}m^{4}}G_{\mu\nu}^{a}\tilde{G}^{b,\mu\nu}G_{\rho\sigma}^{a}\tilde{G}^{b,\rho\sigma}\\
 & +\gamma_{4,5}\frac{g_{S}^{4}}{6!\pi^{2}m^{4}}f^{abe}f^{cde}G_{\mu\nu}^{a}G^{c,\mu\nu}G_{\rho\sigma}^{b}G^{d,\rho\sigma}+\gamma_{4,6}\frac{g_{S}^{4}}{6!\pi^{2}m^{4}}f^{abe}f^{cde}G_{\mu\nu}^{a}\tilde{G}^{c,\mu\nu}G_{\rho\sigma}^{b}\tilde{G}^{d,\rho\sigma}\\
 & +\gamma_{4,7}\frac{g_{S}^{4}}{6!\pi^{2}m^{4}}d^{abcd}G_{\mu\nu}^{a}G^{b,\mu\nu}G_{\rho\sigma}^{c}G^{d,\rho\sigma}+\gamma_{4,8}\frac{g_{S}^{4}}{6!\pi^{2}m^{4}}d^{abcd}G_{\mu\nu}^{a}\tilde{G}^{b,\mu\nu}G_{\rho\sigma}^{c}\tilde{G}^{d,\rho\sigma}\;.
\end{aligned}
\label{eq:EffSUNB}
\end{equation}
Le nombre total de huit opérateurs requis pour $SU(N)$ ainsi que
leur structure tensorielle quartique est en accord avec la référence
\cite{Gracey17_1,Gracey17_2}. Cependant, notons que la définition
de $\Lambda(\mathbf{R})$ est conventionnelle et le choix de la convention
affecte indirectement la définition de tous les opérateurs effectifs
sauf ceux associés à $\gamma_{4,3}$ et $\gamma_{4,6}$. La convention
de l'équation (\ref{eq:ConvLambda}) pour $\Lambda(\mathbf{R})$ adoptée
ici semble optimale car elle assure que $I_{4}(\mathbf{R})=0$ pour
toutes les représentations de $SU(2)$ et de $SU(3)$, comme il se
doit puisque ces algèbres ne possèdent pas de tenseur invariant irréductible
de rang quatre. Tous ces résultats restent valides pour les algèbres
$SO(N)$, à l'exception de $SO(8)$. En effet, comme cela est expliqué
dans l'Annexe \ref{AnnexeC}, $SO(8)$ à la propriété unique de posséder
deux symboles quartiques et en conséquence un terme supplémentaire
apparaît dans l'équation (\ref{eq:D02}). Dans ce cas, deux opérateurs
additionnels sont requis, dont les coefficients de Wilson sont proportionnels
au deuxième symbole quartique de l'équation (\ref{eq:SymTensor}).

\begin{figure}[h]
\begin{center}
\includegraphics[height=1.2272in,width=3.5189in]{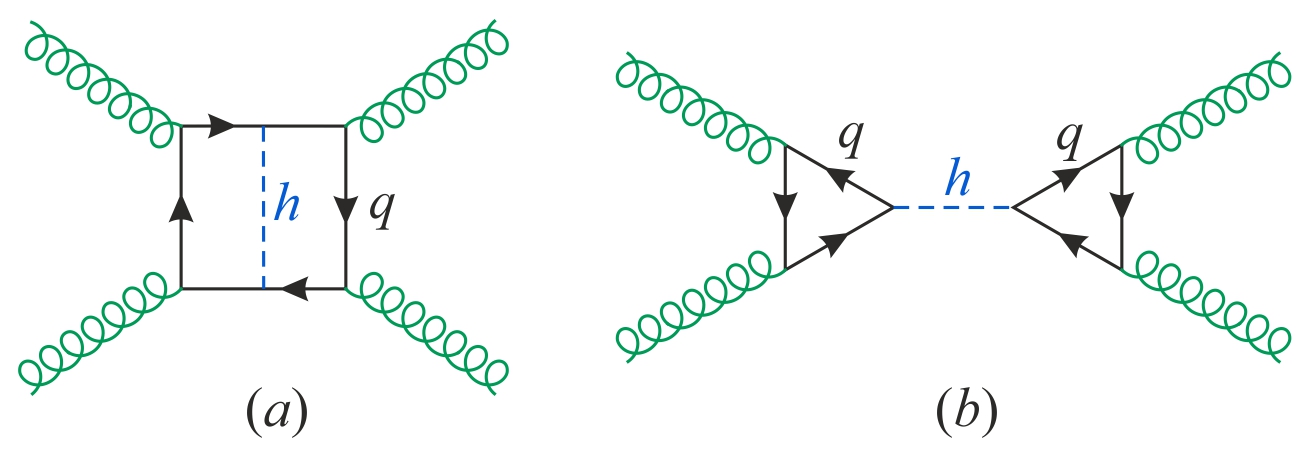}
\caption{Exemples de diagrammes à deux boucles dans le MS préservant ($a$) ou violant ($b$) les prédictions à une boucle de l'équation  (\ref{eq:CoeffCoh}) sur les opérateurs gluoniques. Les particules circulants dans la boucle sont des quarks lourds et les lignes pointillées représentent le boson de Higgs.}
\label{Fig2LH}
\end{center}
\end{figure}

En fait, bien que huit opérateurs indépendants (ou dix pour $SO(8)$)
peuvent être construits dans le cas général, les résultats de nos
calculs ont montrés qu'à une boucle, la plupart de ces opérateurs
découlent d'une même trace totalement symétrisée et sont donc toujours
corrélés. En particulier, quelque soit la représentation et le spin
de la particule circulant dans la boucle, on a:
\begin{equation}
\begin{aligned}\gamma_{4,1} & =\frac{1}{2}\gamma_{4,3}\ ,\\
\gamma_{4,2} & =\frac{1}{2}\gamma_{4,4}\ .
\end{aligned}
\label{eq:CoeffCoh}
\end{equation}
Il s'ensuit alors que deux combinaisons particulières d'opérateurs
ne se produisent jamais dans l'action effective à une boucle. D'un
point de vue de théorie effective, ceci devrait rester vrai dans la
plupart des cas dans la mesure où ce résultat découle de la symétrie
de l'amplitude. Au delà d'une boucle, une condition nécessaire pour
que ce résultat survive est l'absence de diagrammes ayant une discontinuité
du courant de couleur, c'est à dire lorsqu'un produit de traces intervient
au lieu d'une unique trace. Ce cas ne se produit jamais lorsqu'un
seul état lourd est intégré mais pourrait se réaliser dans des configurations
plus générales. Par exemple, dans le MS, si on intègre à la fois des
quarks lourds et le boson de Higgs, les diagrammes de la figure \ref{Fig2LH}
se présentent à deux boucles. 

Puisque le couplage effectif Higgs-gluon-gluon conservant CP est de
la forme $h^{0}G_{\mu\nu}^{a}G^{a,\mu\nu}$, il est clair que l'échange
de boson de Higgs dans la figure \ref{Fig2LH}b contribue à $\gamma_{4,1}$
mais pas à $\gamma_{4,3}$.

Les coefficients de Wilson pour un champ complexe (fermion, scalaire
ou vecteur) circulant dans les boucles sont donnés dans le tableau
\ref{TableSUN}. Si on considère une particule de Majorana (auto-conjuguée),
les valeurs des coefficients de Wilson seront la moitié de ceux rapportés
ici. En effet, quand le propagateur n'est pas orienté, certains diagrammes
de Feynman sont multipliés par un facteur de symétrie $1/2$ supplémentaire
alors que pour les autres, l'impulsion dans la boucle doit circuler
dans un seul sens et ne peut pas être renversée. Cette dernière situation
apporte également un facteur $1/2$ car on a $(T_{\mathbf{R}}^{a})^{T}=-T_{\mathbf{R}}^{a}$
pour une représentation réelle. Par exemple, au lieu de l'équation
(\ref{eq:TrTTT}), les diagrammes triangles sont maintenant proportionnels
à:
\begin{equation}
\left.\operatorname*{Tr}(T_{\mathbf{R}}^{a}T_{\mathbf{R}}^{b}T_{\mathbf{R}}^{c})\right\vert _{self-conjugate}=\frac{1}{2}\operatorname*{Tr}(T_{\mathbf{R}}^{a}[T_{\mathbf{R}}^{b},T_{\mathbf{R}}^{c}])=\frac{1}{2}iI_{2}(\mathbf{R})f^{abc}\ .
\end{equation}
Similairement, les coefficients pour l'amplitude à 4-points satisfont:
\begin{equation}
\left.C_{1}^{abcd}\right\vert _{self-conjugate}=\operatorname*{Tr}(T_{\mathbf{R}}^{a}T_{\mathbf{R}}^{b}T_{\mathbf{R}}^{d}T_{\mathbf{R}}^{c})=\frac{1}{2}(\operatorname*{Tr}(T_{\mathbf{R}}^{a}T_{\mathbf{R}}^{c}T_{\mathbf{R}}^{d}T_{\mathbf{R}}^{b})+\operatorname*{Tr}(T_{\mathbf{R}}^{a}T_{\mathbf{R}}^{c}T_{\mathbf{R}}^{d}T_{\mathbf{R}}^{b}))=\frac{1}{2}C_{1}^{abcd}\ .
\end{equation}
Nous avons vérifié cette propriété des coefficients de Wilson dans
deux cas pratiques pertinents physiquement, à savoir: les contributions
des bosons de Higgs $H_{G}^{a}$ de $SU(5)$ et des gluinos du MSSM
aux coefficients des opérateurs gluoniques, tout deux étant des champs
auto-conjugués dans la représentation adjointe de $SU(3)_{C}$.

\begin{table}[h] \centering 
$ 
\begin{tabular} [c]{ccccccc}\hline 

& $\alpha_{0}$ & $\alpha_{2}$ & $\alpha_{4}$ & $\beta_{2}$ & $\beta_{4,1}$ & $\beta_{4,2}$\smallskip\\\hline 

\multicolumn{1}{r}{Scalar} & $\dfrac{1}{2}I_{2}(\mathbf{R})D_{\varepsilon}$ & $-\dfrac{1}{8}I_{2}(\mathbf{R})$ & $\dfrac{3}{56}I_{2}(\mathbf{R})$ & $\dfrac{1}{24}I_{2}(\mathbf{R})$ & $-\dfrac{1}{14}I_{2}(\mathbf{R})$ & $0\rule[-0.18in]{0in}{0.4in}$\\ 

\multicolumn{1}{r}{Fermion} & $2I_{2}(\mathbf{R})D_{\varepsilon}$ & $-I_{2}(\mathbf{R})$ & $\dfrac{9}{14}I_{2}(\mathbf{R})$ & $-\dfrac{1}{12}I_{2}(\mathbf{R})$ & $\dfrac{1}{7}I_{2}(\mathbf{R})$ & $-\dfrac{3}{2}I_{2}(\mathbf{R})\rule[-0.18in]{0in}{0.4in}$\\ 

\multicolumn{1}{r}{Vector} & $-\dfrac{21D_{\varepsilon}+2}{2}I_{2}(\mathbf{R})$ & $\dfrac{37}{8}I_{2}(\mathbf{R})$ & $-\dfrac{159}{56}I_{2}(\mathbf{R})$ & $\dfrac{1}{8}I_{2}(\mathbf{R})$ & $-\dfrac{3}{14}I_{2}(\mathbf{R})$ & $6I_{2}(\mathbf{R})\rule[-0.18in]{0in}{0.4in}$\\\hline 

\multicolumn{1}{r}{} & $\gamma_{4,1}=\gamma_{4,3}/2$ & $\gamma_{4,2}=\gamma_{4,4}/2$ & $\gamma_{4,5}$ & $\gamma_{4,6}$ & $\gamma_{4,7}$ & $\gamma_{4,8}$\smallskip\\\hline 

\multicolumn{1}{r}{Scalar} & $\dfrac{7}{32}\Lambda(\mathbf{R})$ & $\dfrac {1}{32}\Lambda(\mathbf{R})$ & $\dfrac{1}{48}I_{2}(\mathbf{R})$ & $\dfrac {1}{336}I_{2}(\mathbf{R})$ & $\dfrac{7}{32}I_{4}(\mathbf{R})$ & $\dfrac{1}{32}I_{4}(\mathbf{R})\rule[-0.18in]{0in}{0.4in}$\\ 

\multicolumn{1}{r}{Fermion} & $\dfrac{1}{2}\Lambda(\mathbf{R})$ & $\dfrac {7}{8}\Lambda(\mathbf{R})$ & $\dfrac{1}{48}I_{2}(\mathbf{R})$ & $\dfrac {19}{336}I_{2}(\mathbf{R})$ & $\dfrac{1}{2}I_{4}(\mathbf{R})$ & $\dfrac{7}{8}I_{4}(\mathbf{R})\rule[-0.18in]{0in}{0.4in}$\\ 

\multicolumn{1}{r}{Vector} & $\dfrac{261}{32}\Lambda(\mathbf{R})$ & $\dfrac{243}{32}\Lambda(\mathbf{R})$ & $-\dfrac{3}{16}I_{2}(\mathbf{R})$ & $-\dfrac{27}{112}I_{2}(\mathbf{R})$ & $\dfrac{261}{32}I_{4}(\mathbf{R})$ & $\dfrac{243}{32}I_{4}(\mathbf{R})\rule[-0.18in]{0in}{0.4in}$\\\hline 

\end{tabular} 
$ 

\caption{Coefficients de Wilson des opérateurs effectifs pour les bosons de jauges de $SU(N)$ ou $SO(N\neq8)$, induits par des champs complexes de spin 0, 1/2 et 1 siégeant dans la représentation \textbf{R}. Pour des champs réels, tous les coefficients sont réduits de moitié.} 

\label{TableSUN} 

\end{table}

\subsubsection{Réductions vers les groupes $SU(3)$ et $SU(2)$}

\begin{figure}[t]
\begin{center}
\includegraphics[height=1.9026in,width=6.3261in]{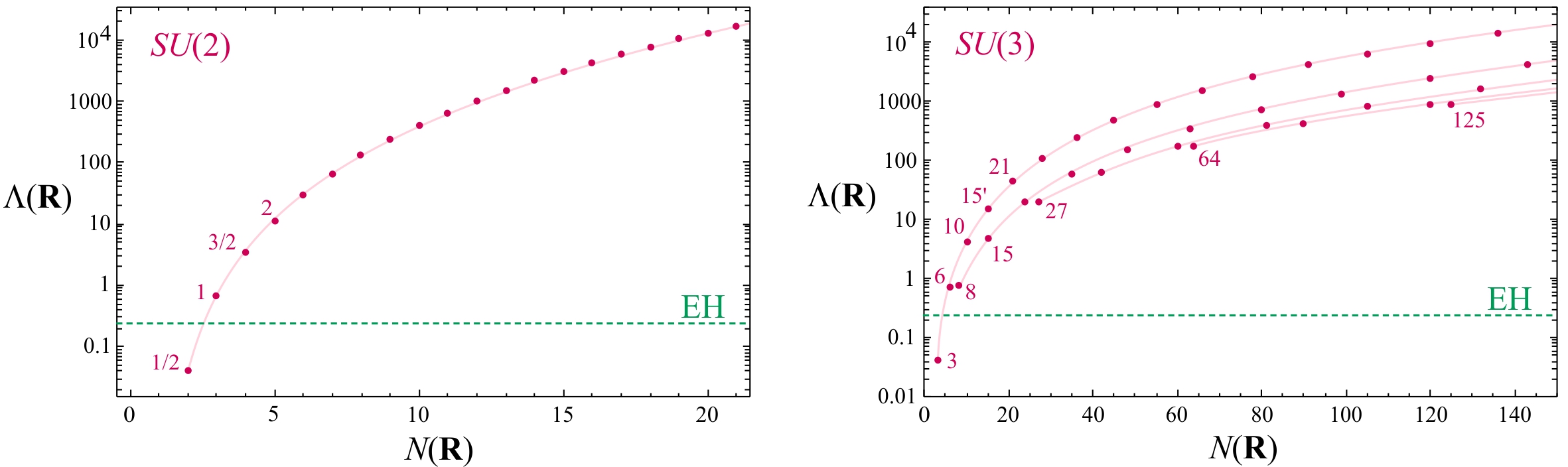}
\caption{Evolution de $\Lambda(\mathbf{R})$ en fonction de la dimension de la représentation $N(\mathbf{R})$ pour $SU(2)$ et $SU(3)$. Dans le premier cas, on note l'isospin correspondant des premières représentations. Dans le cas de $SU(3)$, plusieurs branches sont illustrées, chacunes commençant avec une représentation réelle. Les lignes horizontales en pointillé repèrent la valeur dans le cas d'Euler-Heinsenberg, identifiée a $\Lambda(\mathbf{1})=1/3$ pour une particule de charge unité dans la boucle à partir de l'équation (\ref{eq:ReducU1}).}
\label{LambaR}
\end{center}
\end{figure}

La base générale d'opérateurs effectifs se réduit immédiatement à
$SU(3)$ en retirant les opérateurs proportionnels aux invariants
quartiques, c'est à dire en annulant $\gamma_{4,7}$ et $\gamma_{4,8}$.
Pour la représentation fondamentale, on a $I_{2}^{SU(3)}(\mathbf{F})=1/2$
et $\Lambda^{SU(3)}(\mathbf{F})=1/24$ et on retrouve les résultats
du tableau \ref{TableSU3}. Cependant, un comportement intéressant
apparaît pour des représentations plus générales. A priori, lorsqu'on
augmente la dimension des représentations, on s'attendrait à ce que
l'intensité des couplages effectifs augmente mécaniquement du fait
du plus grand nombre de particules circulant dans la boucle. Toutefois,
nous montrons dans la figure \ref{LambaR} que $\Lambda(\mathbf{R})$
croît plus vite que $N(\mathbf{R})$. Les croissances les plus rapides
concernent les représentations qui sont des produits tensoriels symétriques
des représentations fondamentales, pour lesquelles on a: $\Lambda(\mathbf{R})\sim N(\mathbf{R})^{3}$.
Par exemple, $\Lambda(\mathbf{3})=1/24$ mais $\Lambda(\mathbf{6}=\mathbf{3}\otimes_{S}\mathbf{3})=17/24$,
$\Lambda(\mathbf{10}=\mathbf{3}\otimes_{S}\mathbf{3}\otimes_{S}\mathbf{3})=99/24$
et $\Lambda(\mathbf{15}=\mathbf{3}\otimes_{S}\mathbf{3}\otimes_{S}\mathbf{3}\otimes_{S}\mathbf{3})=371/24$.
La représentation adjointe n'en fait pas partie quoique les interactions
effectives sont quand même plus fortes qu'attendues naïvement par
l'argument de la dimension puisque $\Lambda^{SU(3)}(\mathbf{8})=3/4=18\times\Lambda^{SU(3)}(\mathbf{3})$.
Ceci correspond a des scénarios physiques sensibles comme par exemple
celui des gluinos dans le MSSM pour lesquels (en incluant le facteur
$1/2$ pour les particules auto-conjuguées):
\begin{equation}
\frac{1}{2}\times\frac{g_{S}^{4}}{6!\pi^{2}m_{\tilde{g}}^{4}}\gamma_{4,1}=-\frac{1}{2}\times\dfrac{1}{2}18\frac{g_{S}^{4}}{6!\pi^{2}m_{\tilde{g}}^{4}}=\frac{\alpha_{S}}{10m_{\tilde{g}}^{4}}\ ,
\end{equation}
, ce qui est un ordre de grandeur plus grand que le coefficient des
interactions effectives de photons dans le lagrangien d'Euler-Heisenberg.

Quant à $SU(2)$, le lagrangien effectif se simplifie grâce à l'identité
\begin{equation}
f^{abe}f^{cde}\rightarrow\varepsilon^{abe}\varepsilon^{cde}=\delta^{ac}\delta^{bd}-\delta^{ad}\delta^{bc}\ ,
\end{equation}
ce qui permet d'éliminer deux opérateurs. En exprimant les quatre
opérateurs restants explicitement en termes des états du triplet de
$SU(2)$, noté $\{W_{\mu}^{-},W_{\mu}^{3},W_{\mu}^{+}\}$, on obtient:
\begin{equation}
\begin{aligned}\mathfrak{L}_{eff,SU(2)_{L}}^{(4)} & =\frac{(\gamma_{4,1}+\gamma_{4,3})g^{4}}{6!\pi^{2}m^{4}}(W_{\mu\nu}^{3}W^{3,\mu\nu})^{2}+\frac{(\gamma_{4,2}+\gamma_{4,4})g^{4}}{6!\pi^{2}m^{4}}(W_{\mu\nu}^{3}\tilde{W}^{3,\mu\nu})^{2}\\
 & +\frac{4(\gamma_{4,1}+\gamma_{4,5})g^{4}}{6!\pi^{2}m^{4}}W_{\mu\nu}^{3}W^{3,\mu\nu}W_{\rho\sigma}^{+}W^{-,\rho\sigma}+\frac{4(\gamma_{4,2}+\gamma_{4,6})g^{4}}{6!\pi^{2}m^{4}}W_{\mu\nu}^{3}\tilde{W}^{3,\mu\nu}W_{\rho\sigma}^{+}\tilde{W}^{-,\rho\sigma}\\
 & +\frac{4(\gamma_{4,3}-\gamma_{4,5})g^{4}}{6!\pi^{2}m^{4}}|W_{\mu\nu}^{3}W^{+,\mu\nu}|^{2}+\frac{4(\gamma_{4,4}-\gamma_{4,6})g^{4}}{6!\pi^{2}m^{4}}|W_{\mu\nu}^{3}\tilde{W}^{+,\mu\nu}|^{2}\\
 & +\frac{2(2\gamma_{4,1}+\gamma_{4,3}+\gamma_{4,5})g^{4}}{6!\pi^{2}m^{4}}(W_{\mu\nu}^{+}W^{-,\mu\nu})^{2}+\frac{2(\gamma_{4,4}-\gamma_{4,6})g^{4}}{6!\pi^{2}m^{4}}|W_{\mu\nu}^{+}\tilde{W}^{+,\mu\nu}|^{2}\\
 & +\frac{2(\gamma_{4,3}-\gamma_{4,5})g^{4}}{6!\pi^{2}m^{4}}|W_{\mu\nu}{}^{+}W^{+,\mu\nu}|^{2}+\frac{2(2\gamma_{4,2}+\gamma_{4,4}+\gamma_{4,6})g^{4}}{6!\pi^{2}m^{4}}(W_{\mu\nu}^{+}\tilde{W}^{-,\mu\nu})^{2}\;.
\end{aligned}
\label{eq:LargEffSU2}
\end{equation}
Ces opérateurs et coefficients sont obtenus à partir de l'action effective
et sont indépendants de la masse invariante des particules externes.
Ainsi, ces derniers restent valides pour les bosons de jauge massifs
de l'interaction faible en pattes externes, du moins tant que $m$
est suffisamment grand devant $M_{Z,W}$. Cependant, une importante
mise en garde, pertinente pour le MS, concerne la présence de fermions
chiraux. En effet, ces derniers ne peuvent pas être massifs sans briser
la symétrie de jauge et donc le développement en l'inverse de la masse
n'est défini que dans la phase brisée. Dans ce cas, des opérateurs
non-invariants de jauge peuvent alors être présents aux ordres $\mathcal{O}(m^{0})$
et $\mathcal{O}(m^{-2})$.

Concernant l'intensité des couplages effectifs, ici aussi $\Lambda(\mathbf{R})$
croît bien plus vite que $N(\mathbf{R})$. En fait, comme les représentations
de $SU(2)$ sont plus petites que celles de $SU(3)$, la croissance
est plus prononcée avec $\Lambda(\mathbf{R})\sim N(\mathbf{R})^{5}$
(voir figure \ref{LambaR}). Alors que $\Lambda(\mathbf{F})=1/24$,
on gagne déjà un ordre de grandeur avec la représentation adjointe
$\Lambda(\mathbf{3})=2/3=16\times\Lambda(\mathbf{2})$.

Pour finir, il est instructif de se pencher sur ces résultats d'un
point de vue de théorie des groupes. Jusque là, les lagrangiens effectifs
pour $SU(2)$ et $SU(3)$ sont obtenus simplement en mettant $N=2$
ou $N=3$ dans le résultat général. Néanmoins, si $N$ est assez grand,
de sorte que $SU(N)$ puisse contenir une sous-algèbre $SU(2)$ ou
$SU(3)$, on pourrait alors se demander comment celle-ci se manifesterait
au niveau du lagrangien général $SU(N)$. Plus généralement, considérons
le lagrangien effectif pour une représentation $\mathbf{R}_{M}$ de
$SU(M)$. Ces $N(\mathbf{R}_{M})$ états s'organisent dans des représentations
de $SU(N)\subset SU(M)$, c'est à dire que $\mathbf{R}_{M}$ se décompose
en une somme directe de représentations $\mathbf{R}_{N}$ de $SU(N)$.
Par suite, du point de vue de $SU(N)$, les coefficients de $SU(M)$
encodent la circulation d'une collection d'états dans la boucle. Etant
donné que ces contributions s'ajoutent simplement, les coefficients
de $SU(M)$ doivent être la somme des coefficients de $SU(N)$ pour
toutes les représentations $\mathbf{R}_{N}$ présentes dans la représentation
$\mathbf{R}_{M}$. En repartant de l'équation (\ref{eq:D02}), on
doit alors avoir
\begin{equation}
\begin{aligned}\frac{1}{6}D_{0}^{abcd} & =I_{4}(\mathbf{R}_{M})d_{M}^{abcd}+\Lambda_{N}(\mathbf{R}_{M})(\delta^{ab}\delta^{cd}+\delta^{ac}\delta^{bd}+\delta^{ad}\delta^{bc})\\
 & =\sum_{\mathbf{R}_{N}\subset\mathbf{R}_{M}}I_{4}(\mathbf{R}_{N})d_{N}^{abcd}+\sum_{\mathbf{R}_{N}\subset\mathbf{R}_{M}}\Lambda_{N}(\mathbf{R}_{N})(\delta^{ab}\delta^{cd}+\delta^{ac}\delta^{bd}+\delta^{ad}\delta^{bc})\ ,
\end{aligned}
\end{equation}
où les indices $a,b,c,d$ dénotent les générateurs de $SU(M)$ qui
correspondent à la sous-algèbre $SU(N)$. La difficulté principale
est que même restreint à ces générateurs particuliers, $d_{M}^{abcd}\neq d_{N}^{abcd}$
car la définition de l'invariant quartique implique différentes fonctions
$\Lambda_{N}$ et $\Lambda_{M}$. Pour procéder, supposons que la
représentation fondamentale s'embranche comme ceci $\mathbf{F}_{M}\rightarrow\mathbf{F}_{N}$.
Sachant que $I_{4}(\mathbf{F}{}_{M})=I_{4}(\mathbf{F}_{N})=1$ par
définition, on a
\begin{equation}
\begin{aligned}I_{4}(\mathbf{R}_{M}) & =\sum_{\mathbf{R}_{N}\subset\mathbf{R}_{M}}I_{4}(\mathbf{R}_{N})\ ,\\
I_{4}(\mathbf{R}_{M})(\Lambda_{N}(\mathbf{F}_{N})-\Lambda_{M}(\mathbf{F}_{M}))+\Lambda_{M}(\mathbf{R}_{M}) & =\sum_{\mathbf{R}_{N}\subset\mathbf{R}_{M}}\Lambda_{N}(\mathbf{R}_{N})\ .
\end{aligned}
\label{eq:IdI4I2}
\end{equation}
En utilisant les nombres rapportés dans l'Annexe \ref{AnnexeC} ainsi
que les règles d'embranchements dans \cite{Slansky}, on peut vérifier
que les deux formules sont valides pour $SU(3)\subset SU(4)$ et $SU(4)\subset SU(5)$.
La seconde s'applique également à $SU(2)\subset SU(3)$, auquel cas
elle devient une règle de somme pour les fonctions $\Lambda$ vu que
$I_{4}(\mathbf{R})=0$ dans $SU(3)$. D'un point de vue calculatoire,
une fois que les règles d'embranchements des représentations de $SU(M)$
sont connues, ces équations sont particulièrement puissantes, la deuxième
permettant même de déterminer $I_{4}(\mathbf{R}_{M})$ en fonction
de $\Lambda_{N}$ et $\Lambda_{M}$, c'est à dire exprimé entièrement
en termes des invariants quadratiques $I_{2}(\mathbf{R}_{N})$ et
$I_{2}(\mathbf{R}_{M})$.

Grâce à la convention (\ref{eq:ConvLambda}), la règle d'embranchement
pour l'invariant $I_{4}$ est très simple \cite{Okubo81}, bien qu'il
y ait un prix à payer. Certaines parties des opérateurs $\gamma_{4,7}$
et $\gamma_{4,8}$ de $SU(M)$ sont déplacées dans les opérateurs
de $\gamma_{4,1}$ à $\gamma_{4,4}$ de $SU(N<M)$. Ceci est dû à
la définition première des opérateurs en termes de symboles quartiques
différents et non à la structure de boucle de l'amplitude ou encore
aux règles d'embranchements spécifiques. Par exemple, si pour un groupe
d'unification donné, on trouve un mécanisme spécifique qui génère
seulement $\gamma_{4,7}$ et $\gamma_{4,8}$, les quatre opérateurs
associés à $\gamma_{4,1}$ à $\gamma_{4,4}$ sont en général présents
dès lors que la symétrie est brisée spontanément simplement car le
symbole $d^{abcd}$ est défini différemment au sein de la sous-algèbre
survivante.

\subsubsection{Réduction vers $U(1)$}

En comparant les coefficients $\gamma_{4,i}$ du tableau \ref{TableSUN}
dans le cas de $SU(N)$ et les résultats dans le cas d'Euler-Heisenberg
dans le tableau \ref{TableU1}, il apparaît clairement que ces derniers
sont liés. De façon heuristique, il est simple de comprendre la relation
entre les deux cas en adaptant la décomposition (\ref{eq:Decomp4p})
pour $U(1)$. Lorsqu'un unique générateur intervient, on a $C_{1}=C_{2}=C_{3}=2Q^{4}$.
Ceci assure l'annulation de la divergence UV et plus généralement
l'absence de tous les opérateurs proportionnels aux constantes de
structures. L'amplitude totale est alors proportionnelle à
\begin{equation}
D_{0}\equiv C_{1}+C_{2}+C_{3}=6Q^{4}\ .\label{eq:D03}
\end{equation}
Comme le même facteur 6 apparaît dans le résultat (\ref{eq:D02})
pour $SU(N)$, il est clair que $\gamma_{4,1}^{EH}$ et $\gamma_{4,2}^{EH}$
peuvent être obtenus de façon équivalente à partir de $\gamma_{4,1}^{SU(N)}$
et $\gamma_{4,2}^{SU(N)}$ avec $\Lambda(\mathbf{R})\rightarrow Q^{4}$
ou bien à partir de $\gamma_{4,7}^{SU(N)}$ et $\gamma_{4,8}^{SU(N)}$
avec $I_{4}(\mathbf{R})\rightarrow Q^{4}$, en accord avec les tableaux
\ref{TableSUN} et \ref{TableU1}. Bien entendu, ce raisonnement est
une identification naïve des coefficients des fonctions de boucles
et non une réduction rigoureuse de $SU(N)$ vers un de ses sous-groupes
$U(1)$ basée sur la théorie des groupes.

Afin de réaliser proprement la réduction, notons $T^{a}$ un des générateurs
diagonaux de l'algèbre de Cartan de $SU(N)$. Ce générateur induit
un $U(1)_{\alpha}\subset SU(N)$ pour lequel le lagrangien effectif
$SU(N)$ se réduit à 
\begin{equation}
\begin{aligned}\mathfrak{L}_{eff}^{(4)}(U(1)_{\alpha}\subset SU(N)) & =(\gamma_{4,1}+\gamma_{4,3}+d^{\alpha\alpha\alpha\alpha}\gamma_{4,7})\frac{g_{S}^{4}}{6!\pi^{2}m^{4}}G_{\mu\nu}^{\alpha}G^{\alpha,\mu\nu}G_{\rho\sigma}^{\alpha}G^{\alpha,\rho\sigma}\\
 & +(\gamma_{4,2}+\gamma_{4,4}+d^{\alpha\alpha\alpha\alpha}\gamma_{4,8})\frac{g_{S}^{4}}{6!\pi^{2}m^{4}}G_{\mu\nu}^{\alpha}\tilde{G}^{\alpha,\mu\nu}G_{\rho\sigma}^{\alpha}\tilde{G}^{\alpha,\rho\sigma}\ .
\end{aligned}
\end{equation}
Le résultat dans le cas d'Euler-Heisenberg doit provenir d'une combinaison
de six des huit opérateurs $SU(N)$, incluant ceux qui impliquent
l'invariant quartique. En regardant à nouveau leurs valeurs pour une
représentation donnée $\mathbf{R}$ dans le tableau \ref{TableSUN},
cette réduction correspond aux résultats dans le tableau \ref{TableU1}
pour les cas scalaire, fermion et vecteur à condition de satisfaire
la seule condition suivante:
\begin{equation}
3\Lambda(\mathbf{R})+d^{\alpha\alpha\alpha\alpha}I_{4}(\mathbf{R})=\sum_{q_{\alpha}\in\mathbf{R}}q_{\alpha}^{4}\ .\label{eq:ReducU1}
\end{equation}
La somme dans le membre de droite est réalisée sur tous les états
vivant dans la représentation $\mathbf{R}$. Pour voir que cette condition
est vérifiée en général, il suffit de retourner à la définition première
de l'invariant quartique (voir Annexe \ref{AnnexeC}) qui pour un
unique générateur devient alors:
\begin{equation}
\frac{1}{4!}S\operatorname*{Tr}(T_{\mathbf{R}}^{\alpha}T_{\mathbf{R}}^{\alpha}T_{\mathbf{R}}^{\alpha}T_{\mathbf{R}}^{\alpha})=\operatorname*{Tr}((T_{\mathbf{R}}^{\alpha})^{4})=I_{4}(\mathbf{R})d^{\alpha\alpha\alpha\alpha}+3\Lambda(\mathbf{R})\ .
\end{equation}
Comme $T_{\mathbf{R}}^{\alpha}$ est diagonal, la trace se réduit
à une somme des puissances quartiques ses valeurs propres, c'est à
dire sur les puissances quartiques des charges $U(1)_{\alpha}$ des
états de la représentation $\mathbf{R}$. La dernière étape pour retrouver
le tableau \ref{TableU1} consiste à normaliser correctement le générateur
$T_{\mathbf{R}}^{\alpha}$ de sorte à avoir la charge $U(1)_{\alpha}$
en unités de $Q$. Notons que cette relation peut être trivialement
généralisée à d'autres invariants de Casimir. En particulier, pour
les opérateurs de dimension 4 et 6, $I_{2}(\mathbf{R})=\operatorname*{Tr}((T_{\mathbf{R}}^{\alpha})^{2})=\sum_{q_{\alpha}\in\mathbf{R}}q_{\alpha}^{2}$
montrant que les coefficients $\alpha_{i}$ pour $SU(N)$ se réduisent
à ceux pour la QED moyennant le remplacement naïf $I_{2}(\mathbf{R})\rightarrow Q^{2}$
dans le tableau \ref{TableSUN}.

Des applications numériques qui illustrent cette formule sont en Annexe
\ref{AnnexeC}. Remarquons que pour $SU(2)$ et $SU(3)$, il n'y a
pas d'invariant quartique et les coefficients d'Euler-Heisenberg pour
un état de charge unité sont formellement obtenus en mettant $\Lambda(\mathbf{1})=1/3$
dans l'équation (\ref{eq:ReducU1}). Cette valeur est représentée
sur la figure \ref{LambaR} pour comparaison.

\subsubsection{Réduction vers des groupes produits}

\begin{figure}[h]
\begin{center}
\includegraphics[height=1.1485in,width=3.0346in]{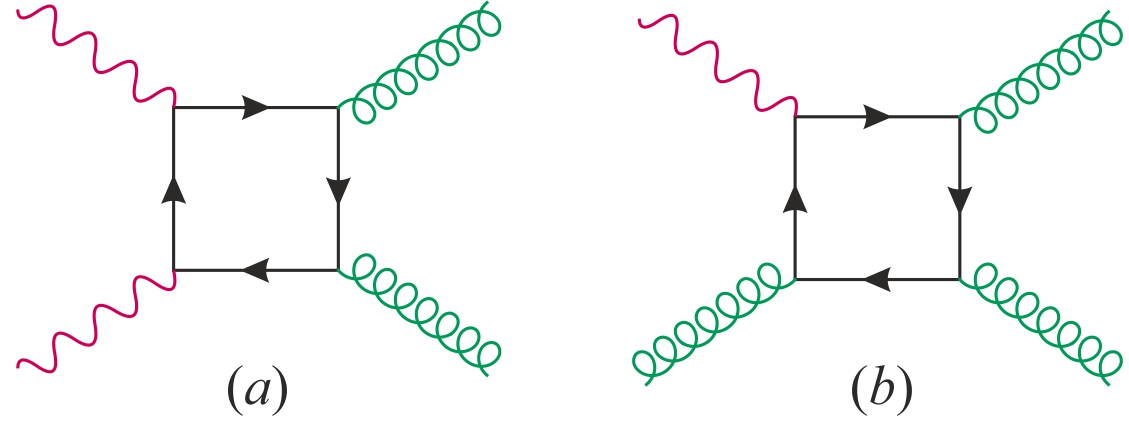}
\caption{Boucles de quarks générant les interactions effectives photon-gluon de dimension 8.}
\label{EHphotglue}
\end{center}
\end{figure}

Le résultat général peut aussi se réduire aux interactions mixtes,
c'est à dire impliquant les bosons de jauge de deux différentes algèbres.
Avant d'examiner cette réduction, calculons les directement à l'aide
des modèles FeynArts. Pour cela, considérons les interactions photon-gluon
induites par des boucles de quarks, de squarks ou de leptoquarks de
$SU(5)$ dans la jauge non-linéaire (voir figure \ref{EHphotglue}).

On peut alors généraliser aisément les résultats obtenus pour la représentation
fondamentale de $SU(3\text{)}_{C}$ à ceux pour des représentations
génériques de $SU(N)$. Les boucles sont convergentes et les interactions
effectives démarrent avec les opérateurs de dimension 8,
\begin{equation}
\begin{aligned}\mathfrak{L}_{eff}^{(4)}(U(1)\otimes SU(N)) & =\alpha_{1}\frac{g_{1}^{2}g_{n}^{2}}{6!\pi^{2}m^{4}}F_{\mu\nu}F^{\mu\nu}G_{\rho\sigma}^{a}G^{a,\rho\sigma}+\alpha_{2}\frac{g_{1}^{2}g_{n}^{2}}{6!\pi^{2}m^{4}}F_{\mu\nu}\tilde{F}^{\mu\nu}G_{\rho\sigma}^{a}\tilde{G}^{a,\rho\sigma}\\
 & +\alpha_{3}\frac{g_{1}^{2}g_{n}^{2}}{6!\pi^{2}m^{4}}F_{\mu\nu}G^{a,\mu\nu}F_{\rho\sigma}G^{a,\rho\sigma}+\alpha_{4}\frac{g_{1}^{2}g_{n}^{2}}{6!\pi^{2}m^{4}}F_{\mu\nu}\tilde{G}^{a,\mu\nu}F_{\rho\sigma}\tilde{G}^{a,\rho\sigma}\\
 & +\beta_{1}\frac{g_{1}g_{n}^{3}}{6!\pi^{2}m^{4}}d^{abc}F_{\mu\nu}G^{a,\mu\nu}G_{\rho\sigma}^{b}G^{c,\rho\sigma}+\beta_{2}\frac{g_{1}g_{n}^{3}}{6!\pi^{2}m^{4}}d^{abc}F_{\mu\nu}\tilde{G}^{a,\mu\nu}G_{\rho\sigma}^{b}\tilde{G}^{c,\rho\sigma}\;,
\end{aligned}
\label{eq:EffU1SUN}
\end{equation}
où $g_{1}$ et $g_{n}$ sont respectivement les constantes de couplages
associées à $U(1)$ et $SU(N)$. Les résultats obtenus pour les valeurs
des coefficients de Wilson sont rapportés dans le tableau \ref{TableMixed}. 

\begin{table}[t] \centering
$
\begin{tabular} [c]{ccccc}\hline
& $\alpha_{1}=\alpha_{3}/2$ & $\alpha_{2}=\alpha_{4}/2$ & $\beta_{1}$ & $\beta_{2}$\smallskip\\\hline
Scalar & $\dfrac{7}{16}Q(\mathbf{R})^{2}I_{2}(\mathbf{R})$ & $\dfrac{1}{16}Q(\mathbf{R})^{2}I_{2}(\mathbf{R})$ & $\dfrac{7}{32}Q(\mathbf{R})I_{3}(\mathbf{R})$ & $\dfrac{1}{32}Q(\mathbf{R})I_{3}(\mathbf{R})\rule[-0.18in]{0in}{0.4in}$\\
Fermion & $Q(\mathbf{R})^{2}I_{2}(\mathbf{R})$ & $\dfrac{7}{4}Q(\mathbf{R})^{2}I_{2}(\mathbf{R})$ & $\dfrac{1}{2}Q(\mathbf{R})I_{3}(\mathbf{R})$ & $\dfrac{7}{8}Q(\mathbf{R})I_{3}(\mathbf{R})\rule[-0.18in]{0in}{0.4in}$\\
Vector & $\dfrac{261}{16}Q(\mathbf{R})^{2}I_{2}(\mathbf{R})$ & $\dfrac{243}{16}Q(\mathbf{R})^{2}I_{2}(\mathbf{R})$ & $\dfrac{261}{32}Q(\mathbf{R})I_{3}(\mathbf{R})$ & $\dfrac{243}{32}Q(\mathbf{R})I_{3}(\mathbf{R})\rule[-0.18in]{0in}{0.4in}$\\\hline
\end{tabular}
$
\caption{Coefficients de Wilson  des opérateurs effectifs pour les interactions mixtes, induit par un champ complexe (scalaire, fermion ou vecteur) dans la représentation \textbf{R} de $SU(N)$ et avec une charge $U(1)$ de $Q(\mathbf{R})$. Les coefficients $\alpha_i$ pour deux bosons de jauge $SU(N)$ et deux bosons de jauge $SU(M)$ sont obtenus en remplaçant $Q(\mathbf{R})^{2}I_{2}(\mathbf{R}) \rightarrow I_{2}^{M}(\mathbf{R}_M) I_{2}^{N}(\mathbf{R}_N)$.}
\label{TableMixed}
\end{table}Ils sont invariants sous l'action de la conjugaison de charge en raison
de $Q(\mathbf{R}^{\ast})=-Q(\mathbf{R})$, $I_{2}(\mathbf{R}^{\ast})=+I_{2}(\mathbf{R})$
ainsi que $I_{3}(\mathbf{R}^{\ast})=-I_{3}(\mathbf{R})$, et ils s'annulent
pour une représentation réelle. Notons en particulier que les leptoquarks
de $SU(5)$ donnent $\beta_{i}<0$ étant donné que la charge électrique
de l'antitriplet est positive, $Q(\mathbf{\bar{3}})=+\sqrt{5/12}$.

Les quatre premières interactions sont immédiatement étendues aux
cas de deux bosons de jauge $SU(N)$ et deux bosons de jauge $SU(M)$.
Dans ce cas, les opérateurs sont alors
\begin{equation}
\begin{aligned}\mathfrak{L}_{eff}^{(4)}(SU(M)\otimes SU(N)) & =\alpha_{1}\frac{g_{m}^{2}g_{n}^{2}}{6!\pi^{2}m^{4}}W_{\mu\nu}^{i}W^{i,\mu\nu}G_{\rho\sigma}^{a}G^{a,\rho\sigma}+\alpha_{2}\frac{g_{m}^{2}g_{n}^{2}}{6!\pi^{2}m^{4}}W_{\mu\nu}^{i}\tilde{W}^{i,\mu\nu}G_{\rho\sigma}^{a}\tilde{G}^{a,\rho\sigma}\\
 & +\alpha_{3}\frac{g_{m}^{2}g_{n}^{2}}{6!\pi^{2}m^{4}}W_{\mu\nu}^{i}G^{a,\mu\nu}W_{\rho\sigma}^{i}G^{a,\rho\sigma}+\alpha_{4}\frac{g_{m}^{2}g_{n}^{2}}{6!\pi^{2}m^{4}}W_{\mu\nu}^{i}\tilde{G}^{a,\mu\nu}W_{\rho\sigma}^{i}\tilde{G}^{a,\rho\sigma}\ ,
\end{aligned}
\end{equation}
où $g_{m}$ et $g_{n}$ sont respectivement les constantes de couplages
associées à $SU(M)$ et $SU(N)$. En regardant la figure \ref{EHphotglue}$a$,
on se convainc aisément que les coefficients sont obtenus à partir
de ceux de $U(1)$ dans le tableau \ref{TableMixed} en remplaçant
$Q(\mathbf{R})^{2}I_{2}(\mathbf{R})\rightarrow I_{2}^{M}(\mathbf{R}_{M})I_{2}^{N}(\mathbf{R}_{N})$
quand les particules circulant dans la boucle sont dans la représentation
$(\mathbf{R}_{M},\mathbf{R}_{N})$ de $SU(M)\otimes SU(N)$.

Pour le MS, le cas $SU(2)_{L}\otimes SU(3)_{C}$ est immédiatement
obtenu dans la base $\{W_{\mu}^{-},W_{\mu}^{3},W_{\mu}^{+}\}$ en
effectuant les substitutions $W_{\mu\nu}^{i}W^{i,\mu\nu}=W_{\mu\nu}^{3}W^{3,\mu\nu}+2W_{\mu\nu}^{+}W^{-,\mu\nu}$
et $g_{n}\rightarrow g$, $g_{m}\rightarrow g_{S}$. Notons cependant
que la même mise en garde que pour les interactions effectives dans
(\ref{eq:LargEffSU2}) s'applique. En effet, en présence de fermions
chiraux, ces interactions ne sont pas dominantes et des opérateurs
de dimension 6 à l'ordre $\mathcal{O}(m^{-2})$ apparaissent, comme
par exemple $\tilde{G}_{\mu\nu}^{a}G^{a,\nu\rho}Z^{\mu\rho}$ ou encore
$Z_{\mu}Z_{\rho}G_{\mu\nu}^{a}G^{a,\rho\nu}$ induisant $Z\rightarrow ggg$
\cite{LaursenMS85} et $gg\rightarrow ZZ$ \cite{ZZgg1,ZZgg2}. Les
seules exceptions sont les interactions $Z\rightarrow gg\gamma$ \cite{LaursenSTS83}
et $Z\rightarrow\gamma\gamma\gamma$ \cite{Z3phot1,Z3phot2} pour
des gluons et des photons on-shell, qui commencent quand même à l'ordre
$\mathcal{O}(m^{-4})$ pour des fermions chiraux car le terme axial
du couplage entre le boson $Z$ et les fermions s'annule. Dans le
cas on-shell, ces interactions effectives sont simplement obtenues
à partir des résultats pour $\gamma\gamma\rightarrow gg$ et $\gamma\gamma\rightarrow\gamma\gamma$
en normalisant un couplage de photons de sorte à ce qu'il corresponde
avec celui du boson $Z$.

En raison du fait que $U(1)\otimes SU(N)\subset SU(M\geqslant N+1)$,
les coefficients $\alpha_{i}$ et $\beta_{i}$ dans le tableau \ref{TableMixed}
sont directement reliés aux $\gamma_{4,i}$ du tableau \ref{TableSUN},
ce qui n'est pas très surprenant en comparant leurs valeurs. Quant
à la réduction vers $U(1)$ de la partie précédente, elle peut être
comprise en analysant les coefficients des fonctions de boucles. Pour
les coefficients $\alpha_{i}$, la décomposition (\ref{eq:Decomp4p})
devient $C_{1}^{ab}=C_{2}^{ab}=C_{3}^{ab}=2I_{2}(\mathbf{R})Q^{2}\delta^{ab}$,
et donc $D_{0}^{ab}=6I_{2}(\mathbf{R})Q^{2}\delta^{ab}$. En comparant
avec l'équation (\ref{eq:D02}), on remarque que $\alpha_{i}=2\gamma_{4,i}$
avec le remplacement $\Lambda(\mathbf{R}_{M})\rightarrow Q(\mathbf{R}_{N})^{2}I_{2}(\mathbf{R}_{N})$
dans le tableau \ref{TableSUN}. Le facteur 2 provient des deux différentes
manières d'identifier les bosons de jauge de $U(1)$ et de $SU(N)$,
e.g. $(G_{\mu\nu}^{a}G^{a,\mu\nu})_{M}^{2}\rightarrow2(F_{\rho\sigma}F^{\rho\sigma})(G_{\mu\nu}^{a}G^{a,\mu\nu})_{N}$.
Un raisonnement similaire peut être mené pour les coefficients $\beta_{i}$.

Pour dépasser l'identification naïve des fonctions de boucles, notons
$T^{a}$ le générateur de Cartan de $SU(M)$ générant $U(1)$ et $\{T^{i}|i=2,...,N^{2}-1\}$
ceux générant $SU(N)$. Puisque $[T^{\alpha},T^{i}]=0$ implique $f^{\alpha ia}=0$,
les contributions UV divergentes disparaissent et les opérateurs associés
à $\gamma_{4,5}^{SU(M)}$ et $\gamma_{4,6}^{SU(M)}$ ne contribuent
pas aux interactions effectives de $U(1)\otimes SU(N)$. Quant aux
autres coefficients, considérons une représentation spécifique de
$SU(M)$ avec une règle d'embranchement $\mathbf{R}_{M}\rightarrow\sum\mathbf{R}_{N}$,
et notons $q_{\alpha}(\mathbf{R}_{N})$ les charges $U(1)_{\alpha}$
des états de la représentation $\mathbf{R}_{N}$. Mathématiquement,
cet embranchement signifie que $N^{2}$ des générateurs $T_{\mathbf{R}_{M}}$
de $SU(M)$ peuvent être ramenés sous une forme diagonale par blocs.
Ceux qui correspondent à $SU(N)$ possèdent des blocs contenant les
générateurs de $SU(N)$ dans la représentation $\mathbf{R}_{N}$,
alors que le générateur $T^{a}$ est une matrice diagonale contenant
toutes les charges $q_{\alpha}(\mathbf{R}_{N})$, ces dernières sont
constantes dans chaque bloc car $[T^{\alpha},T^{i}]=0$. La trace
totalement symétrisée impliquant deux ou trois générateurs de $SU(N)$
prend alors nécessairement le forme suivante:
\begin{equation}
\begin{aligned}\frac{1}{4!}S\operatorname*{Tr}(T_{\mathbf{R}}^{\alpha}T_{\mathbf{R}}^{\alpha}T_{\mathbf{R}}^{i}T_{\mathbf{R}}^{j}) & =\Lambda(\mathbf{R}_{M})\delta^{ij}+d^{\alpha\alpha ij}I_{4}(\mathbf{R}_{M})=\sum_{\mathbf{R}_{N}\subset\mathbf{R}_{M}}q_{\alpha}(\mathbf{R}_{N})^{2}I_{2}(\mathbf{R}_{N})\delta^{ij}\ ,\\
\frac{1}{4!}S\operatorname*{Tr}(T_{\mathbf{R}}^{\alpha}T_{\mathbf{R}}^{i}T_{\mathbf{R}}^{j}T_{\mathbf{R}}^{k}) & =d^{\alpha ijk}I_{4}(\mathbf{R}_{M})=\frac{1}{4}\sum_{\mathbf{R}_{N}\subset\mathbf{R}_{M}}q_{\alpha}(\mathbf{R}_{N})I_{3}(\mathbf{R}_{N})d^{ijk}\ .
\end{aligned}
\label{eq:BranchingI4}
\end{equation}
Ceci montre comment les coefficients $\alpha_{i}$ et $\beta_{i}$
de $U(1)\otimes SU(N)$ proviennent des coefficients $\gamma_{4,i}$
du lagrangien effectif général pour $SU(M\geqslant N+1)$. D'un point
de vue calculatoire, pour vérifier ces identités, il faut d'abord
établir la relation entre les symboles symétriques. En général, tout
ce qu'on peut dire à partir de la structure diagonale par blocs des
générateurs est que $d_{M}^{\alpha\alpha ij}=\eta_{1}\delta^{ij}$
et $d_{M}^{\alpha ijk}=\eta_{2}d_{N}^{ijk}$ (voir Annexe \ref{AnnexeC}),
bien que les facteurs de proportionnalité $\eta_{1}$ et $\eta_{2}$
dépendent de la manière avec laquelle $U(1)\otimes SU(N)$ est intégré
dans $SU(M)$. Ceci est illustré en Annexe \ref{AnnexeC}, où l'équation
(\ref{eq:BranchingI4}) est utilisée pour obtenir l'invariant de Casimir
quartique $I_{4}$ de $SU(5)$ à partir des coefficients d'anomalies
$I_{3}$ de $SU(3)$.

En tant que corollaire intéressant de cette réduction exacte, les
identités dans l'équation (\ref{eq:CoeffCoh}) restent valides et
impliquent que $\alpha_{1,2}=\alpha_{3,4}/2$. D'où le fait qu'il
y ait seulement deux opérateurs indépendants à une boucle, et ce quelque
soit le spin et la représentation de la particule circulant dans la
boucle. Comme précédemment, ceci n'est pas vrai en général si on intègre
plus d'un champ. Par exemple, l'analogue du diagramme d'échange du
boson de Higgs illustré en figure \ref{Fig2LH}$b$ contribue à $\alpha_{1}$
seulement car les couplages effectifs du boson de Higgs avec les photons
et les gluons sont de la forme $h^{0}F_{\mu\nu}F^{\mu\nu}$ et $h^{0}G_{\mu\nu}^{a}G^{a,\mu\nu}$.

\newpage\null\thispagestyle{empty}\newpage

\chapter*{Conclusion
\addstarredchapter{Conclusion}
\markboth{{Conclusion}}{Conclusion}}

Dans cette thèse, la notion de théorie effective des champs a été
mise au service de l'étude de la violation-$\mathcal{CP}$ dans un
premier temps, dans le MS et au-delà, à travers l'analyse des structures
de saveurs des EDMs, puis appliquée à différents secteurs de jauges
dans un deuxième temps, dans l'élaboration de l'action effective pour
des bosons de jauges. \newline

Dans la première étude, basée sur l'article \cite{ST-EDM}, les structures
de saveurs des EDMs dans le MS et au-delà ont été systématiquement
analysées en se reposant sur des outils et techniques inspirés de
MFV. Il est d'usage d'estimer les EDMs ou plus généralement la taille
de la violation-$\mathcal{CP}$ en utilisant des invariants de type
Jarlskog. Cependant, ceci est valide seulement pour des processus
dans lesquels la violation-$\mathcal{CP}$ survient dans une boucle
de fermions. Pour les contributions CKM aux EDMs de quarks, ou les
contributions PMNS aux EDMs de leptons, les diagrammes dominants ont
une topologie rainbow dont la structure ne se réduit pas à des invariants
de saveurs. La symétrie de saveur est bien adaptée pour étudier ces
diagrammes, qui avec l'aide des identités de Cayley-Hamilton, permet
d'identifier leurs structures de saveurs. De plus, l'étude combinée
des structures de type Jarlskog et de type rainbow nous éclaire sur
les possibles corrélations entre les EDMs de quarks et de leptons.
Curieusement, nous trouvons des comportements opposés pour des neutrinos
de Dirac ou de Majorana. Les EDMs de quarks et de leptons sont strictement
proportionnels dans le premier cas alors qu'ils deviennent largement
indépendants dans le second. En conséquence, l'EDM du quark est nécessairement
hors de portée dans le seesaw de type I étant donné que toutes les
grandes structures de saveurs sont de type Dirac. En revanche, les
EDMs de quarks pourraient être notre meilleure fenêtre sur le seesaw
de type II car la structure de saveur amplifiée de neutrino est de
type Majorana.

Tout au long de cette étude, une attention particulière a été portée
aux phases violant-$\mathcal{CP}$ de singlets de saveurs et en particulier
celles associées aux sous-groupes $U(1)$ du groupe de saveur, dont
plusieurs combinaisons s'avèrent être anomales dans le MS. La symétrie
de saveur a été adaptée pour traiter ce type de phases, en gardant
une trace de ces dernières dans les valeurs de fonds des spurions
de Yukawa ou des termes de masses de Majorana. Ceci permet de paramétrer
l'impact de l'interaction de violation-$\mathcal{CP}$ forte sur les
EDMs de quarks et de leptons ou encore d'analyser les relations entre
les phases de Majorana et des éventuelles interactions violant les
nombres leptoniques et/ou baryoniques.

Les méthodes développées dans cette étude peuvent aisément être adaptées
à des modèles plus compliqués, par exemple en présence de plusieurs
neutrinos stériles légers ou encore avec des structures de saveurs
additionnelles non-alignées avec celles des mécanismes de seesaw minimaux.
Dans ces contextes, il est important de ne pas considérer uniquement
les invariants de type Jarlskog mais également les structures de saveurs
non-invariantes. Etant en général bien moins supprimées, ces dernières
sont d'une importance phénoménologique primordiale et représentent
souvent notre seul accès à la physique sous-jacente.\newline

Dans la seconde étude, basée sur l'article \cite{QST-GBEFT}, l'action
effective pour les théories de jauges est revisitée. En intégrant
certains champs lourds chargés, les auto-interactions entre bosons
de jauges sont encodées dans des opérateurs effectifs. En adoptant
une approche diagrammatique, nous construisons explicitement ces interactions
jusqu'aux opérateurs de dimension-8 et nous calculons explicitement
leurs coefficients de Wilson, induits par des boucles de particules
lourdes de spin 0, 1/2 ou 1. Plus spécifiquement,
\begin{itemize}[label=\textbullet]
\item Pour poser le cadre et identifier les possibles complications, nous
revoyons d'abord en détails la construction des couplages effectifs
de photons hors de leur couche de masse (off-shell). Dans l'approche
diagrammatique, intégrer des fermions ou des scalaires ne pose pas
de problèmes particuliers et nous retrouvons le résultat usuel d'Euler-Heinsenberg.
Pour des champs de vecteurs lourds, la procédure de correspondance
(matching) pose bien plus de problèmes. En effet, dans la jauge de
’t Hooft-Feynman, le terme de fixation de jauge requis pour les champs
vectoriels massifs brise l'invariance de jauge $U(1)$. Par conséquent,
l'amplitude à quatre-photons off-shell ne satisfait plus les identités
de Ward de QED et la procédure usuelle pour construire l'action effective
échoue. Afin de résoudre ce problème, nous adoptons la stratégie de
\cite{Boudjema86} en quantifiant le MS en jauge non-linéaire. Le
matching est alors cohérent off-shell et l'approche diagrammatique
suit parallèlement la méthode CDE \cite{Gaillard1985uh,Cheyette1987qz}.
basée sur les intégrales de chemins. Les coefficients de Wilson dans
cette jauge sont calculés.
\item Le calcul de l'EFT pour photons a été ensuite étendu aux cas des gluons
de la QCD. La base la plus générale d'opérateurs gluoniques jusqu'à
la dimension-8 est assez différente de celle dans le cas de la QED
en raison de la nature non-abélienne de la QCD \cite{Gracey17_1,Gracey17_2}.
Nous calculons explicitement les coefficients de Wilson des opérateurs
effectifs pour les cas d'un scalaire, un fermion ou un vecteur dans
la représentation fondamentale. Comme dans le cas des photons, intégrer
des champs vectoriels lourds nécessite de gérer des dépendances de
jauges. Notre stratégie a été d'utiliser le modèle minimal $SU(5)$
GUT, brisé spontanément par un Higgs scalaire dans la représentation
adjointe en gardant la symétrie du MS conservée. Douze des bosons
de jauges de $SU(5\text{)}$ acquièrent une masses dans le processus
et ces derniers possèdent précisément les nombres quantiques désirés
afin d'induire les interactions effectives de gluons. Techniquement,
nous montrons également que la jauge non-linéaire possède la propriété
confortable de réduire drastiquement le nombre de diagrammes pour
un processus donné.
\item Ensuite, nous généralisons l'étude dans le cas de la QCD à des groupes
de Lie génériques comme groupes de jauges, en prenant $SU(N)$, $U(1)\otimes SU(N)$
et $SU(M)\otimes SU(N)$ comme exemples et en autorisant la particule
lourde à siéger dans des représentations arbitraires. Les coefficients
de Wilson pour un champ complexe de spin 0, 1/2 ou 1 circulant dans
les boucles sont calculés pour des bosons de jauges $SU(N)$ d'une
part et pour des groupes de jauges non-simples d'autre part. Suite
à ce calcul, une propriété apparente vaut la peine d'être soulignée:
à une boucle, certains opérateurs sont redondants quelque soit la
représentation ou le spin de la particule circulant dans les boucles.
Nous trouvons que deux combinaisons particulières d'opérateurs ne
se produisent jamais dans l'action effective à une boucle pour des
bosons de jauges $SU(N)$. Cela implique que seulement quatre opérateurs
au lieu de six sont requis en QCD et uniquement deux opérateurs au
lieu de quatre s'avèrent être suffisant pour décrire les interactions
mixtes de deux gluons et deux photons. Enfin, la généralisation du
résultat de QCD à une algèbre de Lie arbitraire nécessite une analyse
minutieuse des invariants quartiques de Casimir. Bien que toutes les
informations nécessaires peuvent être récoltées dans la littérature
\cite{Okubo81,vanRitbergenSV99}, il nous a semblé qu'une courte revue
détaillant toutes les définitions et conventions avec une emphase
particulière sur les calculs à boucles pratiques manquait, ce qui
a été fait en Annexe \ref{AnnexeC}.
\item D'un point de vue plus technique, la relation entre action effective
et correspondance de diagrammes de Feynman a été soigneusement analysée.
En effet, l'action effective peut être établie à partir des amplitudes
1PI à une boucle off-shell. De cette manière, les coefficients de
Wilson de tous les opérateurs, incluant ceux s'annulant sous l'application
de l'équation du mouvement, sont obtenus. Néanmoins, ces coefficients
ne sont pas nécessairement invariants de jauges. En fait, vu que la
correspondance n'est possible qu'en utilisant un terme de fixation
de jauge non-linéaire, ils ne sont bien définis que dans cette jauge.
Ceci est à comparer avec le calcul des coefficients pour des processus
on-shell, où les amplitudes à une boucle on-shell physique correspondent
sur un sous-ensemble d'opérateurs. Ces opérateurs qui s'annulent sous
l'équation du mouvement sont absents donc l'action effective totale
n'est jamais reproduite. En outre, d'un point de vue calculatoire,
effectuer la correspondance avec des processus on-shell nécessite
de gérer à la fois des amplitudes 1PI et non-1PI. Par exemple, le
coefficient de Wilson de l'opérateur à trois gluons $f^{abc}G_{\mu}^{a\;\nu}G_{\nu}^{b\;\rho}G_{\rho}^{c\;\mu}$
ne peut pas être obtenu à partir d'un processus à trois gluons car
il est interdit cinématiquement. A la place, il doit être extrait
avec tous les opérateurs à quatre tenseurs de gluons en les faisant
correspondre sur les amplitudes physiques à quatre gluons.
\end{itemize}
Finalement, la construction du lagrangien effectif pour bosons de
jauges jusqu'aux opérateurs de dimension-8 est maintenant totalement
sous contrôle dans l'approche diagrammatique. Les bases d'opérateurs
sont confirmées, leurs propriétés de théorie des groupes sont clarifiées
et les coefficients de Wilson sont connus pour les scénarios standards
de champs lourds scalaires, fermioniques ou vectoriels. Phénoménologiquement,
bien que les couplages effectifs à quatre bosons faibles ou à quatre
gluons ne seront probablement jamais observés, étant donné la présence
de tels couplages dans le lagrangien à l'arbre, il devrait y avoir
de la place pour le processus $\gamma\gamma\rightarrow gg$. En tout
cas, avoir élaboré une stratégie bien définie pour construire des
actions effectives générales impliquant des bosons de jauges s'avèrera
utile dans le futur.\newline

Depuis la découverte du boson de Higgs en 2012, aucune nouvelle particule
élémentaire n'a été observée au LHC. Néanmoins, de la nouvelle physique
n'implique pas forcément de nouvelles particules et peut émerger sous
forme de déviations par rapport aux prédictions du modèle standard
au fur et à mesure que les précisions des mesures augmentent. Récemment,
des déviations de la sorte ont été observées par les fabriques à mésons
B dans des processus impliquant des mésons beaux. Ces anomalies doivent
être clarifiées en étudiant d'autres canaux de désintégrations dans
des expériences futures, en particulier par Belle II au Japon qui
fonctionnera avec une luminosité plus élevée réduisant alors les incertitudes.
Si ces déviations venaient à être confirmées, elles constitueraient
alors la première observation de physique au-delà du modèle standard
dans des collisionneurs.

D'autre part, beaucoup d'espoirs sont placés dans les futurs grands
collisionneurs comme le \emph{Future Circular Collider (FCC)} qui
atteindra une énergie dans le centre de masse de 100 TeV ou encore
les améliorations du LHC afin d'en augmenter la luminosité \emph{(HL-LHC)}
ou encore l'énergie \emph{(HE-HLC). }D'un côté, l'augmentation de
luminosité permettra de gagner en précision et peut être observer
des déviations par rapport au modèle standard. Le HL-LHC vise en particulier
à étudier les auto-couplages du boson de Higgs, ce qui permettra de
sonder le potentiel scalaire et donc l'origine de la brisure spontanée
de la symétrie électrofaible qui se trouve par définition au-delà
du modèle standard. De l'autre côté, l'augmentation d'énergie permettra
d'explorer des territoires jusque-là complètement inconnus se déroulant
à des échelles d'énergies de plus en plus grande (ou de façon équivalente
des échelles de distances de plus en plus petites). Cette démarche
se trouve dans la continuité de l'aventure initiée il y a 27 siècles
par les penseurs grecs présocratiques concernant la quête des constituants
fondamentaux de la matière. Sur le chemin vers l'échelle de Planck,
de nouvelles particules élémentaires seront très probablement découvertes
et notre vision actuelle sera une nouvelle fois déconstruite pour
laisser place à un nouveau paradigme, comme cela l'a toujours été.
Cependant, mon avis personnel est que cette aventure se verra malheureusement
interrompue par la catastrophe écologique grandissante.

\appendix
\cleardoublepage
\phantomsection
\addcontentsline{toc}{chapter}{Annexes}

\chapter{Théorème de Cayley-Hamilton \label{AnnexeA}}

Le théorème de Cayley-Hamilton affirme que toute matrice $n\times n$
est solution de sa propre équation caractéristique. $\forall\mathbf{X}\in M_{n}(\mathbb{C}),$
\begin{equation}
p(\lambda)=\det\left[\mathbf{X}-\lambda1\right]\Rightarrow p(\mathbf{X})=0_{n},
\end{equation}
où $0_{n}$ est la matrice nulle $n\times n$. A première vue, on
peut penser que c'est trivial. C'est tentant d'écrire $p\left(\mathbf{X}\right)=\det\left[\mathbf{X}-\mathbf{X}\cdot\mathbf{1}\right]=0$,
mais ceci n'est pas correct car $\mathbf{X}
$ n'a du sens uniquement pour $\lambda\in\mathbb{R}$, et $\det\left[\mathbf{X}-\mathbf{X}\cdot\mathbf{1}\right]$
est un scalaire alors que $p\left(\mathbf{X}\right)$ doit être égal
à la matrice nulle. La démonstration dans le cas où $\mathbf{X}$
est diagonalisable est directe. Par définition, une valeur propre
$\lambda$ vérifie:
\begin{equation}
p\left(\lambda\right)=\lambda^{n}+c_{n-1}\lambda^{n-1}+...+c_{0}1=0.
\end{equation}
Par ailleurs, il existe un vecteur $\mathbf{v}$ tel que $\mathbf{X\cdot v}=\lambda\mathbf{v}$,
ce qui signifie:
\begin{equation}
\mathbf{X}^{n}\cdot\mathbf{v}+c_{n-1}\mathbf{X}^{n-1}\cdot\mathbf{v}+...+c_{0}\mathbf{1}\cdot\mathbf{v}=p\left(\lambda\right)\mathbf{v\;,}
\end{equation}
Comme ceci est vrai pour toutes les valeurs propres de $\mathbf{X}$,
nous pouvons en déduire $\mathbf{X}^{n}+c_{n-1}\mathbf{X}^{n-1}+...+c_{0}\mathbf{1}=\mathbf{0}$,
i.e., $p\left(\mathbf{X}\right)=\mathbf{0}$. Le théorème est bien
évidemment valable pour des matrices non diagonalisable mais nous
ne ferons pas la démonstration dans le cas général.

Considérons le cas des matrices $3\times3$ hermitiennes, les trois
valeurs propres $\lambda_{1,2,3}$ de $\mathbf{X}$ peuvent être exprimées
en fonction de traces et du déterminant de $\mathbf{X}$:
\begin{equation}
p\left(\mathbf{X}\right)=\left(\mathbf{X}-\lambda_{1}\mathbf{1}\right)\left(\mathbf{X}-\lambda_{2}\mathbf{1}\right)\left(\mathbf{X}-\lambda_{3}\mathbf{1}\right)=\mathbf{X}^{3}-\langle\mathbf{X}\rangle\mathbf{X}^{2}+\dfrac{1}{2}\mathbf{X}(\langle\mathbf{X}\rangle^{2}-\langle\mathbf{X}^{2}\rangle)-\det\mathbf{X}=0.\label{CH2}
\end{equation}
En prenant la trace de cette équation, on obtient:
\begin{equation}
\det\mathbf{X}=\frac{1}{3}\langle\mathbf{X}^{3}\rangle-\frac{1}{2}\langle\mathbf{X}\rangle\langle\mathbf{X}^{2}\rangle+\dfrac{1}{6}\langle\mathbf{X}\rangle^{3}.\label{eq:A30}
\end{equation}
D'autres identités supplémentaires peuvent être établie en exprimant
$\mathbf{X}=x_{1}
$ et en isolant une puissance donnée de $x_{1}\ensuremath{,}x_{2}$,$...$.
Par exemple, en prenant $\mathbf{X}=x\mathbf{X}
$, et en isolant $x^{2}y$:

\begin{equation}
\begin{aligned}\mathbf{X}^{2}\mathbf{Y}+\mathbf{XYX}+\mathbf{YX}^{2}-\langle\mathbf{X}{}^{2}\mathbf{Y}\rangle-\mathbf{X}^{2}\langle\mathbf{Y}\rangle-\langle\mathbf{X}\rangle\left(\mathbf{XY}+\mathbf{YX-}\langle\mathbf{XY}\rangle\right)\\
=\mathbf{X}\left(\langle\mathbf{XY}\rangle-\langle\mathbf{X}\rangle\langle\mathbf{Y}\rangle\right)-\dfrac{1}{2}(\mathbf{Y-}\langle\mathbf{Y}\rangle)\left(\langle\mathbf{X}\rangle^{2}-\langle\mathbf{X}{}^{2}\rangle\right)
\end{aligned}
.
\end{equation}
En combinant la définition du déterminant, $\varepsilon^{LMN}\mathbf{X}
$, avec le théorème de Cayley-Hamilton on obtient plusieurs identités
utiles. Le point de départ est l'équation (\ref{eq:A30}):

\begin{equation}
\varepsilon^{LMN}\mathbf{X}^{LI}\mathbf{X}^{MJ}\mathbf{X}^{NK}\equiv\varepsilon^{IJK}\det\mathbf{X}=\varepsilon^{IJK}[\;\tfrac{1}{3}\langle\mathbf{X}^{3}\rangle-\tfrac{1}{2}\langle\mathbf{X}\rangle\langle\mathbf{X}^{2}\rangle+\tfrac{1}{6}\langle\mathbf{X}\rangle^{3}\;]\;.\label{eq:A32}
\end{equation}
On peut établir des identités plus simples impliquant des traces et
des contractions antisymétriques en effectuant $\mathbf{X}\rightarrow\mathbf{1}+\mathbf{X}$
puis en développant et ensuite on isole les termes linéaires et quadratiques
en $\mathbf{X}$:
\begin{equation}
\begin{aligned}\varepsilon^{LJK}\mathbf{X}^{LI}+\varepsilon^{ILK}\mathbf{X}^{LJ}+\varepsilon^{IJL}\mathbf{X}^{LK} & =\varepsilon^{IJK}\langle\mathbf{X}\rangle\;,\\
\varepsilon^{LMK}\mathbf{X}^{LI}\mathbf{X}^{MJ}+\varepsilon^{LJM}\mathbf{X}^{LI}\mathbf{X}^{MK}+\varepsilon^{ILM}\mathbf{X}^{LJ}\mathbf{X}{}^{MK} & =\varepsilon^{IJK}\tfrac{1}{2}[\,\langle\mathbf{X}\rangle^{2}-\langle\mathbf{X}^{2}\rangle\,].
\end{aligned}
\end{equation}
D'autres identités utiles sont dérivées en multipliant la définition
du déterminant par $\mathbf{X}^{-1}$,
\begin{equation}
(\mathbf{X}^{-1})^{PK}\varepsilon^{IJP}\det\mathbf{X}=\varepsilon^{LMN}\mathbf{X}^{LI}\mathbf{X}^{MJ}\mathbf{X}^{NP}(\mathbf{X}^{-1})^{PK}=\varepsilon^{LMK}\mathbf{X}^{LI}\mathbf{X}^{MJ}.
\end{equation}
Le membre de gauche peut être simplifié en utilisant le théorème de
Cayley-Hamilton. En multipliant les deux membres de l'équation (\ref{CH2})
par $\mathbf{X}^{-1}$ nous donne: 
\begin{equation}
\varepsilon^{ILM}\mathbf{X}^{LJ}\mathbf{X}^{MK}=\varepsilon^{LJK}[\,\mathbf{X}^{2}-\langle\mathbf{X}\rangle\mathbf{X}+\tfrac{1}{2}\langle\mathbf{X}\rangle^{2}-\tfrac{1}{2}\langle\mathbf{X}^{2}\rangle\,]^{LI}.\label{eq:A35}
\end{equation}
Finalement, il y a aussi des identités impliquant plusieurs matrices.
Par exemple, en injectant $\mathbf{X}\rightarrow\mathbf{X}+\mathbf{Y}$
dans l'équation (\ref{eq:A35}), on peut dériver:
\begin{equation}
\varepsilon^{ILM}(\mathbf{X}^{LJ}\mathbf{Y}^{MK}+\mathbf{Y}^{LJ}\mathbf{X}^{MK})=\varepsilon^{LJK}\left[\,\{\mathbf{X},\mathbf{Y}\}-\langle\mathbf{X}\rangle\mathbf{Y}-\langle\mathbf{Y}\rangle\mathbf{X}+\langle\mathbf{X}\rangle\langle\mathbf{Y}\rangle-\langle\mathbf{XY}\rangle\,\right]^{LI}.
\end{equation}
La relation la plus générale impliquant trois matrices est obtenue
en effectuant $\mathbf{X}
$ dans l'équation (\ref{eq:A32}),
\begin{equation}
\varepsilon^{LMN}\{\mathbf{X},\mathbf{Y},\mathbf{Z}\}^{LI,MJ,NK}=\varepsilon^{IJK}\left[\langle\mathbf{XYZ}+\mathbf{ZYX}\rangle-\langle\mathbf{X}\rangle\langle\mathbf{ZY}\rangle-\langle\mathbf{Y}\rangle\langle\mathbf{XZ}\rangle-\langle\mathbf{Z}\rangle\langle\mathbf{XY}\rangle+\langle\mathbf{X}\rangle\langle\mathbf{Y}\rangle\langle\mathbf{Z}\rangle\right],
\end{equation}
où $\{\mathbf{X},\mathbf{Y},\mathbf{Z}\}^{a,b,c}\equiv\mathbf{X}
$. On peut particulariser cette relation en mettant des matrices à
l'identité ou en en égalant pour obtenir d'autres relations utiles.

\newpage

\chapter{La théorie de grande unification $SU(5)$ GUT quantifiée en jauge
non-linéaire\label{AnnexeB}}

\selectlanguage{english}%
Cette annexe n'est pas déstinée à être une revue du modèle minimal $SU(5)$. Il s'agit plutôt d'un guide de construction du lagrangien de $SU(5)$ brisé en $SU(3)_{C}\otimes SU(2)_{L}\otimes U(1)_{Y}$, quantifié en utilisant un terme de fixation de jauge non-linéaire et sous une forme appropriée pour l'utilisation d'outils de calculs informatiques. Le principal objectif de cette annexe est d'exprimer tous les termes du lagrangien d'une manière cohérente et traçable. Cela nécessite de spécifier un certain nombre de conventions et de définitions que nous avons pensé utiles de détailler ici.

On commence avec les bosons de jauge de $SU(5)$ et on va les exprimer en termes de ceux du groupe de jauge $SU(3)_{C}\otimes SU(2)_{L}\otimes U(1)_{Y}$. Pour cela, nous partons de la règle d'embranchement de la représentation adjointe $\mathbf{24}$:
\begin{equation}
\mathbf{24}=(\mathbf{8},\mathbf{1})_{0}+(\mathbf{1},\mathbf{3})_{0}%
+(\mathbf{3},\mathbf{2})_{5}+(\mathbf{\bar{3}},\mathbf{2})_{-5}+(\mathbf{1}%
,\mathbf{1})_{0}\;.
\end{equation}

En notant $A,B,...=1,...,24$ les indices de la représentation adjointe de $SU(5)$, $a,b,...=1,...,8$ les indices adjoints de couleurs, et $i,j,...=1,2,3$ les indices de la représentation fondamentale de $SU(3)$, les $24$ bosons de jauge $A_{A}^{\mu}$ sont identifiés comme l'octet de gluons $(\mathbf{8}\otimes\mathbf{1})_{0}\sim G_{i}^{\mu}=A_{i}^{\mu
}$, $a=1,...,8$, le triplet de bosons faibles $(\mathbf{1}\otimes\mathbf{3}%
)_{0}\sim W^{\pm\mu}=(A_{9}^{\mu}\mp iA_{10}^{\mu})/\sqrt{2}$, $W_{3}^{\mu
}=A_{11}^{\mu}$, ainsi que le singlet $(\mathbf{1}\otimes\mathbf{1})_{0}\sim
B^{\mu}=A_{24}^{\mu}$. Les champs restants correspondent respectivement aux $12$ leptoquarks et leurs champs conjugués dans les représentations $(\mathbf{\bar{3}}\otimes
\mathbf{2})_{5/3}$ et $(\mathbf{3}\otimes\mathbf{\bar{2}})_{-5/3}$. Nous définissons ces champs par $X_{1}^{\mu\pm
}=(A_{12}^{\mu}\pm iA_{13}^{\mu})/\sqrt{2}$, $Y_{1}^{\mu\pm}=(A_{18}^{\mu}\pm
iA_{19}^{\mu})/\sqrt{2}$, et ainsi de suite. Notons que les leptoquarks sont chargés sous tous les groupes de jauges du MS, et ceux possédant une hypercharge positive se transforment comme des antiquarks sous $SU(3)_{C}$.

Étant donné que la représentation adjointe est contenu dans $\mathbf{5}\otimes\mathbf{\bar{5}%
}=\mathbf{24}\oplus\mathbf{1}$, toutes ces identifications des champs de jauges peuvent être combinées afin de construire une matrice $5\times5$ de trace nulle pour les champs de jauges de  $SU(5)$:
\begin{equation}
\mathbf{A}^{\mu}=A_{A}^{\mu}T^{A}=\left(
\begin{array}
[c]{ccc}%
T_{ij}^{a}G_{\mu}^{a}-\frac{1}{\sqrt{15}}B_{\mu}\delta_{ij} & \frac{1}%
{\sqrt{2}}X_{\mu}^{i-} & \frac{1}{\sqrt{2}}Y_{\mu}^{i-}\\
\frac{1}{\sqrt{2}}X_{\mu}^{j+} & \frac{1}{2}W_{\mu}^{3}+\frac{3}{2\sqrt{15}%
}B_{\mu} & \frac{1}{\sqrt{2}}W_{\mu}^{+}\\
\frac{1}{\sqrt{2}}Y_{\mu}^{j+} & \frac{1}{\sqrt{2}}W_{\mu}^{-} & -\frac{1}%
{2}W_{\mu}^{3}+\frac{3}{2\sqrt{15}}B_{\mu}%
\end{array}
\right)  \,,
\label{SU5Boson}
\end{equation}
où les $T^{A}$ sont les générateurs conventionnels de $SU(5)$ dans la représentation fondamentale, normalisés par $\operatorname*{Tr}(T^{A}T^{B})=\delta^{AB}/2$. Cette identification est compatible avec les états propres de l'opérateur charge électrique,
\begin{equation}
Q=T^{11}+\sqrt{5/3}T^{24}\ ,\ \ \left[  Q,\mathbf{A}_{\mu}\right]  =\frac
{1}{\sqrt{2}}\left(
\begin{array}
[c]{ccc}%
0 & -4/3X_{\mu}^{i-} & -1/3Y_{\mu}^{i-}\\
4/3X_{\mu}^{j+} & 0 & +W_{\mu}^{+}\\
1/3Y_{\mu}^{j+} & -W_{\mu}^{-} & 0
\end{array}
\right)  \ ,
\end{equation}
avec la normalisation de l'opérateur hypercharge $Y=2\sqrt{5/3}T^{24}$. En pratique, nous avons utilisé les librairies Mathematica \textit{FeynArts}~\cite{FeynArts} et \textit{FeynCalc}~\cite{FeynCalc1,FeynCalc2}. Tous les deux permettent de garder implicite la sommation sur les indices $SU(3)$, donc   $\mathbf{A}^{\mu}$ est donné en entrée sous la forme de la matrice de l'équation~(\ref{SU5Boson}). Une fois que toutes les parties du lagrangien sont implémentées, il est relativement aisé d'extraire les règles de Feyman et de les exporter dans \textit{FeynArts}. Penchons nous maintenant sur les termes du lagrangien qui sont pertinents pour nous.

\subsection*{Les interactions de jauges}

Les auto-couplages de jauges sont obtenus à partir du terme cinétique de Yang-Mills
\begin{equation}
\mathcal{L}_{\text{gauge}}=-\frac{1}{2}\langle\mathbf{A}_{\mu\nu}%
\mathbf{A}^{\mu\nu}\rangle=-\frac{1}{4}A_{\mu\nu}^{A}A^{A,\mu\nu}\;,
\end{equation}
avec le tenseur de champs de jauge
\begin{equation}
\mathbf{A}_{\mu\nu}=\partial_{\mu}\mathbf{A}_{\nu}-\partial_{\nu}%
\mathbf{A}_{\mu}-ig_{5}[\mathbf{A}_{\mu},\mathbf{A}_{\nu}]=(\partial_{\mu
}A_{\nu}^{A}-\partial_{\nu}A_{\mu}^{A}+gf^{ABC}A_{\mu}^{B}A_{\nu}^{C})T^{A}\ .
\end{equation}
Les constantes de structures de $SU(5)$ sont définies par $[T^{A},T^{B}]=if^{ABC}T^{C}$. Un calcul explicite montre qu'il y a 68 $f^{ABC}$ non-nulles, avec les permutations  antisymétriques des indices. Neuf d'entre elles sont celles de $SU(3)$ et une constitue l'unique $\varepsilon^{ijk}$ de $SU(2)$, qui reproduisent les auto-interactions de la QCD et de l'interaction électrofaible. Toutes les autres constantes de structures non-nulles sont les $f^{ABC}$ avec $A,B=12,...,23$ et $C=1,...,11,24$. En d'autres termes, elles font intervenir deux fois les champs de leptoquarks, comme on pourrait s'y attendre puisque ces particules sont chargées sous les trois groupes de jauges du MS. La même constante de couplage $g_{5}$ intervient dans toutes les interactions entre bosons de jauges. Explicitement, on a
\begin{align}
\mathcal{L}_{\text{gauge}}  & =-\frac{1}{2}\langle(\partial_{\mu}%
\mathbf{A}_{\nu}-\partial_{\nu}\mathbf{A}_{\mu})(\partial^{\mu}\mathbf{A}%
^{\nu}-\partial^{\nu}\mathbf{A}^{\mu})+4ig_{5}\mathbf{A}_{\mu}\mathbf{A}_{\nu
}(\partial^{\mu}\mathbf{A}^{\nu}-\partial^{\nu}\mathbf{A}^{\mu})-2g_{5}%
^{2}\mathbf{A}_{\mu}\mathbf{A}_{\nu}[\mathbf{A}^{\mu},\mathbf{A}^{\nu}%
]\rangle\nonumber\\
& =-\frac{1}{4}G_{\mu\nu}^{a}G^{a,\mu\nu}-\frac{1}{2}W_{\mu\nu}^{+}W^{-,\mu
\nu}-\frac{1}{4}W_{\mu\nu}^{3}W^{3,\mu\nu}-\frac{1}{4}B_{\mu\nu}B^{\mu\nu
}\nonumber\\
& -\frac{1}{2}(D_{\mu}X_{\nu}^{+}-D_{\nu}X_{\mu}^{+})^{i}(D^{\mu}X^{-\nu
}-D^{\nu}X^{-\mu})^{i}-\frac{1}{2}(D_{\mu}Y_{\nu}^{+}-D_{\nu}Y_{\mu}^{+}%
)^{i}(D^{\mu}Y^{-\nu}-D^{\nu}Y^{-\mu})^{i}\nonumber\\
& \ \ \ +ig_{5}G_{\mu\nu}^{a}(X_{\mu}^{j+}(-T_{ji}^{a})X_{\nu}^{i-}+Y_{\mu
}^{j+}(-T_{ji}^{a})Y_{\nu}^{i-})+i\frac{g_{5}}{\sqrt{2}}(W^{+,\mu\nu}Y_{\mu
}^{i+}X_{\nu}^{i-}+W^{-,\mu\nu}X_{\mu}^{i+}Y_{\nu}^{i-})\nonumber\\
& \ \ \ +i\frac{g_{5}}{2}W^{3,\mu\nu}(X_{\mu}^{i+}X_{\nu}^{i-}-Y_{\mu}%
^{i+}Y_{\nu}^{i-})+ig_{5}\frac{\sqrt{15}}{6}B^{\mu\nu}(X_{\mu}^{i+}X_{\nu
}^{i-}+Y_{\mu}^{i+}Y_{\nu}^{i-})+\mathcal{O}((X,Y)^{4})\ ,
\end{align}
où les tenseurs de champs de jauges des interactions faibles et fortes contiennent implicitement leurs termes non-abéliens respectifs, comme suit
\begin{align*}
G_{\mu\nu}^{a}  & =\partial_{\nu}G_{\mu}^{a}-\partial_{\mu}G_{\nu}^{a}%
+g_{5}f^{abc}G_{\mu}^{b}G_{\nu}^{c}\rightarrow G_{\mu\nu}^{a}T^{a}%
=\partial_{\nu}G_{\mu}^{a}T^{a}-\partial_{\mu}G_{\nu}^{a}T^{a}-ig_{5}[G_{\mu
}^{b}T^{b},G_{\nu}^{c}T^{c}]\ ,\\
W^{i,\mu\nu}  & =\partial_{\nu}W_{\mu}^{i}-\partial_{\mu}W_{\nu}^{i}%
+g_{5}\varepsilon^{ijk}W_{\mu}^{j}W_{\nu}^{k}\ \rightarrow\left\{
\begin{array}
[c]{c}%
W^{3,\mu\nu}=\partial_{\nu}W_{\mu}^{3}-\partial_{\mu}W_{\nu}^{3}+ig_{5}%
(W_{\mu}^{-}W_{\nu}^{+}-W_{\mu}^{+}W_{\nu}^{-})\ ,\\
W^{+,\mu\nu}=\partial_{\nu}W_{\mu}^{+}-\partial_{\mu}W_{\nu}^{+}+ig_{5}%
(W_{\mu}^{+}W_{\nu}^{3}-W_{\mu}^{3}W_{\nu}^{+})\ ,\\
W^{-,\mu\nu}=\partial_{\nu}W_{\mu}^{-}-\partial_{\mu}W_{\nu}^{-}+ig_{5}%
(W_{\mu}^{3}W_{\nu}^{-}-W_{\mu}^{-}W_{\nu}^{3})\ .
\end{array}
\right.
\end{align*}
La dérivée covariante $D^{\mu}=\partial^{\mu}\mathbf{1}-ig_{5}T^{A}%
A_{A}^{\mu}$ agissant sur les douze leptoquarks siégeant dans la représentation $(\mathbf{\bar{3}}\otimes\mathbf{2})_{5/3}$ est donnée par
\begin{align}
(D_{\mu})_{ij}X_{\nu}^{j+}  & =\partial_{\mu}X_{\nu}^{i+}-ig_{5}\left(
X_{\nu}^{j+}(-T_{ji}^{a})G_{\mu}^{a}+\frac{1}{2}W_{\mu}^{3}X_{\nu}^{i+}%
+\frac{1}{\sqrt{2}}W_{\mu}^{+}Y_{\nu}^{i+}+y\frac{5}{6}B_{\mu}X_{\nu}%
^{i+}\right)  \ ,\\
(D_{\mu})_{ij}Y_{\nu}^{j+}  & =\partial_{\mu}Y_{\nu}^{i\pm}-ig_{5}\left(
Y_{\nu}^{j+}(-T_{ji}^{a})G_{\mu}^{a}-\frac{1}{2}W_{\mu}^{3}Y_{\nu}^{i+}%
+\frac{1}{\sqrt{2}}W_{\mu}^{-}X_{\nu}^{i+}+y\frac{5}{6}B_{\mu}Y_{\nu}%
^{i+}\right)  \ ,
\end{align}
où $y=\sqrt{3/5}$ est le facteur de normalisation de l'hypercharge. Enfin, $\mathcal{O}((X,Y)^{4})$ dénote les interactions quartiques entre les bosons de jauge $X$ et $Y$, auquels on ne s'intéressera pas ici. Il est intéressant de remarquer que l'invariance de jauge du MS est satisfaite séparément pour les termes cinétique de $X$ et $Y$ (grâce aux dérivées covariantes), les interactions magnétiques (les      $B_{\mu\nu}X^{\mu}X^{\nu}$ et les termes similaires), ainsi que  les interactions $\mathcal{O}((X,Y)^{4})$. Au niveau du MS, l'intensité des interactions magnétiques et $\mathcal{O}((X,Y)^{4}$ ne sont pas contraintes et celles-ci pourraient même être absentes. En revanche, ici, leurs intensités relatives sont fixées par l'invariance de jauge SU(5) sous-jacente. La situation est similaire dans le MS, avec les intensités relatives des interactions $(D_{\mu}W_{\nu}^{+}-D_{\nu}W_{\mu}^{+})(D^{\mu
}W^{-\nu}-D^{\nu}W^{-\mu})$ et $F_{\mu\nu}W_{\mu}^{+}W^{-\nu}$ qui sont fixées par la symétrie $SU(2)_{L}\otimes U(1)_{Y}$ sous-jacente.

\subsection*{Interactions de scalaires}

Dans cette étude, nous sommes uniquement intéressés par l'étape  initiale de brisure

\begin{equation}
SU(5)\rightarrow SU(3)_{C}\otimes SU(2)_{L}\otimes U(1)_{Y}\ .
\end{equation}
Pour cela, nous avons besoin d'un scalaire dans la représentation adjointe, $\mathbf{\bar{H}%
}_{\mathbf{24}}=\sqrt{2}H_{A}T_{A}$. Notons que $\mathbf{\bar{H}}_{\mathbf{24}%
}=\mathbf{\bar{H}}_{\mathbf{24}}^{\dagger}$, puisque l'adjointe est une représentation réelle, et en supposant une symétrie $\mathbf{\bar{H}}_{\mathbf{24}%
}\rightarrow-\mathbf{\bar{H}}_{\mathbf{24}}$ pour se débarrasser des interactions cubiques, le lagrangien le plus général est donné par
\begin{equation}
\mathcal{L}_{\text{scalar}}=\frac{1}{2}\langle D_{\mu}\mathbf{\bar{H}%
}_{\mathbf{24}}D^{\mu}\mathbf{\bar{H}}_{\mathbf{24}}\rangle+\frac{\mu^{2}}%
{2}\langle\mathbf{H}_{\mathbf{24}}^{2}\rangle-\frac{a}{4}\langle
\mathbf{H}_{\mathbf{24}}^{2}\rangle^{2}-\frac{b}{2}\langle\mathbf{H}%
_{\mathbf{24}}^{4}\rangle\ .\label{LagrScalar}%
\end{equation}
La brisure de la symétrie $SU(5)$ survient lorsque $\mathbf{\bar{H}%
}_{\mathbf{24}}$ prend sa valeur moyenne dans le vide (\textit{vev}) $\langle0|\mathbf{\bar{H}%
}_{\mathbf{24}}|0\rangle\sim v_{5}>0$, ce qui se réalise pour $\mu^{2}>0$. Il y a deux types de minimums dépendant du signe de $b$. D'abord, il est possible de trouver des valeurs de $\mu$, $a$, et $b<0$ tels que le minimum soit de la forme $\langle0|\mathbf{\bar{H}}_{\mathbf{24}}|0\rangle=\operatorname*{diag}%
(v,v,v,v,-4v)$. Ceci correspond à $SU(5)\rightarrow SU(4)\otimes U(1)$. Le second type de minimum se réalise lorsque $b>0$ et est tel que $\langle0|\mathbf{\bar{H}%
}_{\mathbf{24}}|0\rangle$ commute avec les générateurs de  $SU(3)_{C}$, $SU(2)_{L}$, et
$U(1)_{Y}$ :%
\begin{equation}
\mathbf{H}_{\mathbf{24}}^{0}=\langle0|\mathbf{\bar{H}}_{\mathbf{24}}%
|0\rangle=\frac{1}{\sqrt{2}}v_{5}\,\operatorname*{diag}%
(1,1,1,-3/2,-3/2)=-v_{5}\sqrt{15/4}T^{24},\ \;\;\;\;v_{5}^{2}=\dfrac{4\mu^{2}%
}{15a+7b}\;.
\end{equation}
La valeur de $v_{5}$ est déterminée en requièrant que ce soit un minimum global du potentiel, ce qui nécessite $15a+7b>0$.

En injectant cette contrainte dans le potentiel scalaire et en écrivant%
\begin{equation}
\mathbf{H}_{\mathbf{24}}=\mathbf{\bar{H}_{\mathbf{24}}}-\mathbf{H}%
_{\mathbf{24}}^{0}=\sqrt{2}\left(
\begin{array}
[c]{ccc}%
T_{ij}^{a}H_{G}^{a}-\frac{1}{\sqrt{15}}H_{B}^{0}\delta_{ij} & \frac{1}%
{\sqrt{2}}H_{X}^{i-} & \frac{1}{\sqrt{2}}H_{Y}^{i-}\\
\frac{1}{\sqrt{2}}H_{X}^{j+} & \frac{1}{2}H_{W}^{3}+\frac{3}{2\sqrt{15}}%
H_{B}^{0} & \frac{1}{\sqrt{2}}H_{W}^{+}\\
\frac{1}{\sqrt{2}}H_{Y}^{j+} & \frac{1}{\sqrt{2}}H_{W}^{-} & -\frac{1}{2}%
H_{W}^{3}+\frac{3}{2\sqrt{15}}H_{B}^{0}%
\end{array}
\right)  \;,
\end{equation}
les masses des bosons de Higgs sont données par
\[
M_{H_{W}^{i}}^{2}=4M_{H_{G}^{a}}^{2}=5bv_{5}^{2}\ ,\ \ M_{H_{B}}^{2}=2\mu
^{2}\ ,\ \ M_{H_{X,Y}^{i}}^{2}=0\ .
\]
Notons que le $\sqrt{2}$ est conventionnel, il assure que les termes cinétiques soient correctement normalisés étant donné   le lagrangien de l'équation (\ref{LagrScalar}). Des couplages supplémentaires impliquant trois et quatre scalaires sont obtenus à partir du potentiel, ces couplages sont tous proportionnels à $v_{5}$.

Pour obtenir les couplages scalaires aux bosons de jauge, il suffit alors de développer la dérivée covariante, avec pour la représentation adjointe,
\begin{equation}
D^{\mu}\mathbf{\bar{H}}_{\mathbf{24}}=\partial^{\mu}\mathbf{\bar{H}%
}_{\mathbf{24}}-ig_{5}\left[  \mathbf{A}^{\mu},\mathbf{\bar{H}}_{\mathbf{24}%
}\right]  =\partial^{\mu}\mathbf{H}_{\mathbf{24}}-ig_{5}\left[  \mathbf{A}%
^{\mu},\mathbf{H}_{\mathbf{24}}\right]  -ig_{5}\left[  \mathbf{A}^{\mu
},\mathbf{H}_{\mathbf{24}}^{0}\right]  \;.
\end{equation}
Cela donne
\begin{equation}
\frac{1}{2}\langle D_{\mu}\mathbf{\bar{H}}_{\mathbf{24}}D^{\mu}\mathbf{\bar
{H}}_{\mathbf{24}}\rangle\rightarrow\frac{1}{2}\langle\partial_{\mu}%
\mathbf{H}_{\mathbf{24}}\partial^{\mu}\mathbf{H}_{\mathbf{24}}\rangle
+\mathcal{L}_{\text{mass}}+\mathcal{L}_{\text{mix}}+\mathcal{L}%
_{\text{gauge-Higgs}}\ .
\end{equation}
Les couplages $\mathcal{L}_{\text{mass}}$ sont juste les termes de masses des leptoquarks,%
\begin{equation}
\mathcal{L}_{\text{mass}}=-\frac{1}{2}g_{5}^{2}\langle\left[  \mathbf{A}_{\mu
},\mathbf{H}_{\mathbf{24}}^{0}\right]  \left[  \mathbf{A}^{\mu},\mathbf{H}%
_{\mathbf{24}}^{0}\right]  \rangle=\frac{25}{16}g_{5}^{2}v_{5}^{2}\left(
X_{\mu}^{i+}X^{i-\mu}+Y_{\mu}^{i+}Y^{i-\mu}\right)  \ ,
\end{equation}
donc $M_{XY}=5g_{5}v_{5}/4$. La partie $\mathcal{L}_{\text{mix}}$ induit un mélange entre les bosons de jauge $X^{\mu}$ et $Y^{\mu}$  et leurs bosons de Goldstone associés,
\begin{equation}
\mathcal{L}_{\text{mix}}=-ig_{5}\langle\left[  \mathbf{A}_{\mu},\mathbf{H}%
_{\mathbf{24}}^{0}\right]  \partial^{\mu}\mathbf{H}_{\mathbf{24}}%
\rangle=iM_{XY}X_{\mu}^{k-}\partial^{\mu}H_{X}^{k+}+iM_{XY}Y_{\mu}%
^{k-}\partial^{\mu}H_{Y}^{k+}+h.c.\ .
\end{equation}
Les autres couplages impliquent des bosons de jauges et scalaires,%
\begin{equation}
\mathcal{L}_{\text{gauge-Higgs}}=-ig_{5}\langle\left[  \mathbf{A}_{\mu
},\mathbf{H}_{\mathbf{24}}\right]  \partial^{\mu}\mathbf{H}_{\mathbf{24}%
}\rangle-g_{5}^{2}\langle\left[  \mathbf{A}_{\mu},\mathbf{H}_{\mathbf{24}}%
^{0}\right]  \left[  \mathbf{A}^{\mu},\mathbf{H}_{\mathbf{24}}\right]
\rangle-\frac{g_{5}^{2}}{2}\langle\left[  \mathbf{A}_{\mu},\mathbf{H}%
_{\mathbf{24}}\right]  \left[  \mathbf{A}^{\mu},\mathbf{H}_{\mathbf{24}%
}\right]  \rangle\ .
\end{equation}
Les formes explicites peuvent être obtenues aisément et ne seront pas données ici. Remarquons cependant qu'en raison de  la disparition de tous les bosons de jauge du MS de $\left[
\mathbf{A}_{\mu},\mathbf{H}_{\mathbf{24}}^{0}\right]  $, $\mathcal{L}%
_{\text{AAH}}$ ne couplent que les scalaires aux bosons de jauge massifs, avec des couplages proportionnels à leurs masses.

\subsection*{Fixation de jauge et interactions de fantômes}

La prochaine étape pour quantifier cette théorie est de fixer la jauge et d'ajouter les termes fantômes (ghosts) correspondants. L'ansatz général dans la jauge $R_{\xi}$ linéaire est de définir la contrainte en termes des bosons de Goldstone associés comme ceci
\begin{align}
\mathbf{G}  & =\sqrt{2}\partial_{\mu}\mathbf{A}^{\mu}+\xi M_{XY}\left(
\begin{array}
[c]{ccc}%
0 & iH_{X}^{i-} & iH_{Y}^{i-}\\
-iH_{X}^{j+} & 0 & 0\\
-iH_{Y}^{j+} & 0 & 0
\end{array}
\right) \nonumber\\
& =\left(
\begin{array}
[c]{ccc}%
\sqrt{2}T_{ij}^{a}\partial^{\mu}G_{\mu}^{a}-\sqrt{\frac{2}{15}}\partial^{\mu
}B_{\mu}\delta_{ij} & \partial^{\mu}X_{\mu}^{i-}+i\xi M_{XY}H_{X}^{i-} &
\partial^{\mu}Y_{\mu}^{i-}+i\xi M_{XY}H_{Y}^{i-}\\
\partial^{\mu}X_{\mu}^{j+}-i\xi M_{XY}H_{X}^{j+} & \frac{1}{\sqrt{2}}%
\partial^{\mu}W_{\mu}^{3}+\sqrt{\frac{3}{10}}\partial^{\mu}B_{\mu} &
\partial^{\mu}W_{\mu}^{+}\\
\partial^{\mu}Y_{\mu}^{j+}-i\xi M_{XY}H_{Y}^{j+} & \partial^{\mu}W_{\mu}^{-} &
-\frac{1}{\sqrt{2}}\partial^{\mu}W_{\mu}^{3}+\frac{3}{2\sqrt{15}}\partial
^{\mu}B_{\mu}%
\end{array}
\right)  \ ,\label{Gphys}%
\end{align}
de sorte que
\begin{align}
\mathcal{L}_{\text{gf}}\overset{}{=}-\frac{1}{2\xi}\langle\mathbf{G}%
^{2}\rangle &  =-\frac{1}{\xi}|\partial^{\mu}X_{\mu}^{k+}-i\xi M_{XY}%
H_{X}^{k+}|^{2}-\frac{1}{\xi}|\partial^{\mu}Y_{\mu}^{k+}-i\xi M_{XY}H_{Y}%
^{k+}|^{2}\\
&  \ \ \ \ -\frac{1}{\xi}|\partial^{\mu}W_{\mu}^{+}|^{2}-\frac{1}{2\xi
}(\partial^{\mu}W_{\mu}^{3})^{2}-\frac{1}{2\xi}\left(  \partial^{\mu}B_{\mu
}\right)  ^{2}-\frac{1}{2\xi}(\partial^{\mu}G_{\mu}^{a})^{2}\ .
\end{align}
Vu qu'en pratique, tous nos calculs sont effectués dans la jauge de 't Hooft-Feynman, un paramètre commun $\xi$ est introduit pour tous les bosons de jauges. Évidemment, les paramètres pour $G_{\mu}^{a}$, $B_{\mu}$, $W_{\mu}^{3}$, et $W_{\mu}^{\pm}$ peuvent être tous différents étant donné qu'ils apparaissent seulement dans le propagateur respectif et non dans les vertex. Pour $X_{\mu}^{i\pm}$ et $Y_{\mu
}^{i\pm}$, ne pas prendre un paramètre commun compliquerait les choses puisque ces deux champs forment un doublet de $SU(2)_{L}$. Quand la première ligne est développée, les termes linéaires en $M_{XY}$ s'annulent précisément avec les termes dans $\mathcal{L}_{\text{mix}}$, tandis que les termes quadratiques impliquent $M_{H_{XY}}^{2}=\xi M_{XY}^{2}$. Rappelons que les bosons de Goldstone associés (Would-Be Goldstone, WBG) n'acquièrent aucun terme de masse par le potentiel scalaire.

Le but de la fixation de jauge non-linéaire de la Ref.~\cite{GavelaGMS81} est de maintenir les symétries de jauge non brisées de façon explicite. Cela nécessite des dérivées covariantes générales dans les contraintes impliquant les bosons de jauges massifs. Pour permettre l'interpolation entre la jauge linéaire et non-linéaire, nous introduisons les paramètres $\alpha_{G}$, $\alpha_{W}$, $\alpha_{B}$ et utilisons:
\begin{align}
\partial^{\mu}X_{\mu}^{i+}  & \rightarrow\partial^{\mu}X_{\mu}^{i+}%
-ig_{5}\left(  \alpha_{G}X_{\nu}^{j+}(-T_{ji}^{a})G_{\mu}^{a}+\alpha_{W}%
\frac{1}{2}W_{\mu}^{3}X_{\nu}^{i+}+\alpha_{W}\frac{1}{\sqrt{2}}W_{\mu}%
^{+}Y_{\nu}^{i+}+\alpha_{B}y\frac{5}{6}B_{\mu}X_{\nu}^{i+}\right)  \ ,\\
\partial^{\mu}Y_{\nu}^{i+}  & \rightarrow\partial^{\mu}Y_{\mu}^{i\pm}%
-ig_{5}\left(  \alpha_{G}Y_{\nu}^{j+}(-T_{ji}^{a})G_{\mu}^{a}-\alpha_{W}%
\frac{1}{2}W_{\mu}^{3}Y_{\nu}^{i+}+\alpha_{W}\frac{1}{\sqrt{2}}W_{\mu}%
^{-}X_{\nu}^{i+}+\alpha_{B}y\frac{5}{6}B_{\mu}Y_{\nu}^{i+}\right)  \ .
\end{align}
Injecter ceci dans $\mathcal{L}_{\text{gf}}$ génère de nouvelles contributions à $\mathcal{L}_{\text{gauge}}$ et $\mathcal{L}_{\text{gauge-Higgs}}$. A ce stade, un des intérêts de cette jauge devient apparent. Les lagrangiens de jauge et  de jauge-WBG de la partie précédente doivent être invariants sous la symétrie de jauge du MS. Cela signifie que parmi les interactions de bosons WBG-jauge-jauge, il y'a précisément  celles nécessaires pour promouvoir les dérivées dans $\mathcal{L}_{\text{mix}}$ en dérivées covariantes. Cependant, avoir des dérivées covariantes dans $\mathcal{L}_{\text{gf}}$ les annulent. Par conséquent, quand $\alpha_{i}=1$, les couplages $A-A-WBG$ deviennent plus simples. A cette contrainte correspond le lagrangien des fantômes
\begin{equation}
\mathcal{L}_{\text{ghost}}=c^{A\dagger}\left(  \left.  (-g_{5})\frac{\delta
G^{A}}{\delta\lambda^{B}}\right\vert _{\lambda=0}\right)  c^{B}\ .
\end{equation}
Pour obtenir les variations de $G^{A}$ sous une transformation de jauge, nous devons d'abord avoir celle des champs, exprimés dans la même base physique que les bosons de jauges et les scalaires WBG. Pour les champs de jauge, la variation sous une transformation de jauge est
\begin{equation}
\delta\mathbf{A}^{\mu}=\frac{1}{g_{5}}D_{\mu}\mathbf{\lambda}=\frac{1}{g_{5}%
}\partial^{\mu}\mathbf{\lambda}-i\left[  \mathbf{A}^{\mu},\mathbf{\lambda
}\right]  \ ,
\end{equation}
où les paramètres de la base physique sont définis par $\mathbf{\lambda}%
=\lambda^{A}T^{A}$ en complète analogie avec les bosons de jauge. Sous forme explicite, en reconstruisant les transformations des champs individuels, on obtient
\begin{align}
\delta G_{\mu}^{a}  & =\frac{1}{g_{5}}\partial^{\mu}\lambda_{G}^{a}%
+f^{abc}G_{\mu}^{b}\lambda_{G}^{c}+i(X_{\mu}^{i+}T_{ij}^{a}\lambda_{X}%
^{j-}-\lambda_{X}^{i+}T_{ij}^{a}X_{\mu}^{j-}+Y_{\mu}^{i+}T_{ij}^{a}\lambda
_{Y}^{j-}-\lambda_{Y}^{i+}T_{ij}^{a}Y_{\mu}^{j-})\ ,\\
\delta W_{\mu}^{+}  & =\frac{1}{g_{5}}\partial^{\mu}\lambda_{W}^{+}+iW_{\mu
}^{+}\lambda_{W}^{3}-iW_{\mu}^{3}\lambda_{W}^{+}+\frac{i}{\sqrt{2}}%
(\lambda_{X}^{i+}Y_{\mu}^{i-}-\lambda_{Y}^{i-}X_{\mu}^{i+})\ ,\ \delta W_{\mu
}^{-}=(\delta W_{\mu}^{+})^{\dagger}\ ,\\
\delta W_{\mu}^{3}  & =\frac{1}{g_{5}}\partial^{\mu}\lambda_{W}^{3}+iW_{\mu
}^{-}\lambda_{W}^{+}-iW_{\mu}^{+}\lambda_{W}^{-}+\frac{i}{2}(\lambda_{X}%
^{i+}X_{\mu}^{i-}-\lambda_{X}^{i-}X_{\mu}^{i+}-\lambda_{Y}^{i+}Y_{\mu}%
^{i-}+\lambda_{Y}^{i-}Y_{\mu}^{i+})\ ,\\
\delta B_{\mu}  & =\frac{1}{g_{5}}\partial^{\mu}\lambda_{B}+\frac{i}{2}%
\sqrt{\frac{5}{3}}(\lambda_{X}^{i+}X_{\mu}^{i-}-\lambda_{X}^{i-}X_{\mu}%
^{i+}+\lambda_{Y}^{i+}Y_{\mu}^{i-}-\lambda_{Y}^{i-}Y_{\mu}^{i+})\ ,\\
\delta X_{\mu}^{i+}  & =\frac{1}{g_{5}}\partial^{\mu}\lambda_{X}%
^{i+}-i(\lambda_{X}^{k+}T_{ki}^{a}G_{\mu}^{a}-X_{\mu}^{k+}T_{ki}^{a}%
\lambda_{G}^{a})+\frac{i}{\sqrt{2}}(\lambda_{W}^{+}Y_{\mu}^{i+}-\lambda
_{Y}^{i+}W_{\mu}^{+})\nonumber\\
& +\frac{i}{2}(\lambda_{W}^{3}X_{\mu}^{i+}-\lambda_{X}^{i+}W_{\mu}^{3}%
)+\frac{i}{2}\sqrt{\frac{5}{3}}(\lambda_{B}X_{\mu}^{i+}-\lambda_{X}^{i+}%
B_{\mu})\ ,\ \delta X_{\mu}^{i-}=(\delta X_{\mu}^{i+})^{\dagger}\ ,
\end{align}
\begin{align}
\delta Y_{\mu}^{i+}  & =\frac{1}{g_{5}}\partial^{\mu}\lambda_{Y}%
^{i+}+i(\lambda_{Y}^{k+}T_{ki}^{a}G_{\mu}^{a}-Y_{\mu}^{k+}T_{ki}^{a}%
\lambda_{G}^{a})+\frac{i}{\sqrt{2}}(\lambda_{W}^{-}X_{\mu}^{i+}-\lambda
_{X}^{i+}W_{\mu}^{-})\nonumber\\
& -\frac{i}{2}(\lambda_{W}^{3}Y_{\mu}^{i+}-\lambda_{Y}^{i+}W_{\mu}^{3}%
)+\frac{i}{2}\sqrt{\frac{5}{3}}(\lambda_{B}Y_{\mu}^{i+}-\lambda_{Y}^{i+}%
B_{\mu})\ \ ,\ \delta Y_{\mu}^{i-}=(\delta Y_{\mu}^{i+})^{\dagger}\ .
\end{align}
De manière similaire, la transformation des champs scalaires dans la représentation adjointe $\delta H^{A}=f^{ABC}H^{B}\lambda^{C}$ peut être obtenue sous forme matricielle%
\[
g\delta\mathbf{\bar{H}}_{24}=i[\mathbf{\lambda},\mathbf{\bar{H}}%
_{24}]\rightarrow\delta\mathbf{H}_{24}=i[\mathbf{\lambda},\mathbf{H}%
_{24}]+i[\mathbf{\lambda},\mathbf{H}_{24}^{0}]\ .
\]
Nous avons uniquement besoin de la règle de transformation des WGBs, étant donné que les autres champs scalaires ne seront pas introduits dans les contraintes de la jauge:%
\begin{align}
\delta H_{X}^{i+}  & =-iH_{X}^{k+}T_{ki}^{a}\lambda_{G}^{a}+\frac{i}{\sqrt{2}%
}\lambda_{W}^{+}H_{Y}^{i+}+\frac{i}{2}\lambda_{W}^{3}H_{X}^{i+}+\frac{i}%
{2}\sqrt{\frac{5}{3}}\lambda_{B}H_{X}^{i+}+i\frac{5}{4}v_{5}\lambda_{X}^{i+}\nonumber\\
& +i\lambda_{X}^{k+}T_{ki}^{a}H_{G}^{a}-\frac{i}{\sqrt{2}}H_{W}^{+}\lambda
_{Y}^{i+}-\frac{i}{2}H_{W}^{3}\lambda_{X}^{i+}-\frac{i}{2}\sqrt{\frac{5}{3}%
}H_{B}\lambda_{X}^{i+}\ ,\\
\delta H_{Y}^{i+}  & =-iH_{Y}^{k+}T_{ki}^{a}\lambda_{G}^{a}+\frac{i}{\sqrt{2}%
}\lambda_{W}^{-}H_{X}^{i+}-\frac{i}{2}\lambda_{W}^{3}H_{Y}^{i+}+\frac{i}%
{2}\sqrt{\frac{5}{3}}\lambda_{B}H_{Y}^{i+}+i\frac{5}{4}v_{5}\lambda_{Y}^{i+}\nonumber\\
& +i\lambda_{Y}^{k+}T_{ki}^{a}H_{G}^{a}-\frac{i}{\sqrt{2}}H_{W}^{-}\lambda
_{X}^{i+}+\frac{i}{2}H_{W}^{3}\lambda_{Y}^{i+}-\frac{i}{2}\sqrt{\frac{5}{3}%
}H_{B}\lambda_{Y}^{i+}\ .
\end{align}
Notons que ces règles de transformations impliquent que seuls les champs de fantômes associés aux bosons de jauges massifs se couplent à tous les bosons de Higgs, comme attendu par l'absence de couplages directs entre les champs scalaires et les bosons de jauges du MS.

Une fois que $\mathbf{G}$ est exprimée dans la base physique (comme dans l'équation (\ref{Gphys}) pour la jauge linéaire), les paramètres de jauges physiques identifiés à partir de $\mathbf{\lambda}$ et les matrices fantômes définies en complète analogie comme ceci
\begin{equation}
\mathbf{c}=c_{A}T^{A}=\left(
\begin{array}
[c]{ccc}%
T_{ij}^{a}c_{G}^{a}-\frac{1}{\sqrt{15}}c_{B}\delta_{ij} & \frac{1}{\sqrt{2}%
}c_{X}^{i-} & \frac{1}{\sqrt{2}}c_{Y}^{i-}\\
\frac{1}{\sqrt{2}}c_{X}^{j+} & \frac{1}{2}c_{W}^{3}+\frac{3}{2\sqrt{15}}c_{B}
& \frac{1}{\sqrt{2}}c_{W}^{+}\\
\frac{1}{\sqrt{2}}c_{Y}^{j+} & \frac{1}{\sqrt{2}}c_{W}^{-} & -\frac{1}{2}%
c_{W}^{3}+\frac{3}{2\sqrt{15}}c_{B}%
\end{array}
\right)  \,,
\end{equation}
on peut procéder en calculant $-\sqrt{2}g_{5}\langle\mathbf{c}^{\dagger
}\mathbf{\ }\delta\mathbf{G}\rangle$ et en remplaçant chacun des $\lambda$ par le fantôme correspondant, i.e., $\lambda_{B}\rightarrow c_{B}$, $\lambda_{G}%
^{a}\rightarrow c_{G}^{a}$, etc. Etant donné les nombreux couplages possibles dès que la fixation de jauge non-linéaire est imposée, l'expression finale est très longue et ne sera pas explicitée ici. Remarquons juste que seuls les fantômes associés aux leptoquarks deviennent massifs,%
\begin{align}
\mathcal{L}_{\text{ghost}}  & =c_{G}^{a\dagger}(-\partial^{2})c_{G}%
+c_{B}^{\dagger}(-\partial^{2})c_{B}+c_{W}^{3\dagger}(-\partial^{2})c_{W}%
^{3}+c_{W}^{\dagger+}(-\partial^{2})c_{W}^{-}+c_{W}^{\dagger-}(-\partial
^{2})c_{W}^{+}\nonumber\\
& +c_{X}^{\dagger+}(-\partial^{2}-\xi M_{XY}^{2})c_{X}^{-}+c_{X}^{\dagger
-}(-\partial^{2}-\xi M_{XY}^{2})c_{X}^{+}+c_{Y}^{\dagger+}(-\partial^{2}-\xi
M_{XY}^{2})c_{Y}^{-}+c_{Y}^{\dagger-}(-\partial^{2}-\xi M_{XY}^{2})c_{Y}%
^{+}\nonumber\\
& +\mathcal{L}_{\text{CCV}}+\mathcal{L}_{\text{CCH}}+\mathcal{L}_{\text{CCVV}%
}\ .
\end{align}
Néanmoins, les fantômes du MS développent de nouvelles interactions avec des paires d'états lourds (un fantôme, un boson de jauge). Notons également que $\mathcal{L}_{\text{CCVV}}$ provient entièrement de la fixation de jauge non-linéaire.

\selectlanguage{french}%
\newpage\null\thispagestyle{empty}\newpage

\chapter{Invariants de Casimir des algèbres de Lie standard\label{AnnexeC}}

Les constantes de structures d'une algèbre de Lie simple sont définies
par $[T_{\mathbf{R}}^{a},T_{\mathbf{R}}^{b}]=if^{abc}T_{\mathbf{R}}^{c}$,
avec $T_{\mathbf{R}}^{a}$ les générateurs dans la représentation
$\mathbf{R}$. Les invariants de Casimir quadratiques et cubiques
sont définis par la trace totalement symétrisée de deux et de trois
générateurs respectivement
\begin{equation}
\begin{aligned}\frac{1}{2!}S\operatorname*{Tr}(T_{\mathbf{R}}^{a}T_{\mathbf{R}}^{b}) & =\operatorname*{Tr}(T_{\mathbf{R}}^{a}T_{\mathbf{R}}^{b})\equiv I_{2}(\mathbf{R})d^{ab}\ ,\\
\frac{1}{3!}S\operatorname*{Tr}(T_{\mathbf{R}}^{a}T_{\mathbf{R}}^{b}T_{\mathbf{R}}^{c}) & =\frac{1}{2}\operatorname*{Tr}(T_{\mathbf{R}}^{a}\{T_{\mathbf{R}}^{b},T_{\mathbf{R}}^{c}\})\equiv\frac{1}{4}I_{3}(\mathbf{R})d^{abc}\ .
\end{aligned}
\end{equation}
En termes de ces deux invariants, on peut réduire la trace de trois
générateurs comme ceci
\begin{equation}
\operatorname*{Tr}(T_{\mathbf{R}}^{a}T_{\mathbf{R}}^{b}T_{\mathbf{R}}^{c})=\frac{1}{2}\operatorname*{Tr}([T_{\mathbf{R}}^{a},T_{\mathbf{R}}^{b}]T_{\mathbf{R}}^{c})+\frac{1}{2}\operatorname*{Tr}(\{T_{\mathbf{R}}^{a},T_{\mathbf{R}}^{b}\}T_{\mathbf{R}}^{c})=\frac{I_{3}(\mathbf{R})}{4}d^{abc}+\frac{iI_{2}(\mathbf{R})}{2}f^{abc}\ .\label{eq:TriTrace}
\end{equation}

L'invariant quadratique défini une métrique dans l'espace des générateurs.
$\operatorname*{Tr}(T_{\mathbf{R}}^{a}T_{\mathbf{R}}^{b})$ étant
définie positive, il est toujours possible de choisir une base pour
les générateurs telle que $d^{ab}=\delta^{ab}$. Par convention, les
générateurs sont de plus normalisés tels que $I_{2}(\mathbf{F})\equiv c$,
avec $\mathbf{F}$ la représentation définissante de dimension $N(\mathbf{F})=N$
et la constante $c$ usuellement fixée à $1/2$ ou $1$. Notons également
que dès que $d^{ab}=\delta^{ab}$, $T_{\mathbf{R}}^{a}T_{\mathbf{R}}^{a}$
devient proportionnel à l'identité, avec $T_{\mathbf{R}}^{a}T_{\mathbf{R}}^{a}=(N(\mathbf{A})I_{2}(\mathbf{R})/N(\mathbf{R}))\mathbf{1}_{N(\mathbf{R})\times N(\mathbf{R})}$
où $N(\mathbf{R})$ dénote la dimension de la représentation $\mathbf{R}$,
tandis que $\mathbf{A}$ est la représentation adjointe.

Le tenseur complètement symétrique $d^{abc}$ est normalisé tel que
$I_{3}(\mathbf{F})\equiv1$ pour les groupes unitaires. Il est absent
pour les groupes orthogonaux, hormis pour $SO(6)$ qui est isomorphe
à $SU(4)$. Lorsqu'il est défini, le coefficient $I_{3}(\mathbf{R})$
est souvent appelé le coefficient d'anomalie de la représentation
$\mathbf{R}$.

\subsection*{Symbole symétrique quartique}

Pour calculer des traces de quatre générateurs, nous avons besoin
d'étendre la base de sorte à y inclure le symbole symétrique quartique
et son invariant associé (pour plus d'information, voir \cite{vanRitbergenSV99}).
Elles ne sont pas données directement par la trace totalement symétrique
de quatre générateurs car le produit symétrisé de deux symboles symétriques
de second ordre est un tenseur symétrique invariant à quatre indices.
Pour spécifier, la décomposition la plus générale est:
\begin{equation}
\frac{1}{4!}S\operatorname*{Tr}(T_{\mathbf{R}}^{a}T_{\mathbf{R}}^{b}T_{\mathbf{R}}^{c}T_{\mathbf{R}}^{d})=I_{4}(\mathbf{R})d^{abcd}+\Lambda(\mathbf{R})(\delta^{ab}\delta^{cd}+\delta^{ac}\delta^{bd}+\delta^{ad}\delta^{bc})\ .\label{eq:GenQuartic}
\end{equation}
La constante $\Lambda(\mathbf{R})$ est conventionnelle, tandis que
$d^{abcd}$ est normalisé en fixant $I_{4}(\mathbf{F})=c$ pour une
constante arbitraire $c$. Afin de fixer $\Lambda(\mathbf{R})$, on
choisit de définir le tenseur $d^{abcd}$ comme étant orthogonal aux
invariants de rang inférieur, i.e., tel que $d_{ab}d_{cd}d^{abcd}=0$,
\begin{equation}
\begin{aligned}I_{4}(\mathbf{R})d_{ab}d_{cd}d^{abcd} & =\frac{1}{4!}\delta_{ab}\delta_{cd}S\operatorname*{Tr}(T_{\mathbf{R}}^{a}T_{\mathbf{R}}^{b}T_{\mathbf{R}}^{c}T_{\mathbf{R}}^{d})-\delta_{ab}\delta_{cd}\Lambda(\mathbf{R})(\delta^{ab}\delta^{cd}+\delta^{ac}\delta^{bd}+\delta^{ad}\delta^{bc})\\
 & =\operatorname*{Tr}(T_{\mathbf{R}}^{a}T_{\mathbf{R}}^{a}T_{\mathbf{R}}^{b}T_{\mathbf{R}}^{b})+\frac{1}{3}\operatorname*{Tr}(T_{\mathbf{R}}^{a}[T_{\mathbf{R}}^{b},T_{\mathbf{R}}^{a}]T_{\mathbf{R}}^{b})-\Lambda(\mathbf{R})(2+N(\mathbf{A}))N(\mathbf{A})\\
 & =\left(\frac{N(\mathbf{A})I_{2}(\mathbf{R})}{N(\mathbf{R})}-\frac{I_{2}(\mathbf{A})}{6}\right)I_{2}(\mathbf{R})N(\mathbf{A})-\Lambda(\mathbf{R})(2+N(\mathbf{A}))N(\mathbf{A})\ ,
\end{aligned}
\end{equation}
où l'on a utilisé $f^{abc}f^{dbc}=I_{2}(\mathbf{A})\delta^{ad}$,
$f^{abc}f^{abc}=I_{2}(\mathbf{A})N(\mathbf{A})$. Ainsi, $d_{ab}d_{cd}d^{abcd}$
s'annule à condition que
\begin{equation}
\Lambda(\mathbf{R})=\left(\frac{N(\mathbf{A})I_{2}(\mathbf{R})}{N(\mathbf{R})}-\frac{I_{2}(\mathbf{A})}{6}\right)\frac{I_{2}(\mathbf{R})}{2+N(\mathbf{A})}\ .\label{eq:LambdaDef}
\end{equation}
Cette convention assure que $d^{abcd}$ n'a pas de partie restante
proportionnelle au symbole quadratique. Ceci est particulièrement
pratique car $I_{4}(\mathbf{R})$ s'annule alors pour toute représentation
$\mathbf{R}$ de $SU(2)$ et $SU(3)$. On rappelle qu'un tenseur $d^{abcd}$
tel que $d_{ab}d_{cd}d^{abcd}=0$ n'existe pas pour $SU(N\leqslant3)$.
Pour $N=2,3$, $\Lambda(\mathbf{F})=1/24$ et
\begin{equation}
S\operatorname*{Tr}(T_{\mathbf{F}}^{a}T_{\mathbf{F}}^{b}T_{\mathbf{F}}^{c}T_{\mathbf{F}}^{d})\overset{N=2,3}{=}\delta^{ab}\delta^{cd}+\delta^{ac}\delta^{bd}+\delta^{ad}\delta^{bc}\ .
\end{equation}
Cette formule fournit également une identité utile pour les constantes
de structures de $SU(3)$
\begin{equation}
\frac{1}{4!}S\operatorname*{Tr}(T_{\mathbf{8}}^{a}T_{\mathbf{8}}^{b}T_{\mathbf{8}}^{c}T_{\mathbf{8}}^{b})=\frac{1}{4!}\sum_{perm(a,b,c,d)}f^{ax_{1}x_{2}}f^{bx_{2}x_{3}}f^{cx_{3}x_{4}}f^{dx_{4}x_{1}}=\frac{3}{4}(\delta^{ab}\delta^{cd}+\delta^{ac}\delta^{bd}+\delta^{ad}\delta^{bc})\ ,
\end{equation}
sachant que $\Lambda(\mathbf{8})=3/4$.

La formule (\ref{eq:GenQuartic}) est valable pour toute algèbre orthogonale
ou unitaire, hormis pour $SO(8)$. En effet, le symbole de Levi-Civita
de dimension $N$ est un invariant pour $SO(N)$, et lorsque $N$
est pair, il est possible de construire à partir de celui-ci un symbole
symétrique avec $N/2$ indices. Pour le voir, rappelons que la représentation
adjointe $\mathbf{A}$ de $SO(N)$ est obtenue comme étant le produit
tensoriel antisymétrique de la représentation de dimension $N$ définissante,
$\mathbf{F}$, $\mathbf{A}=\mathbf{F}\otimes_{A}\mathbf{F}$. Ainsi
les générateurs $SO(N)$ peuvent être indexées par les combinaisons
de deux indices antisymétriques $i,j=1,...,N$. Si on dénote $a=(i,j)$,
avec $a=1,...,N(N-1)/2$, alors
\begin{equation}
\Theta^{a_{1}...a_{N/2}}=\eta\varepsilon^{i_{1}...i_{N}}\ ,\label{eq:SymTensor}
\end{equation}
avec $\eta$ constante, est un tenseur invariant complètement symétrique
avec $N/2$ indices. Ceci explique un aspect de l'isomorphisme $SO(6)\sim SU(4)$.
Aucune des algèbres orthogonales ne possède un vrai symbole $d^{abc}$,
mais le tenseur invariant additionnel $\Theta^{abc}$ de $SO(6)$
correspond au symbole $d^{abc}$ de $SU(4)$. Pour $SO(8)$, $\Theta^{abcd}$
est un symbole quartique supplémentaire, orthogonal aux deux structures
tensorielles de l'équation (\ref{eq:GenQuartic}). Ainsi, la trace
complètement symétrique de quatre générateurs de $SO(8)$ ne se projète
pas juste sur deux mais trois structures tensorielles.

\subsection*{Réductions de traces d'ordre quatre}

Toute trace de quatre générateurs peut être réduite et exprimée entièrement
en termes des tenseurs invariants. Par exemple, pour $SU(N)$ et $SO(N\neq8)$,
on peut écrire
\begin{equation}
\begin{aligned}\frac{1}{4!}S\operatorname*{Tr}(T_{\mathbf{R}}^{a}T_{\mathbf{R}}^{b}T_{\mathbf{R}}^{c}T_{\mathbf{R}}^{d}) & =\frac{1}{6}\operatorname*{Tr}(T_{\mathbf{R}}^{a}T_{\mathbf{R}}^{b}T_{\mathbf{R}}^{c}T_{\mathbf{R}}^{d})+\frac{1}{6}\operatorname*{Tr}(T_{\mathbf{R}}^{a}T_{\mathbf{R}}^{b}T_{\mathbf{R}}^{d}T_{\mathbf{R}}^{c})+\frac{1}{6}\operatorname*{Tr}(T_{\mathbf{R}}^{a}T_{\mathbf{R}}^{c}T_{\mathbf{R}}^{b}T_{\mathbf{R}}^{d})\\
 & +\frac{1}{6}\operatorname*{Tr}(T_{\mathbf{R}}^{a}T_{\mathbf{R}}^{c}T_{\mathbf{R}}^{d}T_{\mathbf{R}}^{b})+\frac{1}{6}\operatorname*{Tr}(T_{\mathbf{R}}^{a}T_{\mathbf{R}}^{d}T_{\mathbf{R}}^{b}T_{\mathbf{R}}^{c})+\frac{1}{6}\operatorname*{Tr}(T_{\mathbf{R}}^{a}T_{\mathbf{R}}^{d}T_{\mathbf{R}}^{c}T_{\mathbf{R}}^{b})\\
 & =\operatorname*{Tr}(T_{\mathbf{R}}^{a}T_{\mathbf{R}}^{b}T_{\mathbf{R}}^{c}T_{\mathbf{R}}^{d})+\frac{2}{6}if^{dce}\operatorname*{Tr}(T_{\mathbf{R}}^{a}T_{\mathbf{R}}^{b}T_{\mathbf{R}}^{e})+\frac{3}{6}if^{cbe}\operatorname*{Tr}(T_{\mathbf{R}}^{a}T_{\mathbf{R}}^{e}T_{\mathbf{R}}^{d})\\
 & +\frac{2}{6}if^{dbe}\operatorname*{Tr}(T_{\mathbf{R}}^{a}T_{\mathbf{R}}^{c}T_{\mathbf{R}}^{e})+\frac{1}{6}if^{dbe}\operatorname*{Tr}(T_{\mathbf{R}}^{a}T_{\mathbf{R}}^{e}T_{\mathbf{R}}^{c})+\frac{1}{6}if^{dce}\operatorname*{Tr}(T_{\mathbf{R}}^{a}T_{\mathbf{R}}^{e}T_{\mathbf{R}}^{b})\\
 & =\operatorname*{Tr}(T_{\mathbf{R}}^{a}T_{\mathbf{R}}^{b}T_{\mathbf{R}}^{c}T_{\mathbf{R}}^{d})+i\frac{I_{3}(\mathbf{R})}{8}(f^{dce}d^{abe}+f^{cbe}d^{aed}+f^{dbe}d^{ace})\\
 & +\frac{I_{2}(\mathbf{R})}{12}f^{abe}f^{cde}-\frac{I_{2}(\mathbf{R})}{4}f^{ade}f^{bce}+\frac{I_{2}(\mathbf{R})}{12}f^{ace}f^{bde}\ .
\end{aligned}
\end{equation}
Ou, en introduisant l'invariant quartique:
\begin{equation}
\begin{aligned}\operatorname*{Tr}(T_{\mathbf{R}}^{a}T_{\mathbf{R}}^{b}T_{\mathbf{R}}^{c}T_{\mathbf{R}}^{d}) & =I_{4}(\mathbf{R})d^{abcd}-i\frac{I_{3}(\mathbf{R})}{8}(f^{dce}d^{abe}+f^{cbe}d^{aed}+f^{dbe}d^{ace})\\
 & -\frac{I_{2}(\mathbf{R})}{12}(f^{abe}f^{cde}-3f^{ade}f^{bce}+f^{ace}f^{bde})+\Lambda(\mathbf{R})(\delta^{ab}\delta^{cd}+\delta^{ac}\delta^{bd}+\delta^{ad}\delta^{bc})\ .
\end{aligned}
\end{equation}
Comme cas particuliers, on peut poser $I_{3}(\mathbf{R})=0$ pour
$SO(N\neq6)$, $I_{4}(\mathbf{R})=0$ pour $SU(3)$, et $I_{4}(\mathbf{R})=I_{3}(\mathbf{R})=0$
pour $SU(2)$. Notons que les deux derniers termes peuvent être écrit
sous une forme plus simple bien que moins symétrique à l'aide des
identités de Jacobi:
\begin{equation}
\begin{aligned}f^{cde}d^{abe}+f^{ade}d^{bce}+f^{bde}d^{ace} & =0\ ,\\
f^{abe}f^{cde}-f^{ace}f^{bde}+f^{ade}f^{bce} & =0\ .
\end{aligned}
\end{equation}
La première identité provient de $\operatorname*{Tr}(T_{\mathbf{A}}^{a}T_{\mathbf{A}}^{b}T_{\mathbf{A}}^{c})=\operatorname*{Tr}(T_{\mathbf{A}}^{a}[T_{\mathbf{A}}^{b},T_{\mathbf{A}}^{c}])/2$
en raison de $(T_{\mathbf{A}}^{a})^{T}=-T_{\mathbf{A}}^{a}$ pour
une représentation réelle.

Dans le cas de $SU(N)$, il y a une autre façon d'obtenir le symbole
symétrique d'ordre quatre. D'abord, rappelons que,
\begin{equation}
T_{\mathbf{F}}^{a}T_{\mathbf{F}}^{b}=\frac{1}{N}I_{2}(\mathbf{F})\delta^{ab}+\frac{I_{3}(\mathbf{F})}{4I_{2}(\mathbf{F})}d^{abc}T_{\mathbf{F}}^{c}+\frac{i}{2}I_{2}(\mathbf{F})f^{abc}T_{\mathbf{F}}^{c}\;.
\end{equation}
Avec ceci, on peut établir
\begin{equation}
Tr\left[\{T_{\mathbf{F}}^{a}T_{\mathbf{F}}^{b}\}\{T_{\mathbf{F}}^{c}T_{\mathbf{F}}^{d}\}\right]=\frac{4I_{2}(\mathbf{F})^{2}}{N}\delta^{ab}\delta^{cd}+\frac{I_{3}(\mathbf{F})^{2}}{4I_{2}(\mathbf{F})}d^{abe}d^{cde}=\frac{1}{N}\delta^{ab}\delta^{cd}+\frac{1}{2}d^{abe}d^{cde}\ .
\end{equation}
D'autre part, cette trace peut être calculée en utilisant la réduction
générale en termes d'invariants, ce qui donne
\begin{equation}
Tr\left[\{T_{\mathbf{F}}^{a}T_{\mathbf{F}}^{b}\}\{T_{\mathbf{F}}^{c}T_{\mathbf{F}}^{d}\}\right]=4I_{4}(\mathbf{F})d^{abcd}+\frac{1}{3}I_{2}(\mathbf{F})(f^{ace}f^{bde}+f^{ade}f^{bce})+4\Lambda(\mathbf{F})(\delta^{ab}\delta^{cd}+\delta^{ac}\delta^{bd}+\delta^{ad}\delta^{bc})\ .
\end{equation}
En combinant les deux,
\begin{equation}
I_{4}(\mathbf{F})d^{abcd}=\frac{I_{3}(\mathbf{F})^{2}}{16I_{2}(\mathbf{F})}d^{abe}d^{cde}-\frac{I_{2}(\mathbf{F})}{12}(f^{ace}f^{bde}+f^{ade}f^{bce})-\Lambda(\mathbf{F})(\delta^{ab}\delta^{cd}+\delta^{ac}\delta^{bd}+\delta^{ad}\delta^{bc})+\frac{I_{2}(\mathbf{F})^{2}}{N}\delta^{ab}\delta^{cd}\ .\label{eq:D4D3D3}
\end{equation}
Avec la convention $I_{4}(\mathbf{F})=1$, cette identité permet de
calculer le symbole quartique $d^{abcd}$ directement à partir des
invariants de rangs inférieurs. On peut maintenant vérifier que pour
$N=3$, $I_{2}(\mathbf{F})=1/2$, $I_{3}(\mathbf{F})=1$, $I_{4}(\mathbf{F})=0$
et $\Lambda(\mathbf{F})=1/24$,
\begin{equation}
0=\frac{1}{8}d^{abe}d^{cde}-\frac{1}{24}(f^{ace}f^{bde}+f^{ade}f^{bce})-\frac{1}{24}(\delta^{ac}\delta^{bd}-\delta^{ab}\delta^{cd}+\delta^{ad}\delta^{bc})\ ,
\end{equation}
ce qui redonne l'identité (\ref{eq:IddSU3}). Pour $N=2$, $I_{2}(\mathbf{F})=1/2$,
$I_{3}(\mathbf{F})=I_{4}(\mathbf{F})=0$, $\Lambda(\mathbf{F})=1/24$,
on retrouve la formule de réduction usuelle pour le tenseur de Levi-Civita:
\begin{equation}
0=-\frac{1}{24}(\varepsilon^{ace}\varepsilon^{bde}+\varepsilon^{ade}\varepsilon^{bce})-\frac{1}{24}(\delta^{ac}\delta^{bd}-2\delta^{ab}\delta^{cd}+\delta^{ad}\delta^{bc})\ .
\end{equation}

\begin{table}[t] \centering\small
\begin{tabular}[c]{lcccccccccc}\hline
\multicolumn{11}{l}{$SU(2)\rule[-0.06in]{0in}{0.2in}$}\\\hline
$\mathbf{R}$ & $(1)$ & $(\mathbf{2})$ & $(3)$ & $(4)$ & $(5)$ & $(6)$ & $(7)$
& $(8)$ & $(9)$ & $(10)\rule[-0.04in]{0in}{0.17in}$\\
$N$ & $\mathbf{2}$ & $\mathbf{3}$ & $\mathbf{4}$ & $\mathbf{5}$ & $\mathbf{6}$
& $\mathbf{7}$ & $\mathbf{8}$ & $\mathbf{9}$ & $\mathbf{10}$ & $\mathbf{11}%
\rule[-0.04in]{0in}{0.17in}$\\
$I_{2}$ & $1/2$ & $2$ & $5$ & $10$ & $35/2$ & $28$ & $42$ & $60$ & $165/2$ &
$110\rule[-0.04in]{0in}{0.17in}$\\
$\Lambda$ & $\dfrac{1}{24}$ & $\dfrac{2}{3}$ & $\dfrac{41}{12}$ & $\dfrac
{34}{3}$ & $\dfrac{707}{24}$ & $\dfrac{196}{3}$ & $\dfrac{259}{2}$ & $236$ &
$\dfrac{3223}{8}$ & $\dfrac{1958}{3}\rule[-0.12in]{0in}{0.32in}$\\\hline
\multicolumn{11}{l}{$SU(3)\rule[-0.06in]{0in}{0.2in}$}\\\hline
$\mathbf{R}$ & $(10)$ & $(20)$ & $(\mathbf{11})$ & $(30)$ & $(21)$ & $(40)$ &
$(05)$ & $(13)$ & $(22)$ & $(60)\rule[-0.04in]{0in}{0.17in}$\\
$N$ & $\mathbf{3}$ & $\mathbf{6}$ & $\mathbf{8}$ & $\mathbf{10}$ &
$\mathbf{15}$ & $\mathbf{15}^{\prime}$ & $\mathbf{21}$ & $\mathbf{24}$ &
$\mathbf{27}$ & $\mathbf{28}\rule[-0.04in]{0in}{0.17in}$\\
$I_{2}$ & $1/2$ & $5/2$ & $3$ & $15/2$ & $10$ & $35/2$ & $35$ & $25$ & $27$ &
$63\rule[-0.04in]{0in}{0.17in}$\\
$I_{3}$ & $1$ & $7$ & $0$ & $27$ & $14$ & $77$ & $-182$ & $-64$ & $0$ &
$378\rule[-0.04in]{0in}{0.17in}$\\
$\Lambda$ & $\dfrac{1}{24}$ & $\dfrac{17}{24}$ & $\dfrac{3}{4}$ & $\dfrac
{33}{8}$ & $\dfrac{29}{6}$ & $\dfrac{371}{24}$ & $\dfrac{539}{12}$ &
$\dfrac{235}{12}$ & $\dfrac{81}{4}$ & $\dfrac{441}{3}\rule[-0.12in]
{0in}{0.32in}$\\\hline
\multicolumn{11}{l}{$SU(4)\rule[-0.06in]{0in}{0.2in}\rule[-0.06in]
{0in}{0.2in}$}\\\hline
$\mathbf{R}$ & $(100)$ & $(010)$ & $(200)$ & $(\mathbf{101})$ & $(011)$ &
$(020)$ & $(003)$ & $(400)$ & $(201)$ & $(210)\rule[-0.04in]{0in}{0.17in}$\\
$N$ & $\mathbf{4}$ & $\mathbf{6}$ & $\mathbf{10}$ & $\mathbf{15}$ &
$\mathbf{20}$ & $\mathbf{20}^{\prime}$ & $\mathbf{20}^{\prime\prime}$ &
$\mathbf{35}$ & $\mathbf{36}$ & $\mathbf{45}\rule[-0.04in]{0in}{0.17in}$\\
$I_{2}$ & $1/2$ & $1$ & $3$ & $4$ & $13/2$ & $8$ & $21/2$ & $28$ & $33/2$ &
$24\rule[-0.04in]{0in}{0.17in}$\\
$I_{3}$ & $1$ & $0$ & $8$ & $0$ & $-7$ & $0$ & $-35$ & $112$ & $21$ &
$48\rule[-0.04in]{0in}{0.17in}$\\
$I_{4}$ & $1$ & $-4$ & $12$ & $8$ & $-11$ & $-56$ & $69$ & $272$ & $57$ &
$24\rule[-0.04in]{0in}{0.17in}$\\
$\Lambda$ & $\dfrac{29}{816}$ & $\dfrac{11}{102}$ & $\dfrac{23}{34}$ &
$\dfrac{40}{51}$ & $\dfrac{1313}{816}$ & $\dfrac{128}{51}$ & $\dfrac
{1211}{272}$ & $\dfrac{56}{3}$ & $\dfrac{1639}{272}$ & $\dfrac{176}
{17}\rule[-0.12in]{0in}{0.32in}$\\\hline
\multicolumn{11}{l}{$SU(5)\rule[-0.06in]{0in}{0.2in}$}\\\hline
$\mathbf{R}$ & \multicolumn{1}{l}{$(1000)$} & \multicolumn{1}{l}{$(0100)$} &
\multicolumn{1}{l}{$(2000)$} & \multicolumn{1}{l}{$(\mathbf{1001})$} &
\multicolumn{1}{l}{$(0003)$} & \multicolumn{1}{l}{$(0011)$} &
\multicolumn{1}{l}{$(0101)$} & \multicolumn{1}{l}{$(0020)$} &
\multicolumn{1}{l}{$(2001)$} & \multicolumn{1}{l}{$(0110)\rule[-0.04in]
{0in}{0.17in}$}\\
$N$ & $\mathbf{5}$ & $\mathbf{10}$ & $\mathbf{15}$ & $\mathbf{24}$ &
$\mathbf{35}$ & $\mathbf{40}$ & $\mathbf{45}$ & $\mathbf{50}$ & $\mathbf{70}$
& $\mathbf{75}\rule[-0.04in]{0in}{0.17in}$\\
$I_{2}$ & $1/2$ & $3/2$ & $7/2$ & $5$ & $14$ & $11$ & $12$ & $35/2$ & $49/2$ &
$25\rule[-0.04in]{0in}{0.17in}$\\
$I_{3}$ & $1$ & $1$ & $9$ & $0$ & $-44$ & $-16$ & $-6$ & $-15$ & $29$ &
$0\rule[-0.04in]{0in}{0.17in}$\\
$I_{4}$ & $1$ & $-3$ & $13$ & $10$ & $82$ & $-2$ & $-6$ & $-55$ & $79$ &
$-70\rule[-0.04in]{0in}{0.17in}$\\
$\Lambda$ & $\dfrac{47}{1560}$ & $\dfrac{83}{520}$ & $\dfrac{77}{120}$ &
$\dfrac{125}{156}$ & $\dfrac{1841}{390}$ & $\dfrac{1903}{780}$ & $\dfrac
{167}{65}$ & $\dfrac{1589}{312}$ & $\dfrac{11123}{1560}$ & $\dfrac{1075}
{156}\rule[-0.12in]{0in}{0.32in}$\\\hline
\end{tabular}
\caption{Quelques premières représentations de $SU(N)$, $N=2,3,4,5$, indexées par leur indice de Dynkin et leur dimension, les invariants de Casimir quadratiques, cubiques et quartiques avec $\Lambda(\mathbf{R})$ donné par l'équation (\ref{eq:LambdaDef}).}
\label{TableCasimirsSU}
\end{table}

\begin{table}[t] \centering\small
\begin{tabular}[c]{lcccccccccc}\hline
\multicolumn{11}{l}{$SO(5)\rule[-0.06in]{0in}{0.2in}$}\\\hline
$\mathbf{R}$ & $(10)$ & $(01)$ & $(\mathbf{02})$ & $(20)$ & $(11)$ & $(03)$ &
$(30)$ & $(12)$ & $(04)$ & $(21)\rule[-0.04in]{0in}{0.17in}$\\
$N$ & $\mathbf{5}$ & $\mathbf{4}$ & $\mathbf{10}$ & $\mathbf{14}$ &
$\mathbf{16}$ & $\mathbf{20}$ & $\mathbf{30}$ & $\mathbf{35}$ & $\mathbf{35}
^{\prime}$ & $\mathbf{40}\rule[-0.04in]{0in}{0.17in}$\\
$I_{2}$ & $1$ & $1/2$ & $3$ & $7$ & $6$ & $21/2$ & $27$ & $21$ & $28$ &
$29\rule[-0.04in]{0in}{0.17in}$\\
$I_{4}$ & $2$ & $-1/2$ & $-6$ & $26$ & $6$ & $-69/2$ & $162$ & $-6$ & $-132$ &
$91\rule[-0.04in]{0in}{0.17in}$\\
$\Lambda$ & $\dfrac{1}{8}$ & $\dfrac{1}{32}$ & $\dfrac{5}{8}$ & $\dfrac{21}
{8}$ & $\dfrac{13}{8}$ & $\dfrac{133}{32}$ & $\dfrac{153}{8}$ & $\dfrac{77}
{8}$ & $\dfrac{35}{2}$ & $\dfrac{261}{16}\rule[-0.12in]{0in}{0.32in}$\\\hline
\multicolumn{11}{l}{$SO(7)\rule[-0.06in]{0in}{0.2in}$}\\\hline
$\mathbf{R}$ & $(100)$ & $(001)$ & $(\mathbf{010})$ & $(200)$ & $(002)$ &
$(101)$ & $(300)$ & $(110)$ & $(011)$ & $(003)\rule[-0.04in]{0in}{0.17in}$\\
$N$ & $\mathbf{7}$ & $\mathbf{8}$ & $\mathbf{21}$ & $\mathbf{27}$ &
$\mathbf{35}$ & $\mathbf{48}$ & $\mathbf{77}$ & $\mathbf{105}$ &
$\mathbf{112}$ & $\mathbf{112}\rule[-0.04in]{0in}{0.17in}$\\
$I_{2}$ & $1$ & $1$ & $5$ & $9$ & $10$ & $14$ & $44$ & $45$ & $46$ &
$54\rule[-0.04in]{0in}{0.17in}$\\
$I_{4}$ & $2$ & $-1$ & $-2$ & $30$ & $-16$ & $10$ & $220$ & $42$ & $-46$ &
$-126\rule[-0.04in]{0in}{0.17in}$\\
$\Lambda$ & $\dfrac{13}{138}$ & $\dfrac{43}{552}$ & $\dfrac{125}{138}$ &
$\dfrac{111}{46}$ & $\dfrac{155}{69}$ & $\dfrac{889}{276}$ & $\dfrac{1474}
{69}$ & $\dfrac{735}{46}$ & $\dfrac{187}{12}$ & $\dfrac{2007}{92}
\rule[-0.12in]{0in}{0.32in}$\\\hline
\multicolumn{11}{l}{$SO(9)\rule[-0.06in]{0in}{0.2in}\rule[-0.06in]
{0in}{0.2in}$}\\\hline
$\mathbf{R}$ & $(1000)$ & $(0001)$ & $(\mathbf{0100})$ & $(2000)$ & $(0010)$ &
$(0002)$ & $(1001)$ & $(3000)$ & $(1100)$ & $(0101)\rule[-0.04in]
{0in}{0.17in}$\\
$N$ & $\mathbf{9}$ & $\mathbf{16}$ & $\mathbf{36}$ & $\mathbf{44}$ &
$\mathbf{84}$ & $\mathbf{126}$ & $\mathbf{128}$ & $\mathbf{156}$ &
$\mathbf{231}$ & $\mathbf{432}\rule[-0.04in]{0in}{0.17in}$\\
$I_{2}$ & $1$ & $2$ & $7$ & $11$ & $21$ & $35$ & $32$ & $65$ & $77$ &
$150\rule[-0.04in]{0in}{0.17in}$\\
$I_{4}$ & $2$ & $-2$ & $2$ & $34$ & $-18$ & $-50$ & $16$ & $286$ & $106$ &
$-54\rule[-0.04in]{0in}{0.17in}$\\
$\Lambda$ & $\dfrac{17}{228}$ & $\dfrac{10}{57}$ & $\dfrac{245}{228}$ &
$\dfrac{517}{228}$ & $\dfrac{329}{76}$ & $\dfrac{1855}{228}$ & $\dfrac
{376}{57}$ & $\dfrac{5395}{228}$ & $\dfrac{5005}{228}$ & $\dfrac{850}
{19}\rule[-0.12in]{0in}{0.32in}$\\\hline
\multicolumn{11}{l}{$SO(10)\rule[-0.06in]{0in}{0.2in}$}\\\hline
$\mathbf{R}$ & $(10000)$ & $(00001)$ & $(\mathbf{01000})$ & $(20000)$ &
$(00100)$ & $(00002)$ & $(10010)$ & $(00011)$ & $(30000)$ &
$(11000)\rule[-0.04in]{0in}{0.17in}$\\
$N$ & $\mathbf{10}$ & $\mathbf{16}$ & $\mathbf{45}$ & $\mathbf{54}$ &
$\mathbf{120}$ & $\mathbf{126}$ & $\mathbf{144}$ & $\mathbf{210}$ &
$\mathbf{210}^{\prime}$ & $\mathbf{320}\rule[-0.04in]{0in}{0.17in}$\\
$I_{2}$ & $1$ & $2$ & $8$ & $12$ & $28$ & $35$ & $34$ & $56$ & $77$ &
$96\rule[-0.04in]{0in}{0.17in}$\\
$I_{4}$ & $2$ & $-2$ & $4$ & $36$ & $-16$ & $-50$ & $14$ & $-68$ & $322$ &
$144\rule[-0.04in]{0in}{0.17in}$\\
$\Lambda$ & $\dfrac{19}{282}$ & $\dfrac{103}{564}$ & $\dfrac{160}{141}$ &
$\dfrac{104}{47}$ & $\dfrac{770}{141}$ & $\dfrac{2345}{282}$ & $\dfrac
{3791}{564}$ & $\dfrac{1792}{141}$ & $\dfrac{7007}{282}$ & $\dfrac{1168}
{47}\rule[-0.12in]{0in}{0.32in}$\\\hline
\end{tabular}
\caption{Quelques premières représentations de $SO(N)$, $N=5,7,9,10$, indexées par leur indice de Dynkin et leur dimension, les invariants de Casimir quadratiques, cubiques et quartiques avec $\Lambda(\mathbf{R})$ donné par l'équation (\ref{eq:LambdaDef}). L'invariant cubique s'annule pour toutes ces algèbres. La normalisation des générateurs et des symboles quartiques est fixée en termes de celle adoptée pour les algèbres $SU(N)$, à l'aide de l'équation (\ref{eq:InBR}). Les algèbres $SO(4)$ et $SO(6)$ ne sont pas inclues puisqu'elles sont respectivement isomorphes à $SU(2)\otimes SU(2)$ et $SU(4)$. Notons que les normalisations ne correspondent pas nécessairement, avec par exemple $I_2(SU(4)) = I_2(SO(6))$ mais $I_4(SU(4)) = - 2 I_4(SO(6))$.}
\label{TableCasimirsSO}
\end{table}

\begin{table}[t] \centering\small
\begin{tabular}[c]{|c|ccc|ccc|ccc|c|}\hline
\multicolumn{11}{|l|}{$SO(8)$\rule[-0.06in]{0in}{0.2in}\rule[-0.06in]%
{0in}{0.2in}}\\\hline
& \multicolumn{3}{|c|}{$\mathbf{8}\ (I_{2},\Lambda=1,1/12)$} &
\multicolumn{3}{|c|}{$\mathbf{112}\ (I_{2},\Lambda=54,45/2)$} &
\multicolumn{3}{|c|}{$\mathbf{224}\ (I_{2},\Lambda=100,115/3)$} &
$\mathbf{28}(\ I_{2},\Lambda=6,1)$\\
& $(1000)$ & $(0001)$ & $(0010)$ & $(2000)$ & $(0002)$ & $(0020)$ & $(1002)$ &
$(1020)$ & $(2001)$ & $(\mathbf{0100)}\rule[-0.04in]{0in}{0.17in}$\\
$I_{4}$ & $2$ & $-1$ & $-1$ & $252$ & $-126$ & $-126$ & $-40$ & $-40$ & $212$
& $0\rule[-0.04in]{0in}{0.17in}$\\
$I_{4}^{\prime}$ & $0$ & $-1$ & $1$ & $0$ & $-126$ & $126$ & $-128$ & $128$ &
$-44$ & $0\rule[-0.04in]{0in}{0.17in}$\\\cline{2-7}\cline{5-7}\cline{11-11}
& \multicolumn{3}{|c|}{$\mathbf{35}\ (I_{2},\Lambda=10,7/3)$} &
\multicolumn{3}{|c|}{$\mathbf{160}\ (I_{2},\Lambda=60,19)$} &  &  &  &
$\mathbf{300}\ (I_{2},\Lambda=150,65)$\\
& $(2000)$ & $(0002)$ & $(0020)$ & $(1100)$ & $(0101)$ & $(0110)$ & $(0012)$ &
$(0021)$ & $(2010)$ & $(0200)\rule[-0.04in]{0in}{0.17in}$\\
$I_{4}$ & $32$ & $-16$ & $-16$ & $72$ & $-36$ & $-36$ & $-172$ & $-172$ &
$212$ & $0\rule[-0.04in]{0in}{0.17in}$\\
$I_{4}^{\prime}$ & $0$ & $-16$ & $16$ & $0$ & $-36$ & $36$ & $-84$ & $84$ &
$44$ & $0\rule[-0.04in]{0in}{0.17in}$\\\cline{2-11}\cline{5-11}
& \multicolumn{3}{|c|}{$\mathbf{56}\ (I_{2},\Lambda=15,13/4)$} &
\multicolumn{3}{|c|}{$\mathbf{294}\ (I_{2},\Lambda=210,133)$} &
\multicolumn{3}{|c|}{$\mathbf{567}\ (I_{2},\Lambda=324,162)$} & $\mathbf{350}
\ (I_{2},\Lambda=150,55)$\\
& $(0011)$ & $(1001)$ & $(1010)$ & $(4000)$ & $(0004)$ & $(0040)$ & $(2100)$ &
$(0102)$ & $(0120)$ & $(1011)\rule[-0.04in]{0in}{0.17in}$\\
$I_{4}$ & $-18$ & $9$ & $9$ & $1344$ & $-672$ & $-672$ & $864$ & $-432$ &
$-432$ & $0\rule[-0.04in]{0in}{0.17in}$\\
$I_{4}^{\prime}$ & $0$ & $9$ & $-9$ & $0$ & $-672$ & $672$ & $0$ & $-432$ &
$432$ & $0\rule[-0.04in]{0in}{0.17in}$\\\hline
\end{tabular}
\caption{Quelques premières représentations de $SO(8)$, indexées par leur indice de Dynkin. En raison de l'invariance du tenseur de Levi-Civita de dimension huit, cette algèbre possède un second tenseur invariant quartique. Sa normalisation est fixée de sorte à rendre manifeste la relation entre les valeurs des deux invariants de Casimir quartiques, et correspond à $\eta=-1/8$ dans l'équation              (\ref{eq:SymTensor}). Une deuxième propriété de $SO(8)$ est sa symétrie de trialité: les dimensions et les invariants de Casimir quadratiques sont les mêmes sous permutations des première, troisième et quatrième racines simples. Les deux invariants de Casimir quartiques s'annulent quand ils sont sommés sur les représentations liées par la symétrie de permutation \cite{Okubo81}. Ceci signifie en particulier qu'ils s'annulent identiquement pour la $\mathbf{28}$, $\mathbf{300}$, and $\mathbf{350}$.}
\label{TableCasimirsSO8}
\end{table}

\subsection*{Invariants de Casimir pour les groupes simples}

Grâce à la condition d'orthogonalité adoptée pour fixer $\Lambda(\mathbf{R})$
\cite{Okubo81}, la formule usuelle peut être employée pour avoir
les valeurs explicites de l'invariant $I_{4}(\mathbf{R})$ pour plusieurs
représentations,
\begin{equation}
\begin{aligned}I_{n}(\mathbf{R}) & =(-1)^{n}I_{n}(\mathbf{R}^{\dagger})\ ,\\
I_{n}(\mathbf{R}_{1}\oplus\mathbf{R}_{2}) & =I_{n}(\mathbf{R}_{1})+I_{n}(\mathbf{R}_{2})\ ,\\
I_{n}(\mathbf{R}_{1}\otimes\mathbf{R}_{2}) & =I_{n}(\mathbf{R}_{1})N(\mathbf{R}_{2})+I_{n}(\mathbf{R}_{2})N(\mathbf{R}_{1})=\sum I_{n}(\mathbf{R}_{i}^{\prime})\;,
\end{aligned}
\end{equation}
avec $n=2,3,4$ et où $\mathbf{R}_{1}\otimes\mathbf{R}_{2}=\sum_{i}\mathbf{R}_{i}^{\prime}$.
Tout compte fait, ces relations sont plus que suffisantes pour établir
les invariants de Casimir de n'importe quelle algèbre de Lie standard.
Nous donnons dans les tableaux \ref{TableCasimirsSU},~\ref{TableCasimirsSO} et \ref{TableCasimirsSO8}
leurs valeurs pour quelques premières représentations de certaines
algèbres unitaires et orthogonales de rang $r\leq5$, ainsi que $\Lambda(\mathbf{R})$.
Nous avons aussi vérifié ces nombres en calculant $I_{2,3,4}(\mathbf{R})$
directement en utilisant des représentations matricielles explicites
pour les premières représentations de chaque algèbre. Ces nombres
sont compatibles avec la formule explicite en termes d'indices de
Dynkin donnée dans \cite{Okubo81}, aux conventions de normalisations
près.

La normalisation des générateurs adoptée pour les algèbres $SO(N)$
dans les tableaux \ref{TableCasimirsSO} et \ref{TableCasimirsSO8}
n'est pas standard mais inspirée physiquement. Spécifiquement, les
invariants d'une algèbre $M$ peuvent être exprimés en termes de ceux
de sa sous-algèbre $N$. Par exemple, si une représentation $\mathbf{R}_{M}$
se branche sur la somme directe des représentations $\mathbf{R}_{N}$,
on a la règle de somme simple:
\begin{equation}
I_{n}(\mathbf{R}_{M})=\eta\sum_{\mathbf{R}_{N}\subset\mathbf{R}_{M}}I_{n}(\mathbf{R}_{N})\ ,\label{eq:InBR}
\end{equation}
où $\eta$ est une constante reflétant la convention de normalisation
adoptée de $M$ et $N$. Dans le tableau \ref{TableCasimirsSO}, on
choisit de fixer $\eta=1$. Par exemple, les générateurs dans la représentation
définissante de $SO(10)$ sont normalisés tels que
\begin{equation}
I_{2}(\mathbf{10})^{SO(10)}=I_{2}(\mathbf{\bar{5}})^{SU(5)}+I_{2}(\mathbf{5})^{SU(5)}=1\ ,
\end{equation}
étant donné que $\mathbf{10}\rightarrow\mathbf{\bar{5}}+\mathbf{5}$.
De même, la normalisation du symbole quartique de $SO(10)$ est alors
fixée en imposant $I_{4}(\mathbf{10})^{SO(10)}=2I_{4}(\mathbf{5})^{SU(5)}=2$.
Cela fait sens physiquement si on pense à un champ siégeant dans une
représentation donnée de $SO(10)$ circulant dans une boucle. Nos
conventions de normalisations font que le matching de cette amplitude
avec celle calculée en termes des champs de la sous-algèbre soit plus
transparent. Notons que les générateurs ainsi que les symboles quartiques
de toutes les algèbres $SO(N)$ sont fixés dès que ceux de $SO(10)$
le sont, car $SO(N)\subset SO(N+1)$. De plus, nous avons aussi vérifié
que ces conventions soient compatibles avec $SO(3)\otimes SO(7)\subset SO(10)$
et $SO(4)\otimes SO(6)\subset SO(10)$, avec $SO(4)\sim SU(2)\otimes SU(2)$.

D'autres relations entre les invariants d'une algèbre et ceux de ses
sous-algèbres sont données dans le texte, voir en particulier l'équation
(\ref{eq:BranchingI4}) qui donne $I_{n}(\mathbf{R}_{M})$ en fonction
de $I_{n-1}(\mathbf{R}_{N})$, équation (\ref{eq:IdI4I2}) qui fixe
$I_{4}(\mathbf{R}_{M})$ en fonction de $I_{2}(\mathbf{R}_{N})$,
ou l'équation (\ref{eq:ReducU1}) qui donne $I_{n}(\mathbf{R}_{M})$
en termes des charges $U(1)$ des états $\mathbf{R}_{M}$. Pour clore
cette partie, illustrons ces relations avec quelques exemples.

Considérons d'abord la réduction de $SU(2)$ vers un sous-groupe $U(1)$
de $SU(2)$ généré par $T^{3}$. Vu qu'il n'y a pas d'invariant quartique
pour $SU(2)$, l'équation (\ref{eq:ReducU1}) est facile à vérifier.
La représentation fondamentale de $SU(2)$ correspond à deux états
complexes de charges $|T^{3}|=1/2$, alors on peut identifier $2(1/2)^{4}=3\Lambda^{SU(2)}(\mathbf{2})$
car $\Lambda^{SU(2)}(\mathbf{2})=1/24$. De même, la représentation
complexe adjointe de $SU(2)$ contient deux états de charges unités,
ainsi $2=3\Lambda^{SU(2)}(\mathbf{3})$, et l'isospin $3/2$ se décompose
en quatre états tels que $2((1/2)^{4}+(3/2)^{4})=3\Lambda^{SU(2)}(\mathbf{4})$,
ce qui redonne les valeurs correctes $\Lambda^{SU(2)}(\mathbf{3})=2/3$
et $\Lambda^{SU(2)}(\mathbf{4})=41/12$. Le même exercice peut être
répétée pour $SU(3)$, pour lequel l'absence d'invariant quartique
assure que $\operatorname*{Tr}((T_{\mathbf{R}}^{3})^{4})=\operatorname*{Tr}((T_{\mathbf{R}}^{8})^{4})$
si $T^{3}$ et $T^{8}$ sont les générateurs de Cartan conventionnels
(égaux à la moitié des matrices de Gell-Mann correspondantes dans
la représentation fondamentale). Afin d'appliquer la même méthodes
à $SU(5)$, nous avons d'abord besoin de fixer deux paramètres libres.
Pour spécifier, on part de
\begin{equation}
3\Lambda(\mathbf{R})+d^{\alpha\alpha\alpha\alpha}I_{4}(\mathbf{R})=\delta\sum_{q_{\alpha}\in\mathbf{R}}q_{\alpha}^{4}\ ,
\end{equation}
La valeur de $d^{\alpha\alpha\alpha\alpha}$ et la normalisation $\delta$
du générateur de $U(1)$ (qui était accidentellement égale à $1$
dans l'exemple précédent de $SU(2)$) doivent être fixées. Si on identifie
$T^{\alpha}$ avec le générateur de l'hypercharge dans la sous-algèbre
$SU(3)\otimes SU(2)\otimes U(1)\subset SU(5)$, on peut utiliser les
règles d'embranchements \cite{Slansky}
\begin{equation}
\begin{aligned}\mathbf{5} & =(\mathbf{3},\mathbf{1})_{2}+(\mathbf{1},\mathbf{2})_{-3}\ ,\\
\mathbf{10} & =(\mathbf{\bar{3}},\mathbf{1})_{4}+(\mathbf{3},\mathbf{2})_{-1}+(\mathbf{1},\mathbf{1})_{-6}\ ,
\end{aligned}
\label{eq:SU5BR}
\end{equation}
et ces deux constantes sont fixées par
\begin{equation}
\left\{ \begin{array}{c}
I_{4}(\mathbf{5})d^{\alpha\alpha\alpha\alpha}+3\Lambda(\mathbf{5})=\delta(2\times3^{4}+3\times2^{4})\\
I_{4}(\mathbf{10})d^{\alpha\alpha\alpha\alpha}+3\Lambda(\mathbf{10})=\delta(6^{4}+3\times4^{4}+3\times2\times1^{4})
\end{array}\right.\rightarrow\left\{ \begin{array}{l}
\delta=1/60^{2}\ ,\\
d^{\alpha\alpha\alpha\alpha}=-5/156\ .
\end{array}\right.
\end{equation}
On peut alors vérifier que les $I_{4}$ pour les autres représentations
de $SU(5)$ sont correctement reproduits.

Les mêmes règles d'embranchements peuvent être utilisées en rapport
avec l'équation (\ref{eq:BranchingI4}), que nous écrivons
\begin{equation}
I_{4}(\mathbf{R}_{M})=\eta\sum_{\mathbf{R}_{N}\subset\mathbf{R}_{M}}q_{\alpha}(\mathbf{R}_{N})I_{3}(\mathbf{R}_{N})\ .
\end{equation}
La sous-algèbre n'a pas besoin d'être maximale donc on considère $U(1)\otimes SU(3)\subset SU(5)$.
En utilisant les valeurs reportées dans le tableau \ref{TableCasimirsSU},
la première règle de l'équation (\ref{eq:SU5BR}) se traduit en $I_{4}^{SU(5)}(\mathbf{5})=2\eta I_{3}^{SU(3)}(\mathbf{3})$
et fixe $\eta=1/2$. Ainsi, on peut vérifier que cette équation est
valide pour toutes les autres représentations de $SU(5)$ listées
dans le tableau \ref{TableCasimirsSU}.

\newpage\null\thispagestyle{empty}\newpage

\bibliographystyle{JHEP}
\bibliography{Manuscrit2.bib}

\newpage\null\thispagestyle{empty}\newpage

\newpage\null\thispagestyle{empty}\newpage

\thispagestyle{empty}

\includepdf{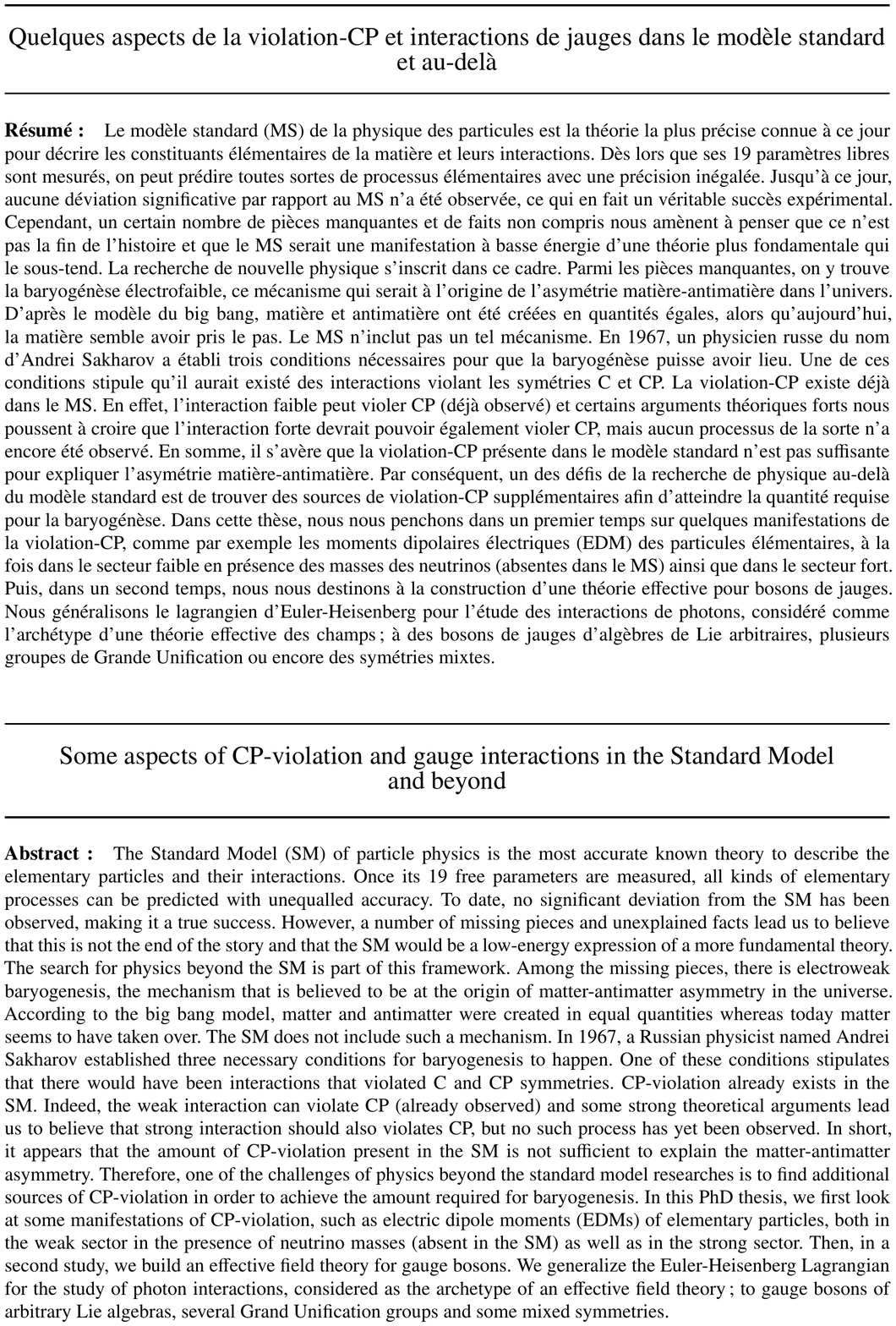}
\end{document}